\documentclass[12pt]{article}
\usepackage{geometry} 
\usepackage{amsmath,amssymb,bm}
\usepackage{textcomp,gensymb}
\usepackage{graphicx}
\usepackage{epstopdf}
\usepackage[comma,authoryear]{natbib}
\usepackage{color}
\usepackage{calrsfs}
\usepackage[english]{babel}
\usepackage{setspace}
\usepackage{gensymb}
\usepackage{xfrac}

\usepackage{mathptmx}
\usepackage{amsmath}
\usepackage{amsfonts}

\numberwithin{equation}{section}

\usepackage[hidelinks,breaklinks]{hyperref}
\usepackage[hyphenbreaks]{breakurl}

\def\bbar{{\mathchar'26\mkern-9mu b}}

\begin{document}

\title{Putting probabilities first. \\ How Hilbert space generates and constrains them.\footnote{This preprint, which we originally posted in October 2019, has since been further developed into the book: \emph{Understanding Quantum Raffles: Quantum Mechanics on an Informational Approach: Structure and Interpretation} (Springer, Cham, 2021, ISBN: \href{https://www.springer.com/gp/book/9783030859381}{978-3-030-85938-1}). The following are the most notable changes between this preprint and the book: Section 5 (now Chapter 6 of \emph{Understanding Quantum Raffles}), especially Section 5.5 on measurement, has been largely rewritten and greatly expanded. The upshot of that Section has also been strengthened: Chapter 6 of \emph{Understanding Quantum Raffles} defends the view that \emph{both} what we call the ``small'' and the ``big'' measurement problems are not bugs but features of quantum mechanics. Section 1 has also been re-written and expanded in the book (where it is now Chapter 1) and parts of Section 2 (which has now been refactored into two separate chapters) have been rewritten to be more easily digestible by general readers. Section 6 (Chapter 7 in the book) now includes a comparison of our interpretation of quantum mechanics with the Everett interpretation (which like our interpretation adds nothing to the quantum-mechanical formalism). Finally, \emph{Understanding Quantum Raffles} includes a foreword by Jeffrey Bub.}$~\;^{\raisebox{-.1em}{\normalsize ,}}\,$\footnote{Dedicated to the memory of Itamar Pitowsky (1950--2010) and William Demopoulos (1943--2017).}}
%Polytopes and elliptopes: classical and quantum constraints on correlations

\author{Michael Janas,\footnote{School of Physics and Astronomy, University of Minnesota; E-mail: jana0030@umn.edu, janss011@umn.edu} \, Michael E. Cuffaro,\footnote{Rotman Institute of Philosophy, University of Western Ontario; Munich Center for Mathematical Philosophy, Ludwig-Maximilians-Universit\"at M\"unchen; E-mail: mike@michaelcuffaro.com} \, Michel Janssen$^\dagger$}
\date{}

\maketitle

\begin{abstract}
\noindent 
We use correlation arrays, the workhorse of Bub's (2016) \emph{Bananaworld}, to analyze the correlations found in an experimental setup due to \citet{Mermin 1981} for measurements on the singlet state of a pair of spin-$\frac12$ particles. Adopting an approach pioneered by \citet{Pitowsky 1989b} and promoted in \emph{Bananaworld}, we geometrically represent the class of correlations allowed by quantum mechanics in this setup as an elliptope in a non-signaling cube.  To determine which of these quantum correlations are allowed by local hidden-variable theories, we investigate which ones we can simulate using raffles with baskets of tickets that have the outcomes for all combinations of measurement settings printed on them. The class of correlations found this way can be represented geometrically by a tetrahedron contained within the elliptope. We use the same Bub-Pitowsky framework to analyze a generalization of the Mermin setup for measurements on the singlet state of two particles with higher spin. The class of correlations allowed by quantum mechanics in this case is still represented by the elliptope; the subclass of those whose main features can be simulated with our raffles can be represented by polyhedra that, with increasing spin, have more and more vertices and facets and get closer and closer to the elliptope. We use these results to advocate for Bubism (not to be confused with QBism), an interpretation of quantum mechanics along the lines of \emph{Bananaworld}. Probabilities and expectation values are primary in this interpretation. They are determined by inner products of vectors in Hilbert space. Such vectors do not themselves represent what is real in the quantum world. They encode families of probability distributions over values of different sets of observables. As in classical theory, these values ultimately represent what is real in the quantum world. Hilbert space puts constraints on possible combinations of such values, just as Minkowski space-time puts constraints on possible spatio-temporal constellations of events. Illustrating how generic such constraints are, the constraint derived in this paper, the equation for the elliptope, is a general constraint on correlation coefficients that can be found in older literature on statistics and probability theory. \citet{Yule 1896} already stated the constraint. De Finetti (1937) already gave it a geometrical interpretation sharing important features with its interpretation in Hilbert space.
\end{abstract}


\tableofcontents

%SECTION 1 (label: 0)
\section{Introduction}\label{0}
%!TEX root =  ./JanasJanssenCuffaro-August2019.tex

%SECTION 1 -- OVERALL INTRO
%\section{Introduction}\label{0}

This paper is a brief for a specific take on the general framework of quantum mechanics.\footnote{This paper deals with \emph{philosophy}, \emph{pedagogy} and \emph{polytopes}. In this introduction, we will explain how these three components are connected, both to each other and to \emph{Bananaworld} \citep{Bub 2016}. Cuffaro's main interest is in philosophy, Janssen's in pedagogy and Janas's in polytopes. Though all three of us made substantial contributions to all six sections of the paper, Janssen had final responsibility for Sections \ref{0}--\ref{1}, Janas for Sections \ref{2}--\ref{3} and Cuffaro for Sections \ref{4}--\ref{5}.} In terms of the usual partisan labels, it is an \emph{information-theoretic} interpretation in which the status of the state vector is \emph{epistemic} rather than \emph{ontic}. On the ontic view, state vectors represent what is ultimately real in the quantum world; on the epistemic view, they are auxiliary quantities for assigning definite values to observables in a world in which it is no longer possible to do so for all observables. Such labels, however, are of limited use for a taxonomy of interpretations of quantum mechanics. A more promising approach might be to construct a genealogy of such interpretations.\footnote{The contemporary literature on quantum foundations has muddied the waters in regards to the classification of interpretations of quantum mechanics, and it is partly for this reason that we prefer to give a genealogy rather than a taxonomy of interpretations. Ours is \emph{not} an epistemic interpretation of quantum mechanics in the sense compatible with the ontological models framework of \citet[]{Harrigan and Spekkens 2010}. In particular it is not among our assumptions that a quantum system has, at any time, a well-defined ontic state. Actually we take one of the lessons of quantum mechanics to be that this view is untenable (see Section \ref{4.3} below). For more on the differences between a view such as ours and the kind of epistemic interpretation explicated in \citet[]{Harrigan and Spekkens 2010}, and for more on why the no-go theorem proved by \citet{PBR} places restrictions on the latter kind of epistemic interpretation but is not relevant to ours, see \citet[]{Ben-Menahem 2017}.}
As this is not a historical paper, however, a rough characterization of the relevant phylogenetic tree must suffice here.\footnote{One of us is working on a two-volume book on the genesis of quantum mechanics, the first of which has recently come out \citep{Duncan and Janssen 2019}.} The main thing to note then is that the mathematical equivalence of wave and matrix mechanics papers over a key difference in what its originators thought their big discovery was. These big discoveries are certainly compatible with one another but there is at least a striking difference in emphasis.\footnote{We will return to this point in Section \ref{4.3a}.} For Erwin Schr\"odinger the big discovery was that a wave phenomenon underlies the particle behavior of matter, just as physicists in the 19th century had discovered that a wave phenomenon underlies geometrical optics \citep{Joas and Lehner 2009}. For Werner Heisenberg it was that the problems facing atomic physics in the 1920s called for a new framework to represent physical quantities just as electrodynamics had called for a new framework to represent their spatio-temporal relations two decades earlier \citep[pp.\ 134--142]{Duncan and Janssen 2007, Janssen 2009}. What are now labeled ontic interpretations---e.g., Everett's many-worlds interpretation, De Broglie-Bohm pilot-wave theory and the spontaneous-collapse theory of Ghirardi, Rimini and Weber (GRW)---can be seen as descendants of wave mechanics; what are now labeled epistemic interpretations---e.g., the much maligned Copenhagen interpretation and Quantum Bayesianism or QBism---as descendants of matrix mechanics.\footnote{David \citet{Wallace 2019} provides an example from the quantum foundations literature showing that the ``big discoveries'' of matrix and wave mechanics are not mutually exclusive. He argues that the Everett interpretation should be seen as a general new framework for physics while endorsing the view that vectors in Hilbert space represent what is real in the quantum world. Wallace and other Oxford Everettians derive the Born rule for probabilities in quantum mechanics from decision-theoretic considerations instead of taking it to be given by the Hilbert space formalism the way von Neumann showed one could (see below). For Berlin Everettians (i.e., at least some of the Christoph Lehners  in their multiverse) state vectors are both ontic and epistemic. They help themselves to the Born rule \emph{\`a la} von Neumann but also use state vectors to represent physical reality (Christoph Lehner, private communication).\label{wallace-recipe}}

The interpretation for which we will advocate in this paper can, more specifically, be seen as a descendant of the (statistical) transformation theory of Pascual \citet{Jordan 1927a, Jordan 1927b} and Paul \citet[amplified in his famous book, Dirac, 1930]{Dirac 1927} and of the ``probability-theoretic construction'' (\emph{Wahrscheinlichkeitstheoretischer Aufbau}) of quantum mechanics in the second installment of the trilogy of papers by John \citet{von Neumann 1927a, von Neumann 1927b, von Neumann 1927c}  that would form the backbone of \emph{his} famous book \citep{von Neumann 1932}. While incorporating the wave functions of wave mechanics, both Jordan's and Dirac's version of transformation theory grew out of matrix mechanics. More strongly than Dirac, Jordan emphasized the statistical aspect. The ``new foundation'' (\emph{Neue Begr\"undung}) of quantum mechanics announced in the titles of Jordan's 1927 papers consisted of some postulates about the probability of finding a value for one quantum variable given the value of another. Von Neumann belongs to that same lineage. Although he proved the mathematical equivalence of wave and matrix mechanics in the process (by showing that they correspond to two different instantiations of Hilbert space), he wrote his 1927 trilogy in direct response to Jordan's version of transformation theory. His \emph{Wahrscheinlichkeitstheoretischer Aufbau} grew out of his dissatisfaction with Jordan's treatment of probabilities. Drawing on work on probability theory by Richard von Mises \citep[soon to be published in book form;][]{von Mises 1928}, he introduced the now familiar density operators characterizing (pure and mixed state) ensembles of quantum systems.\footnote{For historical analysis of these developments, focusing on Jordan and von Neumann, see \citet{Duncan and Janssen 2013} and, for a summary aimed at a broader audience, \citet[pp.\ 142--161]{Janssen 2019}.} He showed that what came to be known as the Born rule for probabilities in quantum mechanics can be derived from the Hilbert space formalism and some seemingly innocuous assumptions about properties of the function giving expectation values \citep[sec.\ 6, pp.\ 246--251]{Duncan and Janssen 2013}. This derivation was later re-purposed for the infamous von Neumann no-hidden variables proof, in which case the assumptions, entirely appropriate in the context of the Hilbert space formalism for quantum mechanics, become highly questionable \citep{Bub 2010, Dieks 2017}. 

A branch on the phylogenetic tree of interpretations of quantum mechanics closer to our own is the one with Jeffrey Bub and Itamar Pitowsky's (2010) ``Two dogmas of quantum mechanics,''  a play on W.\ V.\ O.\ Quine's (1951) celebrated ``Two dogmas of empiricism.'' Bub and Pitowsky presented their paper in the Everettians' lion's den at the 2007 conference in Oxford marking the 50th anniversary of the Everett interpretation.\footnote{The video of their talk can still be watched at 
\textless\url{users.ox.ac.uk/~everett/videobub.htm}\textgreater} It appears in the proceedings of this conference. Enlisting the help of his daughter Tanya, a graphic artist, Bub has since made two valiant attempts to bring his and Pitowsky's take on quantum mechanics to the masses. Despite its title and lavish illustrations, \emph{Bananaworld: Quantum Mechanics for Primates} \citep{Bub 2016} is not really a popular book. Its sequel, however, the graphic novel \emph{Totally Random} \citep{Bub and Bub 2018}, triumphantly succeeds where  \emph{Bananaworld} came up short.\footnote{See, e.g., the review in \emph{Physics World} by Minnesota physicist Jim \citet{Kakalios 2018}, well-known for his use of comic books to explain physics \citep{Kakalios 2009}, and the review in \emph{Physics Today} by philosopher of quantum mechanics Richard \citet{Healey 2019}.} The interpretation promoted overtly in \emph{Bananaworld} and covertly in \emph{Totally Random} has been dubbed \emph{Bubism} by Robert Rynasiewicz (private communication).\footnote{In an essay review of \citet{Ball 2018}, \citet{Becker 2018} and \citet{Freire 2015}, \citet{Bub 2019} gives a concise characterization of his views and places them explicitly in the lineage of Heisenberg sketched above.} 
Like QBism, Bubism is an information-theoretic interpretation but for a Bubist quantum probabilities are objective chances whereas for a QBist they are subjective degrees of belief. Our defense of Bubism builds on the Bubs' two books and on ``Two dogmas \ldots'' as well as on earlier work by (Jeff) Bub and Pitowsky, especially the latter's lecture notes \emph{Quantum Probability---Quantum Logic} and his paper on George Boole's ``conditions of possible experience'' \citep{Pitowsky 1989a, Pitowsky 1994}. We will rely heavily on tools developed by these two authors, Bub's correlation arrays and Pitowsky's correlation polytopes. A third musketeer on whose insights we drew for this paper is William Demopoulos (see, e.g., Demopoulos, 2010, and, especially, Demopoulos, 2018, a monograph he completed shortly before he died, which we fervently hope will be published soon).\footnote{We dedicate our paper to Bill and Itamar. See \citet{Bub and Demopoulos 2010} for a moving obituary of Itamar.} 

In the spirit of \emph{Bananaworld}, \emph{Totally Random} and Louisa Gilder's (2008) lovely \emph{The Age of Entanglement}, we wrote the first part of our paper (i.e., most of Section \ref{1}) with a general audience in mind. We will frame our argument in this part of the paper in terms of a variation of Bub's peeling and tasting of quantum bananas scheme (see Figures \ref{AliceBob-Mermin} and \ref{AliceBob-tasting}). This is not just a gimmick adopted for pedagogical purposes. It is also intended to remind the reader that, on a Bubist view, inspired by Heisenberg rather than Schr\"odinger, quantum mechanics provides a new framework for dealing with arbitrary physical systems, be they waves, particles, or various species of fictitious quantum bananas. The peeling and tasting of bananas also makes for an apt metaphor for the (projective) measurements we will be considering throughout \citep{Popescu 2016}. 

As the title of our paper makes clear, however, we follow Jordan rather than Bub in arguing that quantum mechanics is essentially a new framework for handling \emph{probability} rather than \emph{information}. We are under no illusion that this substitution will help us steer clear of two knee-jerk objections to information-theoretic approaches to the foundations of quantum mechanics: \emph{parochialism} and \emph{instrumentalism} (or \emph{anti-realism}).

What invites complaints of parochialism is the slogan ``Quantum mechanics is all about information,'' which conjures up the unflattering image of a quantum-computing engineer, who, like the proverbial carpenter, only has a hammer and therefore sees every problem as a nail. It famously led John \citet[p.\ 34]{Bell 1990} to object: ``\emph{Whose} information? Information about \emph{what}?'' In \emph{Bananaworld}, \citet[p.\ 7]{Bub 2016} counters that ``we don't ask these questions about a USB flash drive. A 64 GB drive is an information storage device with a certain capacity, and whose information or information about what is irrelevant.'' A computer analogy, however, is probably not the most effective way to combat the lingering impression of parochialism. We can think of two better responses to the parochialism charge.

The first is an analogy with meter rather than memory sticks. Consider the slogan ``Special relativity is all about space-time'' or ``Special relativity is all about spatio-temporal relations." These slogans, we suspect, would not provoke the hostile reactions routinely elicited by the slogan ``Quantum mechanics is all about information.'' Yet, one could ask, parroting Bell: ``spatio-temporal relations of \emph{what}?'' The rejoinder in this case would simply be that \emph{what} could be any physical system allowed by the theory; and that, to qualify as such, it suffices that \emph{what} can consistently be described in terms of mathematical quantities that transform as scalars, vectors, tensors or spinors under Lorentz transformations. When we say that a moving meter stick contracts by such-and-such a factor, we only have to specify its velocity with respect to the inertial frame of interest, not what it is made of. Special relativity imposes certain kinematical constraints on any physical systems allowed by theory. Those constraints are codified in the geometry of Minkowski space-time. There is no need to reify Minkowski space-time. We can think of it in relational rather than substantival  terms \citep{Janssen 2009}. The slogan ``Quantum mechanics is all about information/probability'' can be unpacked in a similar way. Quantum mechanics imposes a kinematical constraint on allowed values and combinations of values of observables. Which observables? Any observable that can be represented by a Hermitian operator on Hilbert space. As in the case of Minkowski space-time, there is no need to reify Hilbert space. So, yes, quantum mechanics is obviously about more than just information, just as special relativity is obviously about more than just space-time. Yet the slogans that special relativity is all about space-time and that quantum mechanics is all about information (or probability) do capture---the way slogans do---what is distinctive about these theories and what sets them apart from the theories they superseded.

In Section \ref{4}, we will revisit this comparison between quantum mechanics and special relativity. We should warn the reader upfront though that the kinematical take on special relativity underlying this comparison, while in line with the majority view among physicists, is not without its detractors. In fact, the defense of the kinematical view by one of us \citep{Janssen 2009} was mounted in response to an alternative dynamical interpretation of special relativity articulated and defended most forcefully by Harvey \citet{Brown 2005}.\footnote{See \citet{Acuna 2014} for an enlightening discussion of the debate over whether special relativity is best understood kinematically or dynamically.} Both \citet[p.\ 228]{Bub 2016} in \emph{Bananaworld} and \citet[p.\ 439]{Bub and Pitowsky 2010} in ``Two dogmas \ldots''  invoked analogies with special relativity to defend their information-theoretic interpretation of quantum mechanics. \citet{Brown and Timpson 2006} have disputed the cogency of these analogies \citep[see also][]{Timpson 2010}.\footnote{What complicates matters here is that the distinction between kinematics and dynamics tends to get conflated with the distinction between constructive and principle theories \citep[p.\ 38; see Section \ref{4.1} below for further discussion]{Janssen 2009}.}

Our second response to the parochialism charge is that the quantum formalism for dealing with intrinsic angular momentum, i.e., spin, laid out in Section \ref{2.1} and used throughout in our analysis of an experimental setup to test the Bell inequalities, is key to spectroscopy and other areas of physics as well. These two responses are not unrelated. In Sections \ref{1.6.2} and \ref{4.3a}, drawing on work on the history of quantum physics by one of us,
%\footnote{See \citet{Duncan and Janssen 2008, Duncan and Janssen 2014, Duncan and Janssen 2015} and \citet{Midwinter and Janssen 2013}.} 
we will give a few examples of puzzles for the old quantum theory that physicists resolved not by altering the dynamical equations but by using key features of the kinematical core of the new quantum mechanics.  

What about the other charge against information-theoretic interpretation of quantum mechanics, \emph{instrumentalism} or \emph{anti-realism}? What invites complaints on this score in the case of Bub and Pitowsky  is their identification of the second of the two dogmas they want to reject: ``the quantum state is a representation of physical reality'' \citep[p.\ 433]{Bub and Pitowsky 2010}. This statement of the purported dogma is offered as shorthand for a more elaborate one: ``[T]he quantum state has an ontological significance analogous to the significance of the classical state as the `truthmaker' for propositions about the occurrence or non-occurrence of events'' (ibid.). Of course, denying that state vectors in Hilbert space represent physical reality in and of itself does not make one an anti-realist. We can still be realists as long as we can point to other elements of the theory's formalism that represent physical reality. The sentence we just quoted from ``Two dogmas \ldots'' suggests that for Bub and Pitowsky ``events'' fit that bill. 

That same sentence also points to an important difference between the role of points in classical phase space and vectors in Hilbert space when it comes to identifying what represents physical reality in classical and quantum mechanics. In fact, their notion of a ``truthmaker'' is particularly useful not just for pinpointing how quantum and classical mechanics differ when it comes to representing physical reality but also---even though this may not have been Bub and Pitowsky's intention---for articulating how they are similar. In both classical and quantum mechanics, reality is ultimately represented by values of observable quantities posited by the theory. How we get from catalogs of values of observable quantities to the notion of some object or system possessing the properties represented by those quantities is a separate issue. Physicists may want to leave that for philosophers to ponder, especially since this is not, we believe, what separates quantum physics from classical physics. In both cases, it seems, catalogs of values of observable quantities are primary and objects carrying properties (be it swarms of particles, fields, bananas, tables and chairs or lions and tigers) are somehow constructed out of those.\footnote{Everettians face the same issue as part of the task of explaining how the seemingly classical (Boolean) world we find ourselves in emerges from their multiverse. Bubists could piggy-back on whatever scheme the Everettians come up with to handle this issue.}  

Where quantum and classical mechanics differ is in how values are assigned to observable quantities. In classical mechanics, observable quantities are represented by functions on phase space. Picking a point in phase space fixes the values of all of these. It is in this sense that points in phase space are ``truthmakers''. In quantum mechanics, observable quantities are represented by Hermitian operators on Hilbert space. The possible values of these quantities are given by the eigenvalues of these operators. Picking a vector in Hilbert space, however, does not fix the value of any observable quantity. It fails to do so in two ways. First, the observable(s) being measured must be selected. Only those selected will be assigned definite values. Quantum mechanics tells us that, once this has happened, it is impossible for any observable represented by an operator that does not commute with those representing the selected ones to be assigned a definite value as well. Second, even after this selection has been made, the state vector will in general only give a probability distribution over the various eigenvalues of the operators for the selected observables. Which of those values is found upon measurement of the observable is a matter of chance. Vectors in Hilbert space thus doubly fail to be ``truthmakers''. \emph{Pace} Bub and Pitowsky, however, it does not follow that classical and quantum states have a different ``ontological significance''. One can maintain that neither vectors in Hilbert space \emph{nor points in phase space} represent physical reality; both can be seen as mathematical auxiliaries for assigning definite values (albeit in radically different ways) to quantities that do.\footnote{In \emph{Wahrscheinlichkeitstheoretischer Aufbau}, von Neumann also resisted the idea that vectors in Hilbert space ultimately represent (our knowledge of) physical reality. He wrote: ``our knowledge  of a system $\mathfrak{S}'$, i.e., of the structure of a statistical ensemble $\{ \mathfrak{S}'_1, \mathfrak{S}'_2,$ $\ldots \}$, is never described by the specification of a state---or even by the corresponding $\varphi$ [i.e., the vector $| \varphi \rangle$]; but usually by the result of measurements performed on the system'' \citep[p.\ 260]{von Neumann 1927b}. He thus wanted to represent ``our knowledge  of a system'' by the values of a set of observables corresponding to a complete set of commuting operators \citep[pp.\ 251--251]{Duncan and Janssen 2013}.} 

This quite naturally leads us to the first dogma \citet[p.\ 433]{Bub and Pitowsky 2010} want to reject: Measurement outcomes should be accounted for in terms of the dynamical interaction between the system being measured and a measuring device. As we will argue in Section \ref{4.4}, rejection of this dogma does not amount to black-boxing measurements. On Bub and Pitowsky's view, any measurement can be analyzed in as much detail as on any other view of quantum mechanics. It does mean, however, that one accepts that there comes a point where no meaningful further analysis can be given of why a measurement gives one particular outcome rather than another. Instead it becomes a matter of irreducible randomness---the ultimate crapshoot.\footnote{Paraphrasing what E.\ M.\ \citet[p.\ 28]{Forster 1942} once said about Virginia Woolf (``[S]he pushed the light of the English language a little further against the darkness''), one might say that quantum mechanics pushes physics right up to the point where total randomness takes over.} 

In the opening sentence of their paper, \citet[p.\ 433]{Bub and Pitowsky 2010} announce that rejection of the two dogmas they identified will expose ``the intractable part of the measurement problem''---which they, with thick irony, call the ``big'' measurement problem---as a pseudo-problem. We agree with Bub and Pitowsky that rejecting the first dogma trivially solves the measurement problem \emph{in its traditional form} of having two different dynamics side-by-side, unitary Schr\"odinger evolution as long as we do not make any measurement, state vector collapse when we do. If one accepts that ultimately measurements do not call for a dynamical account (in the sense just mentioned), the problem \emph{in this particular form} evaporates.

By our reckoning, however, the real problem is still with us, just under a different guise. That the quantum state vector is not a ``truthmaker'' in the two senses explained above raises two questions. First, how does one set of observables rather than another get selected to be assigned definite values? Second, why does an observable, once selected, take on one value rather than another? Rejection of the first dogma makes it respectable to resist the call for a dynamical account to deal with the second question and endorse the ``totally random'' response instead.\footnote{We realize that it is easier to swallow this ``totally random'' response for the observables considered in this paper (where the spin of some particle can be up or down or a banana can taste yummy or nasty) than for others, such as, notably, position (where a particle can be here or on the other side of the universe).}
Though arguments from authority will not carry much weight in these matters, we note that a prominent member of the Copenhagen camp did endorse this very answer. In an essay originally published in 1954, Wolfgang Pauli wrote: ``Like an ultimate fact without any cause, the individual outcome of a measurement is \ldots\  in general not comprehended by laws'' \citep[p.\ 32, quoted by Gilder, 2008, p.\ 169]{Pauli 1994}. This then solves Bub and Pitowsky's ``big'' measurement problem. However, it does not address the first question and thus fails to solve what they, again ironically, call the ``small'' measurement problem, which is closely related to the problem posed by this first question.\footnote{See Section \ref{4.4} for careful discussion of how our profound measurement problem differs from their ``small'' one.}

We will accordingly call  their ``big'' problem the \emph{minor} or \emph{superficial} problem and the problem closely related to their ``small'' one the \emph{major} or \emph{profound} one. The profound problem cannot be solved by a stroke of the pen---crossing out this or that alleged dogma in some quantum catechism. What it would seem to require is some account of the conditions under which one set of observables rather than another acquire (or appear to acquire) definite values (regardless of \emph{which} values). The reader will search our paper in vain for such an account. Instead, we will argue that even in the absence of a solution to the profound problem there are strong indications that Bub and Pitowsky were right to reject the two dogmas they identified (and thereby the Everettian solution to both the profound and the superficial problem).

These indications will come from our analysis---in terms of Bub's correlation arrays and Pitowsky's correlation polytopes---of correlations found in measurements on a special but informative quantum state in a simple experimental setup to test a Bell inequality due to David \citet[see Figure \ref{CA-3set2out-Mermin} for the correlation array for Mermin's example of a violation of this inequality]{Mermin 1981, Mermin 1988}. 

We introduce special raffles to determine which of these quantum correlations can be simulated by local hidden-variable theories (see Figure \ref{raffles-spin32-tickets-mu} for an example of tickets for such raffles and Figures \ref{CA-3set2out-raffles-i-thru-iv} and \ref{CA-3set2out-raffle-mix} for examples of the correlation arrays that raffles with different mixes of these tickets give rise to). These raffles will serve as our models of local hidden-variable theories. They are both easy to visualize and tolerably tractable mathematically (see Section \ref{2.2}). They also make for a natural classical counterpart to the quantum ensembles central to von Neumann's \emph{Wahrscheinlichkeitstheoretischer Aufbau}, which were themselves inspired by von Mises's classical statistical ensembles. Finally, they provide simple examples of theories suffering from the superficial but not the profound measurement problem (see note \ref{minor/major} in Section \ref{1.6.2}). 

The quantum state we will focus on is that of two particles of spin $s$ entangled in the so-called singlet state (with zero overall spin). For most of our argument it suffices to consider entangled pairs of spin-$\frac12$ particles. In Section \ref{1} we will almost exclusively consider this case. Our analysis of this case, however, is informed (and justified) at several junctures by our analysis in Section \ref{2} of cases with larger integer or half-integer values of $s$. In Section \ref{2.1}, we analyze the quantum correlations for these larger spin values; in Section \ref{2.2} we analyze the raffles designed to simulate as many features of these quantum correlations as possible. 

In Section \ref{3} we show how our analysis in Sections \ref{1} and \ref{2} can be adapted to the more common experimental setup used to test the Clauser-Horne-Shimony-Holt (CHSH) inequality. The advantage of the Mermin setup, as we will see in Section \ref{1}, is that in that case the classes of correlations allowed by quantum mechanics and by local hidden-variable theories can be pictured in ordinary three-dimensional space. The corresponding picture for the setup to test the CHSH inequality is four-dimensional. The class of all correlations in the Mermin setup that cannot be used for sending signals faster than light can be represented by an ordinary three-dimensional cube, the so-called \emph{non-signaling cube} for this setup; the class of correlations allowed by quantum mechanics by an elliptope contained within this cube; those allowed by classical mechanics by a tetrahedron contained within this elliptope (see Figures \ref{tetrahedron} and \ref{elliptope}). This provides a concrete example of the way in which Pitowsky and others have used nested polytopes to represent the convex sets formed by these classes and subclasses of correlations (compare the cross-section of the non-signaling cube, the tetrahedron and the elliptope in Figure \ref{elliptope-LQPslice} to the usual Vitruvian-man-like cartoon in Figure \ref{LQP}). Such polytopes completely characterize these classes of correlations whereas the familiar Bell inequalities in the case of local hidden-variable theories or Tsirelson bounds in the case of quantum mechanics only provide partial characterizations. 

As Pitowsky pointed out in the preface of  \emph{Quantum Probability---Quantum Logic}:  
\begin{quote}
The possible range of values of classical correlations is constrained by linear inequalities which can be represented as facets of polytopes, which I call ``classical correlation polytopes.'' These constraints have been the subject of investigation by probability theorists and statisticians at least since the 1930s, though the context of investigation was far removed from physics %The linear constraints in question include Bell's inequalities, Clauser-Horne inequalities and their generalizations 
\citep[p.\ IV]{Pitowsky 1989a}.
\end{quote}
The non-linear constraint represented by the elliptope has likewise been investigated by probability theorists and statisticians before in contexts far removed from physics. As we will see in Section \ref{1.6}, it can be found in a paper by Udny \citet{Yule 1896} on what are now called Pearson correlation coefficients as well as in papers by Ronald A.\ \citet{Fisher 1924} and Bruno de Finetti (1937).
%\citet{De Finetti 1937}. 
Yule, like Pearson, was especially interested in applications in evolutionary biology (see notes \ref{biometrist} and \ref{mendel}). We illustrate the results of these statisticians with a simple example from physics, involving a balance beam with three pans containing different weights (see Figure \ref{3M-balance} in Section \ref{1.6.4}). These antecedents in probability theory and statistics provide us with our strongest argument for the thesis that the Hilbert space formalism of quantum mechanics is best understood as a general framework for handling probabilities in a world in which only some observables can take on definite values. 

In Section \ref{1.5} we show that it follows directly from the geometry of Hilbert space that the correlations found in our simple quantum system are constrained by the elliptope and do not 
 saturate the non-signaling cube. This  derivation of the equation for the elliptope is thus a derivation \emph{from within} quantum mechanics.

\citet{Popescu and Rohrlich 1994} and others have raised the question why quantum mechanics does not allow \emph{all} non-signaling correlations. They introduced an imaginary device, now called a PR box, that exhibits non-signaling correlations stronger than those allowed by quantum mechanics.\footnote{See Figure \ref{CA-PRbox} for the correlation array for a PR box. Figure \ref{raffle-ticket-PRbox} shows that it is impossible to design tickets for a raffle that could simulate the correlations generated by a PR box.} Several authors have looked for information-theoretic principles that would reduce the class of all non-signaling correlations to those allowed by quantum mechanics (see, e.g., Clifton, Bub, and Halvorson, 2003, Bub 2016, Ch. 9, Cuffaro, 2018). Such principles would allow us to derive the elliptope \emph{from without}.\footnote{We took the within/without terminology from the chorus of ``Quinn the Eskimo,'' a song from Bob Dylan's 1967 \emph{Basement Tapes}: ``Come all without, come all within. You'll not see nothing like the mighty Quinn.'' Could ``the mighty Quinn'' be an oblique but prescient reference to a quantum computer?\label{Dylan}} 

What the result of Yule and others shows is that the elliptope expresses a general constraint on the possible correlations between three arbitrary random variables. It has nothing to do with quantum mechanics per se. As such it provides an instructive example of a kinematical constraint encoded in the geometrical structure of Hilbert space, just as time dilation and length contraction provide instructive examples of kinematic constraints encoded in the geometry of Minkowski space-time.  In Sections \ref{4.2}--\ref{4.3}, we return to this and other analogies between quantum mechanics and special relativity. In this context, we take a closer look at the interplay between \emph{from within} and \emph{from without} approaches to understanding fundamental features of quantum mechanics. 

We want to make one more observation before we get down to business. As we just saw, it is not surprising that the correlations found in measurements on a pair of particles of (half-)integer spin $s$ in the singlet state do not saturate the non-signaling cube. No such correlations between three random variables could. What is surprising (see Section \ref{1.6}) is that, even in the spin-$\frac12$ case, these correlations do saturate the elliptope. This is in striking contrast to the correlations that can be generated with the raffles designed to simulate the quantum correlations. In the spin-$\frac12$ case, the correlations allowed by our raffles are all represented by points inside the tetrahedron inscribed in the elliptope. As we will see in Section \ref{1.6}, this is because there are only two possible outcomes in the spin-$\frac12$ case, $\pm \sfrac12$. In the spin-$s$ case, there are $2s+1$ possible outcomes: $-s, -s+1, \ldots, s-1, s$. With considerable help from the computer (see the flowchart in Figure \ref{flowchart} in Section \ref{2.2.2} and the discussion of its limitations in Section \ref{2.2.4}), we generated figures showing that, with increasing $s$, the correlations allowed by the raffles designed to simulate the quantum correlations are represented by polytopes that get closer and closer to the elliptope (see Figures \ref{polytope-spin1}, \ref{SpinThreeHalfFace} and \ref{FacetsSpin2Spin52} for $s = 1, \sfrac32, 2, \sfrac52$). That the quantum correlations already fully saturate the elliptope in the spin-$\frac12$ case is due to a remarkable feature of quantum mechanics: it allows a sum to have a definite value even if the individual terms in this sum do not.

%SECTION 2 (labels: 1, 1.1 through 1.6)
\section{Representing distant correlations by correlation arrays and polytopes} \label{1}
%!TEX root =  ./JanasJanssenCuffaro-August2019.tex
%\section{Representing distant correlations by correlation arrays and polytopes} \label{1}

%SUBSECTION 2.1
\subsection{Taking Mermin to Bananaworld} \label{1.1}

The classical tests of Bell's theorem in the 1970s and 1980s were for a version of the Bell inequality formulated by \citet{CHSH}.\footnote{See \citet[Chs.\ 29--31, pp.\ 250--289]{Gilder 2008} for a concise account, written for a general audience and based on interviews with some of the principals, of how the CHSH inequality was formulated and experimentally tested.} The CHSH inequality, like the one originally proposed by \citet{Bell 1964}, is a bound on the strength of distant correlations allowed by local hidden-variable theories. In such theories, the outcomes of the relevant measurements are predetermined by variables not included in the quantum description (hence: hidden) and cannot be affected by signals traveling faster than light (hence: local). The setup used to test the CHSH inequality involves \emph{two parties}, the ubiquitous Alice and Bob, \emph{two settings per party} of some measuring device (e.g., a polarizer or a Dubois magnet as used in a Stern-Gerlach-type experiment), and \emph{two outcomes per setting} (labeled `0' and `1', `$+$' and `$-$', or `up' and `down'). 

\citet[pp.\ 18--19]{Bell 1964} originally considered three rather than four settings, labeled $\{\hat{a}, \hat{b}, \hat{c}\}$. 
In Bell's setup, one party performs measurements using the pair $\{\hat{a}, \hat{b}\}$ while the other uses $\{\hat{b}, \hat{c}\}$. In the CHSH setup the two parties use two pairs that have no setting in common, $\{\hat{a}, \hat{b}\}$ and $\{\hat{a}', \hat{b}'\}$ in our notation. \citet{Mermin 1981, Mermin 1988} kept Bell's three settings but in his setup both parties use all three settings rather than just two of them. He derived a Bell inequality for this setup, so simple that even those without Mermin's pedagogical skills can explain it to a general audience. 

We use the Mermin setup to illustrate the power of some of the tools in \emph{Bananaworld} \citep{Bub 2016}. We represent the correlations Mermin considered by \emph{correlation arrays}, the workhorse of \emph{Bananaworld}, and parametrize these arrays in such a way that they, in turn, can be represented as points in convex sets in so-called \emph{non-signaling cubes}. This approach was pioneered by \citet{Pitowsky 1989b} in \emph{Quantum Probability---Quantum Logic}.\footnote{\citet[p.\ 120]{Bub 2016} also cites \citet{Pitowsky 2006}, his contribution to a \emph{Festschrift} for Bub, as well as \citet{Pitowsky 1986,Pitowsky 1989a,Pitowsky 1991,Pitowsky 2008}.} 

The representation of classes of correlations in terms of convex sets is well-established in the quantum-foundations literature. Our paper can be seen as another attempt to bring this approach to a broader audience by applying it to Mermin's particularly simple and instructive example. The CHSH setup uses four different settings and its non-signaling cube is a hypercube in four dimensions. The Mermin setup only uses three different settings and its non-signaling cube is an ordinary cube in three dimensions, which makes it easy to visualize. The convex set representing the non-signaling correlations allowed classically is a tetrahedron spanned by four of the eight vertices of the non-signaling cube (see Figure \ref{tetrahedron}); the convex set representing those allowed quantum-mechanically is an elliptope enclosing this tetrahedron (see Figure \ref{elliptope}).  

In \emph{Bananaworld}, \emph{settings} become \emph{peelings}, \emph{outcomes} become \emph{tastes}, and \emph{parties} become characters from \emph{Alice in Wonderland} (Alice stars as Alice, the White Rabbit as Bob). Bananas can be peeled ``from the stem end ($S$)'' or ``from the top end ($T$)'' and can only taste ``\emph{o}rdinary (``o'' or 0)'' or ``\emph{i}ntense, \emph{i}ncredible,  \emph{i}ncredibly delicious (``i'' or 1)'' \citep[pp.\ 8--9, see also p.\ viii]{Bub 2016}.\footnote{Betraying his information-theoretic leanings, \citet{Bub 2016} occasionally refers to \emph{inputs} and \emph{outputs} (both taking on the values 0 and 1) rather than peelings and tastes (see, e.g., p.\ 51, Figure 3.1).} Bub's banana-peeling scheme suffices for the discussion of the CHSH inequality as well as for the analysis of PR boxes, at least those of the original design of their inventors,  \citet{Popescu and Rohrlich 1994}. A PR box is a hypothetical system allowing \emph{superquantum correlations} \citep[p.\ 106]{Bub 2016}, non-signaling correlations that are stronger 
(in some sense to be explicated later)
%(by measures to be introduced below) 
than those allowed by quantum mechanics. Like the CHSH setup, the original design of a PR box involved two parties, two settings per party, and two outcomes per setting. Bub's scheme also works for the analysis of correlations that arise in measurements on so-called GHZ states \citep{GHZ}. While these measurements involve three rather than two parties,\footnote{Bub's illustrator, his daughter Tanya, has the Cheshire Cat (starring as Clio) peel the third GHZ banana \citep[pp.\ 122--123, Clio and Charlie are introduced on p.\ 8]{Bub 2016}.} they still fit the mold of two settings per party and two outcomes per setting. The Mermin setup breaks this mold by using (the same) three settings for both parties. 

\begin{figure}[h]
\centering
    \includegraphics[width=6in]{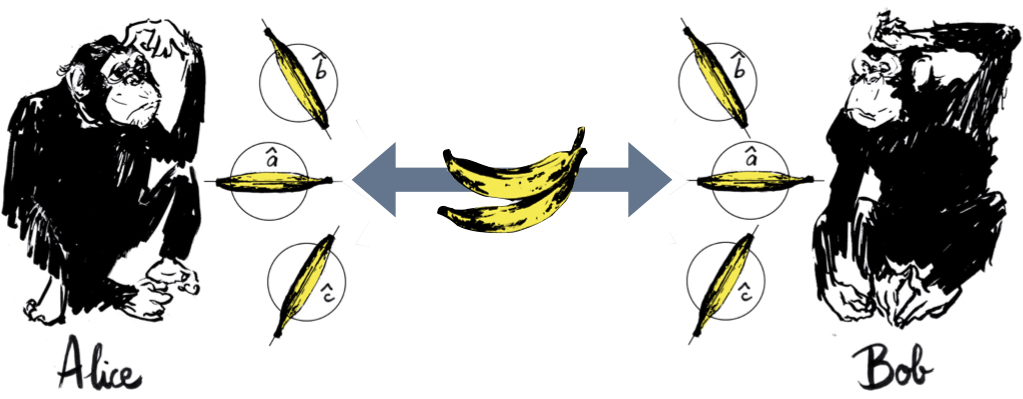}
 \caption{Taking Mermin to \emph{Bananaworld} (I). \emph{Two parties}: the \emph{chimps} Alice and Bob. \emph{Three settings per party}: three \emph{peelings}, ($\hat{a}, \hat{b}, \hat{c}$), given by three unit vectors $(\vec{e}_a, \vec{e}_b, \vec{e}_c)$, in the corresponding \emph{peeling directions} (i.e., the direction of the line going from the top to the stem of the banana while it is being peeled). In Mermin's example, the angles $\varphi_{ab}$ between $\vec{e}_a$ and $\vec{e}_b$, $\varphi_{ac}$ between $\vec{e}_a$ and $\vec{e}_c$, and $\varphi_{bc}$ between $\vec{e}_b$ and $\vec{e}_c$ are all equal to $120\degree$. Drawing by Laurent  Taudin with a nod to Andy Warhol.}
 %(see note \ref{warhol})}
   \label{AliceBob-Mermin}
\end{figure}

\begin{figure}[h]
\centering
    \includegraphics[width=5.5in]{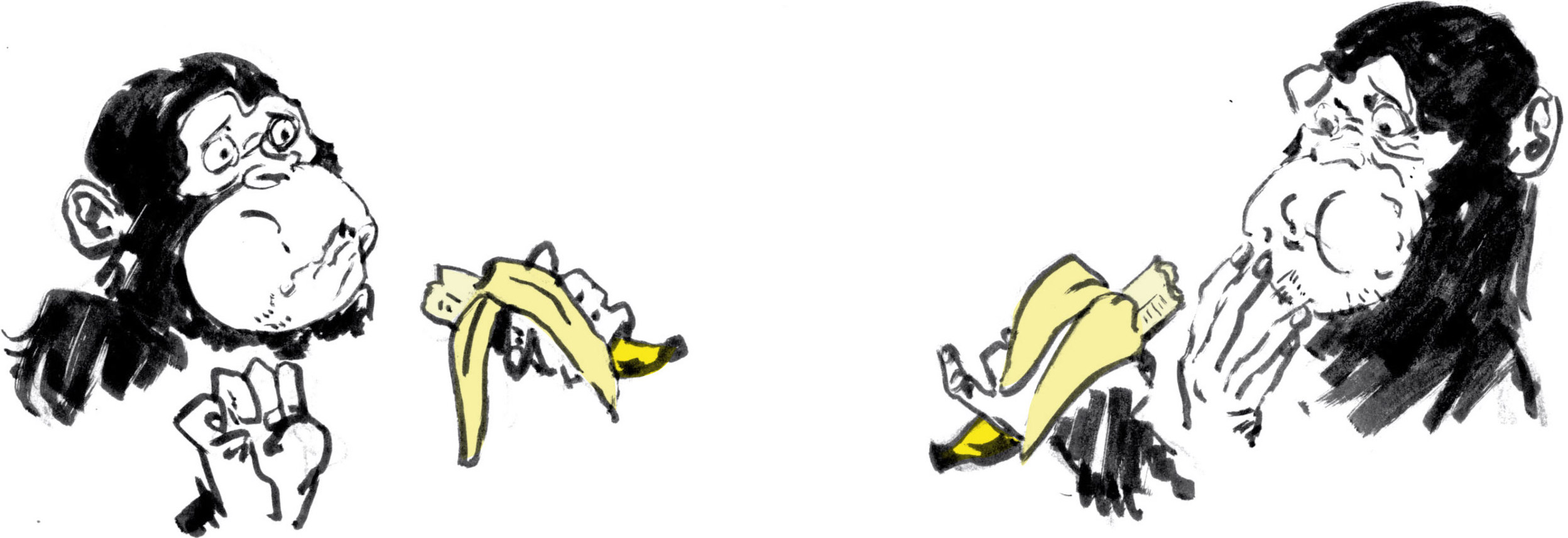}
 \caption{Taking Mermin to \emph{Bananaworld} (II). \emph{Two outcomes per setting}: the \emph{tastes} ``yummy'' ($+$) or ``nasty'' ($-$) for different peeling directions. The peeling and tasting is done by the chimps Alice and Bob. Drawing by Laurent Taudin.}
   \label{AliceBob-tasting}
\end{figure}

To recreate the Mermin setup in \emph{Bananaworld} we thus need a new banana-peeling scheme. Our scheme not only allows infinitely many different settings, it also highlights elements of spherical symmetry in the setups we will examine that turn out to be key to their quantum-mechanical analysis (see Section \ref{2.1}). Figures \ref{AliceBob-Mermin}--\ref{AliceBob-tasting} illustrate our \emph{Bananaworld} version of the Mermin setup.  

We focus on a species of banana that grows in pairs on special banana trees. These bananas can only taste yummy or nasty. Yet we cannot say that they come in two flavors, as they only acquire a definite flavor once they are peeled and tasted. We use these bananas in a long series of peel-and-taste experiments following a protocol familiar from experimental tests of Bell inequalities. We pick a pair of bananas, still joined at the stem, from the banana tree. We separate them and give one each to two chimps, Alice and Bob. Once they have received their respective bananas, they randomly and independently of one another pick a particular \emph{peeling}, defined by the \emph{peeling direction}, i.e., the direction of the line going from the top to the stem of the banana while it is being peeled. Alice and Bob are instructed not to change the orientation of their bananas while peeling so that it is unambiguous which peeling they are using. In the Mermin setup, Alice and Bob get to choose between three peelings, labeled $\hat{a}$, $\hat{b}$ and $\hat{c}$, represented by unit vectors, $\vec{e}_a$, $\vec{e}_b$ and $\vec{e}_c$, in the corresponding peeling directions (see Figure \ref{AliceBob-Mermin}). Once they have randomly chosen one of these three peelings, they point the stem of their banana in the direction of the corresponding unit vector and peel their banana (it does not matter whether they peel from the top or from the stem). When done peeling, Alice and Bob reposition their bananas and take a bite to determine whether they taste yummy or nasty (see Figure \ref{AliceBob-tasting}). The whole procedure is then repeated with a fresh pair of bananas from the banana tree. 

In each run of this peel-and-taste experiment, Alice and Bob record that run's ordinal number, the peeling chosen ($\hat{a}$, $\hat{b}$ or $\hat{c}$) and the taste of their banana, using ``$+$'' for ``yummy'' and ``$-$'' for ``nasty''. Every precaution is taken to ensure that, as long as there are more bananas to be peeled and tasted, Alice and Bob cannot communicate. While they are peeling and tasting, the only contact between them is that the bananas they are given come from pairs originally joined at the stem on the banana tree.

When all bananas are peeled and tasted, Alice and Bob are allowed to compare notes. Just looking at their own records, they see nothing out of the ordinary---just a sequence of pluses and minuses as random as if they had faked their results by tossing a coin for every run. Comparing their records, however, they note that, every time they happened to choose the same peeling (in roughly $33 \%$ of the total number of runs), their results are perfectly anti-correlated. Whenever one banana tasted yummy, its twin tasted nasty. In and of itself, this is not particularly puzzling. Maybe our bananas always grow in pairs in which one is predetermined to taste yummy while its twin is predetermined to taste nasty. This simple explanation, however, is ruled out by another striking correlation our chimps discover while pouring over their data. When they happened to peel differently (in roughly $66 \%$ of the runs), their results were positively correlated, albeit imperfectly. In 75\% of the runs in which they used different peelings, their bananas tasted the same \citep[p.\ 86]{Mermin 1981}.\footnote{In Mermin's (1981, p.\ 86) example, there is a perfect (positive) correlation in runs in which the two parties use the same setting and an imperfect anti-correlation in runs in which they use different settings (see also Mermin, 1988, pp.\ 135--136). To get Mermin's original example, we should have used our pairs of bananas to represent entangled pairs of photons and let ``peel and taste bananas using different peeling directions'' stand for ``measure the polarization of these photons along different axes''. We got our variation on Mermin's example by having pairs of bananas represent pairs of spin-$\frac12$ particles entangled in the singlet state and letting ``peel and taste bananas using different peeling directions''  stand for ``measure spin components of these particles along different axes'' (see Section \ref{1.5}).\label{mermin}} The tastes of two bananas coming from one pair thus depend on the angle between the peeling directions used. This is certainly odd but one could still imagine that our bananas are somehow pre-programmed to respond differently to different peelings and that the set of pre-programmed responses is different for the two bananas in one pair. What Mermin's Bell inequality shows, however, is that it is impossible to pre-program twin bananas in such a way that they would produce the specific correlations found in this case. Such correlations, however, can and have been produced with quantum twins (see Section \ref{1.5}). Given that they persist no matter how far we imagine Alice and Bob to be apart, another explanation of these curious correlations is also unavailing: it would take a signal traveling faster than the speed of light for the taste of one banana peeled a certain way to either affect the way the other banana is peeled or affect its taste when peeled that way. In short, these correlations cannot be accounted for on the basis of any local hidden-variable theory. 

%SUBSECTION 2.2
\subsection{Non-signaling correlation arrays} \label{1.2}

The correlations found in the Mermin setup can be represented in a correlation array consisting of nine cells, one for each of the nine possible combinations of peelings (see Figure \ref{CA-3set2out-Mermin} in Section \ref{1.3}). These cells form a grid with three rows for Alice's three peeling directions and three columns for Bob's. Each cell has four entries, giving the probabilities of the four possible pairs of tastes for that cell's combination of peelings (the entries in one cell thus always sum to 1). 

Since \emph{Bananaworld} focuses on setups with two settings per party, all correlation arrays in it have only four cells. These cells form a $2 \times 2$ grid with rows for Alice peeling from the stem and from the top and columns for Bob peeling from the stem and from the top. Before we turn to the $3 \times 3$ Mermin correlation array we go over some properties of these simpler $2 \times 2$ correlation arrays. 

\begin{figure}[h]
 \centering
   \includegraphics[width=2.5in]{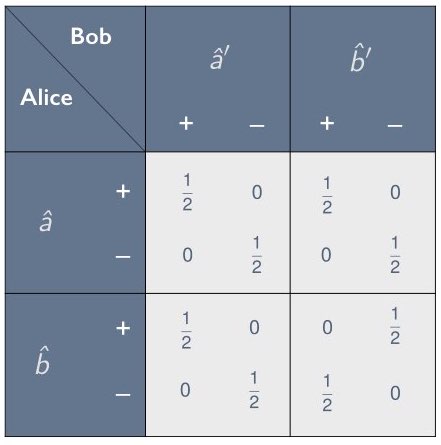} 
   \caption{Correlation array for a Popescu-Rohrlich box.}
   \label{CA-PRbox}
\end{figure}  

The correlation array  in Figure \ref{CA-PRbox} for a PR box in its original design is an example of such an array \citep[p.\ 89, Table 4.2; we switched Alice and Bob to match the convention that the index labeling the rows of a matrix comes before the index labeling its columns]{Bub 2016}. This correlation array plays an important role in \emph{Bananaworld} and is central to its sequel, Tanya and Jeffrey Bub's (2018) enchanting \emph{Totally Random}. A version of it is prominently displayed on many pages of this graphic novel \citep[pp.\ 15, 21, 33, 95, 115, 181, 200 and 227]{Bub and Bub 2018}. The version in \emph{Totally Random} differs in two respects from the version in Figure \ref{CA-PRbox} (which follows \emph{Bananaworld}). First, in Figure \ref{CA-PRbox}, the outcomes found by Alice and Bob are perfectly correlated in three of the four cells, while they are perfectly anti-correlated in the remaining one. In \emph{Totally Random} it is just the other way around. Second, instead of the four entries in each cell in Figure \ref{CA-PRbox}, the cells in \emph{Totally Random} just have ``$=$'' for perfectly correlated and ``$\neq$'' for perfectly anti-correlated. 

In \emph{Bananaworld} the PR-box correlations in Figure \ref{CA-PRbox} are realized with the help of PR bananas growing in pairs on PR banana trees. The settings $\{\hat{a}, \hat{b}\}$ and $\{\hat{a}', \hat{b}'\}$ now stand for Alice and Bob peeling their bananas from the stem ($S$) or from the top ($T$). These peelings could be replaced by two of the peeling directions we introduced. In realizations of this PR box, we can (but do not have to) use the same pair of settings for Alice and Bob (in the case of the CHSH setup we definitely need different pairs of settings; see Section \ref{3}).  

In \emph{Totally Random}, the PR-box correlations  in Figure \ref{CA-PRbox}  are realized with the help of an imaginary device, named for the inventors of the PR box, the ``Superquantum Entangler PR01''. This gadget, which looks like a toaster, has slots for two US quarters. When we insert two ordinary coins, the PR01 turns them into a pair of entangled ``quoins'' \citep[p.\ 7]{Bub and Bub 2018}. The different settings now stand for Alice and Bob holding their quoins heads-up ($\hat{a} = \hat{a}'$) or tails-up ($\hat{b} = \hat{b}'$) when tossing them. The outcomes are the quoins landing heads or tails. What makes this a realization of a PR box with the correlation array shown in Figure \ref{CA-PRbox} is that the quoins invariably land with the same side facing up, except when both are tossed being held tails-up ($\hat{b}, \hat{b}'$), in which case they always land with opposite sides facing up. 

The correlations between the outcomes found in a PR box---be it between the tastes of a pair of PR bananas or the landings of a pair of quoins---are preserved no matter how far its two parts are pulled apart.\footnote{Part of what makes it interesting to contemplate entangled quoins or bananas is that we are free to choose \emph{when} to toss or taste them whereas with entangled photons or spin-$\frac12$ particles we have no choice but to measure their polarization or spin as soon as they arrive at our detectors.} 

An important feature of correlation arrays (no matter how many cells they have or how many entries each cell has) is that they allow us to see at a glance whether or not the correlations they represent can be used for the purposes of instant messaging or superluminal signaling. Suppose Alice wants to use the peeling of a pair of PR bananas to instant-message the answer to some ``yes/no'' question to Bob. They agree ahead of time that Alice will peel $\hat{a}$ if the answer is ``yes'' and $\hat{b}$ if it is ``no''.\footnote{It does not matter in what order Alice and Bob peel their bananas. The correlations in the correlation array in Figure \ref{CA-PRbox} represent constraints on possible combinations of outcomes found by Alice and Bob, not some mechanism through which the outcome of one peeling would cause the outcome of the other.\label{peeling order irrelevant}}  This scheme will not work. No matter how Bob peels his banana, he cannot tell from its taste whether Alice peeled hers $\hat{a}$ or $\hat{b}$. Suppose Bob peels $\hat{b}'$ (essentially the same argument works if Bob peels $\hat{a}'$). In that case, the correlation array in Figure \ref{CA-PRbox} tells us that the marginal probability of Bob finding $+$  if Alice were to peel $\hat{a}$ (trying to transmit ``yes'') is
\begin{equation}
\mathrm{Pr}(+_{\mathrm B}| \hat{a} \,\hat{b}') = \mathrm{Pr}(+\!+| \hat{a} \,\hat{b}') \, + \, \mathrm{Pr}(-\!+| \hat{a} \,\hat{b}') =  \sfrac12 + 0 = \sfrac12,
\label{non-signaling property 1}
\end{equation}
which is the same as the marginal probability of him finding $+$  if Alice were to peel $\hat{b}$ (trying to transmit ``no''):
\begin{equation}
\mathrm{Pr}(+_{\mathrm B}| \hat{b} \,\hat{b}') = \mathrm{Pr}(+\!+| \hat{b} \,\hat{b}') \; + \; \mathrm{Pr}(-\!+| \hat{b} \,\hat{b}') =  0 + \sfrac12 = \sfrac12.
\label{non-signaling property 2}
\end{equation}
Inspection of the correlation array in Figure \ref{CA-PRbox} shows that \emph{all} such marginal probabilities are equal to $\sfrac12$ in this case. PR boxes---whether realized with the help of magic bananas, quoins, or other systems---cannot be used for instant messaging. 

Correlations that do not allow instant messaging are called \emph{non-signaling}. It will be convenient to use this term for their correlation arrays as well. The correlations and correlation arrays for a PR box are always non-signaling. In fact, this is what makes these hypothetical devices intriguing. Even though they would give rise to correlations stronger than those allowed by quantum mechanics, they would not violate special relativity's injunction against superluminal signaling.

Generalizing the results in Eqs.\ (\ref{non-signaling property 1})--(\ref{non-signaling property 2}), we can state the following \emph{non-signaling condition}:
\begin{quote}
\emph{A correlation in a setup with two parties, two settings per party and two outcomes per setting is \emph{non-signaling} if the probabilities in both rows and both columns of all cells in its correlation array add up to $\sfrac12$.}
\end{quote}
The converse is not true. A correlation array with the entries
\begin{equation}
\begin{array}{cccc}
1  & 0  & 0 & 1 \\[.1 cm]
 0 & 0  & 0 & 0 \\[.1 cm]
 0 & 0  & 0 & 0 \\[.1 cm]
1 & 0 & 0 & 1
\end{array}
\end{equation}
is non-signaling even though the entries in half the rows and columns of its cells add up to 1 while the entries in the other half add up to 0. The relevant marginal probabilities, however, are still equal to each other. For instance,
\begin{equation}
\mathrm{Pr}(+_{\mathrm B}|\hat{a} \,\hat{b}') = \mathrm{Pr}(+_{\mathrm B}|\hat{b} \,\hat{b}') = 0 \quad \mathrm{ and} \quad \mathrm{Pr}(-_{\mathrm B}|\hat{a} \,\hat{b}') = \mathrm{Pr}(-_{\mathrm B}|\hat{b} \,\hat{b}') = 1.
\end{equation}
In Section \ref{2}, we will encounter correlation arrays for setups with three outcomes per setting that are non-signaling even though not all rows and columns of its cells add up to the same number (see Figure \ref{CA-3set3out-raffle-vi} in Section \ref{2.2.2}).\footnote{In \emph{Bananaworld}, Bub leaves it to the reader to find examples of correlation arrays that violate the non-signaling condition. Below are the entries for two such correlation arrays:
$$
(a) \quad \begin{array}{cccc}
1  & 0  & 0 & 0 \; \\
 0 & 0  & 0 & 1 \; \\
 0 & 0  & 1 & 0 \; \\
 0 & 1 & 0 & 0,
\end{array}
\quad \quad \quad
(b) \quad \begin{array}{cccc}
\boldsymbol{6/10}  & 1/10  & 2/10 & 1/10 \; \\
1/10  & 2/10  & 1/10 & \boldsymbol{6/10} \; \\
2/10  & 1/10  & \boldsymbol{6/10} & 1/10 \; \\
1/10  & \boldsymbol{6/10}  & 1/10 & 2/10.
\end{array}
$$
If there were a system producing the distant correlations in (a), be it pairs of bananas or pairs of coins, one pair would suffice for Alice and Bob to transmit one bit of information to the other party instantaneously; if there were a system producing the distant correlations in (b), several pairs would be needed to do so with some fidelity. The latter system can be thought of as a noisy version of the former.}

%SUBSECTION 2.3
\subsection{Non-signaling cubes, classical polytopes and the elliptope} \label{1.3}

\begin{figure}[h]
\centering
    \includegraphics[width=3in]{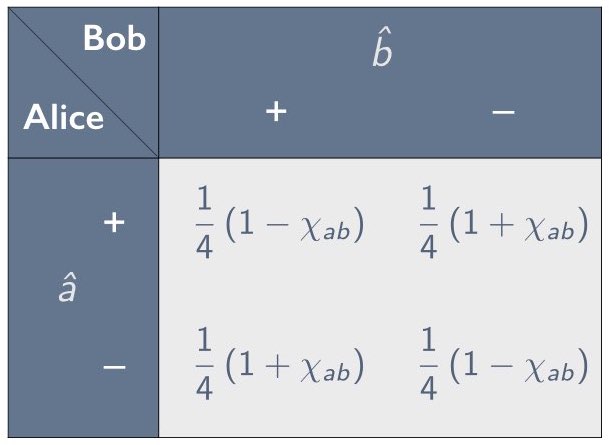}
 \caption{Cell in a non-signaling correlation array parametrized by $-1 \le \chi_{ab} \le 1$.}
   \label{CA-2set2out-cell}
\end{figure}

Any cell in a non-signaling correlation array for any number of settings with two outcomes per setting can be parametrized by a variable with values running from $-1$ to $+1$. Figure  \ref{CA-2set2out-cell} shows such a cell for Alice using setting $\hat{a}$ and Bob using setting $\hat{b}$. Let $-1 \ge \chi_{ab} \ge 1$ be the variable parametrizing this cell. If $\chi_{ab} = 0$, the results of Alice and Bob are uncorrelated; if $\chi_{ab} =-1$, they are perfectly correlated; if $\chi_{ab} =1$, they are perfectly anti-correlated. We will thus call $\chi_{ab}$ an \emph{anti-correlation coefficient.}

Consider the random variable $X_a^A$ measured by Alice using setting $\hat{a}$ and the random variable $X_b^B$ measured by Bob using setting $\hat{b}$. The  \emph{covariance} of these two variables is defined as the expectation value of the product of $X_a^A - \langle X_a^A \rangle$ and $X_b^B - \langle X_b^B \rangle$, where $\langle X \rangle$ is the expectation value of $X$:
\begin{equation}
\mathrm{cov} \! \left( X_a^A, X_b^B \right) \equiv \left\langle \left( X_a^A - \langle X_a^A \rangle \right) \left( X_b^B - \langle X_b^B \rangle \right) \right\rangle.
\label{cov def 0}
\end{equation}
The random variables we will be considering are all \emph{balanced}. That a random variable is \emph{balanced} means that it has the following two properties:
\begin{quote}
A random variable $X$ is balanced IFF
\begin{equation}
\begin{array}{l}
\text{(1) if $x$ is a possible value, then $-x$ is a possible value as well;} \\[.2cm]
\text{(2) the value $x$ is as likely to occur as the value $-x$.} 
\end{array}
\label{def balanced}
\end{equation}
\end{quote}
Such variables have zero expectation value, which means that Eq.\ (\ref{cov def 0}) reduces to:
\begin{equation}
\mathrm{cov} \! \left( X_a^A, X_b^B \right) = \left\langle  X_a^A \, X_b^B \right\rangle.
\label{cov def}
\end{equation}
Bell inequalities (including the CHSH one) are typically expressed in terms of such expectation values. To compute $\langle  X_a^A \, X_b^B \rangle$, we need to assign a numerical value to the taste of a banana. To this end, we introduce the \emph{Bub} or \emph{banana constant} $b$. Yummy ($+$) and nasty ($-$) then correspond to $\pm \bbar/2$, where $\bbar \equiv b/2\pi$ (called \emph{banana split} or \emph{banana bar}). Using the entries in the correlation array in Figure \ref{CA-2set2out-cell}, we evaluate the expectation value of the product of $X_a^A$ and $X_b^B$:
\begin{eqnarray}
\left\langle X_a^A \, X_b^B \right\rangle & = & \frac{\bbar^2}{4} \left(\mathrm{Pr}(+\!+| \hat{a} \,\hat{b}) \, + \, \mathrm{Pr}(-\!-| \hat{a} \,\hat{b})\right) 
- \frac{\bbar^2}{4} \left(\mathrm{Pr}(+\!-| \hat{a} \,\hat{b}) \, + \, \mathrm{Pr}(-\!+| \hat{a} \,\hat{b})\right) \nonumber \\[.2 cm]
& = & \frac{\bbar^2}{4} \left( \frac12 (1 - \chi_{ab}) - \frac12 (1 + \chi_{ab}) \right) \; = \; -\frac{\bbar^2}{4} \chi_{ab}.
\label{prob 2 exp}
\end{eqnarray}
Introducing the standard deviations of $X_a^A$ and $X_b^B$,
\begin{equation}
\begin{array}{c}
\sigma^A_a \equiv \sqrt{ \left\langle (X^A_a)^2 - \langle X_a^A \rangle^2 \right\rangle} = \sqrt{ \left\langle (X^A_a)^2 \right\rangle }= \displaystyle{\frac{\bbar}{2}},    \\[.4cm]
\sigma^B_b \equiv  \sqrt{ \left\langle (X^B_b)^2 - \langle X_b^B \rangle^2 \right\rangle} = \sqrt{ \left\langle (X^B_b)^2 \right\rangle }= \displaystyle{\frac{\bbar}{2}}  
\end{array}
\label{standard deviations a and b}
\end{equation}
where we used that $\langle X_a^A \rangle = \langle X_b^B \rangle = 0$, we can thus write the parameter $\chi_{ab}$  introduced in Figure \ref{CA-2set2out-cell} as
\begin{equation}
\chi_{ab} = - \frac{\left\langle X_a^A \, X_b^B \right\rangle}{\sigma^A_a \sigma^B_b}. 
\label{chi as corr coef}
\end{equation}
This is our formal justification for calling $\chi_{ab}$ (and parameters like it for other cells in this and other correlation arrays) an \emph{anti-correlation coefficient}: it is \emph{minus} what is commonly known as \emph{Pearson's correlation coefficient}. We will return to this information-theoretic interpretation of $\chi_{ab}$ in Section  \ref{1.6}.

\begin{figure}[h]
\centering
    \includegraphics[width=5in]{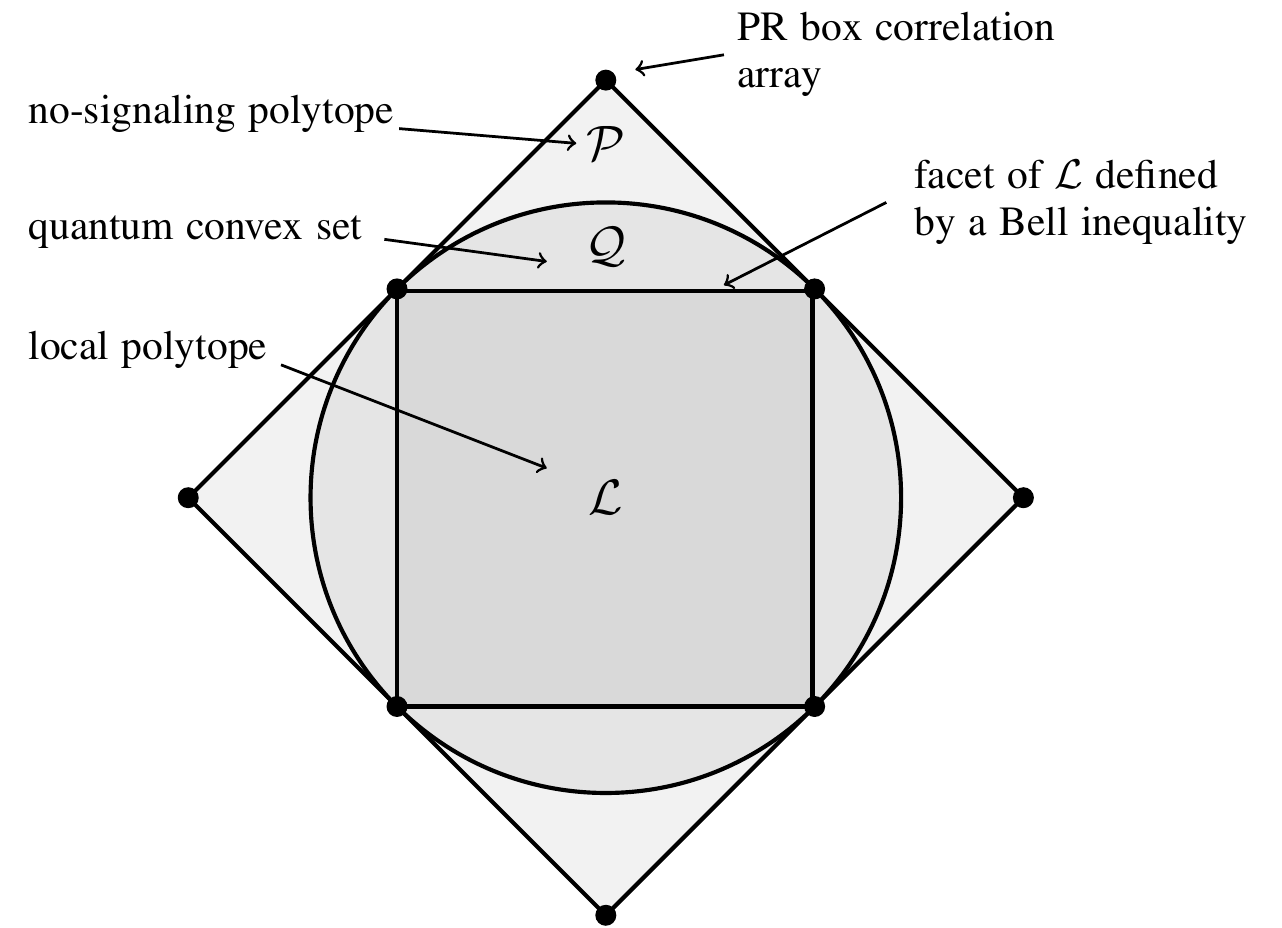}
 \caption{A schematic representation, for some arbitrary experimental setup, of the set $\mathcal{P}$ of all non-signaling correlations, the subset $\mathcal{Q} \subset \mathcal{P}$ of those allowed quantum-mechanically and the subset $\mathcal{L} \subset \mathcal{Q}$ of those allowed classically. One of the facets of $\mathcal{L}$ represents a Bell inequality. The vertex of the non-signaling cube where this Bell inequality is maximally violated represents a PR box for the setup under consideration \citep[p.\ 107, Figure 5.2]{Bub 2016}.}
   \label{LQP}
\end{figure}

A $2 \times 2$ non-signaling correlation array such as the one in Figure \ref{CA-PRbox} for a PR box, with four cells of the form of Figure \ref{CA-2set2out-cell}, can be parametrized by four anti-correlation coefficients
\begin{equation}
-1 \le \chi_{aa'} \le 1, \quad -1 \le \chi_{ab'} \le 1, \quad -1 \le \chi_{ba'} \le 1, \quad -1 \le \chi_{bb'} \le 1.
\label{chi values for PR box}
\end{equation}
Such a correlation array can thus be represented by a point in a hypercube in four dimensions with the anti-correlation coefficients serving as that point's Cartesian coordinates. The correlation array for a PR box is represented by one of the vertices of this hypercube: 
\begin{equation}
(\chi_{aa'}, \chi_{ab'}, \chi_{ba'}, \chi_{bb'}) = (-1, -1, -1, 1).
\label{PR box vertices}
\end{equation}
The four-dimensional hypercube that represents the class of all non-signaling correlations in this setup (two parties, two settings per party, two outcomes per setting) is an example of a so-called \emph{non-signaling polytope}, which can be defined (typically in some higher-dimensional space) for setups with two parties, any number of settings and any number of outcomes per setting. 

\begin{figure}[h]
 \centering
   \includegraphics[width=5in]{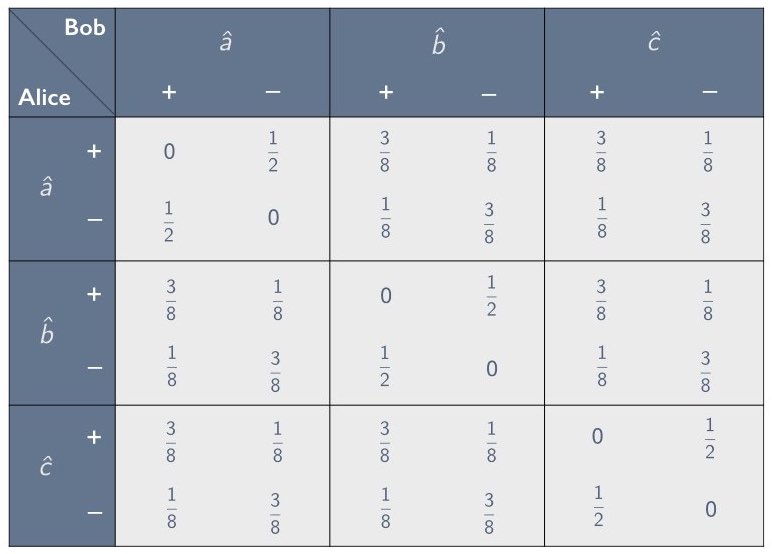} 
   \caption{Correlation array for the correlations found in our variation of the Mermin setup (see Figures \ref{AliceBob-Mermin} and \ref{AliceBob-tasting} and note \ref{mermin}).}
   \label{CA-3set2out-Mermin}
\end{figure}

\begin{figure}[h]
 \centering
   \includegraphics[width=5in]{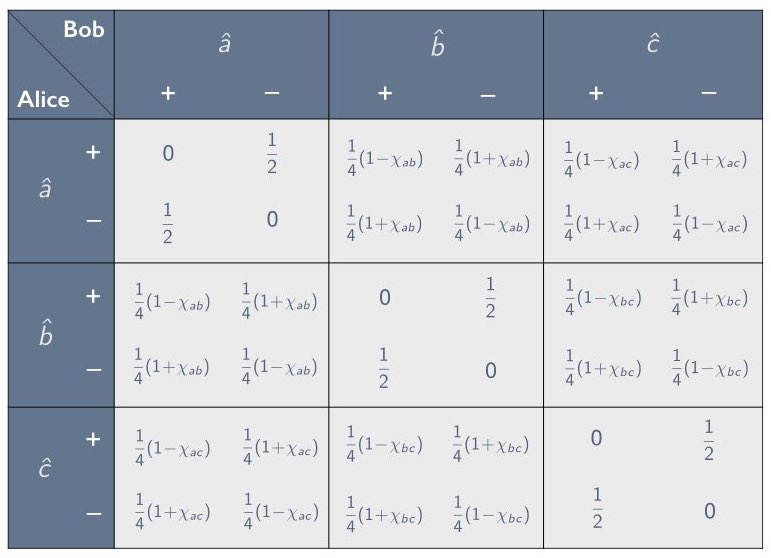} 
   \caption{A non-signaling correlation array for three settings (peelings) and two outcomes (tastes) parametrized by the anti-correlation coefficients $-1 \le \chi_{ab} \le 1$ (for the $\hat{a} \, \hat{b}$ and  $\hat{b} \, \hat{a}$ cells), $-1 \le \chi_{ac} \le 1$ (for the $\hat{a} \, \hat{c}$ and  $\hat{c} \, \hat{a}$ cells) and $-1 \le \chi_{bc} \le 1$ (for the $\hat{b} \, \hat{c}$ and  $\hat{c} \, \hat{b}$ cells).}
   \label{CA-3set2out-non-signaling-chis}
\end{figure}

\begin{figure}[h]
\centering
\includegraphics[width=4.6in]{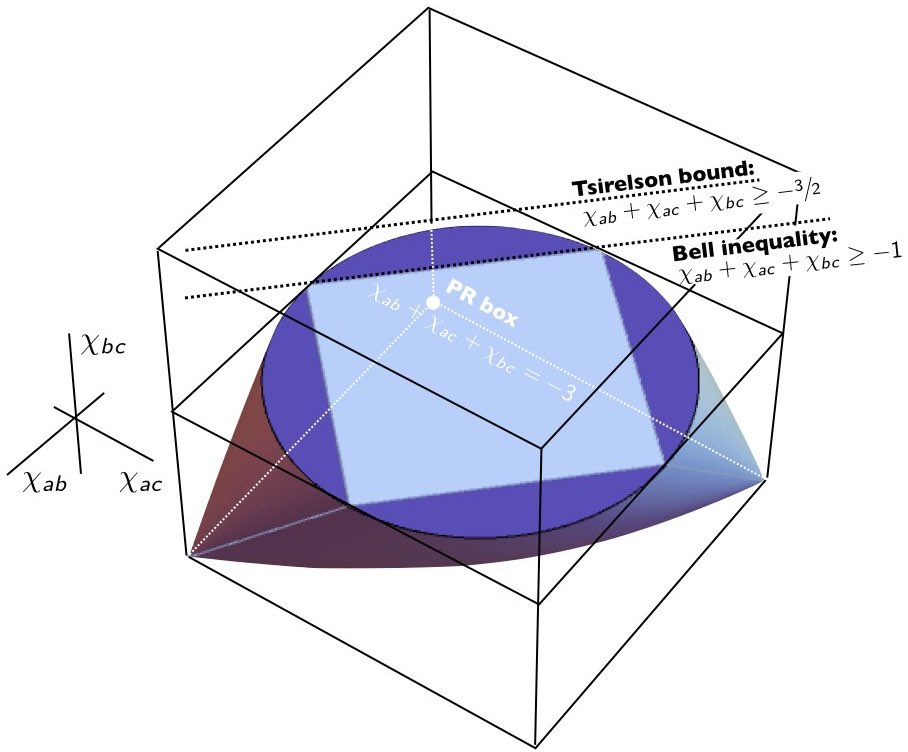}
\caption{Concrete version of the diagram in Figure \ref{LQP} for the correlations in the Mermin setup. The figure shows the cross-section $\chi_{bc} =0$ of the classical tetrahedron and the elliptope in a non-signaling cube in ordinary three-dimensional space (cf.\ Figures \ref{tetrahedron} and \ref{elliptope} below). See Sections  \ref{1.4}--\ref{1.5} for discussion of the two dotted lines representing two inequalities for the sum of the anti-correlation coefficients $\chi_{ab}$, $\chi_{ac}$ and $\chi_{bc}$. These inequalities are maximally violated in the point $(-1, -1, -1)$, which thus represents the PR box for this setup.}
   \label{elliptope-LQPslice}
\end{figure} 

Figure \ref{LQP} gives a schematic representation of the non-signaling polytope for such a setup. The outer square and everything inside of it (the non-signaling polytope $\mathcal{P}$) represents the set of all non-signaling correlations. The inner square and everything inside of it (the local polytope $\mathcal{L}$) represents the set of all non-signaling correlations allowed classically (i.e., by a local hidden-variable theory). The circle in between these two squares and everything inside of it (the quantum convex set $\mathcal{Q}$) represents the set of all correlations allowed quantum-mechanically. One of the facets of $\mathcal{L}$ represents a Bell inequality, a bound on the strength of the correlations allowed classically. The vertex of the non-signaling cube where this bound is maximally violated represents a PR box for the setup under consideration.

Figure \ref{CA-3set2out-Mermin} shows the correlation array for our version of Mermin's example of a quantum correlation violating a Bell inequality. We will refer to it as the \emph{Mermin correlation array}. Its nine cells form a $3 \times 3$ grid. The cells along the diagonal of this grid, when Alice and Bob peel the same way, show a perfect anti-correlation. The six off-diagonal cells, when Alice and Bob peel differently, all show the same imperfect positive correlation. It is easy to see that this correlation array is non-signaling: the entries in both rows and both columns of all nine cells add up to $\sfrac12$. Concisely put, this correlation (array) has \emph{uniform marginals}.

The Mermin correlation array in Figure \ref{CA-3set2out-Mermin} is a special case of the more general correlation array in Figure \ref{CA-3set2out-non-signaling-chis}. The three cells along the diagonal are the same, all showing a perfect anti-correlation (i.e., its diagonal elements are 0 and its off-diagonal elements are $\sfrac12$). Moreover, cells on opposite sides of the diagonal are the same. This correlation array can thus be parametrized by three anti-correlation coefficients of the kind introduced in Figure \ref{CA-2set2out-cell} and Eq.\ (\ref{chi as corr coef}).  In the specific example of the Mermin setup in Figure \ref{CA-3set2out-Mermin}, the three anti-correlation coefficients have the same value:  
\begin{equation}
\chi_{ab} = \chi_{ac} = \chi_{bc} = -\sfrac12.
\label{chi values Mermin example}
\end{equation}

The class of all non-signaling correlations in the Mermin setup can be visualized as a cube in ordinary three-dimensional space with the correlation coefficients, $\chi_{ab}$, $\chi_{ac}$ and $\chi_{bc}$, providing the three Cartesian coordinates of points in this cube. The non-signaling correlations allowed classically can be represented by a tetrahedron spanned by four of the eight vertices of this non-signaling cube (see Figure \ref{tetrahedron} in Section \ref{1.4}); those allowed quantum-mechanically by an elliptope enclosing this tetrahedron (see Figure \ref{elliptope} in Section \ref{1.5}). Figure \ref{elliptope-LQPslice} shows the cross-section $\chi_{bc} =0$ of this non-signaling cube, the classical tetrahedron and the elliptope. This cross-section has exactly the form of the cartoonish rendering in Figure \ref{LQP} of the Vitruvian-man-like structure of the local polytope $\mathcal{L}$ and the quantum convex set $\mathcal{Q}$ inside the non-signaling polytope $\mathcal{P}$. In the next two subsections, we will show in detail how one arrives at the classical tetrahedron and the quantum elliptope in the Mermin setup.

%SUBSECTION 2.4
\subsection{Classical polytopes and raffles to simulate quantum correlations} \label{1.4}

As \citet[p.\ 10]{Bub 2016} explains in the opening chapter of \emph{Bananaworld}, to decide whether or not some correlation array is allowed classically (or quantum-mechanically), he checks whether or not it can be simulated with classical (or quantum-mechanical) resources. Though we will use a more direct approach to find classes of correlations allowed quantum-mechanically (see Sections  \ref{1.5} and \ref{2.1}), we will adopt a variation on Bub's imitation game to find classes of correlations allowed classically (i.e., by some local hidden-variable theory). 

We will use special raffles to simulate the correlations found in our quantum banana peeling and tasting experiments. These raffles involve baskets of tickets such as the ones in Figure \ref{raffle-tickets-3set2out-i-thru-iv}. All tickets list the outcomes for both parties and for all settings in the setup under consideration. We randomly draw a ticket of the appropriate kind from a basket with many such tickets. We tear this ticket in half and randomly decide which side goes to Alice and which side goes to Bob. Alice and Bob then decide, randomly and independently of each other, which setting they will use. They record the outcome for that setting printed on their half of the ticket. We repeat this procedure a great many times. 

Raffles of this kind provide a criterion for determining whether or not a certain correlation is allowed classically:\footnote{In Section \ref{4} we will see that there is an extra bonus to discussing classical theory in terms of such raffles. It makes for a natural comparison between local hidden-variable theories and John von Neumann's (1927b) formulation of quantum theory in terms of statistical ensembles characterized by density operators on Hilbert space. \emph{Single-ticket raffles}, i.e., raffles with baskets of tickets that are all the same, are the classical analogues of pure states in quantum mechanics; \emph{mixed raffles}, i.e., raffles with baskets with different tickets, are the analogues of mixed states. By using the imagery of baskets with different mixes of tickets, we admittedly sweep a mathematical subtlety under the rug: the fractions of different types of tickets in a basket will always be rational numbers. To simulate the quantum correlations we are interested in, however, we need to allow fractions that are real numbers. In Section \ref{2.2.1} we will introduce a different mechanism for selecting tickets that gets around this problem (see Figure \ref{wheelsoffortune}). From a practical point of view, the restriction to rationals is harmless, since the rationals are dense in the reals.\label{dense}}
\begin{quote}
\emph{A correlation array is allowed by a local hidden-variable theory if and only if there is a raffle (i.e., a basket with the appropriate mix of tickets) with which we can simulate that correlation array following the protocol described above.}
\end{quote}

\begin{figure}[h]
 \centering
   \includegraphics[width=4.5in]{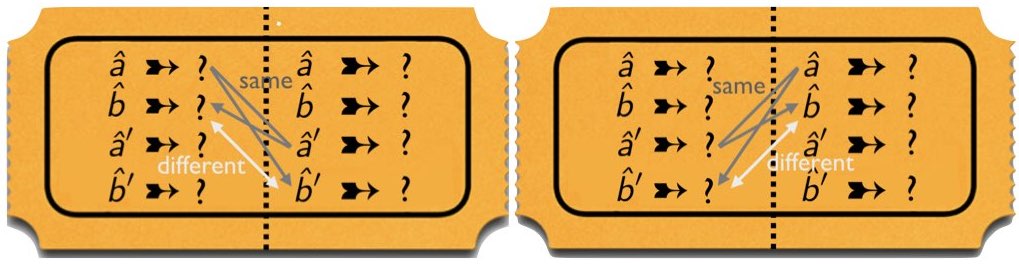} 
   \caption{Trying to design a raffle ticket for a PR box.}
   \label{raffle-ticket-PRbox}
\end{figure}

Invoking this criterion, we can easily show that a PR box with the correlation array in Figure \ref{CA-PRbox} is not allowed classically.\footnote{Essentially the same argument can already be found in \citet[p.\ 970]{Rastall 1995}.} These correlations place impossible demands on the design of the tickets for a raffle that would simulate them (see Figure \ref{raffle-ticket-PRbox}). The perfect positive correlation between the outcomes for three of the four possible combinations of settings ($\hat{a} \, \hat{a}'$, $\hat{a} \, \hat{b}'$ and $\hat{b} \, \hat{a}'$) requires that the outcomes printed on the ticket for $\hat{a}$ and $\hat{b}$ on one side are the same as the outcomes for $\hat{a}'$ and $\hat{b}'$ on the other side. That makes it impossible for the outcomes for $\hat{b}$ and $\hat{b}'$ on opposite sides of the ticket to be different as required by the perfect anti-correlation for the remaining combination of settings ($\hat{b} \, \hat{b}'$). 

\begin{figure}[h]
 \centering
   \includegraphics[width=4.5in]{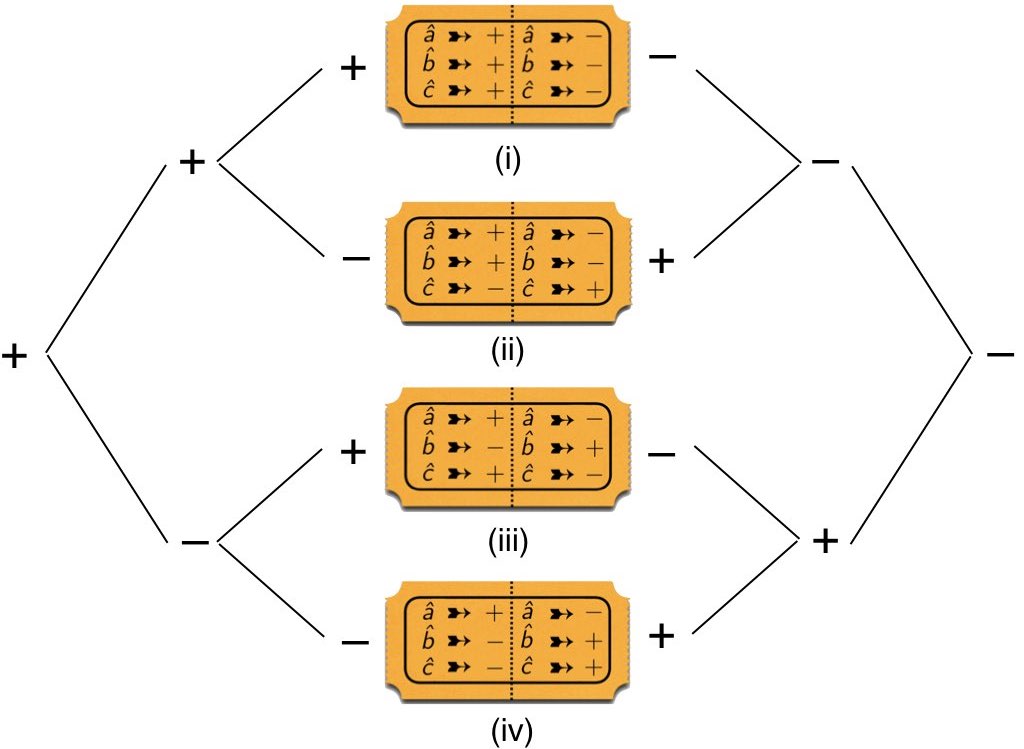} 
   \caption{The four different raffle tickets for three settings and two outcomes. Given the protocol of our raffles, two tickets that differ only in that their left and right sides are swapped are the same ticket.}
   \label{raffle-tickets-3set2out-i-thru-iv}
   \end{figure}

Figure \ref{raffle-tickets-3set2out-i-thru-iv} shows four different types of tickets, labeled (i) through (iv), for raffles meant to simulate correlations found in the Mermin setup in which Alice and Bob choose from the same three settings $(\hat{a}, \hat{b}, \hat{c})$ with two possible outcomes each $(+, -)$. Since in all setups that we will examine Alice and Bob find opposite results whenever they use the same setting, the outcomes on one side of the ticket dictate the outcomes on the other. That reduces the number of different ticket types to $2^3 = 8$. Given that it is decided randomly which side of a ticket goes to Alice and which side to Bob, two tickets that differ only in that the left and the right side are swapped are two equivalent versions of the same ticket type. This further reduces the number of different ticket types to four. As illustrated in Figure \ref{raffle-tickets-3set2out-i-thru-iv}, we chose the ones that have $+$ for the first setting ($\hat{a}$) on the left side of the ticket.

\begin{figure}[h]   
 \centering
   \includegraphics[width=4.5in]{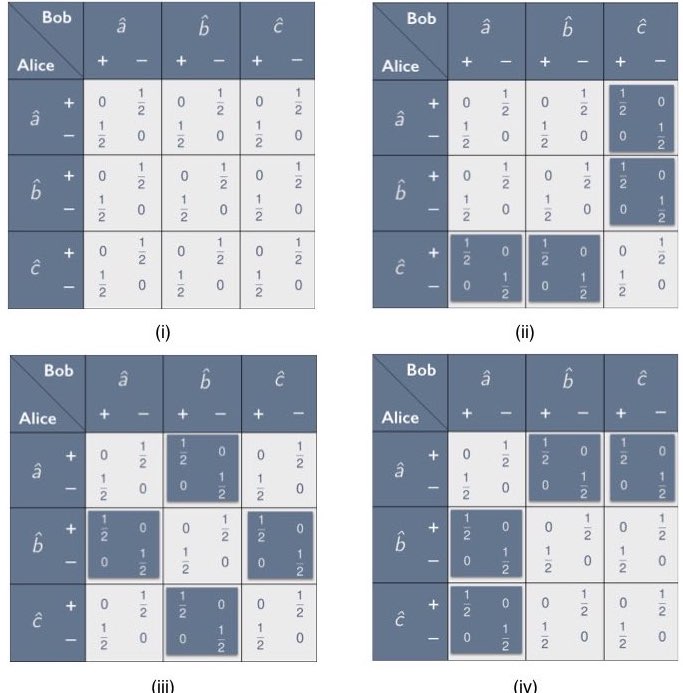} 
   \caption{Correlation arrays for raffles for the four different single-ticket raffles in the Mermin setup. In blue-on-white cells the outcomes are perfectly anti-correlated; in white-on-blue cells they are perfectly correlated.}
   \label{CA-3set2out-raffles-i-thru-iv}
\end{figure}

Figure \ref{CA-3set2out-raffles-i-thru-iv} shows the correlation arrays for raffles with baskets containing only one of the four types of tickets in Figure \ref{raffle-tickets-3set2out-i-thru-iv}. The design of our raffles guarantees that the correlations between the outcomes found by Alice and Bob are non-signaling. This is borne out by the correlation arrays in Figure \ref{CA-3set2out-raffles-i-thru-iv}. The entries in both rows and both columns of all cells in these correlation arrays add up to $\sfrac12$. In other words, these raffles all give uniform marginals. The design of our raffle tickets also guarantees that the outcomes found by Alice and Bob are \emph{balanced} (see the definition in the sentence following  Eq.\ (\ref{cov def 0})).

The entries of correlation arrays like those in Figure \ref{CA-3set2out-raffles-i-thru-iv} form $6 \times 6$ matrices. These matrices are symmetric. This is true both for single-ticket and mixed raffles. All raffles we will consider have this property. This too follows directly from the design of these raffles. It is simply because Alice and Bob are as likely to get the left or the right side of any ticket. 

\begin{figure}[h]   
 \centering
   \includegraphics[width=6in]{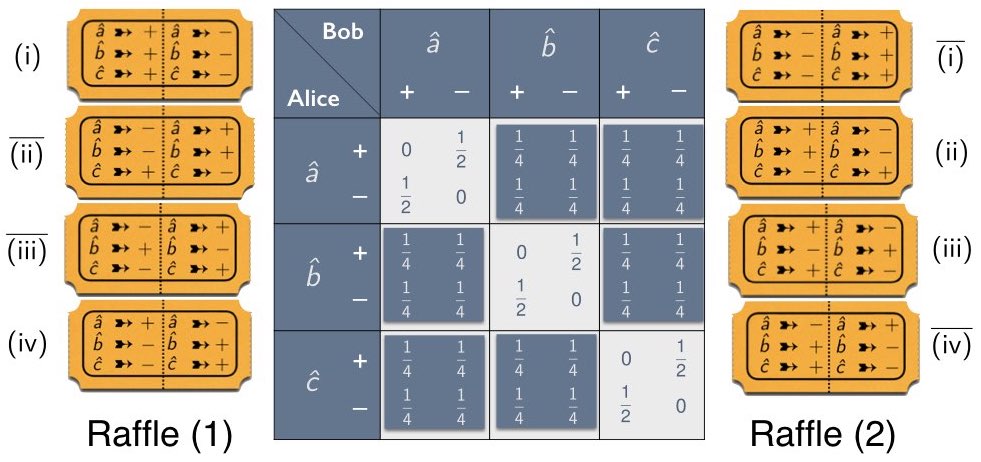} 
   \caption{Two raffles leading to the same correlation array (in blue-on-white cells the outcomes are perfectly anti-correlated; in white-on-blue cells they are completely uncorrelated). In both raffles, whenever a ticket is drawn, Alice gets the left and Bob gets the right side. In addition to tickets (i)--(iv) in Figure \ref{raffle-tickets-3set2out-i-thru-iv} we now have four more tickets, labeled $\overline{(\mathrm{i})}$-$\overline{(\mathrm{iv})}$ and obtained by switching the left and the right side of the tickets (i)--(iv).  Raffle (1) has equal numbers of tickets of type (i), $\overline{(\mathrm{ii})}$, $\overline{(\mathrm{iii})}$ and (iv). Raffle (2) has equal numbers of tickets of type $\overline{(\mathrm{i})}$, (ii), (iii) and $\overline{(\mathrm{iv})}$.}
   \label{CA-3set2out-raffle-25i25ii25iii25iv}
\end{figure}

Before we continue our analysis, we show that changing the protocol of our raffles so that Alice is always given the left side and Bob is always given the right side of any ticket does not give rise to correlation arrays with symmetric associated matrices that cannot be simulated with our more economical protocol---more economical because it requires fewer ticket types. For the alternative protocol, we need four more tickets, labeled $\overline{(\mathrm{i})}$ through $\overline{(\mathrm{iv})}$, that differ from their counterparts (i) through (iv) in that the left and right sides of the ticket have been swapped. Figure \ref{CA-3set2out-raffle-25i25ii25iii25iv} shows two raffles for this alternative protocol. Raffle (1) has equal numbers of tickets of type $\big\{ \mathrm{(i)}, \overline{(\mathrm{ii})}, \overline{(\mathrm{iii})}, \mathrm{(iv)} \big\}$. The matrix associated with the correlation array for this raffle is symmetric. That means that we get the same correlation array if we swap the left and the right sides of all tickets in raffle (1). This turns raffle (1) into raffle (2) with equal numbers of tickets of type $\big\{ \overline{(\mathrm{i})}, \mathrm{(ii)}, \mathrm{(iii)}, \overline{(\mathrm{iv})} \big\}$.  Any raffle mixing raffles (1) and (2) will also give that same correlation array. Consider the special case of a raffle with equal numbers of all eight tickets. This raffle is equivalent to a basket with equal numbers of tickets $\big\{  \mathrm{(i)},  \mathrm{(ii)},  \mathrm{(iii)},  \mathrm{(iv)} \big\}$ with the understanding that it is decided at random which side of the ticket goes to Alice and which side goes to Bob. This construction works for any correlation array with a symmetric associated matrix that we can produce using the protocol in which Alice always get the left side and Bob always get the right side of a ticket. We conclude that we can produce any such correlation array using our more economical protocol. 

\begin{figure}[h]
 \centering
   \includegraphics[width=6in]{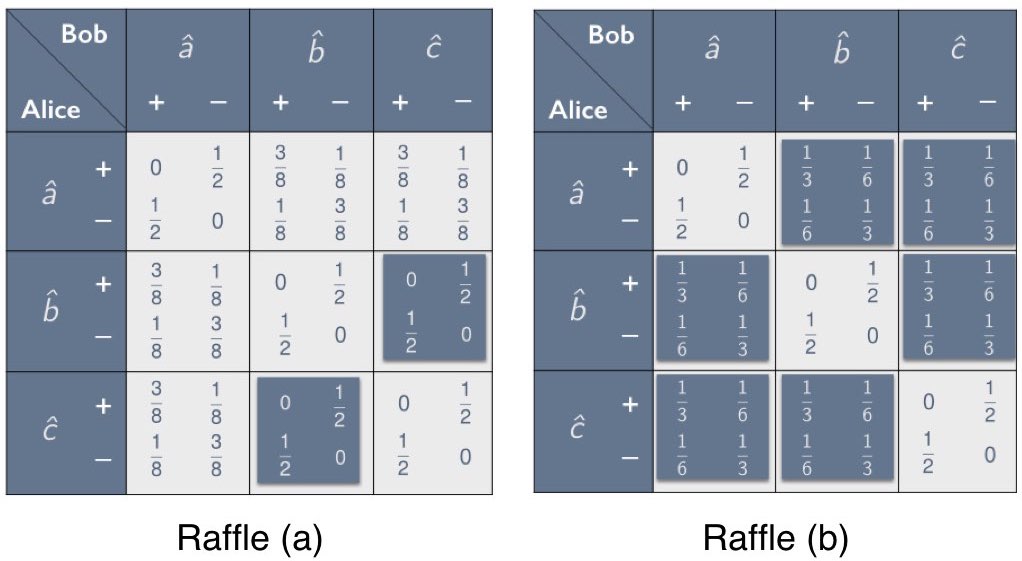} 
   \caption{Correlation arrays for raffles with different mixes of the four tickets in Figure \ref{raffle-tickets-3set2out-i-thru-iv}. Raffle (a) has 25\%  type-(i) tickets and 75\% type-(iv) tickets. Raffle (b) has 33\% each of type-(ii) through type-(iv) tickets. Blue-on-white cells are the same as the corresponding cells in the Mermin correlation array in Figure \ref{CA-3set2out-Mermin}, white-on-blue cells are different.}
   \label{CA-3set2out-raffle-mix}
\end{figure}

There is no mix of tickets (i) through (iv) in Figure \ref{raffle-tickets-3set2out-i-thru-iv} that produces a raffle that can simulate the Mermin correlation array in Figure \ref{CA-3set2out-Mermin}. Figure \ref{CA-3set2out-raffle-mix} shows the results of two unsuccessful attempts to produce one. In the first, we take a basket with 25\% tickets of type (i) and 75\% of type (iv). This results in correlation array (a) in Figure \ref{CA-3set2out-raffle-mix}. This raffle correctly simulates all but two cells of the Mermin correlation array. We get the same result if we replace tickets (iv) by tickets (ii) or (iii), the only difference being that now two other cells will differ from the corresponding ones in the Mermin correlation array. The best we can do overall is to take a basket with 33\% each of tickets (ii) through (iv). This results in correlation array (b) in Figure \ref{CA-3set2out-raffle-mix}. Like the Mermin correlation array we are trying to simulate, this one has the same positive correlation in all six off-diagonal cells but the correlation is weaker ($-\chi_{ab} = -\chi_{ac} = -\chi_{bc}  = \sfrac13$) than in the Mermin case ($-\chi_{ab} =-\chi_{ac}  = -\chi_{bc}  = \sfrac12$).

To prove that there is no raffle that can simulate the Mermin correlation array, we consider the sum $\chi_{ab} + \chi_{ac} + \chi_{bc}$ of the anti-correlation coefficients for a raffle. From the tickets in Figure \ref{raffle-tickets-3set2out-i-thru-iv} we can read off the values of $\chi_{ab}$, $\chi_{ac}$ and $\chi_{bc}$ for the four single-ticket raffles. These values are brought together in Table \ref{values of chi}. 

\begin{table}[h]
\centering
\begin{tabular}{|c||c|c|c|}
\hline
ticket & \quad $\chi_{ab}$ \quad & \quad $\chi_{ac}$ \quad & \quad $\chi_{bc}$ \quad \\[.1cm] 
\hline
 (i) & $+1$ & $+1$ & $+1$ \\[.2cm]
 (ii) & $+1$ & $-1$ & $-1$ \\[.2cm]
 (iii) & $-1$ & $+1$ & $-1$ \\[.2cm]
(iv) & $-1$ & $-1$ & $+1$ \\
 \hline
\end{tabular}
\caption{Values of the anti-correlation coefficients parametrizing the off-diagonal cells of the correlation arrays (i) through (iv) in Figure \ref{CA-3set2out-raffles-i-thru-iv} for single-ticket raffles with tickets (i) through (iv) in Figure \ref{raffle-tickets-3set2out-i-thru-iv}.}
\label{values of chi}
\end{table} 

The cells in the correlation arrays in Figure \ref{CA-3set2out-raffles-i-thru-iv} are all either perfectly anti-correlated or perfectly correlated. The anti-correlation coefficients for these single-ticket raffles can therefore only take on the values $\pm 1$ and their sum can only take on the value 3 (for a raffle with tickets of type (i) only) or $-1$ (for raffles with tickets (ii) or (iii) or (iv) only). For mixed raffles, $\chi_{ab} + \chi_{ac} + \chi_{bc}$ is the weighted average of the value of $\chi_{ab} + \chi_{ac} + \chi_{bc}$ for these four single-ticket raffles, with the weights given by the fractions of each of the four tickets in the raffle.\footnote{For a formal proof of this intuitively plausible result, see Section \ref{2.2.1}.} Hence, for any mix of tickets, this sum must lie between $-1$ and $3$:
\begin{equation}
-1 \le \chi_{ab} + \chi_{ac} + \chi_{bc} \le 3.
\label{Mermin inequality CHSH-like}
\end{equation}
The first of these inequalities, giving the lower bound on $\chi_{ab} + \chi_{ac} + \chi_{bc}$, is the analogue of the CHSH inequality for our variation of the Mermin setup. It is also the form in which \citet{Bell 1964} originally derived the Bell inequality. The CHSH-type Bell inequality is violated by the Mermin correlation array in Figure \ref{CA-3set2out-Mermin}. In that case, $\chi_{ab} = \chi_{ac} = \chi_{bc} = - \sfrac12$ (see Eq.\ (\ref{chi values Mermin example})) and their sum equals $-\sfrac32$. As we will see in Section \ref{1.5}, this is the maximum violation of this inequality allowed by quantum mechanics. Note that the absolute minimum value of $\chi_{ab} + \chi_{ac} + \chi_{bc}$ is $-3$. This value is allowed neither classically nor quantum-mechanically. It is the value reached with the (hypothetical) PR box for this setup.  

\begin{figure}[h]
 \centering
   \includegraphics[width=5in]{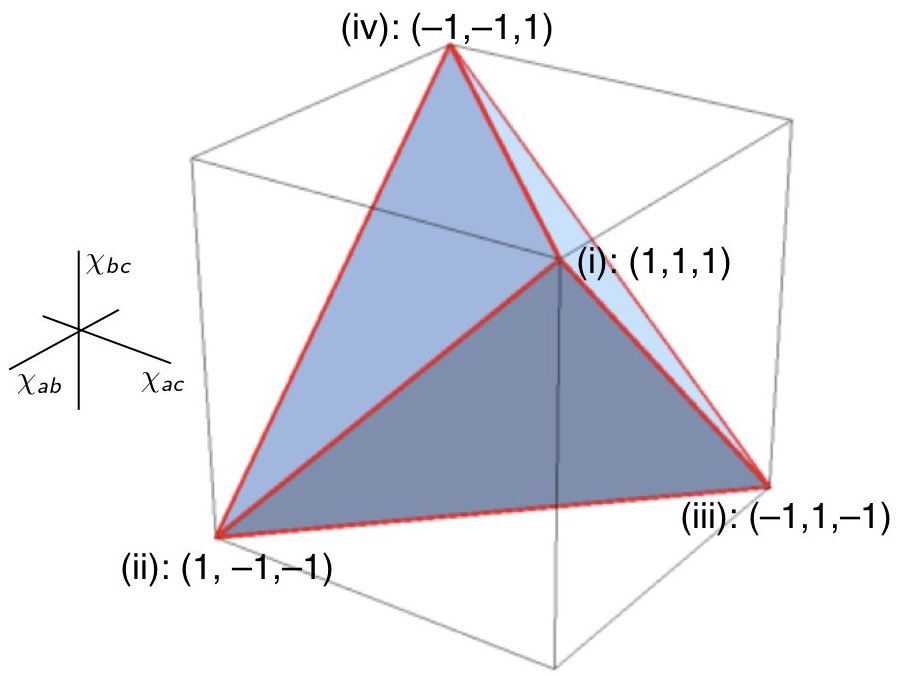} 
   \caption{Tetrahedron of triplets of anti-correlation coeffcients $(\chi_{ab}, \chi_{ac}, \chi_{bc})$ allowed by local hidden-variable theories in our version of the Mermin setup.}
   \label{tetrahedron}
\end{figure}

The values of $\chi_{ab}$,  $\chi_{ac}$ and $\chi_{bc}$ in Table \ref{values of chi} for tickets (i) through (iv) can be used as the Cartesian coordinates of four vertices in the non-signaling cube for the Mermin setup. These are the vertices labeled (i) through (iv) in Figure \ref{tetrahedron}. The vertex $(-1, -1, -1)$ represents the PR box for this setup (see Figure \ref{elliptope-LQPslice}). The vertices (i) through (iv) span a tetrahedron forming the convex set of all raffles that can be obtained by mixing the four types of tickets. The sum $\chi_{ab} + \chi_{ac} + \chi_{bc}$ takes on its maximum value of 3 at the vertex for tickets of type (i) and its minimum value of $-1$ for the facet spanned by the vertices for tickets of types (ii), (iii) and (iv). The inequalities in Eq.\ (\ref{Mermin inequality CHSH-like}) tell us that all correlations that can be simulated with raffles with various mixes of tickets must lie in the region of the non-signaling cube between the vertex (i) and the facet (ii)-(iii)-(iv). 

This is a necessary but not a sufficient condition for a correlation to be allowed by a local hidden-variable theory. As Figure \ref{tetrahedron} shows, there are three forbidden sub-regions in the region between vertex (i) and facet (ii)-(iii)-(iv). A full characterization of the class of correlations allowed classically requires three additional pairs of inequalities like the pair given in Eq.\ (\ref{Mermin inequality CHSH-like}), corresponding to the other three vertices and the other three facets of the tetrahedron. The following four pairs of inequalities do fully characterize the tetrahedron:
\begin{eqnarray}
-1 \le \;\, \chi_{ab} + \chi_{ac} + \chi_{bc} \; \le 3  & \!\!\!\! & \textrm{[between facet (ii)-(iii)-(iv) and vertex (i)]} 
\label{Mermin inequality CHSH-like (i)} \\[.4cm]
-1 \le \;\, \chi_{ab} - \chi_{ac} - \chi_{bc} \; \le 3  & \!\!\!\!  & \textrm{[between facet (i)-(iii)-(iv) and vertex (ii)]}  
\label{Mermin inequality CHSH-like (ii)} \\[.4cm]
-1 \le - \chi_{ab} + \chi_{ac} - \chi_{bc} \le 3 & \!\!\!\!  & \textrm{[between facet (i)-(ii)-(iv) and vertex (iii)]} 
\label{Mermin inequality CHSH-like (iii)} \\[.4cm]
-1 \le - \chi_{ab} - \chi_{ac} + \chi_{bc} \le 3 & \!\!\!\!   & \textrm{[between facet (i)-(ii)-(iii) and vertex (iv)]}.
\label{Mermin inequality CHSH-like (iv)}
\end{eqnarray}
Using the symmetries of the tetrahedron we can easily get from any one of these pairs of inequalities to another.  Another way to see this is to recall that the coordinates $(\chi_{ab}, \chi_{ac}, \chi_{bc})$ are anti-correlation coefficients for different combinations of the measurement settings $(\hat{a}, \hat{b}, \hat{c})$ and to look at what happens when we flip the sign of the outcomes for one of these three settings. If we do this for $\hat{a}$, $\chi_{ab}$ and $\chi_{ac}$ pick up a minus sign and Eq.\ (\ref{Mermin inequality CHSH-like (i)}) turns into Eq.\ (\ref{Mermin inequality CHSH-like (iv)}). If we do this for $\hat{b}$, $\chi_{ab}$ and $\chi_{bc}$ pick up a minus sign and Eq.\ (\ref{Mermin inequality CHSH-like (i)}) turns into Eq.\ (\ref{Mermin inequality CHSH-like (iii)}). Finally, if we do this for $\hat{c}$, $\chi_{ac}$ and $\chi_{bc}$ pick up a minus sign and Eq.\ (\ref{Mermin inequality CHSH-like (i)}) turns into Eq.\ (\ref{Mermin inequality CHSH-like (ii)}).

Mermin formulated a different inequality for this setup, one that implies the lower bound on the sum of anti-correlation coefficients in Eq.\ (\ref{Mermin inequality CHSH-like}) but requires an additional assumption. To derive Mermin's inequality, we have to assume that \emph{Alice and Bob randomly and independently of each other decide which setting to use in any run of the experiment} (whether with raffle tickets, spin-$\frac12$ particles, or quantum bananas). This provision is part of the protocol we described in Section \ref{1.1} but we had no need to invoke it so far. The CHSH-like inequality in Eq.\ (\ref{Mermin inequality CHSH-like}) could be derived without it---and so, for that matter, can the CHSH inequality itself. 

This means that we can test these inequalities without having to change the settings in every run. We can make measurements for one pair of settings at a time, providing data for the correlation array one cell at a time. This is how \citet{Clauser and Freedman 1972} originally tested the CHSH inequality. Changing the orientation of their polarizers was a cumbersome process.\footnote{For a drawing of their apparatus see \citet[p.\ 262]{Gilder 2008}. This drawing is based on a photograph that can be found, for instance, in \citet[p.\ 48]{Kaiser 2011}. For a schematic drawing of the apparatus, see  \citet[p.\ 939, Figure 1]{Clauser and Freedman 1972}.} Because of this limitation of their equipment, the violation of the CHSH inequality they found could conceivably be blamed on the two photons generated as an entangled pair ``knowing'' ahead of time (i.e., the moment they separated) what the orientation of the polarizers would be with which they were going to be measured. To close this loophole, the settings should only be chosen once the photons are in flight. This was accomplished by Aspect and his collaborators later in the 1970s and in the 1980s \citep[Ch.\ 31]{Gilder 2008}. In this paper, we will not be concerned with the extensive experimental efforts to close this and other loopholes.\footnote{David Kaiser alerted us to a paper written by 20 authors (with Kaiser, Alan Guth and Anton Zeilinger listed in 17th, 18th and 20th place, respectively) about one of the latest initiatives in this ongoing effort \citep{Handsteiner 2017}.} 

If we assume that Alice and Bob randomly and independently of each other decide which setting to use in each run,\footnote{We still do not need the stronger assumption that these decisions are made only after they receive their banana, their spin-$\frac12$ particle, or their ticket stub.} the nine possible combinations of settings are equiprobable. Following \citet[pp.\ 86--87]{Mermin 1981}, we ask for the probability, $\mathrm{Pr(opp)}$, that Alice and Bob find opposite results. Consider the Mermin correlation array in Figure \ref{CA-3set2out-Mermin}. For the cells along the diagonal $\mathrm{Pr(opp)} = 1$ (the results are perfectly anti-correlated). For the off-diagonal cells $\mathrm{Pr(opp)} = \sfrac14$, the sum of the off-diagonal entries in those cells. Alice and Bob use the same setting in one out of three runs and different settings in two out of three. Hence, the probability of them finding opposite results is:
\begin{equation}
\mathrm{Pr(opp)} = \sfrac13 \cdot 1 \, + \, \sfrac23 \cdot \sfrac14 = \sfrac12.
\label{Pr opp Mermin}
\end{equation}
Upon inspection of the four correlation arrays in Figure \ref{CA-3set2out-raffles-i-thru-iv}, however, we see that the minimum value for $\mathrm{Pr(opp)}$ in a local hidden variable theory is $\sfrac59$. In correlation array (i), the results in all nine cells are perfectly anti-correlated. In a single-ticket raffle with tickets of type (i), we thus have $\mathrm{Pr(opp)} = 1$.  In each of the other three correlation arrays, there are five cells in which the results are perfectly anti-correlated and four in which they are perfectly correlated. In single-ticket raffles with tickets of type (ii), (iii), or (iv), we thus have $\mathrm{Pr(opp)} = \sfrac59$. For an arbitrary mix of tickets (i) through (iv), we therefore have the inequality
\begin{equation}
\mathrm{Pr(opp)} \ge \sfrac59.
\label{Mermin inequality probs}
\end{equation}
This is the form in which Mermin states the Bell inequality for the setup we are considering. It implies the lower bound in Eq.\ (\ref{Mermin inequality CHSH-like}). Consider, once again, the general non-signaling correlation array in Figure \ref{CA-3set2out-non-signaling-chis} parametrized by the anti-correlation coefficients $\chi_{ab}$, $\chi_{ac}$ and  $\chi_{bc}$. Adding the off-diagonal elements in every cell and dividing by 9, as we are assuming that Alice and Bob use the settings of all nine cells with equal probability, we find
\begin{eqnarray}
\mathrm{Pr(opp)} & \!\! = \!\! & \sfrac39 \, + \, \sfrac29 \cdot \sfrac12 \, \Big(1+ \chi_{ab} \Big) \, + \, \sfrac29 \cdot \sfrac12 \, \Big(1+ \chi_{ac} \Big) \, + \,  \sfrac29 \cdot \sfrac12 \, \Big(1+ \chi_{bc} \Big) \nonumber \\[.2cm]
 & \!\! = \!\!  & \sfrac{2}{3} \, + \, \sfrac{1}{9} \, \Big(\chi_{ab} + \chi_{ac} + \chi_{bc} \Big). 
 \label{Pr opp general} 
\end{eqnarray}
If $\mathrm{Pr(opp)}$ must at least be $\sfrac59$, then $\chi_{ab} + \chi_{ac} + \chi_{bc}$ cannot be smaller than $-1$. Conversely, if $\chi_{ab} + \chi_{ac} + \chi_{bc} \ge -1$ \emph{and} all nine combinations of the settings $\hat{a}$, $\hat{b}$ and $\hat{c}$ are equiprobable, then 
$\mathrm{Pr(opp)} \ge \sfrac59$. 
 
Mermin's lower bound on the probability of finding opposite results may be easier to grasp for a general audience than a lower bound on a sum of expectation values. The latter, however, does have its own advantages. First, as we just saw, it can be derived from weaker premises. Second, it immediately generates inequalities corresponding to other facets of the polyhedron of classically allowed correlations in the Mermin setup (see Eqs. (\ref{Mermin inequality CHSH-like (i)})--(\ref{Mermin inequality CHSH-like (iv)})). Third, as we will show in detail in Section \ref{3}, it makes it easier to see the connection with the CHSH inequality.

%SECTION 2.5
%!TEX root =  ./JanasJanssenCuffaro-August2019.tex

%SUBSECTION 2.5
\subsection{The singlet state, the Born rule and the elliptope} \label{1.5}

The correlation array in Figure \ref{CA-3set2out-Mermin} for our variation of the Mermin setup  can readily be produced in our quantum banana peeling and tasting experiment. In the end, the bananas, the peelings and the tastes are window-dressing for spin-$\frac12$ particles, measurements of their spin with Dubois magnets and outcomes of such measurements. Picking a pair of quantum bananas from our quantum-banana tree corresponds to preparing two spin-$\frac12$ particles in the singlet state, so called because its overall spin is zero. In the Dirac notation used from here on out we will designate this state as $|0, 0\rangle_{12}$ (see Eq.\ (\ref{singlet state s=1/2 simple}) below and Eq.\ (\ref{singlet state}) and note \ref{singlet note} in Section \ref{2.1}). The different peeling directions, $\vec{e}_a$, $\vec{e}_b$ and $\vec{e}_c$, correspond to different orientations of the axis of the Dubois magnets. The tastes, `yummy' and `nasty', corresponds to the spin values `up' and `down'. 

\begin{figure}[h!]
 \centering
   \includegraphics[width=3.5in]{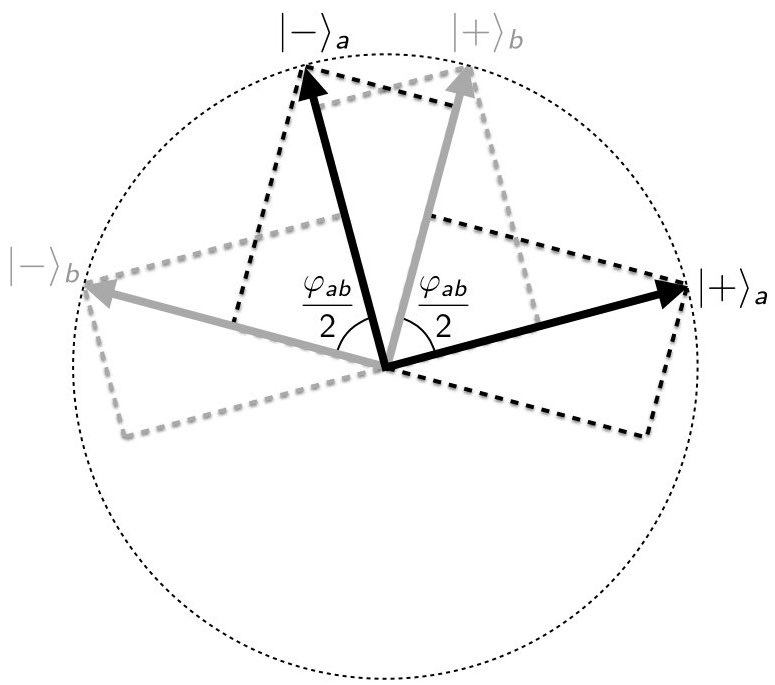} 
   \caption{Eigenvectors for the operators $\hat{S}_a$ and $\hat{S}_b$ acting on a one-banana Hilbert space and representing the tastes of the bananas when peeled being held in the directions $\vec{e}_a$ and $\vec{e}_b$, respectively.}
   \label{vectors}
\end{figure}

Sticking with the conceit of quantum bananas for now, we can say that the state $|0, 0\rangle_{12}$ is a vector in the Hilbert space for two bananas. This Hilbert space is the tensor product of two (identical) Hilbert spaces for one banana. The one-banana Hilbert space can be represented by the set of all vectors lying in the same plane and originating in the same point (see Figure \ref{vectors}). The properties `taste when peeled while being held in the direction of the unit vector $\vec{e}_a$' and `taste when peeled while being held in the direction of the unit vector $\vec{e}_b$' are represented by two operators, $\hat{S}_a$ and $\hat{S}_b$, acting in one-banana Hilbert space, respectively. Let $\{ | + \rangle_a, | - \rangle_a \}$ and  $\{ | + \rangle_b, | - \rangle_b \}$ be unit vectors that are eigenvectors of these operators with eigenvalues $\pm \bbar/2$:
\begin{equation}
\begin{array}{ccc}
\hat{S}_a \, | + \rangle_a = \displaystyle{\frac{\bbar}{2}} \, | + \rangle_a, & & \hat{S}_b \, | + \rangle_b = \displaystyle{\frac{\bbar}{2}}  \, | + \rangle_b,   \\[.4cm]
\hat{S}_a \, | - \rangle_a = - \displaystyle{\frac{\bbar}{2}}  \, | - \rangle_a, & \quad & \hat{S}_b \, | - \rangle_b = - \displaystyle{\frac{\bbar}{2}}  \, | - \rangle_b.   
\end{array}
\label{eigenvectors}
\end{equation}
Both sets of eigenvectors form an orthonormal basis for the one-banana Hilbert space. These eigenvectors are shown in Figure  \ref{vectors}. With malice aforethought, we have chosen the angle between  $|+\rangle_a$ and $|+\rangle_b$ to be half the angle $\varphi_{ab}$ between the unit vectors $\vec{e}_a$ and $\vec{e}_b$ in ordinary three-dimensional space.  

Using the orthonormal basis $\{ |+ \rangle_a, |- \rangle_a \}$ of one-banana Hilbert space, we can write the singlet state $| 0, 0 \rangle_{12}$  in two-banana Hilbert space as  
 \begin{equation}
| 0, 0 \rangle_{12} = \frac{1}{\sqrt{2}} \Big( |+ \rangle_{1a} |- \rangle_{2a} \, - \; |-  \rangle_{1a} | + \rangle_{2a} \Big)
\label{singlet state s=1/2 simple}
\end{equation}
(where the subscripts 1 and 2 refer to the bananas given to Alice and Bob, respectively). This is an \emph{entangled state}: the state vector for the composite system in the two-banana Hilbert space cannot be written as a tensor product of state vectors for its components in the two one-banana Hilbert spaces.

As we will see in Section \ref{2.1}, the singlet state, not just for a pair of spin-$\frac12$ particles but for a  pair of particles of arbitrary spin, is invariant under rotation. This means that $|0, 0 \rangle_{12}$ has the same form regardless of whether we use $\{ |+ \rangle_a, |- \rangle_a \}$ or $\{ |+ \rangle_b, |- \rangle_b \}$ as our orthonormal basis for the one-banana Hilbert space. Here we provide an intuitive proof of this property in the spin-$\frac12$ case. In Section \ref{2.1}, we prove this more rigorously for arbitrary spin.

Upon inspection of Figure \ref{vectors}, one sees that\footnote{What makes the proof in this section intuitive and dubious at the same time is that we take the coefficients in Eq.\ (\ref{QM2}) to be real whereas Hilbert space is a vector space over the complex rather than the real numbers. In Section \ref{2.1} (see Eqs.\ (\ref{Pauli matrices})--(\ref{b +/- trans law 2})), we will show that we can always write the eigenvectors of one spin operator as linear combinations \emph{with real coefficients} of the eigenvectors of another spin operator.}
\begin{equation}
\begin{array}{c}
|+ \rangle_a = \cos{\! \left( {\displaystyle \frac{\varphi_{ab}}{2}} \right)} \, |+ \rangle_b - \sin{\! \left( {\displaystyle \frac{\varphi_{ab}}{2}} \right)} \, |- \rangle_b, \\[.6cm]
|- \rangle_a = \sin{\! \left( {\displaystyle \frac{\varphi_{ab}}{2}} \right)} \, |+ \rangle_b + \cos{\! \left( {\displaystyle \frac{\varphi_{ab}}{2}} \right)} \, |- \rangle_b.
\end{array}
\label{QM2}
\end{equation}
Inserting these expressions into Eq.\ (\ref{singlet state s=1/2 simple}), we arrive at
\begin{eqnarray}
|0, 0 \rangle_{12} & \! \!\! = \! \!\! & \frac{1}{\sqrt{2}} \Big\{ \! \left(  \cos{\! \left( \frac{\varphi_{ab}}{2} \right)}  |+ \rangle_{1b} - \sin{\! \left(  \frac{\varphi_{ab}}{2} \right)} |- \rangle_{1b} \! \right) 
\!\! \left(  \sin{\! \left(  \frac{\varphi_{ab}}{2}  \right)} |+ \rangle_{2b} + \cos{\! \left(  \frac{\varphi_{ab}}{2} \right)} |- \rangle_{2b} \!  \right) 
 \nonumber \\[.3cm]
  &  & \; - \left( \sin{\! \left(  \frac{\varphi_{ab}}{2}  \right)}  |+ \rangle_{1b} + \cos{\! \left(  \frac{\varphi_{ab}}{2}  \right)} |- \rangle_{1b} \right) 
 \!\! \left( \cos{\! \left(  \frac{\varphi_{ab}}{2}  \right)} |+ \rangle_{2b} - \sin{\! \left(  \frac{\varphi_{ab}}{2}  \right)} |- \rangle_{2b}  \right) \! \Big\}.  
 \nonumber
 \end{eqnarray}
Terms with $\sin{\!\left( \varphi_{ab}/{2} \right)} \cos{\!\left( \varphi_{ab}/{2} \right)}$ in this expression cancel; terms with $\cos^2{\!\left( \varphi_{ab}/{2} \right)}$ and $\sin^2{\!\left( \varphi_{ab}/{2} \right)}$ add up to:
\begin{equation}
|0, 0 \rangle_{12} = \frac{1}{\sqrt{2}} \Big( |+ \rangle_{1b} |- \rangle_{2b}  - |-  \rangle_{1b} | + \rangle_{2b} \Big), 
\label{QM3}
\end{equation}
 which has the exact same form as Eq.\ (\ref{singlet state s=1/2 simple}).   

To find the probabilities given by quantum mechanics for the four possible combinations of tastes found when Alice uses peeling direction $\vec{e}_a$  and Bob uses peeling direction $\vec{e}_b$, we need to write out $| 0, 0 \rangle_{12}$ in components with respect to the orthonormal basis:
\begin{equation}
\Big\{ |+ \rangle_{1a} |+ \rangle_{2b}, \; |+ \rangle_{1a} |- \rangle_{2b}, \; | - \rangle_{1a} | + \rangle_{2b}, \; |- \rangle_{1a} |- \rangle_{2b} \Big\}.
\label{QM4}
\end{equation}
These four basis vectors correspond to the four combinations of tastes Alice and Bob can find upon peeling their bananas. Using Eq.\ (\ref{QM2}) to express $|+ \rangle_{2a}$ and $|- \rangle_{2a}$ in Eq.\ (\ref{singlet state s=1/2 simple}) in terms of $|+ \rangle_{2b}$ and $|- \rangle_{2b}$, we find the singlet state in this new orthonormal basis:
\begin{eqnarray}
|0, 0 \rangle_{12}  & \! \! = \! \! & \frac{1}{\sqrt{2}} \Big( \sin{\! \left( \frac{\varphi_{ab}}{2} \right)} \, |+ \rangle_{1a}  |+ \rangle_{2b}  
\; + \;   \cos{\! \left( \frac{\varphi_{ab}}{2} \right)} \, |+ \rangle_{1a} |- \rangle_{2b} \nonumber \\
 &  & \quad \quad \quad \; - \;   \cos{\! \left( \frac{\varphi_{ab}}{2} \right)} \, | -  \rangle_{1a} | + \rangle_{2b}  
\; + \;   \sin{\! \left( \frac{\varphi_{ab}}{2} \right)} \, |-  \rangle_{1a} |- \rangle_{2b} \Big).
\label{expansion in ab}
\end{eqnarray}
We now use the \emph{Born rule}, the basic rule for probabilities in quantum mechanics, which in this case says that, when Alice peels $\hat{a}$ and Bob peels $\hat{b}$, the probabilities of them finding the combination of tastes `$++$', `$+-$', `$-+$' and `$--$', respectively, are the squares of the coefficients of the corresponding terms in the expansion of $| 0, 0 \rangle_{12}$ in Eq.\ (\ref{expansion in ab}).  It does not matter whose banana is peeled first (cf.\ note \ref{peeling order irrelevant}).

To find the probabilities for combinations of peeling directions other than $(\vec{e}_a, \vec{e}_b)$, we simply relabel $a$ and $b$ in Eq.\ (\ref{expansion in ab}) accordingly.  We thus see immediately that quantum mechanics correctly reproduces  the main features of the statistics of our banana-tasting experiment. If $ab$ is replaced by $aa$, $\varphi_{ab} = \varphi_{aa} = 0$ and Eq.\ (\ref{expansion in ab}) reduces to Eq.\ (\ref{singlet state s=1/2 simple}). If $ab$ is replaced by $bb$, Eq.\ (\ref{expansion in ab}) similarly reduces to Eq.\ (\ref{QM3}). Changing $a$ to $c$ in Eq.\ (\ref{singlet state s=1/2 simple}) or $b$ to $c$ in Eq.\ (\ref{QM3}), we likewise find the expansion of $|0, 0\rangle_{12}$ using the orthonormal basis $\{ | + \rangle_c,  | - \rangle_c \}$ of the one-banana Hilbert space. Applying the Born rule if the peeling combinations are $\hat{a}_A\hat{a}_B$, $\hat{b}_A\hat{b}_B$, or $\hat{c}_A\hat{c}_B$, we thus recover the perfect anti-correlation between the outcomes found by Alice and Bob whenever they peel their bananas the same way. Since $\varphi_{ab} = \varphi_{ba}$, $\varphi_{ac} = \varphi_{ca}$ and $\varphi_{bc} = \varphi_{cb}$, the Born rule also correctly predicts that having Alice and Bob swap peeling directions does not affect the probabilities of the various combinations of tastes. 

Thinking in terms of spin-$\frac12$ particles rather than bananas for a moment, we can now also present an intuitive argument as to why the angle between, say, $|+\rangle_a$ and $|+\rangle_b$ in Hilbert space is half the angle between the directions $\vec{e}_a$ and $\vec{e}_b$ in real space. Imagine we place two Dubois magnets in a beam of spin-$\frac12$ particles, one right after the other, with the second one rotated $180^{\mathrm o}$ with respect to the first. In this setup we would only find two possible outcomes $(+_1, -_2)$ and $(-_1, +_2)$ (where $\pm$ refers to spin up/down and 1 and 2 refer to the two Dubois magnets). The probability of finding $(+_1, +_2)$ and $(-_1, -_2)$, in other words, vanishes. For the Born rule to reproduce this result, the angle between the eigenvectors $|+\rangle_a$ and $|+\rangle_b$ should be $90^{\mathrm o}$ if the angle between the vectors $\vec{e}_a$ and $\vec{e}_b$ specifying two orientations of the Dubois magnet is $180^{\mathrm o}$. In Section 2.1, we will give a more general derivation of the relation between these two angles in the spin-$\frac12$ case (see Eqs.\ (\ref{rot proof 1})--(\ref{b +/- trans law 2})).

This argument, incidentally, reveals the limitations of the banana metaphor in two ways. First, we can only peel and taste a banana once. As Sandu \citet[p.\ vi]{Popescu 2016} put it in his foreword to \emph{Bananaworld}: ``once it's peeled it's peeled \ldots\ once it's eaten it's eaten.'' Secondly, there is an intrinsic difference between the tastes `yummy' and `nasty' of a banana (or, for that matter, between `heads' and `tails' of a quoin) whereas there is no intrinsic difference between `spin up' and `spin down'. `Spin up' and `spin down' are defined with respect some preferred axis, given, for instance, by the orientation of a Dubois magnet. 

This second complication also provides a simple argument for why we should not expect to succeed, using only elementary quantum systems as our components, in building a device such as the Superquantum Entangler PR01 of \citet{Bub and Bub 2018} realizing the PR box with the correlation array in Figure \ref{CA-PRbox}. Recall that the cells along the diagonal of the correlation array for this PR box are different. Now one can easily imagine that tossing two coins initially facing heads and tossing two coins initially facing tails would give different results. But this cannot be true for the spin-components of spin-$\frac12$ particles in the singlet state. Suppose we measure the spin-in-the-$z$-direction of both particles with two Dubois magnets. Spherical symmetry requires that the statistics for this experiment do not change if we rotate both Dubois magnets to measure, say, spin-in-the-$x$-direction. Measurements on the singlet state of two spin-$\frac12$ particles can thus not be used to produce a PR box with the correlation array in Figure \ref{CA-PRbox}. We will see below that they also cannot be used to produce a PR box for the Mermin setup with a correlation array represented by the point $\chi_{ab} = \chi_{ac} = \chi_{bc} = -1$ in Figure \ref{elliptope-LQPslice}.

\begin{figure}[h]
 \centering
   \includegraphics[width=5.5in]{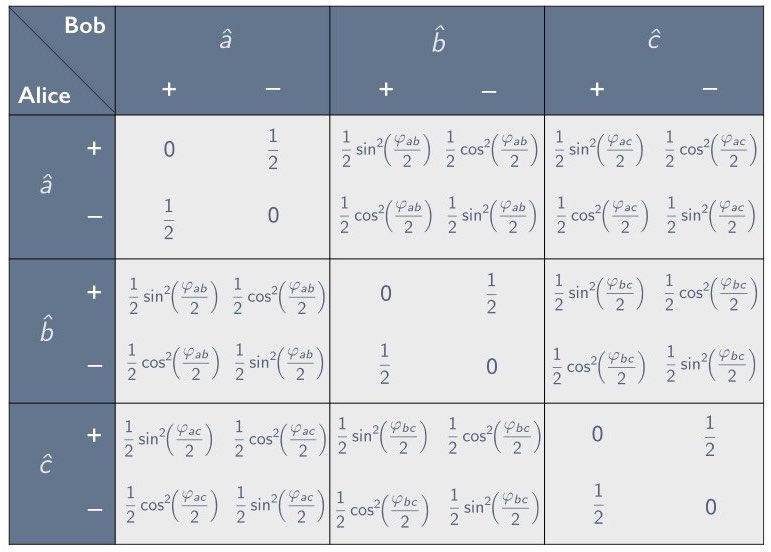} 
   \caption{Correlation array given by quantum mechanics for two parties, one using settings $\hat{a}$ and one using setting $\hat{b}$, to perform a measurement on two spin-$\frac12$ particles in the singlet state.}
   \label{CA-3set2out-non-signaling-halfangles}
\end{figure}

With the help of Eq.\ (\ref{expansion in ab}) and the Born rule, we can fill out the correlation array for the Mermin setup. The result is shown in Figure \ref{CA-3set2out-non-signaling-halfangles}.  Note that the rows and columns of all cells add up to $\sfrac12$. The correlation array thus has uniform marginals. As we saw in Section \ref{1.2}, this is a sufficient condition for it to be non-signaling. 

In Mermin's setup (see Figure \ref{AliceBob-Mermin}), the peeling directions $\vec{e}_a$, $\vec{e}_b$ and  $\vec{e}_c$ corresponding to the settings $\hat{a}$, $\hat{b}$ and $\hat{c}$ are such that $\varphi_{ab} = \varphi_{ac} = \varphi_{bc} = 120\degree$. Inserting
\begin{equation}
\frac12 \sin^2{\! \left( \frac{\varphi_{ab}}{2}\right)} = \frac12 \sin^2{\! 60\degree} = \frac{3}{8}, 
\quad \frac12 \cos^2{\! \left( \frac{\varphi_{ab}}{2}\right)}  = \frac12 \cos^2{\! 60\degree} = \frac{1}{8}, 
\quad \mathrm{etc.,}
\end{equation}
in the correlation array in Figure \ref{CA-3set2out-non-signaling-halfangles}, we recover the Mermin correlation array in Figure \ref{CA-3set2out-Mermin}.

Using the trigonometric identities
\begin{equation}
\cos{\alpha} = \cos^2{\!\left( \frac{\alpha}{2} \right)} - \sin^2{\!\left( \frac{\alpha}{2} \right)} = 2 \cos^2{\!\left( \frac{\alpha}{2} \right)} -1 = 1 - 2 \sin^2{\!\left( \frac{\alpha}{2} \right)},
\label{trigonometry}
\end{equation}
we can replace the squares of sines and cosines of half the angles between peeling directions by cosines of the full angle. For instance,  
\begin{equation}
\frac12 \sin^2{\!\left(\frac{\varphi_{ab}}{2}\right)} = \frac{1}{4} \left( 1 - \cos{ \varphi_{ab}} \right), \quad
\frac12 \cos^2{\!\left(\frac{\varphi_{ab}}{2}\right)} = \frac{1}{4} \left( 1 + \cos{ \varphi_{ab}} \right).
\label{angle2halfangle}
\end{equation}
Comparison with the way we wrote the entries of a cell in a non-signaling array in Figure \ref{CA-2set2out-cell} then leads us to identify anti-correlation coefficients with these cosines:
\begin{equation}
\chi_{ab} = \cos{\varphi_{ab}}, \quad \chi_{ac} = \cos{\varphi_{ac}}, \quad \chi_{bc} = \cos{\varphi_{bc}}.
\label{chi values repeat}
\end{equation}

We can verify directly that $\chi_{ab} = \cos{\varphi_{ab}}$ by evaluating the expectation value $\langle \hat{S}_{1a} \, \hat{S}_{2b} \rangle_{00}$ of the product of the outcomes found by Alice and Bob when they use settings $\hat{a}$ and $\hat{b}$, respectively (cf.\ Eq.\ (\ref{prob 2 exp}); the subscript $00$ indicates that this expectation values is for systems in the singlet state $|0, 0 \rangle_{12}$). Recalling that the taste of a banana is $\pm \bbar/2$ and using the probabilities in the correlation array in Figure \ref{CA-3set2out-non-signaling-halfangles}, we find
\begin{eqnarray}
\langle \hat{S}_{1a} \, \hat{S}_{2b} \rangle_{00} & \!\! = \!\! & \frac{\bbar^2}{4} \left(\mathrm{Pr}(+\!+| \hat{a} \,\hat{b}) \, + \, \mathrm{Pr}(-\!-| \hat{a} \,\hat{b})\right) 
- \frac{\bbar^2}{4} \left(\mathrm{Pr}(+\!-| \hat{a} \,\hat{b}) \, + \, \mathrm{Pr}(-\!+| \hat{a} \,\hat{b})\right) \nonumber \\
&  \!\! = \!\! & \frac{\bbar^2}{4} \left(  \sin^2{\!\left(\frac{\varphi_{ab}}{2}\right)} - \cos^2{\!\left(\frac{\varphi_{ab}}{2}\right)} \right) \; = \; -\frac{\bbar^2}{4}\cos{\varphi_{ab}}.
\label{prob 2 exp b}
\end{eqnarray}
Using the standard deviations 
\begin{equation}
\sigma_{1a} = \sqrt{\langle \hat{S}^2_{1a} \rangle} = \bbar/2, \quad \sigma_{2b} = \sqrt{\langle \hat{S}^2_{2b} \rangle} = \bbar/2
\label{standard deviations a b}
\end{equation}
(cf.\ Eq.\ (\ref{standard deviations a and b})) and Eq.\ (\ref{chi as corr coef}) for $\chi_{ab}$, we find that 
\begin{equation}
\chi_{ab} \equiv  - \frac{\langle \hat{S}_{1a} \, \hat{S}_{2b} \rangle_{00}}{\sigma_{1a} \sigma_{2b}}  = \cos{\varphi_{ab}}. 
\label{chi2angle}
\end{equation}    

If the angles $\varphi_{ab}$, $\varphi_{ac}$ and $\varphi_{bc}$ could be chosen independently of one another the class of non-signaling correlation in our banana-tasting experiment allowed by quantum mechanics would saturate the non-signaling cube and we could realize the PR box represented by $\chi_{ab} = \chi_{ac} = \chi_{bc} = -1$. These angles, however, cannot be chosen independently of one another. 

To derive the constraint on the angles $\varphi_{ab}$, $\varphi_{ac}$ and $\varphi_{bc}$ , we introduce the \emph{matrix of anti-correlation coefficients} or \emph{anti-correlation matrix} $\chi$ for our version of the Mermin setup:
\begin{equation}
\chi
\equiv  
\begin{pmatrix}
\; 1 \; & \; \chi_{ab} \; & \;  \chi_{ac} \; \\[.2cm]
\; \chi_{ab} & \; 1 \; & \;  \chi_{bc} \; \\[.2cm]
 \; \chi_{ac} \; & \; \chi_{bc} \; & \;  1 
\end{pmatrix}
=
\begin{pmatrix}
\cos{\varphi_{aa}} & \cos{\varphi_{ab}} & \cos{\varphi_{ac}} \\[.2cm]
\cos{\varphi_{ab}} & \cos{\varphi_{bb}} & \cos{\varphi_{bc}} \\[.2cm]
\cos{\varphi_{ac}} & \cos{\varphi_{bc}} & \cos{\varphi_{cc}} 
\end{pmatrix}.
\label{chi matrix}
\end{equation}
Since the anti-correlation coefficients are given by these cosines, they can be written as inner products of the unit vectors in the associated peeling directions:
\begin{equation}
\chi_{ab} = \cos{\varphi_{ab}} = \vec{e}_a \cdot \vec{e}_b, \quad
\chi_{ac} = \cos{\varphi_{ac}} = \vec{e}_a \cdot \vec{e}_c, \quad
\chi_{bc} = \cos{\varphi_{bc}} = \vec{e}_b \cdot \vec{e}_c. \quad
\label{chis2angles2innerproducts}
\end{equation}
Such a matrix of inner products is called a \emph{Gram matrix}. Writing the components of the three unit vectors as
\begin{equation}
\vec{e}_a = (a_x, a_y, a_z),  \quad \vec{e}_b = (b_x, b_y, b_z), \quad \vec{e}_c = (c_x, c_y, c_z),
\label{comps of unit vectors e_abc}
\end{equation}
we can write this Gram matrix as
\begin{equation}
\chi
= 
\begin{pmatrix}
\vec{e}_a \! \cdot  \vec{e}_a &  \vec{e}_a \! \cdot  \vec{e}_b  &   \vec{e}_a \! \cdot  \vec{e}_c  \\[.2cm]
\vec{e}_b \! \cdot  \vec{e}_a & \vec{e}_b \! \cdot  \vec{e}_b & \vec{e}_b \! \cdot  \vec{e}_c \\[.2cm]
\vec{e}_c \! \cdot  \vec{e}_a & \vec{e}_c \! \cdot  \vec{e}_b  & \vec{e}_c \! \cdot  \vec{e}_c 
\end{pmatrix}
=
\begin{pmatrix}
\; a_x \; & \; a_y \; & \;  a_z \; \\[.2cm]
\; b_x \; & \; b_y \; & \;  b_z \; \\[.2cm]
 \; c_x \; & \; c_y \; & \;  c_z 
\end{pmatrix}
\begin{pmatrix}
\; a_x \; & \; b_x \; & \;  c_x \; \\[.2cm]
\; a_y \; & \; b_y \; & \;  c_y \; \\[.2cm] 
 \; a_z \; & \; b_z \; & \;  c_x 
\end{pmatrix}.
\label{QM10}
\end{equation}
Introducing
\begin{equation}
L \equiv
\begin{pmatrix}
\; a_x \; & \; b_x \; & \;  c_x \; \\[.2cm]
\; a_y \; & \; b_y \; & \;  c_y \; \\[.2cm]
 \; a_z \; & \; b_z \; & \;  c_x 
\end{pmatrix},
\label{QM11}
\end{equation}
we can write Eq.\ (\ref{QM10}) more compactly as
\begin{equation}
\chi = L^\top L.
\label{QM12}
\end{equation}
It follows that the determinant of $\chi$ is non-negative: 
%\citep[p.\ 515]{Deza and Laurent 1997}:
\begin{equation}
\det{\chi} = \det{\!(L^\top L)} = (\det{L^\top})(\det{L}) = (\det{L})^2 \ge 0. 
\label{QM13}
\end{equation}
Using Eq.\ (\ref{chi matrix}) to evaluate the determinant of $\det{\chi}$, we can rewrite this condition as
\begin{equation}
1 - \chi_{ab}^2 - \chi_{ac}^2 - \chi_{bc}^2 + 2 \, \chi_{ab} \, \chi_{ac} \, \chi_{bc} \ge 0.
\label{QM14}
\end{equation}

\begin{figure}[h]
 \centering
   \includegraphics[width=5in]{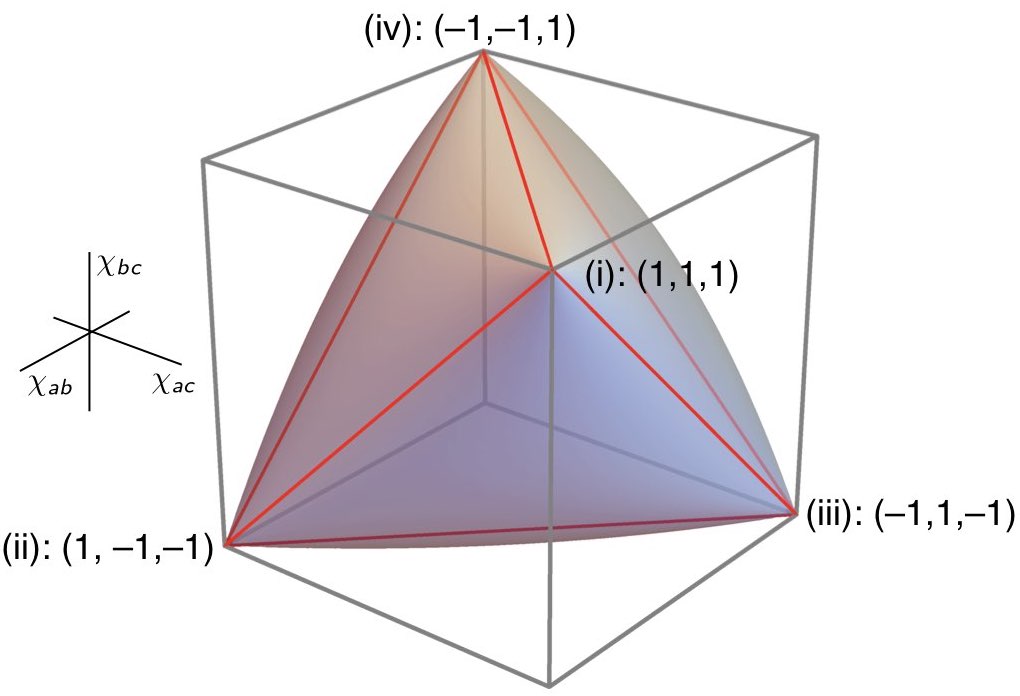} 
   \caption{Elliptope of triplets of anti-correlation coefficients $(\chi_{ab}, \chi_{ac}, \chi_{bc})$ allowed by quantum mechanics in our version of the Mermin setup.}
   \label{elliptope}
\end{figure}

Eq.\ (\ref{QM14}) is the constraint we were looking for. Quantum mechanics only allows those non-signaling correlation arrays in Figure \ref{CA-3set2out-non-signaling-halfangles}  for which the anti-correlation coefficients that can be used to parametrize them (see Eqs.\ (\ref{angle2halfangle})--(\ref{chi values repeat})) satisfy this constraint. The region of the non-signaling cube picked out by Eq.\ (\ref{QM14})  is the elliptope in Figure \ref{elliptope}, an ``inflated'' version of the tetrahedron of classically allowed triplets of correlation coefficients in Figure \ref{tetrahedron}.\footnote{In Section \ref{1.3} we already showed the cross-section $\chi_{bc} = 0$ of the non-signaling cube, the elliptope and the classical tetrahedron (see Figure \ref{elliptope-LQPslice}).} 

It is easy to verify that the elliptope contains the classical tetrahedron. The vertices (i) through (iv) with the values
\begin{equation}
(1, 1, 1), \quad (1, -1, -1), \quad (-1, 1, -1), \quad (-1, -1, 1),
\label{vertices a}
\end{equation}
for $(\chi_{ab}, \chi_{ac}, \chi_{bc})$ all satisfy Eq.\ (\ref{QM14}) with an equality sign:
\begin{equation}
1 - \chi_{ab}^2 - \chi_{ac}^2 - \chi_{bc}^2 + 2 \, \chi_{ab} \, \chi_{ac} \, \chi_{bc} = 0.
\label{QM14a}
\end{equation}
So do the six lines connecting these four vertices. Consider the top and bottom face of the non-signaling cube, where $\chi_{bc} =1$ and $\chi_{bc} = -1$, respectively. Eq.\ (\ref{QM14a}) then reduces to:
\begin{equation}
- \chi_{ab}^2 - \chi_{ac}^2 \pm 2\, \chi_{ab} \, \chi_{ac} = 0.
\label{QM14b}
\end{equation}
So the line $\chi_{ab} = \chi_{ac}$ on the $\chi_{bc} =1$ face of the cube and the line $\chi_{ab} = -\chi_{ac}$ on the $\chi_{bc} =-1$ face of the cube satisfy Eq.\ (\ref{QM14a}). These are the lines connecting the vertices on these two faces of the cube. We similarly find that the other four lines connecting vertices of the tetrahedron satisfy Eq.\ (\ref{QM14a}): the $\chi_{ac} = \chi_{bc}$ line on the $\chi_{ab}=1$ face, the $\chi_{ac}=-\chi_{bc}$ line on the $\chi_{ab}=-1$ face, the $\chi_{ab}= \chi_{bc}$ line on the $\chi_{ac}=1$ face and the $\chi_{ab}=-\chi_{bc}$ line on the $\chi_{ac}=-1$ face.
 
The volumes of the tetrahedron and the elliptope are $\sfrac83$ and $\pi^2/2$, respectively, which means that their ratio is $.54$. By this metric, the class of correlations allowed quantum-mechanically in this setup is thus almost twice as large as the class of correlations allowed by local hidden-variable theories.     

Once again consider Eq.\ (\ref{QM14}). The constraint it expresses has a simple geometrical interpretation. Note that $\det{L} = \pm V$, where $V$ is the volume of the parallelepiped spanned by the three unit vectors $\vec{e}_a$, $\vec{e}_b$ and $\vec{e}_c$. If this triplet is positively oriented, $\det{L} = V$; if it is negatively oriented, $\det{L} = -V$. Consider the case that $\det{\chi} = (\det{L})^2 = 0$, which means that $V=0$. This, in turn, means that $\vec{e}_a$, $\vec{e}_b$ and $\vec{e}_c$ are coplanar. Hence, the statement that $\det{\chi} =0$ is equivalent to the statement that either the three angles $(\varphi_{ab}, \varphi_{ac}, \varphi_{bc})$ add up to $360\degree$ or one of them is the sum of the other two \citep[p.\ 515]{Deza and Laurent 1997}.

\begin{figure}[h]
 \centering
   \includegraphics[width=4.5in]{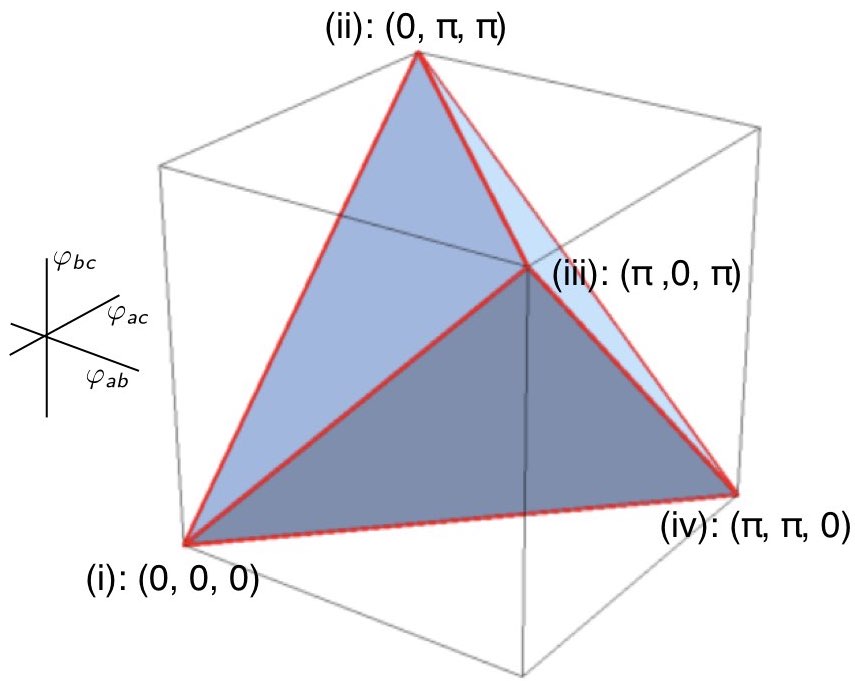} 
   \caption{Tetrahedron of triplets of angles $(\varphi_{ab}, \varphi_{ac}, \varphi_{bc})$ allowed by quantum mechanics in our version of the Mermin setup.}
   \label{tetrahedron-angles}
\end{figure}

To verify this equivalence we use the trigonometric identity
\begin{eqnarray}
(2 \cos{\alpha}) (2 \cos{\beta}) (2 \cos{\gamma}) & \! \!  = \! \!  & 2 \cos{(\alpha + \beta + \gamma)}
+ 2 \cos{(- \alpha + \beta + \gamma)} \nonumber \\
 & & \quad  + \; 2 \cos{(\alpha - \beta + \gamma)} + 2 \cos{(\alpha + \beta - \gamma)}, 
\label{QM15}
\end{eqnarray}
which can easily be checked with the help of the Euler formula, $2 \cos{\alpha} = e^{i \alpha} + e^{-i \alpha}$. In all cases in which one of the angles $(\alpha, \beta, \gamma)$ is either the sum of the other two ($\alpha = \beta + \gamma$, $\beta = \alpha + \gamma$, or $\gamma = \alpha + \beta$) or is $360\degree$ minus this sum ($\alpha+\beta+\gamma=360\degree$), the right-hand side of Eq.\ (\ref{QM15}) reduces to
\begin{equation}
2 + 2 \cos{2 \alpha} + 2 \cos{2 \beta} + 2 \cos{2 \gamma}
= 4 \cos^2{\! \alpha} + 4 \cos^2{\!\beta} + 4 \cos^2{\!\gamma} - 4.
 \label{QM16}
\end{equation}
Eq.\ (\ref{QM15}) can thus be rewritten as
\begin{equation}
2 \cos{\alpha} \cos{\beta} \cos{\gamma} = \cos^2{\!\alpha} + \cos^2{\!\beta} + \cos^2{\!\gamma} - 1.
\label{QM17}
\end{equation}
Substituting $(\varphi_{ab}, \varphi_{ac}, \varphi_{bc})$ for  $(\alpha, \beta, \gamma)$ and $(\chi_{ab}, \chi_{ac}, \chi_{bc})$ for  $(\cos{\varphi_{ab}}, \cos{\varphi_{ac}}, \cos{\varphi_{bc}})$, 
%(see Eq.\ (\ref{chis2angles2innerproducts})), 
we see that Eq.\ (\ref{QM17}) reduces to Eq.\ (\ref{QM14}) with an `$=$' sign rather than a `$\ge$' sign. This shows that the equality $\det{\chi} =0$ is indeed equivalent to the statement that the angles $(\varphi_{ab}, \varphi_{ac}, \varphi_{bc})$ either add up to $360\degree$ or that one of them is the sum of the other two. More generally, the \emph{inequality} $\det \chi\geq 0$ is equivalent to the statement that either their sum is no larger than $360\degree$ or the sum of any two of them is no smaller than the third, i.e.,  
\begin{equation}
\begin{array}{c}
 \varphi_{ab} +  \varphi_{ac} + \varphi_{bc} \le 360\degree  \\[.2cm]
\mathrm{or} \;\; \big( \; \varphi_{ab} \le  \varphi_{ac} + \varphi_{bc} \;\; \mathrm{and} \;\; \varphi_{ac} \le  \varphi_{ab} + \varphi_{bc}  \;\;  \mathrm{and} \;\; \varphi_{bc} \le  \varphi_{ab} + \varphi_{ac} \; \big) 
\end{array}
\label{angle inequalities}
\end{equation}
\citep[p.\ 515; cf.\ the triangle on the right in Figure \ref{vectors4elliptope} in Section \ref{1.6}]{Deza and Laurent 1997}.
%\citep[for proof, see][p.\ 515]{Deza and Laurent 1997}. 
These angle inequalities can be represented geometrically by the tetrahedron in Figure \ref{tetrahedron-angles}.\footnote{These inequalities can also be found in \citet[remark on p.\ 170]{Accardi and Fedullo 1982}. These authors also give Eq.\  (\ref{QM14}) for the elliptope in terms of \emph{the cosines of} these angles, which are just our anti-correlation coefficients $(\chi_{ab}, \chi_{ac}, \chi_{bc})$ (ibid., p.\ 166, Eq.\ 19; p.\ 168, Eq.\ 33; p.\ 169, Eq.\ 37):
$$
2 \, \cos{\varphi_{ab}} \, \cos{\varphi_{ac}} \, \cos{\varphi_{bc}} \ge  \cos^2{\!\varphi_{ab}} + \cos^2{\!\varphi_{ac}} + \cos^2{\!\varphi_{bc}} - 1.
%\label{QM17a}
$$
Citing \citet{Accardi and Fedullo 1982}, Bas \citet[pp.\ 120--121]{Van Fraassen 1991} states this same inequality in terms of probabilities rather than correlation coefficients, just as Mermin did with the Bell inequality (see Section \ref{1.4}, Eqs.\ (\ref{Pr opp Mermin})--(\ref{Pr opp general})).\label{Accardi Fedullo}}  

In the special case that Mermin considered, as we saw in Section \ref{1.3} (see Eq.\ (\ref{chi values Mermin example})),  
\begin{equation}
\chi_{ab} = \chi_{ac} = \chi_{bc} = -\sfrac12.
\label{chi values Mermin example repeat}
\end{equation}
In terms of the corresponding angles, this is the point
\begin{equation}
\varphi_{ab} = \varphi_{ac} = \varphi_{bc} = 120\degree
\end{equation}
at the center of the facet (ii)-(iii)-(iv) of the tetrahedron in Figure \ref{tetrahedron-angles}. 

At this point, the sum of the anti-correlation coefficients has its minimum value of $-\sfrac32$. This sum thus satisfies the inequality,
\begin{equation}
\chi_{ab} + \chi_{ac} + \chi_{bc} \ge -\sfrac32.
\label{Mermin Tsirelson bound on chis}
\end{equation}
This is the quantum analogue of the CHSH-type Bell inequality for local hidden-variable theories, which says that this quantity cannot be smaller than $-1$ (see Eq.\ (\ref{Mermin inequality CHSH-like})). The inequality in Eq.\ (\ref{Mermin Tsirelson bound on chis}) gives the \emph{Tsirelson bound} for this setup, named for Boris \citet{Cirel'son 1980}. Both the Tsirelson bound and the CHSH-type Bell inequality for this setup are represented by dotted lines in the cross-section of the quantum elliptope and the classical tetrahedron in Figure \ref{elliptope-LQPslice}.\footnote{The line marked ``Tsirelson bound'' in Figure \ref{elliptope-LQPslice} does not touch the circle representing the quantum convex set because the point $(\chi_{ab}, \chi_{ac}, \chi_{bc}) = (-\sfrac12, -\sfrac12, -\sfrac12)$ for which $\chi_{ab} + \chi_{ac} + \chi_{bc}$ reaches the minimum value quantum mechanics allows, lies below the plane $\chi_{bc}=0$ (cf. Figure \ref{elliptope}).} 

The point where the sum of the anti-correlation coefficients reaches the minimum value allowed by quantum mechanics in the Mermin setup is also the point where the probability of Alice and Bob finding opposite results reaches the minimum value allowed. Substituting the values for the anti-correlation coefficients in Eq.\ (\ref{chi values Mermin example repeat}) into Eq.\ (\ref{Pr opp Mermin}), we find
\begin{equation}
\mathrm{Pr(opp)} = \sfrac{2}{3} \, + \, \sfrac{1}{9} \, \Big(\chi_{ab} + \chi_{ac} + \chi_{bc} \Big) = \sfrac23 \, - \, \sfrac16 = \sfrac12.
\label{Pr opp Mermin (double)}
\end{equation}

As in the classical case (see Eqs.\ (\ref{Mermin inequality CHSH-like (i)})--(\ref{Mermin inequality CHSH-like (iv)})), we can write down three pairs of inequalities like the one for which the lower bound is given in Eq.\ (\ref{Mermin Tsirelson bound on chis}):  
\begin{equation}
-\sfrac32 \le \;\, \chi_{ab} - \chi_{ac} - \chi_{bc} \; \le 3, 
\label{Mermin Tsirelson bound on chis (ii)}
\end{equation}
\begin{equation}
-\sfrac32 \le - \chi_{ab} + \chi_{ac} - \chi_{bc} \le 3, 
\label{Mermin Tsirelson bound on chis (iii)}
\end{equation}
\begin{equation}
-\sfrac32 \le - \chi_{ab} - \chi_{ac} + \chi_{bc} \le 3.
\label{Mermin Tsirelson bound on chis (iv)}
\end{equation}
However, while the classical counterparts of these four linear inequalities sufficed to fully characterize the classical tetrahedron, we needed the non-linear inequality in Eq.\ (\ref{QM14}) to characterize the elliptope with its curved surface. 

%SECTION 2.6
%!TEX root =  ./JanasJanssenCuffaro-August2019.tex

%SUBSECTION 2.6
\subsection{The elliptope and the geometry of correlations} \label{1.6}

In Section \ref{1.5} we derived the equation for the elliptope using elements of quantum mechanics. In particular, we used the Born rule to show that the anti-correlation coefficients $(\chi_{ab}, \chi_{ac}, \chi_{bc})$ in the Mermin setup are given by the cosines of the angles $(\varphi_{ab}, \varphi_{ac}, \varphi_{bc})$ between the peeling directions $(\vec{e}_a, \vec{e}_b, \vec{e}_c)$. That meant that the anti-correlation matrix $\chi$ is a Gram matrix:
\begin{equation}
\chi \equiv \begin{pmatrix}
1 & \chi_{ab} & \chi_{ac} \\[.2 cm]
\chi_{ab} & 1 & \chi_{bc} \\[.2 cm]
\chi_{ac} & \chi_{bc} & 1
\end{pmatrix} =
\begin{pmatrix}
\vec{e}_a \!  \cdot  \vec{e}_a &  \vec{e}_a \! \cdot  \vec{e}_b  &   \vec{e}_a \! \cdot  \vec{e}_c  \\[.2cm]
\vec{e}_b \!  \cdot  \vec{e}_a & \vec{e}_b \! \cdot  \vec{e}_b & \vec{e}_b \! \cdot  \vec{e}_c \\[.2cm]
\vec{e}_c \!  \cdot   \vec{e}_a & \vec{e}_c \! \cdot  \vec{e}_b  & \vec{e}_c \! \cdot  \vec{e}_c 
\end{pmatrix}.
\label{gram matrix reprise}
\end{equation}
Eq.\ (\ref{QM14}) for the elliptope then follows from the observation that the determinant of a Gram matrix is non-negative. 

Our derivation of this equation was thus a derivation \emph{from within} quantum mechanics. It can, however, also be derived \emph{from without} (cf.\ note \ref{Dylan}).
%\footnote{We took the within/without terminology from the chorus of ``Quinn the Eskimo,'' a song from Bob Dylan's 1967 \emph{Basement Tapes}: ``Come all without, come all within. You'll not see nothing like the mighty Quinn.'' Could ``the mighty Quinn'' be an oblique but prescient reference to a quantum computer?\label{Dylan}} 
The constraint on (anti-) correlation coefficients expressed in the elliptope equation, it turns out, has nothing to do with quantum mechanics per se. It is a general constraint on correlations between three random variables. A paper on the early history of least-squares estimates by \citet{Aldrich 1998} led us to a paper by Udny \citet[p.\ 487]{Yule 1896} on what today are called Pearson correlation coefficients (see Eq.\ (\ref{chi as corr coef})), in which this constraint can already be found.\footnote{Yule was an associate of Karl Pearson and is remembered by historians of biology today for his role in bridging the divide between Mendelians and Darwinian biometrists, which would eventually result in the modern synthesis \citep[p.\ 329]{Bowler 2003}. In his 1897 paper, Yule refers to work by  Auguste \citet{Bravais 1846} half a century earlier. In a historical note in Part III of his seminal paper on the mathematics of Darwinian evolutionary theory as conceived by biometrists such as Francis Galton and Raphael Weldon, \citet{Pearson 1896} likewise writes that the correlation coefficient ``appears in Bravais' work, but a single symbol is not used for it'' (quoted in Hald, 1998, p.\ 622). Pearson later ``bitterly regretted this unbalanced evaluation'' of Bravais's contribution and tried to set the record straight in another historical note, ``equally unbalanced, but in the other direction'' \citep[p.\ 623; see Pearson, 1921, p.\ 191]{Hald 1998}.\label{biometrist}} 

In this section, we derive the elliptope equation from general statistical considerations (Section \ref{1.6.1}), discuss the application of this general result both to our quantum banana peeling and tasting experiment and to the raffles designed to simulate them (Sections \ref{1.6.2}--\ref{1.6.3}) and provide a geometrical perspective on the general result (Section \ref{1.6.3}), following some remarkable papers by two famous statisticians a generation after Pearson and Yule, Ronald A. \citet{Fisher 1915, Fisher 1924} and  Bruno \citet{De Finetti 1937}.\footnote{Fisher is remembered among many other things for his claim that Mendel faked his data \citep{Franklin et al 2008}; de Finetti mainly for his advocacy of Bayesian personalism \citep{McGrayne 2011}. De Finetti's first paper, however, was on population genetics and De Finetti diagrams are still used in that field.\label{mendel}}

%SUBSECTION 2.6.1
\subsubsection{The elliptope equation as a general constraint on correlations} \label{1.6.1}

Consider three random variables $X_a$, $X_b$ and $X_c$. They can be discrete or continuous but we assume that they are \emph{balanced} (see the definition numbered (\ref{def balanced}) in Section \ref{1.3}). In that case, their expectation values $\langle X_a \rangle$, $\langle X_b \rangle$ and $\langle X_c \rangle$ vanish; their variances (the square of their standard deviations) are given by $\sigma_a^2 \equiv \langle X_a^2 \rangle$, $\sigma_b^2 \equiv \langle X_b^2 \rangle$ and $\sigma_c^2 \equiv \langle X_c^2 \rangle$ (cf.\ Eq.\ (\ref{standard deviations a and b})); and their covariances by $\langle X_a X_b \rangle$, $\langle X_a X_c \rangle$ and $\langle X_b X_c \rangle$ (cf.\ Eq.\ (\ref{cov def})). For any triplet of real numbers $(v_a, v_b, v_c)$, which can be thought of as the components of some vector $\vec{v}$, we have the following inequality: 
\begin{equation}
\Big\langle \Big( v_a \frac{X_a}{\sigma_a} + v_b \frac{X_b}{\sigma_b} + v_c \frac{X_c}{\sigma_c} \Big)^{\!2} \Big\rangle \ge 0.
\label{inf the 1}
\end{equation}
This expands to
\begin{equation}
v_a^2 + v_b^2 + v_c^2 + 2 v_a v_b \frac{\langle X_a X_b \rangle}{\sigma_a \sigma_b} + 2 v_a v_c \frac{\langle X_a X_c \rangle}{\sigma_a \sigma_c} + 2 v_b v_c \frac{\langle X_b X_c \rangle}{\sigma_b \sigma_c} \ge 0,
\label{inf the 3}
\end{equation}
where we used the definition of the standard deviations $\sigma_a$, $\sigma_b$ and $\sigma_c$. Introducing the Pearson correlation coefficients
\begin{equation}
\overline{\chi}_{ab} \equiv \frac{\langle X_a X_b \rangle}{\sigma_a \sigma_b}, \quad
\overline{\chi}_{ac} \equiv \frac{\langle X_a X_c \rangle}{\sigma_a \sigma_c}, \quad
\overline{\chi}_{bc} \equiv \frac{\langle X_b X_c \rangle}{\sigma_b \sigma_c}
\label{corr coeff gen}
\end{equation}
(with overbars to distinguish them from the \emph{anti}-correlation coefficients introduced in Eq.\ (\ref{chi as corr coef})), we can rewrite Eq.\ (\ref{inf the 3}) as
\begin{equation}
v_a^2 + v_b^2 + v_c^2 + 2 v_a v_b \overline{\chi}_{ab} + 2 v_a v_c \overline{\chi}_{ac}  +  2 v_b v_c \overline{\chi}_{bc} \ge 0.
\label{inf the 4a}
\end{equation}
The expression before the $\ge$ sign has the form of a matrix multiplied by $\vec{v} \equiv (v_a, v_b, v_c)$ both from the left and from the right. Eq.\ (\ref{inf the 4a}) can thus be written as
\begin{equation}
\Big(v_a, v_b, v_c \Big) \, \left( 
\begin{array}{ccc}
\; 1 \; & \; \overline{\chi}_{ab} \; & \;  \overline{\chi}_{ac} \; \\
\; \overline{\chi}_{ab} & \; 1 \; & \;  \overline{\chi}_{bc} \; \\
 \; \overline{\chi}_{ac} \; & \; \overline{\chi}_{bc} \; & \;  1 
\end{array}
\right) \, 
\left( \begin{array}{c}
v_a \\
v_b \\
v_c
\end{array}
\right) \ge 0,
\label{inf the 4}
\end{equation}
or, more compactly, as $\vec{v}^{\top}  \overline{\chi} \, \vec{v} \ge 0$, where $\overline{\chi}$ is the correlation matrix, defined in analogy with the anti-correlation matrix $\chi$ in Eq.\ (\ref{chi matrix}) in Section \ref{1.5}. A matrix satisfying this inequality for arbitrary $\vec{v}$ is called \emph{positive semi-definite}. The (real) eigenvalues of such a matrix are non-negative. Let $\vec{u}$ be an eigenvector of $\overline{\chi}$ with some (real) eigenvalue $\lambda$, i.e., $\overline{\chi} \vec{u} = \lambda \vec{u}$. From
\begin{equation}
\vec{u}^{\top}  \overline{\chi} \, \vec{u} = \lambda \, \vec{u}^{\top} \vec{u} \ge 0,
\end{equation}
it follows that $\lambda \ge 0$. A square matrix with non-negative eigenvalues has a non-negative determinant. So $\det{\overline{\chi}} \ge 0$. Evaluation of the determinant of $\overline{\chi}$ gives
\begin{equation}
1 - \overline{\chi}_{ab}^2 - \overline{\chi}_{ac}^2 - \overline{\chi}_{bc}^2 + 2 \, \overline{\chi}_{ab} \, \overline{\chi}_{ac} \, \overline{\chi}_{bc} \ge 0
\label{inf the 5}
\end{equation}
and Bub's your uncle: Eq.\ (\ref{inf the 5}) has the exact same form as Eq.\ (\ref{QM14}) for elements of the anti-correlation matrix $\chi$.  It can thus be represented by the same elliptope that characterizes the class of correlations allowed by quantum mechanics in our banana peeling and tasting experiment. 

As in Section  \ref{1.5}, we can write the correlation coefficients $(\overline{\chi}_{ab}, \overline{\chi}_{ac}, \overline{\chi}_{bc})$ as the cosines of angles $(\overline{\varphi}_{ab}, \overline{\varphi}_{ac}, \overline{\varphi}_{bc})$. These angles will satisfy the angle inequality in Eq.\ (\ref{angle inequalities}) but, in general, will be unrelated to angles in either ordinary or Hilbert space. In Section \ref{1.6.4}, we return to the geometrical interpretation of random variables and angles between them. But first we apply the general formalism in Eqs.\ (\ref{inf the 1})--(\ref{inf the 5}) to the random variables in our quantum banana peeling and tasting experiment.

%SUBSECTION 2.6.2
\subsubsection{Why there are no further restrictions on the quantum correlations} \label{1.6.2}

Let the random variables $X_a$, $X_b$ and $X_c$ be the tastes of Alice's bananas in pairs of such bananas peeled and tasted by Alice and Bob when Alice peels it $\hat{a}$, $\hat{b}$ or $\hat{c}$, corresponding to peeling directions $\vec{e}_a$, $\vec{e}_b$ and $\vec{e}_c$.  These variables are balanced: they can only take on the values $\pm \bbar/2$ and both values occur with equal probability. On the face of it, however, there are three serious obstacles to the application of Eqs.\ (\ref{inf the 1})--(\ref{inf the 5}) to these variables, all three related to the issue we already encountered in Section \ref{1.5}: a banana can only be peeled and eaten once. Or to put it in terms of spin measurements with Dubois magnets in a Stern-Gerlach-type experiment: it can only be checked for one orientation of the axis of the magnet where any one particle lands on the photographic plate behind the magnet. Fortunately, the three hurdles this creates are not insurmountable, though clearing the third calls for some elements of quantum mechanics that we have not introduced yet.  

The first hurdle is how to make sense of covariances such as $\langle X_a X_b \rangle_{00}$ if $X_a$ and $X_b$ refer to the taste of one and the same banana peeled $\hat{a}$ or $\hat{b}$ (for the remainder of this section we drop the subscript $00$ that indicates we are peeling and tasting pairs of bananas in the singlet state). To get over this hurdle, we consider runs in which Alice peels $\hat{a}$ and Bob peels $\hat{b}$. We use the taste of Alice's banana as the value for $X_a$ and the \emph{opposite} of the taste of Bob's banana as the value for $X_b$. After all, we know that, if two bananas in one pair are peeled the same way, their tastes are always opposite. We can thus use $-X_b^B$ as a proxy for $X_b^A$ and $- \langle X_a^A X_b^B \rangle$ as a proxy for $\langle X_a^A X_b^A \rangle$. Because of the extra minus sign, the correlation coefficient $ \overline{\chi}_{ab}$ in Eq.\ (\ref{inf the 5}) gets replaced by the anti-correlation coefficients $\chi_{ab}$ introduced in Eq.\ (\ref{chi as corr coef}) in Section \ref{1.3}:   
\begin{equation}
 \overline{\chi}_{ab} = \frac{\langle X^A_a X^A_b \rangle}{\sigma_a \sigma_b} = - \frac{\langle X^A_a X^B_b \rangle}{\sigma_a \sigma_b} = \chi_{ab}.
 \label{chi 4 bar-chi}
\end{equation}
With the substitution of $(\chi_{ab}, \chi_{ac}, \chi_{bc})$ for $(\overline{\chi}_{ab}, \overline{\chi}_{ac},  \overline{\chi}_{bc})$, the constraint in Eq.\ (\ref{inf the 5}), found \emph{from without}, reduces to the constraint in Eq.\ (\ref{QM14}), found \emph{from within} (cf.\ note \ref{Dylan}).

There are, however, two more hurdles remaining. One is that, in forming linear combinations of $\chi_{ab}$, $\chi_{ac}$ and $\chi_{bc}$, we are combining data from different runs of the experiment.  When Alice peels $\hat{a}$ and Bob peels $\hat{b}$ to give us a value for $X^A_a X^A_b$, neither of them can, in the same run, peel $\hat{c}$ to give us values of $X^A_a X^A_c$ and $X^A_b X^A_c$ as well. The way around this problem is to note that our pairs of bananas start out in the same (singlet) state in every run. This is what makes it meaningful to consider expressions that combine, say, $\chi_{ab}$ and $\chi_{ac}$, the former based on data obtained in runs in which one banana is peeled $\hat{a}$ and the other one is peeled $\hat{b}$, the latter on data obtained in runs in which one is peeled $\hat{a}$ and the other one $\hat{c}$. In deriving the CHSH-type inequality in Eq.\ (\ref{Mermin inequality CHSH-like}) and the Tsirelson bound in Eq.\ (\ref{Mermin Tsirelson bound on chis}), we tacitly made the assumption that data from different runs of the experiment (randomly drawing a ticket from a basket in the case of these raffles) can be combined in this way. In the kind of local hidden-variable theories for which our raffles provide a model, this assumption can easily be avoided. We can simply change the protocol for our raffles and have Alice and Bob record the tastes for all three peelings on their halves of each ticket rather than pick just one. We trust that a moment's reflection will convince the reader that such a change of protocol does not change the correlation arrays for any of our raffles. In the quantum case, the assumption is unavoidable but equally innocuous.\footnote{Incidentally, this demonstrates the point we made in Section \ref{0} that our raffles provide a (toy) model of a theory that suffers from the \emph{superficial} but not the \emph{profound} measurement problem. Whenever a ticket is drawn from the basket, it is totally random who gets which half of the ticket. That means that it is totally random what outcome Alice and Bob will find when they check their ticket stub for the value for the setting they decide to check. So our toy theory runs afoul of the superficial measurement problem. The ticket stubs, however, have values for all settings, so there is nothing like the profound measurement problem in our toy theory.\label{minor/major}}

The third and final hurdle, however, looks more serious than the other two and to clear it we need to take an advance on our coverage of  the quantum-mechanical formalism for spin in Section \ref{2.1.1}.\footnote{We are extremely grateful to Wayne Myrvold for identifying this third hurdle when we presented a preliminary version of this paper in Viterbo in May 2019.\label{Myrvold 1}} 
%It is hard to overestimate the importance of his intervention for the final form of this (critical) section of our paper. 
The reason the expression in Eq.\ (\ref{inf the 5}) is greater or equal than zero is that it is simply a rewritten version of the expectation value of the square of a linear combination of $X_a$, $X_b$ and $X_c$ (see Eq.\ (\ref{inf the 1})). How do we find a value of this linear combination in any given run of the experiment? In one run, as we just saw,
only one of Alice's variables  $(X_a^A, X_b^A, X_c^A)$ can be measured directly and only one of the other two can be inferred from a measurement of one of Bob's variables $(X_a^B, X_b^B, X_c^B)$. Quantum mechanics does not allow us to assign values to all three observables in one run. Classically, it is inconceivable to have a situation in which a sum of three terms has a definite value but one of those three terms does not! Quantum mechanics, however, routinely allows such situations and this is what gets us over our third hurdle.\footnote{This response to Wayne Myrvold's request for clarification (see the preceding note) was inspired by a footnote in \emph{Wahrscheinlichkeitstheoretischer Aufbau} 
\citep[p.\ 249, note 9]{von Neumann 1927b}. 
%(1927b, p.\ 249, note 9).
In this footnote, von Neumann points out that the Hamiltonian for the harmonic oscillator $\hat{H}$ has a discrete spectrum even though it is the sum of two terms, $\hat{p}^2/2m$ and $\alpha \, \hat{q}^2$, that both have a continuous spectrum. The value of  $\hat{H}$ will not be the sum of the values of $\hat{p}^2/2m$ and $\alpha \, \hat{q}^2$ even though the expectation value $\big\langle \hat{H} \big\rangle$ will be the sum of the expectation values $\left\langle \hat{p}^2/2m \right\rangle$ and $\left\langle \alpha \, \hat{q}^2 \right\rangle$. We are grateful to Christoph Lehner for emphasizing the importance of this footnote back in 2009.\label{Myrvold 2}}  
%\citet[p.\ 249, note 9]{von Neumann 1927b} 

Recall that the values of the random variable $X_a$ in this case are the eigenvalues of the operator $\hat{S}_a$ (see Eq.\ (\ref{eigenvectors})). Therefore, introducing the notation
\begin{equation}
\lambda_a \equiv v_a/\sigma_a, \quad \lambda_b \equiv v_b/\sigma_b, \quad \lambda_c \equiv v_c/\sigma_c,
\label{lambda=v/sigma}
\end{equation}
we can write Eq.\ (\ref{inf the 1}), applied to our banana tasting experiment, as  
\begin{equation}
\big\langle \big( \lambda_a \, \hat{S}_a + \lambda_b \, \hat{S}_b + \lambda_c \, \hat{S}_c \big)^{\!2} \big\rangle \ge 0.
\label{exp sum S}
\end{equation}
Recall that $\sigma_a \! = \! \sigma_b \! = \! \sigma_c \! = \! \bbar/2$ in this case (see Eq.\ (\ref{standard deviations a and b})), so the vector $\vec{\lambda} \equiv (\lambda_a, \lambda_b, \lambda_c)$ is a vector in the same direction as $\vec{v}$ with the dimension of  $b^{-1}$. 

In Section \ref{2.1.1}, we will introduce the vector $\hat{\vec{S}}$ (see Eq.\ (\ref{def S vector})). The operators $(\hat{S}_a, \hat{S}_b, \hat{S}_c)$ can be written as inner products of this vector with the unit vectors $(\vec{e}_a, \vec{e}_b, \vec{e}_c)$ (see Eq.\ (\ref{spin op})). The inequality above can then be rewritten as
 \begin{equation}
\big\langle \big( \hat{\vec{S}} \! \cdot \! \left( \lambda_a \, \vec{e}_a + \lambda_b \, \vec{e}_b + \lambda_c \, \vec{e}_c \right) \! \big)^{\!2} \big\rangle \ge 0.
\label{exp sum S inner prod}
\end{equation}
The minimum value of 0 is reached whenever 
\begin{equation}
\lambda_a \, \vec{e}_a + \lambda_b \, \vec{e}_b + \lambda_c \, \vec{e}_c = 0.
\label{vecs add to 0}
\end{equation}
This can only happen when the three unit vectors $(\vec{e}_a, \vec{e}_b, \vec{e}_c)$ are coplanar. Recall that the anti-correlation coefficients $(\chi_{ab}, \chi_{ac}, \chi_{bc})$ for our quantum banana peeling and tasting experiment can be written as inner products of the unit vectors in the corresponding peeling directions (see Eq.\ (\ref{gram matrix reprise})). As noted in the paragraph leading up to Eq.\ (\ref{QM15}) in Section \ref{1.5}, these vectors must be coplanar for values of $(\chi_{ab}, \chi_{ac}, \chi_{bc})$ on the surface of the elliptope. For the four vertices the elliptope shares with the tetrahedron (see Figure \ref{elliptope}), they are not just coplanar but collinear as illustrated in Table \ref{collinear peeling directions}.

\begin{table}[h]
\centering
\begin{tabular}{|c||c|c|c||c|c|c|}
\hline
$\!$point$\!$ & $\cos{\varphi_{ab}} $ & $\cos{\varphi_{ac}} $ & $\cos{\varphi_{bc}} $ & $\; \vec{e}_a \;$ & $\; \vec{e}_b \;$ & $\; \vec{e}_c \;$ \\[.1cm] 
\hline
 (i) & $+1$ & $+1$ & $+1$ & $\uparrow$ & $\uparrow$ & $\uparrow$ \\[.2cm]
 (ii) & $+1$ & $-1$ & $-1$ & $\uparrow$ & $\uparrow$ & $\downarrow$ \\[.2cm]
 (iii) & $-1$ & $+1$ & $-1$ & $\uparrow$ & $\downarrow$ & $\uparrow$ \\[.2cm]
(iv) & $-1$ & $-1$ & $+1$ & $\uparrow$ & $\downarrow$ & $\downarrow$ \\
 \hline
\end{tabular}
\caption{Unit vectors $(\vec{e}_a, \vec{e}_b, \vec{e}_c)$, with $\vec{e}_a$ chosen as $\uparrow$, for the triplets of peeling directions corresponding to the points labeled (i) through (iv) of the elliptope in Figure \ref{elliptope}.}
\label{collinear peeling directions}
\end{table} 

\begin{figure}[h!]
 \centering
   \includegraphics[width=4.5in]{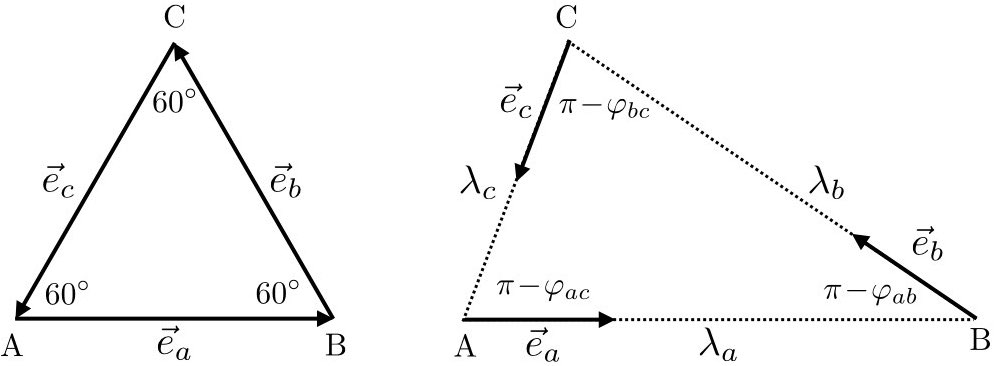} 
   \caption{Coplanar peeling directions $(\vec{e}_a, \vec{e}_b, \vec{e}_c)$ resulting in triplets of correlation coeffients $(\chi_{ab}, \chi_{ac}, \chi_{bc})$ coordinatizing points on the surface of the elliptope in Figure \ref{elliptope}}
   \label{vectors4elliptope}
\end{figure}

Figure \ref{vectors4elliptope} shows two examples of triplets of coplanar (but not collinear) unit vectors $(\vec{e}_a, \vec{e}_b, \vec{e}_c)$ such that some linear combination of them adds up to zero. The triplet on the left is for the situation in which the angles between all three peeling directions is $120\degree$ (cf.\ Figure \ref{AliceBob-Mermin}). This is the combination of peeling directions that gives rise to the Mermin correlation array in Figure \ref{CA-3set2out-Mermin}. In this case, the vectors $(\vec{e}_a, \vec{e}_b, \vec{e}_c)$ form an equilateral triangle with sides of length equal to 1. The vectors $(\vec{e}_a, \vec{e}_b, \vec{e}_c)$ on the right in Figure \ref{vectors4elliptope} are for a generic choice of three peeling directions such that a triangle with sides of length $(\lambda_a, \lambda_b, \lambda_c)$ in the directions of these three unit vectors can be formed. Eq.\ (\ref{vecs add to 0}) is thus satisfied. 

As long as $(\vec{e}_a, \vec{e}_b, \vec{e}_c)$ are coplanar but no two of them are collinear, it will be possible for any triplet of values for 
\begin{equation}
\chi_{ab} = \cos{\varphi_{ab}} =  \vec{e}_a \! \cdot  \vec{e}_b, \quad \chi_{ac} =    \cos{\varphi_{ac}} =   \vec{e}_a \! \cdot \vec{e}_c, \quad \chi_{bc} =    \cos{\varphi_{bc}} \! = \!  \vec{e}_b \! \cdot \vec{e}_c
\label{chi's as inner prods of e's}
\end{equation}
on the surface of the elliptope to construct a triangle with sides of length $(\lambda_a, \lambda_b, \lambda_c)$ in the directions of the triplet of corresponding coplanar unit vectors $(\pm \vec{e}_a, \pm \vec{e}_b, \pm \vec{e}_c)$. If we need to flip one of the three unit vectors to form a triangle, we need to take minus the length of the corresponding side to satisfy Eq.\ (\ref{vecs add to 0}). If we can use $(\vec{e}_a, \vec{e}_b, \vec{e}_c)$ to form a triangle, the angles $(\varphi_{ab}, \varphi_{ab}, \varphi_{ab})$ will add up to $360\degree$. If we have to flip one of the unit vectors to do so, one of the angles will be the sum of the other two (cf.\ our discussion in Section \ref{1.5} of the angle inequalities in Eq.\ (\ref{angle inequalities})).  

This construction shows that by choosing the appropriate peeling directions in our quantum banana peeling and tasting experiment we can reach all points on the surface of the elliptope. In particular we can reach the point $(\chi_{ab}, \chi_{ac}, \chi_{bc}) = (-\sfrac12, -\sfrac12, -\sfrac12)$ at the center of the facet (ii)-(iii)-(iv) of the tetrahedron in Figure \ref{tetrahedron-angles} corresponding to the elliptope. For $v_a = v_b = v_c = 1$, Eq.\ (\ref{inf the 1}) reduces to 
\begin{equation}
\Big\langle \Big(  \frac{X_a}{\sigma_a} +  \frac{X_b}{\sigma_b} + \frac{X_c}{\sigma_c} \Big)^{\!2} \Big\rangle \ge 0.
\label{Tsirelson reprise 1}
\end{equation}
Eq.\ (\ref{inf the 4a}), with $\chi$'s substituted for $\overline{\chi}$'s (see Eq.\ (\ref{chi 4 bar-chi})), then turns into
\begin{equation}
3 + 2 \chi_{ab} + 2 \chi_{ac}  +  2 \chi_{bc} \ge 0.
\label{Tsirelson reprise 2}
\end{equation}
or
\begin{equation}
\chi_{ab} + \chi_{ac}  +  \chi_{bc} \ge -\sfrac32.
\label{Tsirelson reprise 3}
\end{equation}
The value $-\sfrac32$, which is just the Tsirelson bound we found in Section \ref{1.5} (see Eq.\ (\ref{Mermin Tsirelson bound on chis})), is reached at $(\chi_{ab}, \chi_{ac}, \chi_{bc}) = (-\sfrac12, -\sfrac12, -\sfrac12)$.

Note that the vector $\vec{\lambda} = (1, 1, 1)$ is an eigenvector with eigenvalue 0 of the anti-correlation matrix $\chi$ in Eq.\ (\ref{chi matrix}) with $(\chi_{ab}, \chi_{ac}, \chi_{bc}) = (-\sfrac12, -\sfrac12, -\sfrac12)$. If $\varphi_{ab} = \varphi_{ac} = \varphi_{bc} = 120\degree$, then
\begin{equation}
\chi = \begin{pmatrix}
1 & \chi_{ab} & \chi_{ac}  \\[.1cm]
\chi_{ab} & 1 & \chi_{bc} \\[.1cm]
\chi_{ac} & \chi_{bc} & 1
\end{pmatrix}
=
\begin{pmatrix}
1 & \cos{\varphi_{ab}} & \cos{\varphi_{ac}} \\[.1cm]
\cos{\varphi_{ab}} & 1 & \cos{\varphi_{bc}} \\[.1cm]
\cos{\varphi_{ac}} & \cos{\varphi_{bc}} & 1
\end{pmatrix}
=
\begin{pmatrix}
1 & -\sfrac12 & -\sfrac12 \\[.1cm]
-\sfrac12 & 1 & -\sfrac12 \\[.1cm]
 -\sfrac12 & -\sfrac12 & 1
\end{pmatrix}.
\label{chi 120 deg}
\end{equation}
Hence, for $\vec{\lambda} = (1, 1, 1)$, $\chi \, \vec{\lambda} = 0 \, \vec{\lambda}$:
\begin{equation}
\begin{pmatrix}
1 & \chi_{ab} & \chi_{ac}  \\[.1cm]
\chi_{ab} & 1 & \chi_{bc} \\[.1cm]
\chi_{ac} & \chi_{bc} & 1
\end{pmatrix} \!
\begin{pmatrix}
1 \\[.1cm]
1 \\[.1cm]
1
\end{pmatrix}
=
\begin{pmatrix}
0 \\[.1cm]
0 \\[.1cm]
0
\end{pmatrix}.
\label{eigenvector (111)}
\end{equation}

This is just one example of a general property. Let $(\chi_{ab}, \chi_{ac}, \chi_{bc})$ be the coordinates of any point on the surface of the elliptope that is not one of the four points the elliptope shares with the tetrahedron. Let $(\vec{e}_a, \vec{e}_b, \vec{e}_c)$ be a triplet of coplanar (but not collinear) unit vectors whose inner products give $(\chi_{ab}, \chi_{ac}, \chi_{bc})$ (see Eq.\ (\ref{chi's as inner prods of e's})). Let the coefficients $(\lambda_a, \lambda_b, \lambda_c)$, chosen in such a way that Eq.\ (\ref{vecs add to 0}) is satisfied, be the components of a vector $\vec{\lambda}$. This vector will be an eigenvector with eigenvalue 0 of the anti-correlation matrix $\chi$ for the values of $(\chi_{ab}, \chi_{ac}, \chi_{bc})$ in Eq.\ (\ref{chi's as inner prods of e's}). This is a direct consequence of  $\vec{\lambda}$ being an eigenvector with eigenvalue 0 of the matrix $L$ introduced in Section \ref{1.5} to write $\chi = L^\top L$ (see Eqs.\ (\ref{QM12})--(\ref{QM13})). That $L \, \vec{\lambda} = 0 \, \vec{\lambda}$, in turn, simply expresses the linear dependence of the three coplanar vectors $(\vec{e}_a, \vec{e}_b, \vec{e}_c)$. The columns of $L$ are just the components of $(\vec{e}_a, \vec{e}_b, \vec{e}_c)$. Using Eq.\ (\ref{comps of unit vectors e_abc}) for these components, one readily verifies that $L \, \vec{\lambda}$ vanishes:
\begin{eqnarray}
\begin{pmatrix}
a_x & b_x & c_x  \\
a_y & b_y & c_y  \\
a_z & b_z & c_x
\end{pmatrix} \!\!
\begin{pmatrix}
\lambda_a \\
\lambda_b \\
\lambda_c
\end{pmatrix}
& \! \! = \! \! &
\lambda_a \!
\begin{pmatrix}
a_x  \\
a_y \\
a_z
\end{pmatrix}
+
\lambda_b \!
\begin{pmatrix}
b_x  \\
b_y \\
b_z
\end{pmatrix}
+
\lambda_c \!
\begin{pmatrix}
c_x  \\
c_y \\
c_z
\end{pmatrix} \nonumber \\[-.2cm]
 & & \label{eigenvector of L} \\
 & \! \! = \! \! & \lambda_a \, \vec{e}_a + \lambda_b \, \vec{e}_b + \lambda_c \, \vec{e}_c = 0, \nonumber
\end{eqnarray}
where in the last step we used Eq.\ (\ref{vecs add to 0}). It follows that $\chi \vec{\lambda}$ also vanishes:
\begin{equation}
\begin{pmatrix}
\vec{e}_a \! \cdot   \vec{e}_a &  \vec{e}_a \! \cdot  \vec{e}_b  &   \vec{e}_a \! \cdot  \vec{e}_c  \\[.2cm]
\vec{e}_b \! \cdot  \vec{e}_a & \vec{e}_b \! \cdot  \vec{e}_b & \vec{e}_b \! \cdot  \vec{e}_c \\[.2cm]
\vec{e}_c \! \cdot  \vec{e}_a & \vec{e}_c \! \cdot  \vec{e}_b  & \vec{e}_c \! \cdot  \vec{e}_c
\end{pmatrix} \!
\begin{pmatrix}
\lambda_a \\[.2cm]
\lambda_b \\[.2cm]
\lambda_c
\end{pmatrix}
= \begin{pmatrix}
\, \vec{e}_a \! \cdot \! \big(\lambda_a \, \vec{e}_a + \lambda_b \, \vec{e}_b  + \lambda_c \, \vec{e}_c \big) \, \\[.2cm]
\, \vec{e}_b \! \cdot \! \big(\lambda_a \, \vec{e}_a + \lambda_b \, \vec{e}_b  + \lambda_c \, \vec{e}_c \big)  \, \\[.2cm]
\, \vec{e}_c \! \cdot \! \big(\lambda_a \, \vec{e}_a + \lambda_b \, \vec{e}_b  + \lambda_c \, \vec{e}_c \big)  \,
\end{pmatrix} =
\begin{pmatrix}
0 \\[.2cm]
0 \\[.2cm]
0
\end{pmatrix}.
\label{eigenvector lambda}
\end{equation}

One conclusion of the analysis in this subsection is that it is unsurprising that quantum mechanics does not allow any non-signaling correlations that violate the Tsirelson bound---or, more generally, any non-signaling correlations represented by points inside the non-signaling cube but outside the elliptope. This is not because of some elusive physical principle over and above non-signaling but simply because of the general constraint on (anti-)correlation coefficients in Eq.\ (\ref{inf the 5}). What \emph{is} surprising in light of this general constraint, is that the correlations allowed in our quantum banana peeling and tasting experiment get beyond the classical tetrahedron and fill out the entire elliptope. The taste of these bananas, after all, can only be $\pm \bbar/2$, regardless of what peeling is chosen, so it is impossible to find values for $X_a$, $X_b$ and $X_c$, representing the taste of a banana peeled $\hat{a}$, $\hat{b}$ and $\hat{c}$, respectively, such that $X_a + X_b + X_c = 0$. One would therefore expect the minimum value of the square of this sum to be $\bbar^2/4$ rather than 0. In that case, Eq.\ (\ref{Tsirelson reprise 1}) changes to
\begin{equation}
\Big\langle \Big(  \frac{X_a}{\sigma_a} +  \frac{X_b}{\sigma_b} + \frac{X_c}{\sigma_c} \Big)^{\!2} \Big\rangle \ge 1,
\label{CHSH reprise 1}
\end{equation} 
where we used that $\sigma_a = \sigma_b = \sigma_c = \bbar/2$ (see Eq.\ (\ref{standard deviations a and b})). Eq.\ (\ref{Tsirelson reprise 3}) would accordingly change to:
\begin{equation}
\chi_{ab} + \chi_{ac}  +  \chi_{bc} \ge - \sfrac32 + \sfrac12 = -1,
\label{CHSH reprise 2}
\end{equation}
which is just the CHSH-type inequality we found in Section \ref{1.4} (see Eq.\ (\ref{Mermin inequality CHSH-like})). As we saw above, the reason quantum mechanics is less restrictive is that it allows something no classical theory would allow, namely to assign a value to the sum of three variables without assigning a value to all three of them individually.

As illustrated by a simple example in \emph{Totally Random}, the speedup of a quantum computer comes from the ability to assign a truth value to a conjunction without having to assign a truth value to its conjuncts \citep[pp.\ 186--215]{Bub and Bub 2018}. Long before anybody was thinking about quantum computing, however, physicists had already run into a version of the conundrum encountered here and resolved by quantum mechanics. In their popular introductory textbook on quantum physics, \citet[p.\ 258]{Eisberg and Resnick 1985} highlight the peculiar behavior of angular momentum in quantum mechanics. That the wave function for a one-electron atom ``does not describe a state with a definite $x$ and $y$ component of orbital angular momentum,'' they note, ``is mysterious from the point of view of classical mechanics'' (and is equally mysterious for intrinsic or spin angular momentum). In quantum mechanics, they continue, this is required by the uncertainty principle. If the $z$-component has a definite value, the $x$- and $y$-components cannot have definite values.
 
\begin{figure}[h]
 \centering
   \includegraphics[width=2.6in]{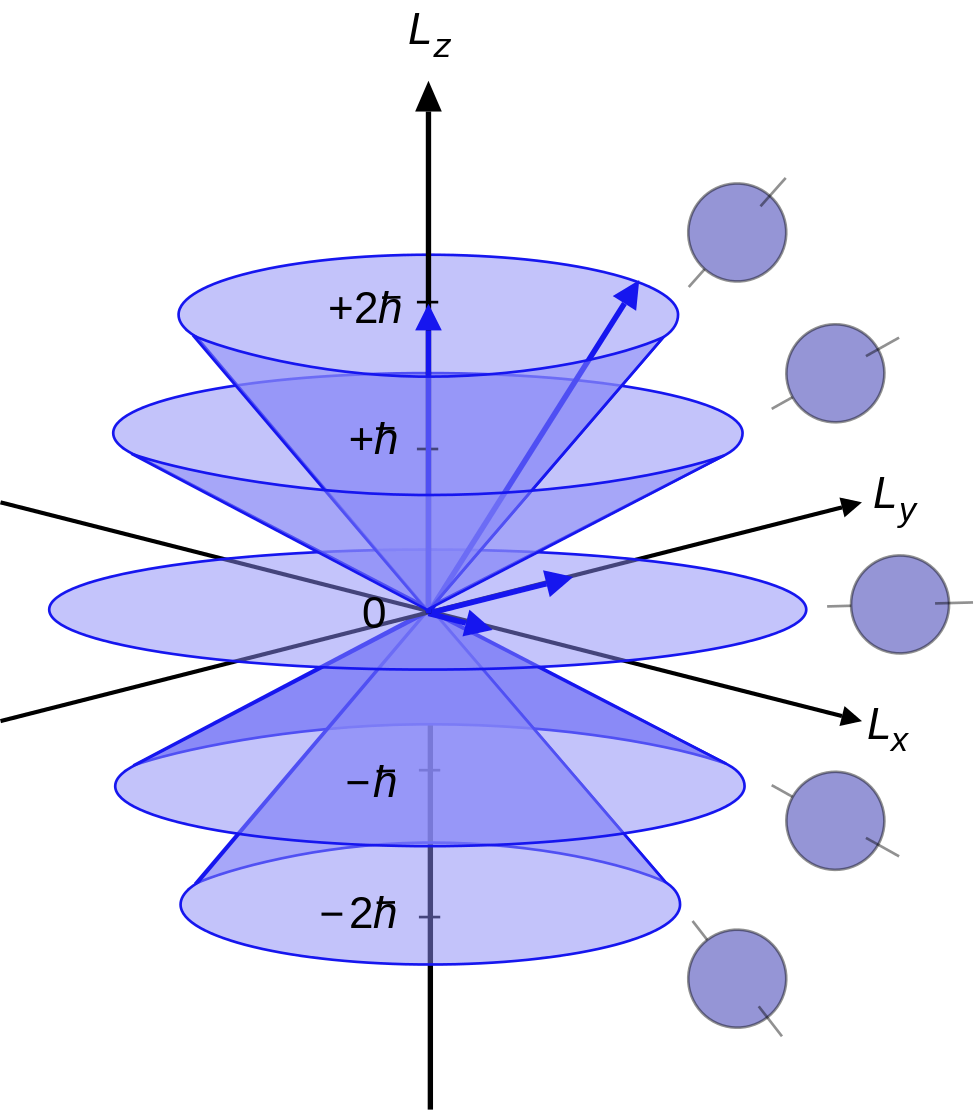} 
   \caption{Vector model of orbital angular momentum in the old quantum theory \citep[Wikimedia Commons; cf.][p.\ 258, Fig.\ 7-12]{Eisberg and Resnick 1985}. }
   \label{Vector_model_of_orbital_angular_momentum}
\end{figure}

This behavior of angular momentum,  \citet[pp.\ 258--259]{Eisberg and Resnick 1985} note in the next paragraph, ``can be conveniently represented by a \emph{vector model},'' i.e., the one shown in Figure \ref{Vector_model_of_orbital_angular_momentum}. In this model, the angular momentum vector precesses around the $z$ axis at a fixed angle determined by the value of its $z$-component. While the $z$-component remains fixed, the $x$- and $y$-components are constantly changing. This model, of course, still does not capture the true state of affairs in quantum mechanics where it is impossible to assign a definite value to all three component at any instant. The vector model is a left-over from the old quantum theory, where it was introduced to deal with problems posed by multiplet spectra and the anomalous Zeeman effect \citep[Sec.\ 1.3.6 and Ch.\ 7]{Duncan and Janssen 2019}.

The quantum-mechanical treatment of angular momentum not only solved these problems in spectroscopy, it also restored order in a completely different field. As John H.\ Van Vleck, the author of an authoritative book on the subject noted in the opening sentence of its preface: ``The new quantum mechanics  is perhaps most noted for its triumphs in the field of spectroscopy, but its less heralded successes in the theory of electric and magnetic susceptibilities must be regarded as one of its great achievements'' \citep[p.\ vii]{Van Vleck 1932}. Since we inserted this digression to preempt charges of parochialism against the information-theoretic interpretation of quantum mechanics we are championing (see Section \ref{0}), it is amusing to note that Van Vleck in the aftermath of the quantum revolution of the mid-1920s felt that he and his colleagues had been too focused on spectroscopy. In an article on the new quantum mechanics in a chemistry journal, he wryly noted that ``[t]he chemist is apt to conceive of the physicist as some one who is so entranced in spectral lines that he closes his eyes to other phenomena'' \citep[p.\ 493]{Van Vleck 1928}.\footnote{Both quotations are taken from \citet[p.\ 137]{Midwinter and Janssen 2013}. Van Vleck's solution to the problem of susceptibilities hinges on the correct quantum-mechanical treatment of angular momentum, in this case of diatomic molecules such as hydrogen chloride (ibid., p.\ 199). We will return to this episode in Section \ref{4.3a} as an example of a problem solved by the new kinematics of quantum mechanics.\label{Van Vleck}}

%SUBSECTION 2.6.3
\subsubsection{Further restrictions on the correlations generated by raffles meant to simulate the quantum correlations} \label{1.6.3}

We can find CHSH-type inequalities like the one in Eq.\ (\ref{CHSH reprise 2}) for raffles we will design in Section \ref{2.2} to simulate correlations found in measurements on pairs of entangled particles of spin $s  = 1, \sfrac32, 2, \sfrac52, \ldots$ in the singlet state. The spin in any direction (corresponding to the taste of a banana peeled in that peeling direction) will take on $2s \! + \! 1$ different values $m \hbar$, where $m = -s, -s+1, \ldots, s-1, s$ and $\hbar$ is Planck's constant divided by $2\pi$. We will analyze these quantum correlations in Section \ref{2.1}. 

\begin{figure}[h]
 \centering
   \includegraphics[width=2.5in]{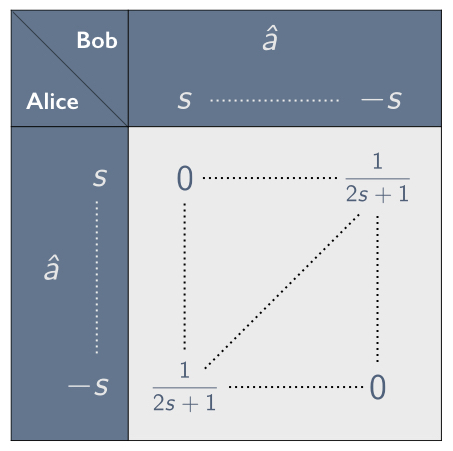} 
   \caption{Cell along the diagonal of a non-signaling correlation array with $2s+1$ outcomes per setting.}
   \label{diag-cell-sxs}
\end{figure}

Figure \ref{diag-cell-sxs} shows a cell along the diagonal of the correlation array for the quantum correlations for arbitrary spin $s$. It tells us that, when Alice and Bob use the same setting, all $2s \! + \! 1$ ways in which they can find opposite results (including $m = 0$ if $s$ is an integer) occur with equal probability. These quantum correlation arrays are non-signaling by virtue of having uniform marginals: in every row and every column of every cell, on or off the diagonal of the correlation array, the entries add up to $1/(2s \! + \! 1)$. 

In designing raffles for these higher-spin cases, we are immediately faced with a complication compared to the spin-$\frac12$ case (see Section \ref{2} for further discussion). Regardless of how many possible outcomes per setting there are, we can only put two outcomes per setting on any given ticket. This simple observation, unfortunately, has serious consequences for the design of our raffles. The reason we did not have to worry about this so far is that, for $s = \sfrac12$, there are only two possible outcomes per setting in the quantum experiment we are trying to simulate with our raffles. For $s > \sfrac12$, however, there are $2s+1$ possible outcomes per setting. Here is why this a problem. As we just saw, for any value of $s$, the quantum correlations are such that if Alice and Bob use the same setting, they are both expected to find all $2s+1$ possible outcomes in equal proportion. There is no way we can simulate this feature of the quantum correlations with single-ticket raffles as these tickets can at most have two of these $2s+1$ outcomes printed on them for each setting  (for integer values of $s$ it is possible that they have one and the same value, i.e., $0$, on both sides). In other words, for $s > \sfrac12$, the correlations we can produce with single-ticket raffles, while non-signaling by construction, do not give uniform marginals (see, e.g., Figure \ref{CA-3set3out-raffle-vi} in Section \ref{2.2}). To make sure that our raffles at least simulate the cells along the diagonal of the quantum correlation arrays correctly, we thus need to restrict ourselves to mixed raffles that do give uniform marginals. In Section \ref{2.2}, we will show how to construct those. All we need at this point is that any admissible raffle gives rise to a correlation array in which the cells along its diagonal have the form shown in Figure \ref{diag-cell-sxs}, which implies that they give uniform marginals.

Consider an arbitrary (single-ticket or mixed) raffle with tickets with opposite outcomes on the left and the right side for the three settings $\hat{a}$, $\hat{b}$ and $\hat{c}$ and with $2s \! + \! 1$ possible outcomes for each of these three settings. As in the case of two outcomes per setting considered so far, we can write:
\begin{equation}
\langle \left( X_a^A + X_b^A + X_c^A \right)^2 \rangle
= \sigma_a^2 + \sigma_b^2 + \sigma_c^2 - 2 \big( \langle X_a^A X_b^B \rangle + \langle X_a^A X_c^B \rangle + \langle X_b^A X_c^B \rangle \big)
\label{sum of X sq 4 any raffle}
\end{equation}
where, as before, we use $-X_a^A X_b^B$ as a proxy for $X_a^A X_b^A$, etc. 

The standard deviations are the square roots of variances computed for the cells along the diagonals of the relevant correlation arrays. Since these cells will all be of the form of the one in Figure \ref{diag-cell-sxs}, the three standard deviations in Eq.\ (\ref{sum of X sq 4 any raffle}) will have the same value $\sigma_s$, given by  
\begin{equation}
 \sigma_s^2 = \sigma_a^2 = \langle (X^A_a)^2 \rangle  \big|_{\mathrm{UM}}  = - \langle X^A_a X^B_a \rangle   \big|_{\mathrm{UM}}, 
\label{SD for adm raffle 1}
\end{equation}
where $ \big|_{\mathrm{UM}}$ indicates that the (co-)variance be evaluated for a raffle giving uniform marginals. Inspection of Figure \ref{diag-cell-sxs} and the well-known sum-of-squares formula gives
\begin{equation}
\sigma_s^2 = - \sum_{m=-s}^s \! \frac{(m \, \hbar) (-m \, \hbar)}{2s+1} = \frac{\hbar^2}{2s+1} \sum_{m=-s}^s \!\! m^2 =  \frac13 s(s+1) \, \hbar^2.
\label{SD for adm raffle 2}
\end{equation}

For raffles giving uniform marginals, we have
\begin{equation}
\chi_{ab}   \big|_{\mathrm{UM}} = \left. - \left( \frac{\langle X_a^A X_b^B \rangle}{\sigma_a \sigma_b} \right) \right|_{\mathrm{UM}} = - \frac{1}{\sigma_s^2} \langle X_a^A X_b^B \rangle   \big|_{\mathrm{UM}}.\label{chi and covariance adm raffle}
\end{equation}
Similar expressions obtain for $\chi_{ac}$ and $\chi_{bc}$. For such raffles, Eq.\ (\ref{sum of X sq 4 any raffle}) can be rewritten as
\begin{equation}
\langle \left( X_a^A + X_b^A + X_c^A \right)^2 \rangle  \big|_{\mathrm{UM}}
= \sigma_s^2 \left( 3 + 2 \! \left. \big( \chi_{ab} + \chi_{ac} + \chi_{bc} \big) \right|_{\mathrm{UM}} \right).
\label{sum of X sq 4 UM raffle}
\end{equation}

For any half-integer value of $s$, $|X_a^A + X_b^A + X_c^A|$ cannot be made smaller than $\hbar/2$, hence
\begin{equation}
\langle \left( X_a^A + X_b^A + X_c^A \right)^{\!2} \rangle \ge \frac{\hbar^2}{4} \quad {\mathrm{for \; half}}\mbox{-}{\mathrm{integer \;}} s.
\label{De Finetti half integer s}
\end{equation}
For any integer value of $s$, $X_a + X_b + X_c$ can be made to vanish, hence
\begin{equation}
\langle \left( X_a^A + X_b^A + X_c^A \right)^{\!2} \rangle \ge 0 \quad {\mathrm{for \; integer \;}} s.
\label{De Finetti integer s}
\end{equation} 
Restricting ourselves to raffles giving uniform marginals, in which case we can use Eq.\ (\ref{sum of X sq 4 UM raffle}), we thus find the following CHSH-type lower bounds on the sum of the anti-correlation coefficients in the Mermin setup:
\begin{equation}
\big( \chi_{ab} + \chi_{ac} + \chi_{bc} \big)  \big|_{\mathrm{UM}} \ge \frac{\hbar^2}{8\sigma_s^2} - \frac32 \quad {\mathrm{for \; half}}\mbox{-}{\mathrm{integer \;}} s,
\label{Mermin CHSH half-integer spin}
\end{equation}
\begin{equation}
\big( \chi_{ab} + \chi_{ac} + \chi_{bc} \big)  \big|_{\mathrm{UM}}  \ge - \frac32 \quad {\mathrm{for \; integer \;}} s.
\label{Mermin CHSH integer spin}
\end{equation}
For $s=\sfrac12$, $\sigma_s^2 = \hbar^2/4$ and Eq.\ (\ref{Mermin CHSH half-integer spin}) reduces to Eq.\ (\ref{CHSH reprise 2}), the CHSH-like inequality for this setup (since all raffles for the spin-$\frac12$ case give uniform marginals, the restriction $\big|_{\mathrm{UM}}$ can be dropped).

Eq.\ (\ref{Mermin CHSH integer spin}) tells us that, for integer values of $s$, we can (at least in principle) always reach the Tsirelson bound for a Mermin setup with $2s+1$ outcomes per setting (in Section \ref{2.2} we will design raffles that do indeed reach this bound). Eq.\ (\ref{Mermin CHSH half-integer spin}) tells us that, for half-integer values of $s$, this is true only in the limit that $s$ goes to infinity, in which case the number of outcomes $2s+1$ and the standard deviation $\sigma_s$ also go to infinity. 

As we will show in detail for the $s=1$ case in Section \ref{2.2}, even though we \emph{can} reach the Tsirelson bound on the sum of (anti-)correlation coefficients with our raffles, these raffles still \emph{cannot} reproduce all individual entries in the correlation arrays for the quantum correlations they are supposed to simulate (see Eq.\ (\ref{off diag cell quantum v raffle})). The reason for this will also become clear in Section \ref{2}. In Section \ref{2.1}, we will show that, regardless of the spin $s$ of the two particles involved, the probabilities in any given cell in these quantum correlation arrays can still be parametrized by the angle between measuring directions. In Section \ref{2.2}, we will see that in our raffles this is true only for the simple case $s = \sfrac12$ of two outcomes per setting that we have been considering so far. Even in the case of $s=1$, we already need two parameters to specify the entries in any off-diagonal cell in our correlation arrays (see Figure \ref{symmetry-spin-1-32}). We will give a simple example of measurements one could in principle do on two spin-1 particles entangled in the singlet state where quantum mechanics predicts results that cannot be accounted for in any local hidden-variable theory \emph{even though these results do not violate the relevant Bell inequality}.

%SUBSECTION 2.6.4
\subsubsection{The geometry of correlations: from Pearson and Yule to Fisher and de Finetti} \label{1.6.4}

\begin{figure}[h!]
 \centering
   \includegraphics[width=5in]{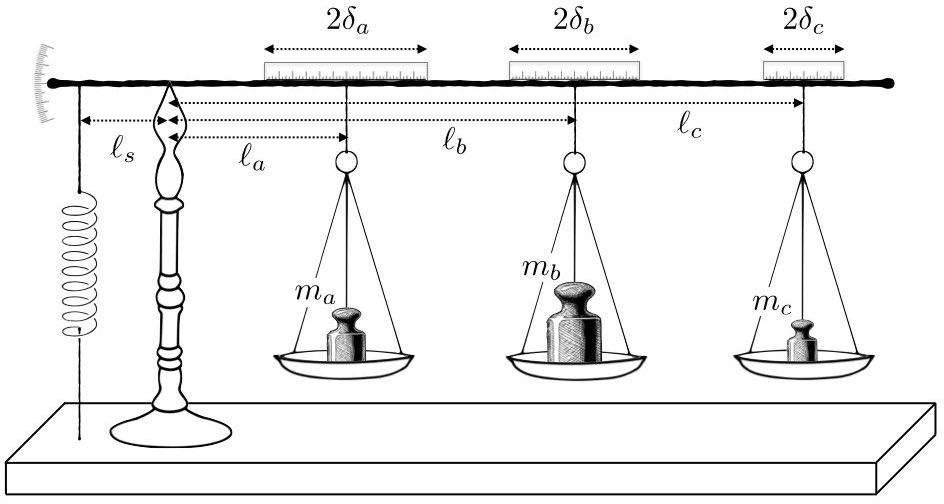} 
   \caption{3$m$ balance: an elementary example of a correlation between three random variables $(X_a, X_b, X_c) \equiv (\Delta \ell_a, \Delta \ell_b, \Delta \ell_c)$ (with $| \Delta \ell_a | \le \delta_a$, $| \Delta \ell_b | \le \delta_b$ and $| \Delta \ell_c | \le \delta_c$).}
   %The values for the displacements $(\Delta \ell_a, \Delta \ell_b, \Delta \ell_c)$ from the points $(\ell_a, \ell_b,  \ell_c)$ are subject to the constraint that the torque coming from the three scale pans on the right (with masses $m_a$, $m_b$ and $m_c$) in them exactly cancels the opposing torque coming from the spring on the left, so that the beam remains horizontal as it is for $(\Delta \ell_a, \Delta \ell_b, \Delta \ell_c) = (0, 0, 0)$.
   \label{3M-balance}
\end{figure}

In this subsection, we indicate how \citet{Yule 1896} found the general constraint on correlation coefficients in Eq.\ (\ref{inf the 5}) in the context of regression theory (i.e., finding the straight lines best approximating correlations between variables) and how \citet{Fisher 1915,Fisher 1924} and \citet{De Finetti 1937} recovered the result Yule found algebraically by treating random variables as vectors and correlations in terms of angles between those vectors. The importance of this geometric approach was emphasized by \citet{Pearson 1916}:
\begin{quote}
It is greatly to be desired that the ``trigonometry'' of higher dimensioned plane space should be fully worked out, for all our relations between multiple correlation and partial correlation coefficients of $n$ variates are properties of the ``angles,'' ``edges'' and ``perpen\-diculars'' of sphero-polyhedra in multiple space. It would be a fine task for an adequately equipped pure mathematician to write a treatise on ``spherical polyhedrometry''; he need not fear that his results would be without practical application for they embrace the whole range of problems from anatomy to medicine and from medicine to sociology and ultimately to the doctrine of evolution \citep[p.\ 237]{Pearson 1916}.\footnote{Linear algebra has proven to be much more convenient than spherical geometry for dealing with the relevant problems in statistics. As \citet[p.\ 73]{Aldrich 1998} notes, ``[t]he treatise on the trigonometry of correlations \ldots\ never materialized. [Kendall, 1961] is a partial offering but it appeared just as a new approach was taking off. This drew on the Hilbert space theory developed in the early part of the century and assembled by [Stone 1932].''}
\end{quote}
Rather than discussing the geometry of correlations in the abstract, we use the concrete example of three balanced random variables $(X_a, X_b, X_c)$ illustrated in Figure \ref{3M-balance}. 

This figure shows a device we will call a \emph{3m balance}, so named for the three masses $m_a$, $m_b$ and $m_c$ in the scale pans hanging from its beam. In the configuration shown in the figure, the three pans are at distances $\ell_a$, $\ell_b$ and $\ell_c$ from the beam's pivot point. In this configuration, the torque coming from the three pans on the right exactly cancels the torque coming from the spring on the left:   
\begin{equation}
F_s \ell_s = g \big( m_a \, \ell_a + m_b \, \ell_b + m_c \, \ell_c \big).
\label{torques}
\end{equation}
Here $F_s$ is the force exerted by the spring, $\ell_s$ its distance to the pivot point and $g$ the acceleration of gravity. Now imagine we take these three pans off the beam and put them back on (with the same masses as before) such that (a) the system is once again perfectly balanced (i.e., the beam is horizontal) and (b) the pans with $m_a$, $m_b$ and $m_c$ are somewhere in the intervals  $\ell_a \pm \delta_a$, $\ell_b \pm \delta_b$ and $\ell_c \pm \delta_c$, respectively. Suppose we end up putting the pans at the displaced positions
\begin{equation}
\ell_a + \Delta \ell_a, \quad \ell_b + \Delta \ell_b, \quad \ell_c + \Delta \ell_c,
\end{equation}
where $| \Delta \ell_a | \le \delta_a$, $| \Delta \ell_b | \le \delta_b$ and $| \Delta \ell_c | \le \delta_c$. The displacements $(\Delta \ell_a, \Delta \ell_b, \Delta \ell_c)$ that characterize this new perfectly balanced configuration of the system will satisfy the linear equation
\begin{equation}
m_a \, \Delta \ell_a + m_b \, \Delta\ell_b + m_c \, \Delta \ell_c  = 0.
\label{lin rel delta l}
\end{equation}
Suppose we repeat this many times.
%experiment of taking the three pans off the beam and putting them back on many times. 
The triplets of displacements $(\Delta \ell_a, \Delta \ell_b, \Delta \ell_c)$ found in consecutive runs of this experiment, all satisfying Eq.\ (\ref{lin rel delta l}), can then be treated as triplets of values for the three random variables
\begin{equation}
(X_a, X_b, X_c) \equiv (\Delta \ell_a, \Delta \ell_b, \Delta \ell_c).
\label{X = delta l}
\end{equation}
Since we are equally likely to find $\Delta \ell_a$ as $-\Delta \ell_a$, $\Delta \ell_b$ as $-\Delta \ell_b$ and $\Delta \ell_c$ as $-\Delta \ell_c$, these variables are 
balanced (see the definition numbered (\ref{def balanced}) in Section \ref{1.3}).\footnote{Since $X_a$ is balanced, its expectation value is zero and its variance is given by
\begin{equation*}
\langle (X_a - \langle X_a \rangle)^2 \rangle = \langle X_a^2 \rangle = \frac{1}{2\delta_a} \int_{-\delta_a}^{\delta_a} \!\! X_a^2 \, dX_a =  \frac{1}{6\delta_a} X_a^3 \Big|_{-\delta_a}^{\delta_a} = \frac{\delta_a^2}{3}.
\label{variance 3M variable} 
\end{equation*}
The corresponding standard deviation is: $\sigma_a \! = \! \sqrt{\langle X_a^2 \rangle} \! = \! \delta_a/\sqrt{3}$. Similarly, $\sigma_b  \! = \!  \delta_b/\sqrt{3}$ and $\sigma_c  \! = \!   \delta_c/\sqrt{3}$.\label{3M balance SDs}} They are correlated: if we vary the position of one of the masses, we need to vary the position of at least one of the other two to satisfy Eq.\ (\ref{lin rel delta l}). This is expressed by the linear relation
\begin{equation}
m_a \, X_a + m_b \, X_b + m_c \, X_c  = 0
\label{lin rel X_abc}
\end{equation}
between the three variables, which is just Eq.\ (\ref{lin rel delta l}) with $\Delta \ell$ replaced by $X$.

Note that in this simple example there is no problem obtaining values for all three variables in every run. In our quantum banana peeling and tasting experiment, we could only ascertain the values (tastes) for two of the three variables (peelings). The experiment with our 3$m$ balance, however, does share several features with both our quantum banana experiment and the raffles meant to simulate them. As with any three balanced random variables the allowed values for the corresponding three correlation coefficients $(\chi_{ab}, \chi_{ac}, \chi_{bc})$ are bound by the elliptope. The correlation will be represented by a point on the surface of the elliptope if some linear combination of the three variables (or, in the quantum case, \emph{the operator representing} the three variables) gives zero. Eq.\ (\ref{lin rel X_abc}) gives this linear combination in the case of the 3$m$ balance. This relation will hold as long as we can ignore errors in our measurement of the displacements $(\Delta \ell_a, \Delta \ell_b, \Delta \ell_c)$ and as long as there are no other masses pulling on the balance's beam. We can represent both of these complicating factors by an additional pan containing some unknown mass pulling on the beam at some unknown location. In that case, the right-hand side of Eq.\ (\ref{lin rel X_abc}) is no longer zero. That, in turn, means that the correlation between these three variables will be represented by a point inside the elliptope rather than on its surface.\footnote{This provides a simple way to understand a comment by Richard Holt, the second H in CHSH, in an interview in 2001 about finding a result in an early test of the CHSH inequality that did not agree with the quantum-mechanical prediction: ``whenever you're looking for a stronger correlation, any kind of systematic error you can imagine typically weakens it and moves it toward the hidden-variable range'' \citep[p.\ 286]{Gilder 2008}.} 

If we set $m_a = m_b = m_ c =m$ and $\delta_a = \delta_b = \delta_c = \delta$  and only allow the values $\pm \delta/2$ for the variables $X_a$, $X_b$ and $X_c$, their sum can no longer be made to vanish. To balance the beam in this case we need to put our thumb on the scale. This too can be represented by an additional pan on the beam. This pan must provide a torque of $mg\delta/2$ to compensate for the smallest torque possible coming from the other three pans combined. The quantity $\left(X_a + X_b + X_c \right)^2$ can never be less than $\delta^2/4$ in this case and we can no longer reach the point $(\chi_{ab}, \chi_{ac}, \chi_{bc}) = 
%(-\delta/2, -\delta/2, -\delta/2)$ 
(-\sfrac12, -\sfrac12, -\sfrac12)$ corresponding to the Tsirelson bound in our quantum banana experiment. Instead we find ourselves right back where we were when we tried to design a raffle to simulate the correlations found in the quantum banana peeling and tasting experiment (cf.\ Eqs.\ (\ref{Tsirelson reprise 1})--(\ref{CHSH reprise 2}) above). 

In our discussion of the 3$m$ balance so far, we have tacitly assumed that we know the masses in the three pans and thus the coefficients in the linear relation between $X_a$, $X_b$ and $X_c$ in Eq.\ (\ref{lin rel X_abc}) that tells us how these three variables are correlated. Typically we will not know those coefficients. Pearson, Yule and Fisher were especially interested in biological variables. These will seldom satisfy a simple linear relation such as the one Eq.\ (\ref{lin rel X_abc}). Suppose, however, that we have reason to believe that three balanced random variables do satisfy a linear relation of this form, with unknown coefficients and a non-zero right-hand side. In terms of our 3$m$ balance this corresponds to the situation in which we do not know the masses in the three pans and cannot rule out that there is a fourth pan somewhere on the beam with another unknown mass. We run the same experiment as before in which we repeatedly put the three pans on the beam somewhere within the allowed intervals always making sure the beam is balanced. Say, we repeat this experiment $n \gg 1$ times. That gives us $n$ triplets of values for $(X_a, X_b, X_c)$. We now plot, say, $X_b$ against $X_a$. Eq.\ (\ref{lin rel X_abc}) with the right-hand side changed from 0 to $U$ (for ``unknown torque'') tells us that
\begin{equation}
X_b  = - \left(\frac{m_a }{m_b} \right) X_a - \left( \frac{m_c}{m_b} \right) X_c \, + \; \frac{U}{m_b}.   
\end{equation}
On the assumption that $U$ does not change much over the course of the $n$ runs and given that for any value of $X_c$ we find we are as likely to find its opposite, the slope of the straight line in our plot of $X_b$ against $X_a$ that best fits the data is a good estimate of the ratio $m_a/m_b$. Plotting $X_c$ against $X_a$, we likewise obtain a good estimate of the ratio $m_a/m_c$. If we are told the value of $m_a$, our experiment thus gives us good estimates of $m_b$ and $m_c$. This is a simple example of regression analysis. It is in this context that Yule studied correlations. Instead of following Yule's algebraic approach, however, we switch to the geometric approaches of Fisher and de Finetti.

For our use of \citet{Fisher 1915, Fisher 1924}, we rely on the paper mentioned at the beginning of this section by \citet[sec.\ 10, pp.\ 72--74; see also Kendall, 1961, pp.\ 55--57]{Aldrich 1998}. Think of the values for the random variables $X_a$, $X_b$ and $X_c$ found in runs $1, 2, \ldots n$ ($n \gg 1$) of our experiment with the 3$m$ balance as the components of the $n$-dimensional vectors
\begin{equation}
\vec{X}^{(n)}_a \equiv \big( X_a^1, \dots X_a^n \big), \; \vec{X}^{(n)}_b \equiv \big( X_b^1, \dots X_b^n \big), \; 
\vec{X}^{(n)}_c \equiv \big( X_c^1, \dots X_c^n \big).
\label{sample vectors}
\end{equation}
We will call such vectors \emph{representative sample vectors}. A sample vector is representative if its components form a balanced sample, i.e., if (a) $\pm X_a, \pm X_b, \pm X_c$ occur with equal frequency and (b) $\sum_{k=1}^n X_a^k = \sum_{k=1}^n X_b^k = \sum_{k=1}^n X_c^k = 0$.\footnote{Here is how we construct a representative sample vector of dimension $n = 2m$ if we only perform a relatively small number $m$ runs of the experiment. For the first $m$ components of the sample vectors $(\vec{X}^{(n)}_a, \vec{X}^{(n)}_b, \vec{X}^{(n)}_c)$, we use the values for $X_a$, $X_b$ and $X_c$ found in these runs; for the last $m$ components, we use \emph{minus} these values. This construction guarantees that the samples from which the sample vectors are constructed are balanced.}

The standard dot product of $\vec{X}^{(n)}_a$ with itself gives $n$ times the variance of $X_a$:
\begin{equation}
\vec{X}^{(n)}_a \! \cdot \! \vec{X}^{(n)}_a = \sum_{k=1}^n \big(X_a^k)^2 = n \langle X_a^2 \rangle = n \, \sigma_a^2,
\label{dot product aa}
\end{equation}
where $\sigma_a$ is the standard deviation. Similar results hold for the dot products of $\vec{X}^{(n)}_b$ and $\vec{X}^{(n)}_c$ with themselves. Hence 
\begin{equation}
\| \vec{X}^{(n)}_a \| = \sqrt{n} \, \sigma_a, \quad \| \vec{X}^{(n)}_b \| = \sqrt{n} \, \sigma_b, \quad \| \vec{X}^{(n)}_c \| = \sqrt{n} \, \sigma_c.
\label{dot products aa bb cc}
\end{equation}
The dot product of $\vec{X}^{(n)}_a$ and $\vec{X}^{(n)}_b$ gives $n$ times the covariance of $X_a$ and $X_b$:
\begin{equation}
\vec{X}^{(n)}_a \! \cdot \! \vec{X}^{(n)}_b = \sum_{k=1}^n  X_a^k  X_b^k = n \langle X_a X_b \rangle.
\label{dot product ab}
\end{equation}
We can use these dot products to define angles between sample vectors. The cosines of these angles are just the Pearson correlation coefficients in Eq.\ (\ref{corr coeff gen})). We verify this for the cosine of the angle $\vartheta_{ab}$ between $\vec{X}^{(n)}_a$ and $\vec{X}^{(n)}_b$:
\begin{equation}
\cos{\vartheta_{ab}} = \frac{\vec{X}^{(n)}_a \! \cdot \! \vec{X}^{(n)}_b}{\| \vec{X}^{(n)}_a \| \| \vec{X}^{(n)}_b \|} = \frac{n  \langle X_a X_b \rangle}{\sqrt{n} \, \sigma_a \, \sqrt{n} \, \sigma_b} =
 \frac{\langle X_a X_b \rangle}{ \sigma_a \sigma_b} = \overline{\chi}_{ab}.
\label{sample vectors a b angle}
\end{equation}

\begin{figure}[h!]
 \centering
   \includegraphics[width=4in]{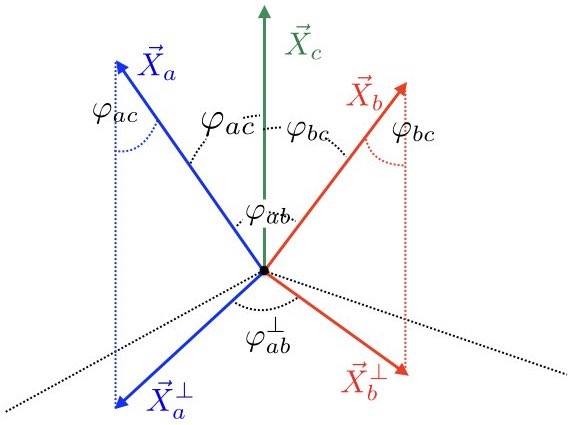} 
   \caption{Vectors representing balanced samples of triplets of values of the random variables $(X_a, X_b, X_c)$.}
   \label{vectors4yule}
\end{figure}

Suppressing the superscripts $(n)$, we now decompose the sample vectors $\vec{X}_a$ and $\vec{X}_b$ into components parallel and perpendicular to the sample vector $\vec{X}_c$:
\begin{equation}
\vec{X}_a = \vec{X}_a^\parallel + \vec{X}_a^\perp, \quad \vec{X}_b = \vec{X}_b^\parallel + \vec{X}_b^\perp
\label{split per par}
\end{equation}
This decomposition is illustrated in Figure \ref{vectors4yule}. The parallel components can be seen as measures of the correlation between the random variables $X_a$ and $X_c$ and the correlation between $X_b$ and $X_c$; the perpendicular components as measures of the so-called partial or residual correlation between $X_a$ and $X_b$ \citep[p.\ 73]{Aldrich 1998}.\footnote{As we saw in Section \ref{1.5}, the tastes two bananas, one peeled $\hat{a}$ by Alice, the other peeled $\hat{b}$ by Bob, are completely \emph{uncorrelated} if the peeling directions $\vec{e}_a$ and $\vec{e}_b$ are orthogonal. It follows that the geometry of these sample vectors differs from the geometry of state vectors in Hilbert space in quantum mechanics. We will return to this point in the context of our discussion of \citet{De Finetti 1937} below.\label{orthogonal preview}} As Fisher put it (translated into our notation): 
\begin{quote}
[T]he correlation between [$X_a$] and [$X_b$] \ldots\ will be the cosine of the angle between  [$\vec{X}_a$] and [$\vec{X}_b$] \ldots\
[T]he partial correlation between [$X_a$] and [$X_b$] is the cosine of the angle between the projections of [$\vec{X}_a$] and [$\vec{X}_b$] upon the region perpendicular to [$\vec{X}_c$]
\citep[pp.\ 329--330]{Fisher 1924}.
\end{quote}
The lengths of the parallel and perpendicular components of $\vec{X}_a$ and $\vec{X}_b$  are
\begin{equation}
\begin{array}{c}
 \| \vec{X}_a^\parallel \| \! = \!  \|  \vec{X}_a  \| \cos{\vartheta_{ac}} \! = \!  \sqrt{n} \, \sigma_a \cos{\vartheta_{ac}}, \quad
\| \vec{X}_a^\perp \| \! = \!  \|  \vec{X}_a  \| \sin{\vartheta_{ac}} \! = \!   \sqrt{n} \, \sigma_a \sin{\vartheta_{ac}},    \\[.4cm]
\| \vec{X}_b^\parallel \| \! = \!  \|  \vec{X}_b  \| \cos{\vartheta_{bc}} \! = \!   \sqrt{n} \, \sigma_b \cos{\vartheta_{bc}}, \quad
\| \vec{X}_b^\perp \| \! = \!  \|  \vec{X}_b  \| \sin{\vartheta_{bc}} \! = \!  \sqrt{n} \, \sigma_b \sin{\vartheta_{bc}}, 
\end{array}
\label{lengths per par}
\end{equation}
where we used Eq.\ (\ref{dot products aa bb cc}) for $ \|  \vec{X}_a  \|$ and $ \|  \vec{X}_b  \|$. 

We can rewrite the dot product in Eq.\ (\ref{sample vectors a b angle}) as
\begin{equation}
\cos{\vartheta_{ab}} = \frac{\big( \vec{X}_a^\parallel + \vec{X}_a^\perp \big) \! \cdot \! \big(  \vec{X}_b^\parallel + \vec{X}_b^\perp \big)}{n \sigma_a \sigma_b}
= \frac{\vec{X}_a^\parallel \! \cdot \! \vec{X}_b^\parallel + \vec{X}_a^\perp \! \cdot \! \vec{X}_b^\perp}{n \sigma_a \sigma_b}.
\label{cos theta ab}
\end{equation}
With the help of Eq.\ (\ref{lengths per par}), the parallel components of $\vec{X}_a$ and $\vec{X}_b$ can be written as
\begin{equation}
\begin{array}{c}
\vec{X}_a^\parallel = \| \vec{X}_a \| \cos{\vartheta_{ac}} \displaystyle{\frac{\vec{X}_c}{\| \vec{X}_c \|}} = \displaystyle{\frac{\sigma_a}{\sigma_c}} \cos{\vartheta_{ac}} \, \vec{X}_c, 
\\[.5cm]
\vec{X}_b^\parallel = \| \vec{X}_b \| \cos{\vartheta_{bc}} \displaystyle{\frac{\vec{X}_c}{\| \vec{X}_c \|}} = \displaystyle{\frac{\sigma_b}{\sigma_c}} \cos{\vartheta_{bc}} \, \vec{X}_c.
\end{array}
\label{ab par vecs and vec c} 
\end{equation}
Using these expressions along with $\vec{X}_c \! \cdot \! \vec{X}_c = \| \vec{X}_c \|^2 = n \, \sigma_c^2$, we can write the dot product of the parallel components of $\vec{X}_a$ and $\vec{X}_b$ as
\begin{equation}
\vec{X}_a^\parallel \! \cdot \! \vec{X}_b^\parallel = n \, \sigma_a \sigma_b \cos{\vartheta_{ac}} \cos{\vartheta_{bc}}.
\label{dot par comps}
\end{equation}
The dot product of the perpendicular components can be written as
\begin{equation}
\vec{X}_a^\perp \! \cdot \! \vec{X}_b^\perp = \| \vec{X}_a^\perp \| \| \vec{X}_b^\perp \|   \cos{\vartheta_{ab}^\perp} =
n \, \sigma_a \sigma_b \sin{\vartheta_{ac}} \sin{\vartheta_{bc}} \cos{\vartheta_{ab}^\perp},
\label{dot perp comps}
\end{equation}
where, once again, we used Eq.\ (\ref{lengths per par}). Inserting Eqs.\ (\ref{dot par comps})--(\ref{dot perp comps}) into Eq.\ (\ref{cos theta ab}), we arrive at
\begin{equation}
\cos{\vartheta_{ab}} = \cos{\vartheta_{ac}} \cos{\vartheta_{bc}} + \sin{\vartheta_{ac}} \sin{\vartheta_{bc}} \cos{\vartheta_{ab}^\perp}.
\label{cos theta ab 2}
\end{equation}
Solving for $\cos{\vartheta_{ab}^\perp}$, we find that\footnote{If all vectors in Figure \ref{vectors4yule} are turned into unit vectors so that there tips all lie on a unit sphere, one easily recognizes that Eq.\ (\ref{cos theta ab perp}) is nothing but the spherical law of cosines. This illustrates the observation by Pearson quoted at the beginning of this subsection. The relation between the ``full'' and the partial correlation between $X_a$ and $X_b$ is thus fairly trivial. Borrowing a comment by \citet[p.\ 73]{Aldrich 1998} on a closely related aspect of Fisher's analysis, we can say that Fisher's derivation of this formula for the partial correlation amounts to ``something of a self-annihilating insight.''}
\begin{equation}
\cos{\vartheta_{ab}^\perp} = \frac{\cos{\vartheta_{ab}} - \cos{\vartheta_{ac}} \cos{\vartheta_{bc}}}{\sin{\vartheta_{ac}} \sin{\vartheta_{bc}}}. 
\label{cos theta ab perp}
\end{equation}
If $\overline{\chi}$'s are substituted for cosines of $\vartheta$'s and expressions of the form $\sqrt{\big( 1 - \overline{\chi}^2 \big)}$ for sines of $\vartheta$, this turns into:
\begin{equation}
\cos{\vartheta_{ab}^\perp} = \frac{\overline{\chi}_{ab} - \overline{\chi}_{ac} \overline{\chi}_{bc}}{\sqrt{\big( 1 - \overline{\chi}_{ac}^2 \big)} \sqrt{\big( 1 - \overline{\chi}_{bc}^2 \big)}}. 
\label{elliptope Yule 1}
\end{equation}
The quantity on the right-hand side can be found in \citet[p.\ 485]{Yule 1896}, who denotes it as $\rho_{12}$ and calls it the ``net coefficient of correlation between $x_1$ and $x_2$'' ($X_a$ and $X_b$ in our notation; Yule's $x_3$, likewise, is $X_c$ in our notation). Squaring both sides of Eq.\ (\ref{elliptope Yule 1}) we find:
\begin{equation}
\cos^2{\! \vartheta_{ab}^\perp} = \frac{ \big( \overline{\chi}_{ab} - \overline{\chi}_{ac} \overline{\chi}_{bc} \big)^2}{\big(1 - \overline{\chi}_{ac}^2 \big) \big( 1 - \overline{\chi}_{bc}^2 \big)}.
\label{elliptope Yule 2}
\end{equation}
Since $\cos^2{\! \vartheta_{ab}^\perp} \le 1$, we have the inequality \citep[p.\ 486]{Yule 1896}
\begin{equation}
  \big(\overline{\chi}_{ab} - \overline{\chi}_{ac} \overline{\chi}_{bc} \big)^2 \; \le \; \big( 1 - \overline{\chi}_{ac}^2 \big) \big( 1 - \overline{\chi}_{bc}^2 \big);
  \label{lastminute}
\end{equation}
or, expanding both sides,  
\begin{equation}
\overline{\chi}_{ab}^2 - 2 \overline{\chi}_{ab} \overline{\chi}_{ac} \overline{\chi}_{bc} + \overline{\chi}_{ac}^2 \overline{\chi}_{bc}^2
\; \le \; 1 - \overline{\chi}_{ac}^2 - \overline{\chi}_{bc}^2 + \overline{\chi}_{ac}^2 \overline{\chi}_{bc}^2.
\label{elliptope deja vu}
\end{equation}
The terms $\overline{\chi}_{ac}^2 \overline{\chi}_{bc}^2$ before and after the `$\le$' cancel and we see that this inequality is equivalent to the one in Eq.\ (\ref{inf the 5}) that defines the elliptope.

The construction above with sample vectors is problematic as probabilities only coincide with relative frequencies in the $n\to\infty$ limit. 
%{\michael There should probably be more elaboration about why it's objectionable.}
To remedy this, we consider the random variables themselves as vectors. This is actually a common approach in modern probability theory \citep[see, e.g.,][]{Fristedt and Gray 1997}. An early instance of it can be found in a paper of \citet{De Finetti 1937}.\footnote{We used the translation by Luca Barone and Peter Laurence, who also co-authored a commentary on the paper \citep{Laurence 2008}.}

De Finetti begins his geometric interpretation by observing that ``since we can consider linear combinations of random variables, we can interpret them as vectors in an `abstract space'\,'' \citep[p.\ 5]{De Finetti 1937}.\footnote{Unless noted otherwise, page references in the remainder of this section are to \citet{De Finetti 1937} } That is to say, linear combinations of random variables are also random variables and so form a vector space. Standard deviations can be interpreted as the norm of vectors in this space (cf.\ Eqs.\ (\ref{dot product aa})--(\ref{dot products aa bb cc}) for Fisher's sample vectors), 
\begin{equation}
\| X\| \equiv \sqrt{\langle X^2 \rangle}.
\end{equation}
%which we may recognize as the $L^2$-norm with respect to the probability measure $P$. 
The distance (or metric) between two random variables $X$ and $Y$ can then be defined as:\footnote{One subtlety of this definition is that $d(X,Y)=0$ when $X$ and $Y$ differ by a constant random variable (i.e., a random variable guaranteed to have the same value every time it is measured). Since we assume our random variables to have zero mean, this amounts to two random variables being equivalent if they are equal with probability 1, i.e., almost surely equal. This point is emphasized by de Finetti earlier in the paper, with him commenting that such 'coinciding' random variables are represented by the same vector. Since  random variables that are almost surely equal agree in their expectation values, we will not distinguish them in our discussion of de Finetti's paper (p.\ 5).}
\begin{equation}
d(X,Y) \equiv \|X-Y\|.
\end{equation}
Similarly, covariances can be interpreted as inner products of two random variables in this vector space (p.\ 6; cf.\ Eq.\ (\ref{dot product ab}) for Fisher's sample vectors):\footnote{The notation $\langle \cdot \,, \cdot \rangle$ is ours: De Finetti does not introduce a special notation for this inner product.}
\begin{equation}
 \langle X,Y\rangle \equiv \langle X Y\rangle. 
\end{equation}
Hence de Finetti deduces that this vector space of random variables is not only a \emph{metric space} but also a (normed) \emph{inner product space}. One can go further and complete this space with respect to the above metric, resulting in a Hilbert space of random variables. 

De Finetti in fact gives two such Hilbert space interpretations, one (on p.\ 6) where the inner product is the covariance $\langle (X-\langle X\rangle) (Y-\langle Y\rangle)\rangle$, the other (on p.\ 8) where the inner product is $\langle X Y\rangle$ (even if $\langle X \rangle$ and/or $\langle Y \rangle$ are non-zero). The first Hilbert space can be viewed as the subspace of the second, consisting of vectors orthogonal to all constant random variables, and so it is the second Hilbert space which is more common to the literature \citep[see again][]{Fristedt and Gray 1997}.  For the purposes of this section, however, we only consider random variables with $\langle X \rangle = \langle Y \rangle = 0$ and so will not distinguish the two spaces.

De Finetti's purpose in his 1937 paper is not to define a Hilbert space but to use his vector space to bolster geometric intuition about random variables.\footnote{While de Finetti does not consider issues of completeness in this paper, he cites a prior note of his (not on probability theory) in which he references Hilbert space and appears to directly address the issue of completeness \citep[p.\ 248 and p.\ 254, respectively]{De Finetti 1930}. It thus seems reasonable to assume that de Finetti understood perfectly well that his vector space of random variables is a Hilbert space. Joseph L.\ \citet{Doob 1934} appears to have been the first to use Hilbert space in the context of general probability theory. Applications to probability theory are not mentioned in the history of functional analysis by \citet{Birkhoff and Kreyszig 1984}. A promising source for filling this gap in the literature on the history of mathematics is \citet{Bingham 2000}.} To this end, he uses his inner product to conclude---like \citet{Fisher 1915,Fisher 1924} before him but now treating the random variables themselves as vectors---that ``the correlation coefficient is the cosine of the angle $\alpha(X,Y)$ between vectors $X$ and $Y$" (p.\ 6). He elaborates:
%Contextualize with reference to Section \ref{1.5}:
\begin{quote}
Zero correlation means orthogonality; positive or negative correlation means that $\alpha$ is acute or obtuse, respectively; for the extreme cases \ldots\ $\alpha = 0$ and $\alpha = \pi$, respectively, the two vectors \ldots\ only differ by a multiplicative constant, positive or negative, respectively \citep[p.\ 6]{De Finetti 1937}.
\end{quote}
In our quantum banana peeling and tasting experiment the taste $X_a^A$ found by Alice peeling $\hat{a}$ is perfectly correlated to the taste $X_b^B$ found by Bob peeling $\hat{b}$ if the angle $\varphi_{ab}$ between the unit vectors $\vec{e}_a$ and $\vec{e}_b$ for the corresponding peeling directions is $180\degree$. As we noted in Section \ref{1.5}, the angle between the corresponding eigenvectors $|+\rangle_a$ and $|+\rangle_b$ in that case is $90\degree$. De Finetti's angle $\alpha$ thus corresponds to the angle between the vectors  $\vec{e}_a$ and $\vec{e}_b$  in ordinary space rather than to the angle between $|+\rangle_a$ and $|+\rangle_b$ in Hilbert space (note that the perfect correlation for $\varphi_{ab} = 180\degree$ turns into a perfect anti-correlation once $-X_b^B$ is used as a proxy for $X_b^A$; see Section \ref{1.6.2}). In de Finetti's formalism, we can thus think of $(X^A_a/\sigma_a)$ as a vector that coincides with $\vec{e}_a$. This vector, in turn, is directly related to the operator for the spin component being measured, $\hat{S}_a \equiv \hat{\vec{S}} \cdot \vec{e}_a$. Though we will not pursue this any further in this paper, we note that the Hilbert space encountered here is not the familiar Hilbert space of state vectors but a Hilbert space of (spin) operators (where the inner product is simply the expectation value of the product of the operators in the quantum state of the system).

Like Fisher (see Eq.\ (\ref{split per par})), de Finetti notes that ``every random variable $Y$ can be decomposed into two components, one correlated with $X$ (the parallel component) and one uncorrelated with $X$ (the orthogonal component)" (p.\ 6). Finally, de Finetti considers the angles between a triplet of random variables $(X, Y, Z)$.
\begin{quote}
If $\alpha(X, Y)$ is the angle between two random variables $X$ and $Y$, a third random variable $Z$ cannot form two arbitrary angles $\alpha(X, Z)$ and $\alpha(Y, Z)$, but we must have (obviously, if we think of the geometric picture) $\alpha (X, Y)$ $\leq \alpha(X, Z) + \alpha(Y, Z) \leq 2 \pi - \alpha (X, Y)$; we have the extreme case $\alpha(X, Z) + \alpha(Y, Z) = \alpha(X, Y)$ if and only if $Z = aX + bY$, $a > 0$, $b > 0$ (the coplanar vector, included in the concave angle between the two vectors), and the other $\alpha(X, Z) + \alpha(Y, Z) = 2 \pi - \alpha(X, Y)$ if and only if $Z = -(aX + bY)$, $a> 0$, $b > $0 (the aforementioned condition applied to minus the same vector). This shows that there are some constraints for the degrees of pairwise correlation among different random variables  \citep[p.\ 6]{De Finetti 1937}.
\end{quote}
These angle inequalities are exactly those in Eq.\ (\ref{angle inequalities}) in Section \ref{1.5}. De Finetti considers the following special case:
\begin{quote}
In particular, if three random variables are all equally correlated, since pairwise they are unable to form an angle bigger than $\sfrac{2\pi}{3}$ (i.e., $120\degree$), the correlation coefficient cannot be smaller than $-\sfrac12$ \citep[pp.\ 6--7]{De Finetti 1937}.
\end{quote}
The observant reader will recognize the $120\degree$ case as the scenario where the Tsirelson bound is fulfilled. De Finetti identified this scenario more than forty years before either \citet{Cirel'son 1980} or \citet[see note \ref{Accardi Fedullo}]{Accardi and Fedullo 1982} published on the subject!

%SECTION 3 (label: 3}
\section{Generalization to the singlet state of two particles with higher spin} \label{2}
%!TEX root =  ./JanasJanssenCuffaro-August2019.tex

%SECTION 3 -- INTRO
%\section{Generalization to the singlet state of two particles with higher spin} \label{2}
%Three settings and more than two outcomes per setting: 
%Generalization to pairs of higher-spin particles entangled in the singlet state

In Section \ref{1} we studied pairs of bananas that behave like pairs of spin-$\frac12$ particles in the singlet state. This is just one type of banana in Bananaworld. There are many more. In Section \ref{1.2}, for instance, we came across Popescu-Rohrlich bananas. \citet[Ch.\ 6]{Bub 2016} has made extensive study of other exotic species, such as Aravind-Mermin bananas and Klyachko bananas. One could add to this taxonomy by introducing pairs of bananas that behave like pairs of particles of arbitrary integer or half-integer spin $s \ge 1$ in the singlet state. For $s=1$, bananas would taste yummy ($+$), nasty ($-$), or meh ($0$). As we move to higher spin, we would have to invent more refined taste palettes. The banana imagery thus starts to feel forced for higher spin and we will largely dispense with it in the remainder of this paper. Instead we will phrase our analysis directly in terms of spin. So rather than have Alice and Bob peel and taste pairs of bananas, we imagine them sending pairs of particles through Dubois magnets and measuring a component of their spin, choosing between three different directions, represented, as before, by unit vectors $(\vec{e}_a, \vec{e}_b, \vec{e}_c)$, corresponding to settings $(\hat{a}, \hat{b}, \hat{c})$. In Section \ref{2.1} we present the quantum-mechanical analysis of the correlations found in these measurements, extending our discussion of the spin-$\frac12$ case in Section \ref{1.5}. In Section \ref{2.2} we design raffles like the ones we introduced in Section \ref{1.4} to simulate these correlations. 

For the quantum-mechanical analysis we rely on the standard treatment of rotation in quantum mechanics (see, e.g., Messiah 1962, Vol.\ 2, Appendix C, or Baym 1969, Ch.\ 17). This will lead us to the so-called Wigner d-matrices (see Eq.\ (\ref{wigner})) with which we can readily compute the probabilities entering into the correlation arrays for measurements on the singlet state of two particles with arbitrary (half-) integer spin $s$. After showing how the results we found in Section \ref{1.4} for the spin-$\frac12$ case are recovered (and justified) in this more general formalism, we use it to find the entries of a typical cell in the correlation array for the spin-1 case (see Figure \ref{CA-cell-spin1-chi}). We then prove that the correlation arrays for higher-spin cases share some key properties with those for the spin-$\frac12$ and spin-1 cases (Sections \ref{2.1.1}--\ref{2.1.3}). 

In Section \ref{2.1.4}, we show that all such correlation arrays have uniform marginals and are therefore non-signaling. In Section \ref{2.1.5}, we show that they can still be parametrized by the anti-correlation coefficients for three of their off-diagonal cells\footnote{In Section \ref{2}, we referred cells on and off the diagonal of various correlation arrays. From now on, we will use the more economical designations `diagonal cells' and `off-diagonal cells'.\label{diag cells economy}} and that these coefficients are still given by the cosines of the angles between measuring directions,
\begin{equation}
\chi_{ab} = \cos{\varphi_{ab}}, \quad \chi_{ac} = \cos{\varphi_{ac}},  \quad \chi_{bc} = \cos{\varphi_{bc}},
\label{intro sec 2a}
\end{equation} 
and subject to the same constraint we found in the spin-$\frac12$ case  in Section \ref{1.4} (see Eq.\ (\ref{QM14})): 
\begin{equation}
1 - \chi_{ab}^2 - \chi_{ac}^2 - \chi_{bc}^2 + 2 \, \chi_{ab} \, \chi_{ac} \, \chi_{bc} \ge 0.
\label{intro sec 2b}
\end{equation}
It follows that the class of correlations allowed by quantum mechanics in measurements on the singlet state of two particles with  (half-)integer spin $s$ can be represented by the elliptope in Figure \ref{elliptope} regardless of the value of $s$.

In Section \ref{2.1.6}, we turn to a property of the correlation arrays for these higher-spin cases we did not pay much attention to in the spin-$\frac12$ case: the symmetries of their cells. We show that, for arbitrary (half)-integer $s$, any cell in the correlation array for measurements on the singlet state of particles with (half-)integer spin $s$ is \emph{centrosymmetric}, \emph{symmetric} and \emph{persymmetric}, i.e., it is unchanged under reflection about its center, across its main diagonal and across its main anti-diagonal (see Eq.\ (\ref{Prob sym}) for what this means in terms of the probabilities that form the entries of such cells). 

In Section \ref{2.2}, we design raffles to simulate these quantum correlations. In preparation for this, we formalize the description and analysis of the raffles we used in Section \ref{1.4} for the spin-$\frac12$ case (Section \ref{2.2.1}). In Sections \ref{2.2.2}--\ref{2.2.4}, we adapt the formalism developed for this case to design raffles that simulate---to the extent that this at all possible---the correlation arrays for measurements on the singlet state of two particles with higher spin. In doing so, we run into four main complications compared to the spin-$\frac12$ case. 

First, as we already noted in Section \ref{1.6}, we can no longer admit single-ticket raffles. Our raffle tickets only have two outcomes per setting. If there are more than two possible outcomes (and for spin $s$ there are $2s+1$), it is therefore impossible to have diagonal cells of the form shown in Figure \ref{diag-cell-sxs}. Such single-ticket raffles not only fail to reproduce the quantum correlations; without identical diagonal cells, the anti-correlation coefficients for the off-diagonal cells no longer suffice to characterize the correlations generated by these raffles. To get around this problem we need to restrict ourselves to mixed raffles that give uniform marginals. These consist of combinations of tickets such that all $2s+1$ outcomes occur with the same frequency. This will guarantee that the diagonal cells in the resulting correlation arrays all have the form of the cell in Figure \ref{diag-cell-sxs}. Since our raffles are non-signaling by construction, this ensures uniform marginals for the entire correlation array.   

The second complication has to do with the relationship between probabilities and expectation values (or anti-correlation coefficients). In the spin-$\frac12$ case, the probabilities 
in any cell of the correlation array could be parametrized by the anti-correlation coefficient for that cell (see Figures \ref{CA-2set2out-cell} and \ref{CA-3set2out-non-signaling-chis} in Section \ref{1.3}). In the quantum correlation arrays, as noted above, this remains true for arbitrary (half-) integer spin $s$. As soon as there are more than two possible outcomes, however, it fails for our raffles. This severely limits our ability to simulate the quantum correlation arrays for higher-spin cases. We can design mixed raffles that simulate the diagonal cells of the quantum correlation arrays and that give the correct values for the \emph{anti-correlation coefficients} of the off-diagonal cells; yet these raffles will, in general, still not give the right values for the \emph{probabilities} in the off-diagonal cells. In Section \ref{2.2.2}, we will encounter a striking example of this complication. We will construct a raffle for the spin-1 case for which the sum of the anti-correlation coefficients is $-\sfrac32$, the Tsirelson bound for this setup (see Eq.\ (\ref{Mermin CHSH integer spin}) in Section \ref{1.6}). Yet the off-diagonal cells of the correlation array for this raffle are different from those in the quantum correlation array it was meant to simulate (see Eq.\ (\ref{off diag cell quantum v raffle})). 

The third (less serious) complication has to do with the symmetries of the off-diagonal cells in the correlation array. The design of our raffles guarantees that all cells in their correlation arrays are centrosymmetric. The condition for centrosymmetry of such cells, say the  $\hat{a} \, \hat{b}$ one,  is (see Eq.\ (\ref{Prob sym})): 
\begin{equation}
\mathrm{Pr}(m_1 m_2| \hat{a} \,\hat{b}) = \mathrm{Pr}(-m_1 -\!m_2| \hat{a} \,\hat{b}).
\label{raffle cell centro-symmetry} 
\end{equation}
This condition is automatically satisfied by our raffles: it simply expresses that Alice and Bob are as likely to get one side of the ticket as the other. The entries in cells of correlation arrays in the spin-$\frac12$ case form $2 \times 2$ matrices. In that case, centrosymmetry trivially implies both symmetry and persymmetry. For the $(2s +1) \times (2s +1)$ matrices formed by the entries in cells of correlation arrays in the spin-$s$ case with $s \ge 1$, this is no longer true (although any two of these symmetries still imply the third). Cells in the quantum correlation arrays, as noted above, have all three symmetries, regardless of the spin of the particles in the singlet state on which Alice and Bob perform their measurements. To correctly simulate this feature of the quantum correlations we thus need to impose additional symmetry conditions on our raffles. Fortunately, this can be done without too much trouble.
%\footnote{It does have the unfortunate consequence, however, that we cannot formally prove that the polyhedra we will construct for the higher-spin cases asymptotically approach the elliptope in the limit that $s$ goes to infinity, as strongly suggested by looking at these polyhedra (see note \ref{no-convergence-proof}).} 

The fourth complication is perhaps the most obvious one. As the number of outcomes increases, so does the number of different ticket types in our raffles. Figure \ref{raffle-tickets-3set3out-i-xiv} shows the $(3^3 + 1)/2 = 14$ different ticket types for raffles in the spin-1 case. Figures \ref{raffles-spin32-tickets-mu} and \ref{raffles-spin32-tickets-nu} show some of the $4^3/2 = 32$ different ticket types for the spin-$\frac32$ case. In dealing with these higher-spin cases, we therefore turned to the computer for guidance.

As in the spin-$\frac12$ case, we will represent the class of triplets of anti-correlation coefficients $(\chi_{ab}, \chi_{ac}, \chi_{bc})$ for \emph{admissible raffles} in the spin-$s$ case (i.e., raffles that give uniform marginals and meet the symmetry requirements) by a polyhedron in the same 3-dimensional non-signaling cube as before. Henceforth, we will call this  the \emph{anti-correlation polyhedron}. In the spin-$\frac12$ case, the anti-correlation polyhedron doubles as the local polytope (see Figures \ref{LQP} and \ref{elliptope-LQPslice}). In the spin-$s$ case (with $s \ge 1$), the anti-correlation polyhedron is a particular (highly informative) projection of (a restricted version of) a now higher-dimensional local polytope (restricted by our admissibility conditions) to three dimensions (cf.\ the discussion above about complications in the relationship between probabilities and anti-correlation coefficients). The flowchart in Figure \ref{flowchart} shows how we get from the local polytope to the anti-correlation polyhedron in the spin-1 case. With considerable help from the computer, we were able to construct anti-correlation polyhedra for $s = 1, \sfrac32,  2, \sfrac52$ (see Figures \ref{polytope-spin1}, \ref{SpinThreeHalfFace} and \ref{FacetsSpin2Spin52}). 

We pay special attention to admissible raffles for which the sum $\chi_{ab} + \chi_{ac} + \chi_{bc}$ takes on its minimum value (see Eqs.\ (\ref{Mermin CHSH half-integer spin}) and (\ref{Mermin CHSH integer spin}) in Section \ref{1.6.3}). In constructing these raffles, we take advantage of the insight that they will involve tickets for which the sum of the outcomes on both sides is either zero (for integer spin) or $\sfrac12$ (for half-integer spin) (see Figures \ref{admissible-raffles-spin1}, \ref{raffles-spin32-tickets-mu}, \ref{raffles-spin32-tickets-nu} and Table \ref{Spin2TicketGroups} for the tickets in such raffles). Comparing the anti-correlation polyhedra for higher spin values $s$  to the tetrahedron for $s = \sfrac12$, we see that these polyhedra get closer and closer to the elliptope as the number of outcomes $2s + 1$ increases (see Figure \ref{polytopevolume}).\footnote{We have not been able to construct a formal proof of this convergence  (see note \ref{no-convergence-proof}).} This is just what we would expect given what we learned in Section \ref{1.6.3}, viz.\ that Eq.\ (\ref{intro sec 2b}) for the elliptope determines the broadest conceivable class of triplets of (anti-)correlation coefficients.
%Slogan: there is no physics beyond the elliptope. 

%SUBSECTION 3.1 (labels: 3.1, 3.1.1 through 3.1.X)
\subsection{The quantum correlations} \label{2.1}
%!TEX root =  ./JanasJanssenCuffaro-August2019.tex

%SUBSECTION 3.1
%\subsection{The quantum correlations} \label{2.1}

%SUBSUBSECTION 3.1.1
\subsubsection{Quantum formalism for one spin-$s$ particle} \label{2.1.1}

We review the formalism for spin angular momentum of particles of arbitrary integer or half-integer spin $s$, starting with the one-particle case.\footnote{Our presentation roughly follows the treatment of angular momentum in \citet[Vol.\ 2, Appendix C]{Messiah 1962}.} The state of a spin-$s$ particle with component $m\hbar$ in the $z$-direction is represented by a state vector $|s,m\rangle_z$ in a one-particle Hilbert space (where $\hbar \equiv h/2\pi$ and $h$ is Planck's constant). These vectors are simultaneous eigenvectors of the Hermitian operators 
\begin{equation}
\hat{S}_z, \quad \hat{S}^2\equiv \hat{S}_x^2+\hat{S}_y^2+\hat{S}_z^2,
\label{Sz and S^2}
\end{equation}
with eigenvalues $m \hbar$ (with $m \in \{ -s, \ldots, s\}$) and $s(s+1) \hbar^2$, respectively:
%\footnote{For the counterpart of Planck's constant in Bananaworld, see the paragraph following Eq.\ (\ref{cov def}).}
\begin{equation} 
\hat{S}_z |s,m\rangle_z =  m  \hbar \, |s,m\rangle_z,\quad \hat{S}^2 |s,m\rangle_z= s(s+1)   \hbar^2 \, |s,m\rangle_z,
\label{state dfn}
\end{equation} 
We will follow the common practice of setting $\hbar =1$. The operators $\hat{S}_x$, $\hat{S}_y$ and $\hat{S}_z$ can be thought of as components of a vector
\begin{equation}
\hat{\vec{S}} \equiv \left(\hat{S}_x, \hat{S}_y, \hat{S}_z \right).
\label{def S vector}
\end{equation}
The operator $\hat{S}^2$ represents the length of this vector. 

The simultaneous eigenvectors of $\hat{S}_z$ and $ \hat{S}^2$ form an orthonormal basis of the $(2s+1)$-dimensional one-particle Hilbert space,
\begin{equation}
\big\{|s,m \rangle_{z}\big\}_{m=-s}^s \;\; \mathrm{with} \;\; {_{z\!}}\langle s,  m | s,  m' \rangle_z = \delta_{mm'},
\label{onb one-particle Hilbert space}
\end{equation}
where $\delta_{mm'}$ is the Kronecker delta ($\delta_{mm'} = 1$ if $m=m'$ and $\delta_{mm'} = 0$ if $m \neq m'$). The operators $\hat{S}_x$, $\hat{S}_y$ and $\hat{S}_z$ satisfy the commutation relations 
\begin{equation}
[\hat{S}_x,\hat{S}_y]=i \hat{S}_z,\quad [\hat{S}_y,\hat{S}_z]=i \hat{S}_x,\quad [\hat{S}_z,\hat{S}_x]=i \hat{S}_y.
\label{ops dfn}
\end{equation}

We will also use the raising/lowering operators $\hat{S}_{\pm} \equiv \hat{S}_x\pm i \hat{S}_y$. The action of $\hat{S}_{\pm}$ on one of the orthonormal basis vectors $|s,m\rangle_z$ is given by
\begin{equation}
\hat{S}_\pm |s,m\rangle_z = C_\pm(s,m)|s,m \pm1\rangle_z.
\label{raising operator}
\end{equation}
Using that $|s,m\rangle_z$ and $|s,m\pm1\rangle_z$ are unit vectors, we can find expression for the constants $C_\pm(s,m)$. We do this for $C_+$. We start from
\begin{equation}
C_+(s,m)^2 = C_+(s,m)^2 {_z}\langle s,m+1|s,m+1\rangle_z = {_z\!}\langle s,m|\hat{S}_+^\dagger \hat{S}_+|s,m\rangle_{\!z}.
\label{raising norm 1}
\end{equation}
Since
\begin{equation}
\hat{S}_+^\dagger \hat{S}_+ = (\hat{S}_x-i \hat{S}_y)(\hat{S}_x+i \hat{S}_y) = \hat{S}_x^2+\hat{S}_y^2 +i[\hat{S}_x,\hat{S}_y]=\hat{S}^2 - \hat{S}_z^2-\hat{S}_z,
\label{raising}
\end{equation}
we can rewrite the right-hand side as
\begin{equation}
{_z\!}\langle s,m|\big(\hat{S}^2-\hat{S}_z^2-\hat{S}_z\big)|s,m\rangle_{\!z} = (s(s+1)-m(m+1)) \, {_z\!}\langle s,m|s,m\rangle_{\!z}.
\label{raising norm 2}
\end{equation}
Choosing $C_+(s,m)$ to be positive and real (the so-called Condon-Shortley convention), we conclude that
\begin{equation}
C_+(s,m)=\sqrt{s(s+1)-m(m+1)}.
\label{coef raising}
\end{equation}
A corollary of this phase convention is that the operator $i\hat{S}_y = \frac12 \hat{S}_{+}-\frac12 \hat{S}_-$ has real matrix elements in the $|s,m\rangle_{\!z}$ basis.

\begin{figure}[h]
 \centering
   \includegraphics[width=1.4in]{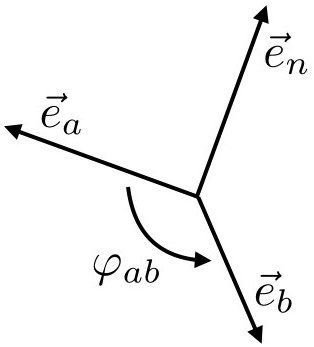} 
   \caption{Rotation by an angle $\varphi_{ab}$ about the direction $\vec{e}_n$, mapping $\vec{e}_z$ to $\vec{e}_a$.}
   \label{rotation}
\end{figure}

As we already saw in Section \ref{1.6}, one can also define spin operators associated with directions other than the Cartesian axes. The spin operator associated with the direction given by the unit vector $\vec{e}_a=(a_x,a_y,a_z)$ is defined as
\begin{equation}
\hat{S}_a \equiv \hat{\vec{S}} \cdot \vec{e}_a = \hat{S}_x a_x+\hat{S}_y a_y+\hat{S}_z a_z.
\label{spin op}
\end{equation}

Such spin operators generate rotations in Hilbert space. Let $\hat{S}_a$ and $\hat{S}_b$ be spin operators associated with the directions given by the unit vectors $\vec{e}_a$ and $\vec{e}_b$.  Let $\vec{e}_n$ be a unit vector in the direction of the cross product $\vec{e}_a\times \vec{e}_b$, so that we get from $\vec{e}_a$ to $\vec{e}_b$ by rotating around $\vec{e}_n$ by the angle $\varphi_{ab}$ between them (see Figure \ref{rotation}). This rotation is implemented in Hilbert space by the rotation operator $e^{-i \varphi_{ab} \hat{S}_n}$. It can be shown that the transformation from the spin operator $\hat{S}_a$ associated with $\vec{e}_a$ to the spin operator $\hat{S}_b$ associated with $\vec{e}_b$ is given by:
\begin{equation}
 \hat{S}_b = e^{-i\varphi_{ab} \hat{S}_n}\hat{S}_a e^{i\varphi_{ab}  \hat{S}_n}
\label{ops rot}
\end{equation}
\citep[Vol.\ 2, pp.\ 530--533; see also Baym 1969, pp.\ 305--307]{Messiah 1962}. The transformation of the eigenvectors of $\hat{S}_a$ to the eigenvectors of $\hat{S}_b$ is accordingly given by:\footnote{Whereas it takes some effort to prove the transformation law in Eq.\ (\ref{ops rot}), it is easy to see that this transformation law for operators entails the one for state vectors in Eq.\ (\ref{state rot}). Using Eq.\ (\ref{ops rot}), we can show that, whenever $|s, m\rangle_a$ is an eigenvector of $\hat{S}_a$ with eigenvalue $m$, $|s, m\rangle_b$ in Eq.\ (\ref{state rot}) is, in fact, an eigenvector of $\hat{S}_b$ with that same eigenvalue, as the notation suggests:
\begin{eqnarray*}
\hat{S}_b |s, m\rangle_b & \!\!\! = \!\!\! & \left( e^{-i\varphi_{ab} \hat{S}_n}\hat{S}_a e^{i\varphi_{ab}  \hat{S}_n} \right) \left( e^{-i \varphi_{ab} \hat{S}_n}|s,m\rangle_a\right) \\[.3 cm]
 & \!\!\! = \!\!\! & e^{-i\varphi_{ab} \hat{S}_n}\hat{S}_a |s,m\rangle_a = e^{-i\varphi_{ab} \hat{S}_n} m |s,m\rangle_a = m |s,m\rangle_b.
\end{eqnarray*}}
\begin{equation}
|s,m\rangle_b = e^{-i \varphi_{ab} \hat{S}_n}|s,m\rangle_a.
\label{state rot}
\end{equation}

We now show that, in the special case that $s = \sfrac12$ and $\vec{e}_a$ to $\vec{e}_b$ are in the $xz$-plane, the transformation law in Eq.\ (\ref{state rot}) reduces to the one in Eq.\ (\ref{QM2}) in Section \ref{1.5}. For $s = \sfrac12$, the elements of the matrices representing the spin operators $\hat{S}_x$, $\hat{S}_y$ and $\hat{S}_z$  in the orthonormal basis  in Eq.\ (\ref{onb one-particle Hilbert space}) of eigenvectors of $\hat{S}_z$ are $\sfrac12$ times the Pauli matrices (as long as the Condon-Shortley convention mentioned above is adopted):
\begin{equation}
\begin{array}{c}
\hat{S}_x = \sfrac12 \, \hat{\sigma}_x = \frac12 \begin{pmatrix} 
0  \! & \! 1 \\
1 \! & \! 0
\end{pmatrix},  \\[.5cm]
\hat{S}_y = \sfrac12 \, \hat{\sigma}_y = \frac12 \begin{pmatrix} 
0  \! & \! - i \\
i \! & \! 0
\end{pmatrix}, \\[.5cm]
\hat{S}_z = \sfrac12 \, \hat{\sigma}_z = \frac12 \begin{pmatrix} 
1 \! & \! 0 \\
0 \! & \! -1
\end{pmatrix}.
\end{array}
\label{Pauli matrices}
\end{equation}
One readily verifies that these matrices satisfy the commutation relations in Eq.\ (\ref{ops dfn}).

For $s = \sfrac12$, the rotation operator $e^{-i\vartheta \hat{S}_n}$ has a particularly simple form, which we can find using 
%\citep[pp.\ 306--307]{Baym 1969}. 
the completeness of the orthonormal basis of eigenvectors 
\begin{equation}
%\big\{| {\textstyle \frac12,\frac12} \rangle_{n}, | {\textstyle \frac12, -\frac12} \rangle_{n} \big\}
\left\{ \left|  \sfrac12,\sfrac12 \right\rangle_{n}, \left|  \sfrac12, -\sfrac12 \right\rangle_{n} \right\}
\label{onb spin S_n}
\end{equation}
of the spin operator $\hat{S}_n$ associated with the unit vector $\vec{e}_n$. For the purposes of this calculation, we revert to the notation $| \pm \rangle_n$ of Section \ref{1.5} for these eigenvectors. With the help of the resolution of unity in the basis of these eigenvectors,
\begin{equation}
\hat{1} = |+\rangle_{\!n} \, _{n\!}\langle +| \; + \; |-\rangle_{\!n} \, _{n\!}\langle -|
\label{rot proof 1}
\end{equation}
we can write the spectral decomposition of the spin operator $\hat{S}_n$ as
\begin{equation}
\hat{S}_n = \textstyle{\frac12} |+\rangle_{\!n} \, _{n\!}\langle + | \, - \, \textstyle{\frac12} |-\rangle_{\!n} \, _{n\!}\langle -|,
\label{rot proof 2}
\end{equation}
and the spectral decomposition of the rotation operator $e^{-i\vartheta \hat{S}_n}$ as
\begin{equation}
e^{-i\vartheta \hat{S}_n} = e^{-i\vartheta/2} |+\rangle_{\!n} \, _{n\!}\langle + | \; + \; e^{i\vartheta/2} |-\rangle_{\!n} \, _{n\!}\langle -|.
\label{rot proof 3}
\end{equation}
Collecting terms with cosines and sines and using Eqs.\ (\ref{rot proof 1})--(\ref{rot proof 2}), we find that
\begin{eqnarray}
e^{-i\vartheta \hat{S}_n} & \!\! = \!\! & \Big( \cos{\!(\vartheta/2)} - i \, \sin{\!(\vartheta/2)} \Big) |+\rangle_{\!n} \, _{n\!}\langle + | \nonumber \\
 & & \quad \quad \quad \quad + \; \Big(\cos{\!(\vartheta/2)} + i \, \sin{\!(\vartheta/2)} \Big) | -\rangle_{\!n} \, _{n\!}\langle - | \nonumber \\
 & \!\! = \!\!  & \cos{\!(\vartheta/2)} \, \hat{1} - i \sin{\!(\vartheta/2)} \, 2\hat{S}_n
 \label{rot proof 4}
\end{eqnarray}
Note that we got from the angle $\vartheta$ to the half-angle $\vartheta/2$ because of the factor of $\sfrac12$ in the eigenvalues $\hbar/2$ of the spin operator $\hat{S}_n$. 

To recover Eq.\ (\ref{QM2}), we choose $\vec{e}_n$ in the $y$-direction and $\vec{e}_a$ in the $z$-direction. Hence, $\hat{S}_n = \hat{S}_y$ and $\hat{S}_z = \hat{S}_a$. In that case, as we saw in Eq.\ (\ref{Pauli matrices}), the matrix representing $2\hat{S}_y$  in the basis $\big\{| + \rangle_{a}, | - \rangle_{a} \big\}$ is the Pauli matrix $\hat{\sigma}_y$. The matrix elements representing the rotation operator $e^{-i\vartheta \hat{S}_n}$ in this basis is then given by:
\begin{equation}
\cos{\! \left(\frac\vartheta2\right)}  
\begin{pmatrix} 
1  \! & \! 0 \\[.2 cm]
0 \! & \! 1
\end{pmatrix} 
\, - \, i \sin{\! \left(\frac\vartheta2\right)} 
\begin{pmatrix} 
0  \! & \! - i \\[.2 cm]
i \! & \! 0
\end{pmatrix}
= 
\begin{pmatrix} 
\cos{\!(\vartheta/2)} \! & \! - \sin{\!(\vartheta/2)} \\[.2 cm]
\sin{\!(\vartheta/2)} \! & \! \cos{\!(\vartheta/2)}
\end{pmatrix}.
\label{spin 1/2 rotation}
\end{equation}
From this matrix we can read off the components of the eigenvectors of $\hat{S}_b$ associated with a unit vector $\vec{e}_b$ in the $xz$-plane obtained by rotating the eigenvectors of $\hat{S}_a$ associated with a unit vector $\vec{e}_a = \vec{e}_z$ (cf.\ Eq.\ (\ref{state rot})):
\begin{equation}
| \pm \rangle_b = e^{-i \varphi_{ab} \hat{S}_y} | \pm \rangle_a.
\end{equation}
Using the resolution of unity in the $\{| \pm \rangle_a \}$ basis, we find:
\begin{eqnarray}
| \pm \rangle_b & \!\!\! =  \!\!\! & \Big( |+\rangle_{\!a} \, _{a\!}\langle +| \; + \; |-\rangle_{\!a} \, _{a\!}\langle -| \Big) e^{-i \varphi_{ab} \hat{S}_y} | \pm \rangle_a
\nonumber \\
 & \!\!\! =  \!\!\! & \Big(\!\,  _{a\!}\langle +| e^{-i \varphi_{ab} \hat{S}_y} | \pm \rangle_a \! \Big) |+\rangle_a 
 \; + \; \Big(\!\,  _{a\!}\langle -| e^{-i \varphi_{ab} \hat{S}_y} | \pm \rangle_a \! \Big) |-\rangle_a 
\label{b +/- trans law}
\end{eqnarray}
Using Eq.\ (\ref{spin 1/2 rotation}) with $\vartheta = \varphi_{ab}$ for the matrix elements of the rotation operator in the $\{ | \pm \rangle_a \}$ basis,
\begin{equation}
_{a\!}\langle \pm | e^{-i \varphi_{ab} \hat{S}_y} | \pm \rangle_a 
=
\begin{pmatrix} 
\cos(\varphi_{ab}/2) \! & \! -\sin(\varphi_{ab}/2)\\[.3 cm]
\sin(\varphi_{ab}/2) \! & \! \cos(\varphi_{ab}/2)
\end{pmatrix},
\label{spin 1/2 rotation 2}
\end{equation}
we arrive at  
\begin{eqnarray}
|+\rangle_b &\!\!=\!\!& \cos\left(\frac{\varphi_{ab}}{2}\right)|+\rangle_a+\sin\left(\frac{\varphi_{ab}}{2}\right)|-\rangle_a, \nonumber \\[.3 cm]
|-\rangle_b &\!\!=\!\!& - \sin\left(\frac{\varphi_{ab}}{2}\right)|+\rangle_a+\cos\left(\frac{\varphi_{ab}}{2}\right)|-\rangle_a.
\label{b +/- trans law 2}
\end{eqnarray}
This is just the inverse of the transformation from $|\pm \rangle_b$ to $|\pm\rangle_a$ in Eq. (\ref{QM2}). We used Eq.\ (\ref{QM2}) to show that the singlet state for a pair of spin-$\frac12$ particles has the same form in orthonormal bases $\{ | \pm \rangle_a \}$ and $\{ | \pm \rangle_b \}$ (see Eqs.\ (\ref{singlet state s=1/2 simple})--(\ref{QM3})). We concluded that this singlet state has the same form in \emph{any} orthonormal basis  $\{ | \pm \rangle_n \}$. Strictly speaking, our derivation of Eq.\ (\ref{b +/- trans law 2}) only allows us to claim this for orthonormal bases of eigenvectors of spin operators associated with unit vectors in the $xz$-plane. However, since for any two unit vectors $\vec{e}_a$ and $\vec{e}_b$, we can choose the plane spanned by those two vectors to be the $xz$ plane, this result holds for \emph{any} orthonormal basis  $\{ | \pm \rangle_n \}$.

We now turn to the special case $s=0$. Eq.\ (\ref{state dfn}) tells us that $\hat{S}^2 |0,m\rangle_z=0$. Using that $\hat{S}^2 = \hat{S}_y^2 + \hat{S}_x^2 +\hat{S}_z^2$, we thus have
\begin{eqnarray}
0 &\!\!=\!\!& {_z} \langle 0,m|\hat{S}_x^2 |0,m\rangle_z + {_z}\langle 0,m|\hat{S}_y^2 |0,m\rangle_z + {_z}\langle 0,m|\hat{S}_z^2 |0,m\rangle{_z} \nonumber \\[.3 cm]
&\!\!=\!\!& \Big|\hat{S}_x|0,m\rangle_z \Big|^2 + \Big|\hat{S}_y|0,m\rangle_z \Big|^2 +\Big|\hat{S}_z|0,m\rangle_z \Big|^2.
\label{zero spin}
\end{eqnarray}
This last expression, being a sum of squared absolute values, can only vanish if 
\begin{equation}
\hat{S}_x |0,m\rangle_z = \hat{S}_y |0,m\rangle_z = \hat{S}_z |0,m\rangle_z = 0. 
\label{zero comps}
\end{equation}
This enforces that $m=0$ (see Eq.\ (\ref{ops dfn})). More generally, it means that the singlet state has zero spin angular momentum along any direction. Hence we must have $\hat{S}_n|0,0\rangle_z = 0$. This implies that the singlet state is invariant under rotation through an angle $\vartheta$ with respect to any direction $\vec{e}_n$:
\begin{equation}
e^{-i\vartheta \hat{S}_n }|0,0\rangle_{z} = \sum_{k=0}^\infty \frac{(-i\vartheta)^k}{k!}\hat{S}^k_n|0,0\rangle_z = |0,0\rangle_z,
\label{singlet rot}
\end{equation}
as the only contribution to the sum comes from the $k=0$ term.

%SUBSUBSECTION 3.1.2
\subsubsection{Quantum formalism for two spin-$s$ particles in the singlet state} \label{2.1.2}

In Section \ref{1.5}, we considered the singlet state for a pair of spin-$\frac12$ particles (see Eq.\ (\ref{singlet state s=1/2 simple})). For the rest of this section, we consider the singlet state for two particles of any (half-)integer spin $s$. Alice performs measurements on particle 1, with a one-particle Hilbert space spanned by $\{|s,m_1\rangle_{1z}\}_{m_1=-s}^s$ and one-particle spin operators $\hat{S}_{1x},\hat{S}_{1y},\hat{S}_{1z}$. Similarly, Bob performs measurements on particle 2, with a one-particle Hilbert space spanned by $\{|s,m_2\rangle_{2z}\}_{m_2=-s}^s$ and one-particle spin operators $\hat{S}_{2x}$, $\hat{S}_{2y}$ and $\hat{S}_{2z}$. The two-particle Hilbert space therefore has dimension $(2s+1)^2$ and is spanned by the orthonormal basis
\begin{equation}
\big\{|s,m_1\rangle_{1z} \, |s,m_2\rangle_{2z}\big\}_{m_1, m_2=-s}^s.
\label{2 particle basis}
\end{equation}
The spin operators on this two-particle Hilbert space are 
\begin{equation}
\hat{S}_{n}=\hat{S}_{1n}+\hat{S}_{2n}.
\label{total spin ops}
\end{equation}
%(see Eq.\ (\ref{spin op}) for the definition of $\hat{S}_n$ in the one-particle case). 
The action of these operators on the basis vectors in Eq.\ (\ref{2 particle basis}) is given by:
\begin{equation}
\hat{S}_{n} \, |s,m_1\rangle_{1z} \, |s,m_2\rangle_{2z} = \Big( \hat{S}_{1n} \, |s,m_1\rangle_{1z} \Big) |s,m_2\rangle_{2z}
+ |s,m_1\rangle_{1z} \Big( \hat{S}_{2n} |s,m_2\rangle_{2z} \Big).
\label{total spin ops action}
\end{equation}

We now show that the two-particle state 
\begin{equation}
|0,0\rangle_{\!12} = \sum_{m=-s}^s \frac{(-1)^{s-m}}{\sqrt{2s+1}} |s,m\rangle_{1z}|s,-m\rangle_{2z}
\label{singlet state}
\end{equation}
is the singlet state of two particles of any (half-)integer spin $s$.\footnote{For $s=0$, Eq.\ (\ref{singlet state}) trivially reduces to $|0, 0 \rangle_{\!12} = |0, 0 \rangle_{1z} |0, 0 \rangle_{2z}$. For $s=\frac12$, it reduces to
$$\textstyle |0, 0 \rangle_{\!12} = \dfrac{1}{\sqrt{2}} 
\left( \,
\left| \frac12, \frac12 \right\rangle_{\!1z} \left| \frac12, -\frac12 \right\rangle_{\!2z} \, - \; 
\left| \frac12, -\frac12 \right\rangle_{\!1z} \left| \frac12, \frac12 \right\rangle_{\!2z} \,
\right)
$$
which can be written more compactly in the familiar form given in Eq.\ (\ref{singlet state s=1/2 simple}) (with $z=a$, i.e., with $\vec{e}_a$ in the $z$-direction).\label{singlet note}}  By definition, the singlet state must be annihilated by $\hat{S}^2$. Given Eq.\ (\ref{raising}), to show that $\hat{S}^2$ annihilates $|0,0\rangle_{12}$, it suffices to show that $\hat{S}_z$ and $\hat{S}_+$ both annihilate $|0,0\rangle_{12}$. For the former, we note that each product state in Eq. (\ref{singlet state}) has $m_1=-m_2=m$ and therefore $\hat{S}_z = \hat{S}_{1z} + \hat{S}_{2z}$ annihilates $|0,0\rangle_{12}$ term by term. For the latter, we first need to write out how $\hat{S}_+ = \hat{S}_{1+}+\hat{S}_{2+}$ acts on $|0,0\rangle_{12}$:
\begin{eqnarray}
\hat{S}_+ |0,0\rangle_{12}
&\!\!=\!\!&  \sum_{m=-s}^s \frac{(-1)^{s-m}}{\sqrt{2s+1}}
\Big[\big(\hat{S}_{1+}|s,m\rangle_{1z}\big)|s,-m\rangle_{2z}+|s,m\rangle_{1z}\big(\hat{S}_{2+}|s,-m\rangle_{2z}\big)\Big] \nonumber \\[.2 cm] 
&\!\!=\!\!&  \sum_{m=-s}^s \frac{(-1)^{s-m}}{\sqrt{2s+1}}
\Big[C_+(s,m)|s,m\!+\!1\rangle_{1z}|s,-m\rangle_{2z} \nonumber \\
&&\hspace{4cm}+C_+(s,-m)|s,m\rangle_{1z}|s,-m\!+\!1\rangle_{2z})\Big].
\label{raising kills}
\end{eqnarray}
Shifting the summation index in the second term on the right-hand side and noting that $C_+(s,m)=C_+(s,-m-1)$, which follows from the expression for $C_+(s,m)$ in Eq.\ (\ref{coef raising}), we readily verify that the sum cancels term by term. So both $\hat{S}_z$ and $\hat{S}_+$ annihilate $|0,0\rangle_z$. We conclude that $|0,0\rangle_{12}$ is indeed the singlet state. 

%% If we pass to the density matrix |00><00| and trace out over the second particle, then the first particle is in the completely mixed state (and vice versa). 

Using the rotational invariance of the singlet state (see Eq.\ (\ref{singlet rot})), we can rewrite the action of any rotation operator as
\begin{equation}
|0,0\rangle_{\!12} 
= e^{-i\vartheta \hat{S}_{n}} |0,0\rangle_{\!12} = \sum_{m=-s}^s \frac{(-1)^{s-m}}{\sqrt{2s+1}} \, e^{-i\vartheta \hat{S}_{1n}} |s,m\rangle_{1z} e^{-i\vartheta  \hat{S}_{2n}}|s,-m\rangle_{2z},
\label{spin11} 
\end{equation}
which, given Eq.\ (\ref{state rot}), reduces to: 
\begin{equation}
|0,0\rangle_{\!12} = \sum_{m=-s}^s \frac{(-1)^{s-m}}{\sqrt{2s+1}} |s,m\rangle_{1b} |s,-m\rangle_{2b}.
\label{spin11a}
\end{equation}
It follows that the singlet state has the same form in all bases. In Section \ref{1.5} we showed this for the special case of the singlet state of two spin-$\frac12$ particles (see Eqs.\ (\ref{singlet state s=1/2 simple})--(\ref{QM3}) and the comment following Eq.\ (\ref{b +/- trans law 2}) in Section \ref{2.1.1}).

%SUBSUBSECTION 3.1.3
\subsubsection{Wigner d-matrices and correlation arrays in the spin-$s$ case} \label{2.1.3}
%Wigner d-matrices and correlation arrays for measurements on a pair of particles with spin $s$ in the singlet state

Suppose that Alice measures the spin component of particle 1 in the direction $\vec{e}_a$ and that Bob measures the spin component of particle 2 in the direction $\vec{e}_b$. Using the Born rule, we can compute the probability that Alice finds $m_1$ and Bob finds $m_2$ by taking the absolute square of the inner product of the singlet state with the product state $|s, m_1 \rangle_{1a} |s, m_2 \rangle_{2b}$ (cf.\ Eq.\ (\ref{expansion in ab}) in Section \ref{1.5}): 
\begin{equation}
\mathrm{Pr}(m_1\, m_2 | \hat{a}\,\hat{b} ) = \Big| \Big(  \,\! _{1a\!}\langle s, m_1| \, _{2b\!}\langle s, m_2 | \Big)| 0, 0\rangle_{\! 12}\, \Big|^{2\!}.
\label{prob 2}
\end{equation} 
Using the expansion of the singlet state in Eq.\ (\ref{singlet state}), we can write the inner product on the right-hand side of Eq.\ (\ref{prob 2}) as
\begin{eqnarray}
\Big(  \,\! _{1a\!}\langle s, m_1| \, _{2b\!}\langle s, m_2 | \Big)| 0, 0\rangle_{\! 12}
&\!\!\!=\!\!\!& \sum_{m=-s}^{s} \frac{(-1)^{s-m}}{\sqrt{2s+1}} \,
_{1a\!}\langle s, m_1|s,m\rangle_{1a} \,
_{2b\!}\langle s, m_2 |s,-m\rangle_{2a} \nonumber \\
&\!\!\!=\!\!\!& \frac{(-1)^{s-m_1}}{\sqrt{2s+1}} \,
_{2b\!}\langle s, m_2 |s,-m_1\rangle_{2a}
\label{inner product 1}
\end{eqnarray}
where we used that $_{1a \!}\langle s, m_1 | s, m \rangle_{1a} = \delta_{mm_1}$. To evaluate $_{2b\!}\langle s, m_2 |s,-m_1\rangle_{2a}$ we use the invariance of the singlet state under rotation to choose, without loss of generality, the unit vectors $\vec{e}_a$ and $\vec{e}_b$ as
\begin{equation}
\vec{e}_a=\vec{e}_z,\quad \vec{e}_b=\vec{e}_z \cos\varphi_{ab}+\vec{e}_x\sin\varphi_{ab}.
\label{vector dirs 2}
\end{equation}
%where $\varphi_{ab}$ is the angle between $\vec{e}_a$ and $\vec{e}_b$.
The unit vector $\vec{e}_b$ is obtained from $\vec{e}_z$ by rotating around the $y$-axis through the angle $\varphi_{ab}$ (see Figure \ref{rotation} with $\vec{e}_n$ and $\vec{e}_a$ relabeled $\vec{e}_y$ and $\vec{e}_z$, respectively). So we can express the states of the second particle as
\begin{equation}
|s,m_2\rangle_{2b} = e^{-i \varphi_{ab} \hat{S}_{2y}}|s,m_2\rangle_{2z}.
\label{rotated 2nd particle states}
\end{equation}
As such, the desired probability may be written as
\begin{eqnarray}
\mathrm{Pr}(m_1\, m_2 | \hat{a}\,\hat{b} )
&\!\!=\!\!& \frac{1}{2s+1} \Big| {_{2b\!}}\langle s, m_2 |s,-m_1\rangle_{2a}\Big|^{\!2} \nonumber \\[.3 cm]
&\!\!=\!\!& \frac{1}{2s+1} \Big| {_{2a\!}}\langle s, -m_1 |s,m_2\rangle_{2b}\Big|^{\!2} \nonumber \\[.3 cm]
&\!\!=\!\!& \frac{1}{2s+1} \Big| {_{2z\!}}\langle s, -m_1| e^{-i \varphi_{ab}\hat{S}_{2y}}|s,m_2\rangle_{2z}\Big|^{\!2}.
\label{inner product 2}
\end{eqnarray}
The inner product in the final expression in Eq.\ (\ref{inner product 2}) is an element of the so-called \emph{Wigner d-matrix}, introduced in \citet[Ch.\ XV]{Wigner 1931}:\footnote{The `d' in d-matrix stands for \emph{Darstellung} (representation) rather than \emph{Drehung} (rotation). \citet[Vol.\ 2, p.\ 1070, Eq.\ C55]{Messiah 1962} uses the notation $r^{(s)}(\beta)$ for this matrix and writes its elements as
$$
r^{(s)}_{mm'}(\beta) \equiv {_{z\!}}\langle s, m | e^{\displaystyle -i \beta \hat{S}_y}   | s, m' \rangle_z.
$$
The only difference between our expressions and Messiah's is that he is considering the total angular momentum, the sum of intrinsic and orbital angular momentum, whereas we are focusing on spin, i.e., intrinsic angular momentum. As Messiah notes, the elements of the Wigner d-matrix are always real (ibid., p. 1071).

Philip W.\ Anderson recalls taking a class on group theory in the late 1940s with John H.\ Van Vleck, using Wigner's book in the original German \citep[p.\ 148]{Midwinter and Janssen 2013}. During World War II, it had been reprinted in facsimile by, as it says on the title page, ``Authority of the Alien Property Custodian.''

Those who have ever taken a course on quantum mechanics covering elements of group theory may vaguely recognize our Wigner d-matrices. They appear on the same page as  Clebsch-Gordan coeffients in the Particle Data Group's \emph{Review of Particle Physics} \citep[p. 564]{Tanabashi et al 2018}. This page makes for a convenient formula sheet for exams.
\label{Messiah}}
\begin{equation}
d^{\,(s)}_{mm'}(\vartheta) \equiv {_{z\!}}\langle s, m |e^{-i \vartheta \hat{S}_{y}}| s, m' \rangle_{z}.
\label{wigner}
\end{equation}
Setting $m = -m_1$, $m' = m_2$ and $\vartheta = \varphi_{ab}$, we  obtain Eq.\ (\ref{prob 2}) in terms of these Wigner d-matrix elements:
\begin{equation}
\mathrm{Pr}(m_1\, m_2 | \hat{a}\,\hat{b} ) = \Big| \Big( {_{1a\!}}\langle s, m_1| \, {_{2b\!}}\langle s, m_2 | \Big)| 0, 0\rangle_{\! 12}\, \Big|^{2\!} = \frac{1}{2s+1} \Big( \, d^{\,(s)}_{-m_1m_2}(\varphi_{ab}) \, \Big)^{2\!}.
\label{prob 3}
\end{equation} 

We examine this result for the special cases of $s= \frac12$ and $s=1$. In Sections \ref{2.1.4}--\ref{2.1.6} we will do so for arbitrary (half-)integer values of $s$. We actually already encountered the Wigner d-matrix for $s=\frac12$ in Section \ref{2.1.1} (see Eq.\ (\ref{spin 1/2 rotation} with $\vartheta = \varphi_{ab}$):
\begin{equation}
d^{(\frac12)}(\varphi_{ab}) = 
\left( 
\begin{array}{cc}
\cos{(\varphi_{ab}/2)} & -\sin{(\varphi_{ab}/2)}  \\[.4 cm]
\sin{(\varphi_{ab}/2)}  & \cos{(\varphi_{ab}/2)} 
\end{array}
\right).
\label{Messiah 1/2} 
\end{equation}
%\citep[Vol.\ 2, p.\ 1073, Eq.\ (C.74), with $\alpha = \gamma = 0$ and $\beta = \varphi_{ab}$; cf.\ note \ref{Messiah}]{Messiah 1962}.
Squaring these matrix elements, we recover the probabilities in the $\hat{a}\hat{b}$ cell of the correlation array for the spin-$\frac12$ case in Figure \ref{CA-3set2out-non-signaling-halfangles}.

For $s=1$, the Wigner d-matrix is
\begin{equation}
d^{(1)}(\varphi_{ab}) = 
\left( 
\begin{array}{ccc}
\frac12 (1 + \cos{\varphi_{ab}}) & -\frac12 \sqrt{2} \sin{\varphi_{ab}} & \frac12 (1 - \cos{\varphi_{ab}}) \\[.4 cm]
\frac12 \sqrt{2} \sin{\varphi_{ab}}  & \cos{\varphi_{ab}} & -\frac12 \sqrt{2} \sin{\varphi_{ab}}  \\[.4 cm]
\frac12 (1 - \cos{\varphi_{ab}}) & \frac12 \sqrt{2} \sin{\varphi_{ab}}  & \frac12 (1 + \cos{\varphi_{ab}}) 
\end{array}
\right)
\label{Messiah 1}
\end{equation}
\citep[Vol.\ 2, p.\ 1073, Eq.\ (C.75) with $\alpha = \gamma = 0$ and $\beta = \varphi_{ab}$; see also Wigner, 1931, p.\ 182, Eq.\ 29]{Messiah 1962}.
%\footnote{See also   in \citet[p.\ 128, Eq.\ 29]{Wigner 1931}.} 
The elements of this matrix are $d^{(1)}_{ij}(\varphi_{ab})$ where $i = 1, 0, -1$ labels the rows and $j= 1, 0, -1$ labels the columns. 

\begin{figure}[h]
 \centering
   \includegraphics[width=4in]{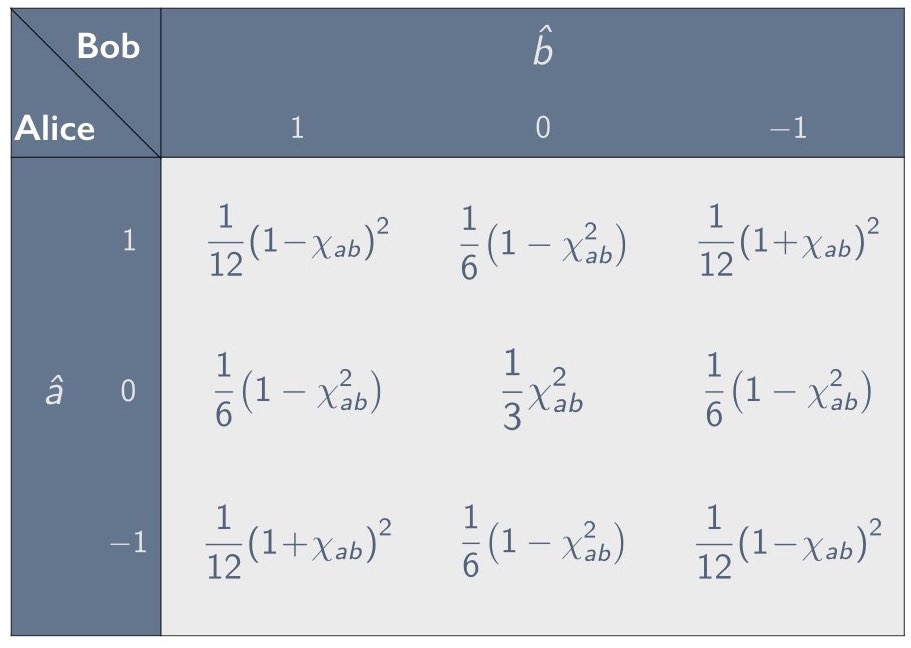} 
   \caption{Cell in a correlation array given by quantum mechanics for measurements on the singlet state of two spin-1 particles ($\chi_{ab} = \cos{\varphi_{ab}}$).}
    \label{CA-cell-spin1-chi}
\end{figure}

Squaring these matrix elements and introducing (with malice aforethought) the notation
\begin{equation}
\chi_{ab} \equiv \cos\varphi_{ab}\label{spin-s cosines}
\end{equation}
(and similarly $\chi_{bc}\equiv \cos\varphi_{bc}$ and $\chi_{ac}\equiv \cos\varphi_{ac}$) we find the probabilities in the cell of a correlation array in Figure \ref{CA-cell-spin1-chi}. 
We verify this for two of these probabilities. The probability that Alice (using setting $\hat{a}$) and Bob  (using setting $\hat{b}$) both find the outcome 1 is given by Eq.\ (\ref{prob 3}) with $s=1$ and the Wigner d-matrix in Eq.\ (\ref{Messiah 1}):
\begin{equation}
\mathrm{Pr}(m_1 = 1, m_2 = 1| \hat{a}\, \hat{b} ) = \frac13 \Big( d^{(1)}_{-11}( \varphi_{ab}) \Big)^{\!2} 
= \frac{1}{12} (1-\cos{\varphi_{ab}})^2 = \frac{1}{12} (1-\chi_{ab})^2.
\label{prob 3aa}
\end{equation}
Similarly, the probability that Alice finds $1$ for $\hat{a}$ and Bob finds $0$ for $\hat{b}$ is given by:
\begin{equation}
\mathrm{Pr}(m_1 = 1, m_2 = 0| \hat{a}\, \hat{b} ) = \frac13 \Big( d^{(1)}_{-10}( \varphi_{ab}) \Big)^{\!2} = \frac16 \sin^{2}{\!\varphi_{ab}} = \frac16 (1-\chi_{ab}^2).
\label{prob 3a}
\end{equation}

The $\hat{a} \, \hat{c}$ and $\hat{b} \, \hat{c}$ cells of the correlation arrays will have the exact same structure as $\hat{a} \, \hat{b}$ one, with $\varphi_{ab}$ replaced by $\varphi_{ac}$ and $\varphi_{bc}$. The correlation arrays for measurements on the singlet state of two spin-$\frac12$ or two spin-1 particles can thus be parametrized by the cosines of the angles $\varphi_{ab}$, $\varphi_{ac}$ and $\varphi_{bc}$ between the directions $\vec{e}_a$, $\vec{e}_b$ and $\vec{e}_c$ in which the spin is being measured. This is also true for the singlet state of two particles with higher spin \citep[Vol.\ 2, p.\ 1072, Eq.\ (3.72)]{Messiah 1962}. In Section \ref{2.1.5}, we will show that these cosines can be interpreted as anti-correlation coefficients, thereby justifying the notation introduced in Eq.\ (\ref{spin-s cosines}).

%SUBSUBSECTION 3.1.4
\subsubsection{Non-signaling in the spin-$s$ case} \label{2.1.4}
%Showing that the correlations found in measurements on two particles with spin $s$ in the singlet state are non-signaling}

We now show that the correlations found in measurements on the singlet state of two particles of any (half-)integer spin $s$ have uniform marginals and are therefore non-signaling. Consider the correlation array in Figure \ref{CA-cell-spin1-chi} and the marginal probability of Alice finding the outcome $m_1 = 1$ when she uses setting $\hat{a}$ and Bob uses setting $\hat{b}$:
\begin{eqnarray} 
\mathrm{Pr}(+1_A| \hat{a}\, \hat{b}) \nonumber
& \!\!\! = \!\!\! &\sum_{m_2=-1}^1 \mathrm{Pr}(+1\,m_2| \hat{a}\, \hat{b}) \nonumber \\[.3 cm]
&\!\!\! = \!\!\!  &\; \frac{1}{12}(1+ \chi_{ab})^2+\frac{1}{6}(1- \chi_{ab}^2)+\frac{1}{12}(1- \chi_{ab})^2 = \frac13.
\label{nonsignal 1}
\end{eqnarray}
For $m_1=0$ and $m_1 = -1$ we similarly find that 
\begin{equation}
\mathrm{Pr}(0_A| \hat{a}\, \hat{b})=\mathrm{Pr}(-1_A| \hat{a}\, \hat{b})= \frac13.
\end{equation}
None of these marginal probabilities---and this observation is key---depend on $\varphi_{ab}$. They are thus unaltered if Bob's settings are changed from $\hat{b}$ to $\hat{a}$ or $\hat{c}$. The same is true for the marginal probabilities of Bob measuring $m_2 = (1, 0, -1)$ using any of these three settings. In every cell of the correlation array that the cell in  Figure \ref{CA-cell-spin1-chi} is part of, all three rows and all three columns add up to $\sfrac13$. Like the singlet state for a pair of spin-$\frac12$ particles, the singlet state for a pair of spin-1 particles thus cannot be used for superluminal signaling.

The same is true for the singlet state of two particles of higher spin. The marginal probability of Alice finding $m_1$ for $\hat{a}$ when Bob uses $\hat{b}$ is:
\begin{equation}
\mathrm{Pr}(m_{1}| \hat{a}\, \hat{b}) = \sum_{m_2=-s}^s \mathrm{Pr}(m_1 \,m_2| \hat{a}\, \hat{b}).
\end{equation}
Using the first line of Eq.\ (\ref{inner product 2}), we find that
\begin{eqnarray} 
\mathrm{Pr}(m_{1}| \hat{a}\, \hat{b}) 
& \!\!\! = \!\!\! &\frac{1}{2s+1} \! \sum_{m_2=-s}^s \! \Big|{_{2b}}\langle s, m_2|s, -m_1\rangle{_{2a}}\Big|^2 \label{nonsignal 2} \nonumber \\[.3 cm]
& \!\!\! = \!\!\! &\frac{1}{2s+1} \! \sum_{m_2=-s}^s \!
{_{2a}}\langle s, -m_1|s,m_2\rangle{_{2b}}\; {_{2b}}\langle s, m_2|s,-m_1\rangle{_{2a}}.
\end{eqnarray}
Evaluating this sum, using the completeness relation
\begin{equation}
\hat{1}_2 = \sum_{m=-s}^s |s,m \rangle_{\!2b}\; {_{2b\!}}\langle s, m|,
\label{completeness}
\end{equation}
%i.e., the resolution of unity in the orthonormal basis $\{|s,m \rangle_{2b}\}_{m=-s}^s$ 
%(see Eq.\ (\ref{onb one-particle Hilbert space})) 
%for the second particle,
we arrive at
\begin{equation}
\mathrm{Pr}(m_{1}| \hat{a}\, \hat{b}) = \frac{1}{2s+1} \;
{_{2a}}\langle s, -m_1|s,-m_1\rangle{_{2a}} = \frac{1}{2s+1}.
\label{nonsignal 3}
\end{equation}
This same formula holds if we substitute $m_2$ for $m_1$ or any two of the triplet of settings $(\hat{a}, \hat{b}, \hat{c})$ for $\hat{a}\, \hat{b}$. It follows that the correlations found in the measurements on the singlet state of two spin-$s$ particles are indeed non-signaling. In every cell of the corresponding correlation array, all $2s+1$ rows and all $2s+1$ columns sum to $1/(2s+1)$.   

%SUBSUBSECTION 3.1.5
\subsubsection{Anti-correlation coefficients in the spin-$s$ case} \label{2.1.5}

We now turn our attention to the quantities $\chi_{ab}$ introduced in Eq.\ (\ref{spin-s cosines}) and the analogous quantities $\chi_{ac}$ and $\chi_{bc}$. We show that these can be interpreted as anti-correlation coefficients just as in the case of $s = \sfrac12$ (see Eqs.\ (\ref{chi values repeat})--(\ref{chi2angle}) in Section \ref{1.5}). Consider the expectation value
\begin{equation}
\langle \hat{S}_{1a}\hat{S}_{2b}\rangle_{00}\equiv {_{12\!}}\langle 0,0|\hat{S}_{1a}\hat{S}_{2b}|0,0\rangle_{12}.
\label{exp def}
\end{equation}
Recalling our choice of $\vec{e}_a$ and $\vec{e}_b$ in Eq.\ (\ref{vector dirs 2}), we have
\begin{equation}
\hat{S}_{1a} = \hat{S}_{1z}, \quad \hat{S}_{2b} = \hat{S}_{2z}  \cos{\varphi_{ab}} + \hat{S}_{2x}  \sin{\varphi_{ab}}. 
\label{rewriting Sa and Sb}
\end{equation}
Inserting these expressions into Eq.\ (\ref{exp def}), we arrive at:
\begin{equation}
\langle \hat{S}_{1a}\hat{S}_{2b}\rangle_{00} = \langle \hat{S}_{1z}\hat{S}_{2z}\rangle_{00} \cos\varphi_{ab}+\langle \hat{S}_{1z}\hat{S}_{2x}\rangle_{00} \sin \varphi_{ab}.
\label{exp value 1}
\end{equation}
The quantity $\langle \hat{S}_{1z}\hat{S}_{2z}\rangle_{00}$ in the first term on the right-hand side is minus the square of the standard deviation $\sigma_s$ (see Eq.\ (\ref{SD for adm raffle 1}) in Section \ref{1.6}). This quantity is thus given by
\begin{equation}
\langle \hat{S}_{1z}\hat{S}_{2z}\rangle_{00} = \sigma_s^2 = - \frac13 s(s+1).
\label{SD for singlet state}
\end{equation}
This same result can be derived directly from properties of the singlet state. The expectation value of the product of $\hat{S}_{1z}$ and $\hat{S}_{2z}$ in the singlet state is given by
\begin{equation}
\langle \hat{S}_{1z} \hat{S}_{2z} \rangle_{00}  = {_{12\!}}\langle 0,0| \hat{S}_{1z}\hat{S}_{2z}|0,0\rangle_{\!12}.
\label{aa product prob 1}  
\end{equation}
Using that $\hat{S}_{2z} = \hat{S}_z - \hat{S}_{1z}$ and that $\hat{S}_z  |0,0\rangle_{\!12} = 0$, we can rewrite this as
\begin{equation}
\langle \hat{S}_{1z} \hat{S}_{2z} \rangle_{00} = -{_{12\!}}\langle 0,0| \hat{S}_{1z}^2|0,0\rangle_{12}.
\label{aa product prob 2} 
\end{equation}
Rotational invariance requires that 
\begin{equation}
{_{12\!}}\langle 0,0| \hat{S}_{1x}^2|0,0\rangle_{12} = {_{12\!}}\langle 0,0| \hat{S}_{1y}^2|0,0\rangle_{12} = {_{12\!}}\langle 0,0| \hat{S}_{1z}^2|0,0\rangle_{12}.
\label{aa product prob 3}  
\end{equation}
Hence
\begin{equation}
\langle \hat{S}_{1z} \hat{S}_{2z} \rangle_{00} = - \frac13 \, {_{12\!}}\langle 0,0| \big( \hat{S}_{1x}^2+\hat{S}_{1y}^2+\hat{S}_{1z}^2 \big) |0,0\rangle_{\!12}.
\label{aa product prob 4}  
\end{equation}
Substituting $\hat{S}_{1}^2$ for $\hat{S}_{1x}^2+\hat{S}_{1y}^2+\hat{S}_{1z}^2$ and using that $\hat{S}_{1}^2 |0,0\rangle_{\!12} = s(s+1) |0,0\rangle_{12}$, we recover Eq.\ (\ref{SD for singlet state}): 
\begin{equation}
\langle \hat{S}_{1z} \hat{S}_{2z} \rangle_{00} = -\frac13 \, {_{12\!}}\langle 0,0| \hat{S}_{1}^2|0,0\rangle_{\!12}  = -\frac13 s(s+1).
\label{aa product prob 5} 
\end{equation}
 
Again using the rotational invariance of the singlet state, we can show that the second term in Eq.\ (\ref{exp value 1}) vanishes. Consider a rotation of the singlet state over $180\degree$ around the $z$-axis. Since the singlet state is invariant under arbitrary rotation, the action of the operator $e^{-i\pi \hat{S}_z}$ implementing this rotation (see Eq.\ (\ref{ops rot})) on the singlet state simply reproduces the singlet state. It follows that
\begin{equation}
\langle \hat{S}_{1z}\hat{S}_{2x}\rangle_{00} = {_{12\!}}\langle 0,0| \hat{S}_{1z}\hat{S}_{2x}|0,0\rangle_{\!12} =  {_{12\!}}\langle 0,0|e^{i\pi \hat{S}_z} \hat{S}_{1z}\hat{S}_{2x}e^{-i\pi \hat{S}_z}|0,0\rangle_{\!12}.
\label{exp value 2 a}
\end{equation}
Inserting $\hat{S}_z = \hat{S}_{1z} + \hat{S}_{2z}$ and using that $\hat{S}_{1z}$ commutes with both $\hat{S}_{2z}$ and $\hat{S}_{2x}$, we can rewrite this as:
\begin{equation}
\langle \hat{S}_{1z}\hat{S}_{2x}\rangle_{00} = {_{12\!}}\langle 0,0| \hat{S}_{1z} \, e^{i\pi \hat{S}_{2z}}\hat{S}_{2x}e^{-i\pi \hat{S}_{2z}} |0,0\rangle_{\!12}.
\label{exp value 2 b}
\end{equation}
Recalling the transformation law for spin operators in Eq.\ (\ref{ops rot}), we note that:
\begin{equation}
\quad e^{i\pi \hat{S}_{2z}}\hat{S}_{2x}e^{-i\pi \hat{S}_{2z}} = -\hat{S}_{2x}.
\label{exp value 2 c}
\end{equation}
Inserting this expression into Eq.\ (\ref{exp value 2 b}), we conclude that
\begin{equation}
\langle \hat{S}_{1z}\hat{S}_{2x}\rangle_{00} = - {_{12\!}}\langle 0,0|  \hat{S}_{1z}\hat{S}_{2x} |0,0\rangle_{\!12} =  - \langle \hat{S}_{1z}\hat{S}_{2x} \rangle_{00} = 0.
\label{exp value 2 d}
\end{equation}
So the only contribution to Eq.\ (\ref{exp value 1}) comes from the first term: 
\begin{equation}
\langle \hat{S}_{1a}\hat{S}_{2b}\rangle_{00} = \langle \hat{S}_{1a}\hat{S}_{2a} \rangle_{00} \cos\varphi_{ab} = -\frac13 s(s+1) \, \chi_{ab},
\label{exp value 3}
\end{equation} 
where we used Eq.\  (\ref{aa product prob 5})) for $ \langle \hat{S}_{1a}\hat{S}_{2a} \rangle_{00}$ and the definition of $\chi_{ab}$ in Eq.\ (\ref{spin-s cosines}). Using that the standard deviations 
\begin{equation}
\sigma_{1a} \equiv \sqrt{\langle \hat{S}_{1a}^2 \rangle_{00} } \quad {\mathrm{and}} \quad
\sigma_{2b} \equiv \sqrt{ \langle \hat{S}_{2b}^2 \rangle_{00} }
\label{sigma s > 1/2 a} 
\end{equation}
are both given by Eq.\ (\ref{SD for singlet state}), we can rewrite Eq.\ (\ref{exp value 3}) as
\begin{equation}
\chi_{ab} = - \frac{\langle \hat{S}_{1a}\hat{S}_{2b}\rangle_{00}}{\sigma_a \sigma_b}
\end{equation}
which we recognize as the definition of the anti-correlation coefficient $\chi_{ab}$ (see Section \ref{1.3}, Eq.\ (\ref{chi as corr coef}), and Section \ref{1.5}, Eq.\ (\ref{chi2angle})). This justifies the use of the $\chi_{ab}$ notation in Eq.\ (\ref{spin-s cosines}). This identification moreover means that the main conclusion of our examination of the special case of spin-$\frac12$ particles in Section \ref{1.5} carries over to the general case considered in this section: the set of values for $(\chi_{ab},\chi_{ac},\chi_{bc})$ that can be obtained by measurements on the singlet state of two particles of higher spin is the elliptope in Figure \ref{elliptope}.

%SUBSUBSECTION 3.1.6
\subsubsection{Cell symmetries in the spin-$s$ case} \label{2.1.6}

\begin{figure}[h]
 \centering
   \includegraphics[width=5in]{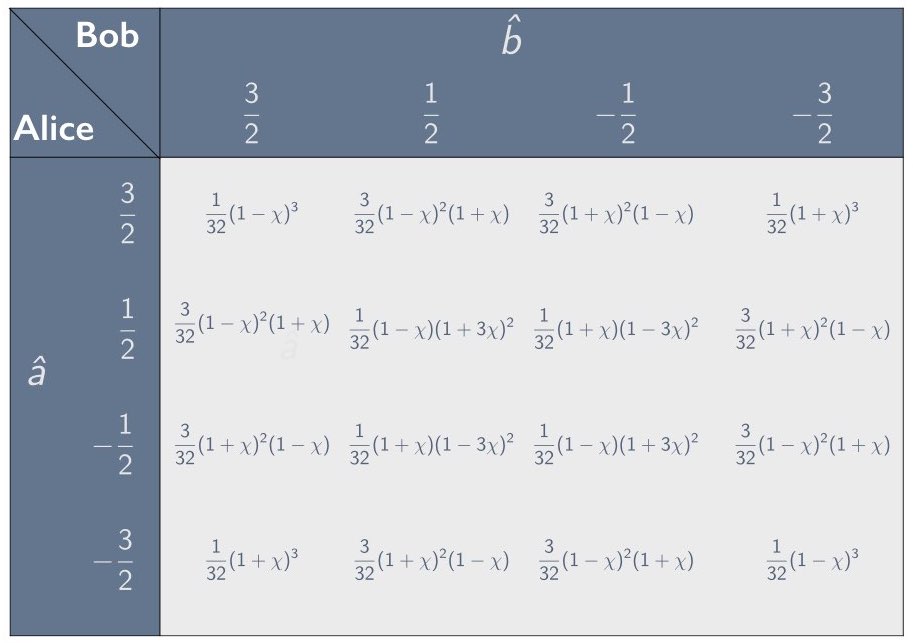} 
   \caption{Cell in a correlation array given by quantum mechanics for measurements on the singlet state of two spin-$\frac32$ particles ($\chi = \cos{\varphi_{ab}}$).}
   \label{CA-2set4out}
\end{figure}

In this subsection, we will consider various symmetries of cells in correlation arrays for measurements on particles of arbitrary (half-)integer spin $s$.  Figures \ref{CA-3set2out-non-signaling-halfangles} and \ref{CA-cell-spin1-chi} show such cells for the $s = \sfrac12$ and $s = 1$ cases. Figure \ref{CA-2set4out} shows it for the $s = \sfrac32$ case.

Cells on the diagonals in all these cases are particularly simple. Since for those cells $\chi_{ab} = \cos{\varphi_{ab}} =1$, elements on the skew-diagonal are equal and sum to 1 while the other elements are 0. This is true not just for  $s = \sfrac12$ and $s = 1$, but for any (half-)integer value of $s$. Whenever the angle between the measuring directions used by Alice and Bob is zero,  Eq.\ (\ref{wigner}) for the Wigner d-matrix element reduces to   
\begin{equation}
d^{\,(s)}_{-m_1m_2}(0) = {_z} \langle s, -m_1  | s, m_2 \rangle_z=\delta_{-m_1m_2}.
\label{wigner zero}
\end{equation}
Hence, the probability for Alice to measure $m_1$ and Bob to measure $m_2$ along a common direction, say $\vec{e}_a$, is given by (cf.\ Eq.\ (\ref{prob 3})):
\begin{equation}
\mathrm{Pr}(m_1\, m_2 | \hat{a}\,\hat{a} ) = \frac{1}{2s+1} \Big( \, d^{\,(s)}_{-m_1m_2}(0) \, \Big)^{2\!}=\frac{1}{2s+1} \, \delta_{-m_1m_2}.
\label{prob 4}
\end{equation}
Any cell on the diagonal of a correlation array for measurements on the singlet state of two particles of any integer or half-integer spin $s$  thus has values $1/(2s+1)$ on the skew-diagonal and zeros everywhere else (see Figure \ref{diag-cell-sxs} in Section \ref{1.6}).  

The off-diagonal cells, while not as simple as the diagonal ones, also exhibit features that are the same for all values of $s$. Note, for instance, that the cells in Figure \ref{CA-3set2out-non-signaling-halfangles} (for $s = \sfrac12$) and Figures \ref{CA-cell-spin1-chi}--\ref{CA-2set4out} (for $s = 1$ and $s = \sfrac32$) are all symmetric across both the diagonal and the skew-diagonal. This is true not just for $s = \sfrac12$ and $s = 1$ but for any (half-)integer $s$. This follows directly from the following three symmetry properties of the Wigner d-matrix in Eq.\ (\ref{wigner}):
%, i.e., from symmetries of the matrix elements for the one-particle rotation operator $e^{-i\vartheta \hat{S}_y}$:
\begin{eqnarray}
d^{\,(s)}_{mm'}(\vartheta) 
&\!\!=\!\!& (-1)^{m-m'} d^{\,(s)}_{m' m}(\vartheta) \label{WignerD sym 1} \quad \\[.3 cm] 
&\!\!=\!\!& (-1)^{m-m'} d^{\,(s)}_{-m-m'}(\vartheta) \label{WignerD sym 2} \\[.3 cm]
&\!\!=\!\!& d^{\,(s)}_{-m'-m}(\vartheta). \label{WignerD sym 3}
\end{eqnarray}

To establish the first of these symmetries, consider a rotation of the one-particle system through an angle of $180\degree$ around the $z$-axis, implemented as $e^{-i \pi \hat{S}_z}$. Such a rotation leaves the $z$-axis unchanged and as such the states are only changed up to an overall phase factor, i.e.,
\begin{equation}
    e^{-i\pi \hat{S}_z}| s, m \rangle_{z} = e^{-i \pi m}| s, m \rangle_{z}. 
    \label{WignerD sym 1 proof a}
\end{equation}
By contrast, this rotation flips the $y$-axis, which means that the operator $\hat{S}_y$ transforms as
\begin{equation}
    e^{-i\pi \hat{S}_z} \hat{S}_y e^{i\pi \hat{S}_z} = -\hat{S}_y.
\end{equation}
The same is true for any function of $\hat{S}_y$:
\begin{equation}
 e^{-i\pi \hat{S}_z} f(\hat{S}_y) \, e^{i\pi \hat{S}_z} = f(-\hat{S}_y).
\end{equation}
Hence, we can rewrite the Wigner d-matrix element in Eq.\ (\ref{wigner}) as
\begin{equation}
 d^{\,(s)}_{mm'}(\vartheta) = {_{z\!}}\langle s, m |e^{-i \vartheta \hat{S}_{y}}| s, m' \rangle_{z} = {_{z\!}}\langle s, m |e^{-i\pi \hat{S}_z}e^{i \vartheta \hat{S}_{y}}e^{i\pi \hat{S}_z}| s, m' \rangle_{z}.
\end{equation}
On account of Eq.\ (\ref{WignerD sym 1 proof a}), this reduces to: 
\begin{equation}
 d^{\,(s)}_{mm'}(\vartheta) = {_{z\!}}\langle s, m |e^{-i\pi m}e^{i \vartheta \hat{S}_{y}}e^{i\pi m'}| s, m' \rangle_{z} = (-1)^{m-m'} d^{\,(s)}_{mm'}(-\vartheta).
  \label{symm of W d-matrices 1}
\end{equation}
%As we noted in the discussion of raising and lowering operators, 
Given the Condon-Shortley phase convention, the operator $i\hat{S}_y$ has real matrix elements in the $|s,m\rangle_{\!z}$ basis (see Eqs.\ (\ref{raising operator})--(\ref{coef raising})). Since the elements of the Wigner d-matrix are matrix elements of a function of $i\hat{S}_y$  in this basis, it follows that they too must be real.
We can thus rewrite $d^{\,(s)}_{mm'}(-\vartheta)$ in Eq.\ (\ref{symm of W d-matrices 1}) as:
\begin{eqnarray}
    d^{\,(s)}_{mm'}(-\vartheta) 
    &\!\!=\!\!& d^{\,(s)}_{mm'}(-\vartheta) ^* \nonumber\\[.3 cm]
    &\!\!=\!\!& {_{z\!}}\langle s, m |e^{i \vartheta \hat{S}_{y}}| s, m' \rangle_{z}^* \nonumber\\[.3 cm]
    &\!\!=\!\!& {_{z\!}}\langle s, m' |e^{-i \vartheta \hat{S}_{y}}| s, m \rangle_{z} \nonumber \\[.3 cm]
    &\!\!=\!\!& d^{\,(s)}_{m'm}(\vartheta).
    \label{symm of W d-matrices 2}
\end{eqnarray}
Inserting this expression in Eq.\ (\ref{symm of W d-matrices 1}), we arrive at the symmetry stated in Eq.\ (\ref{WignerD sym 1}). 

To establish the second symmetry of the Wigner d-matrix, the one in Eq.\ (\ref{WignerD sym 2}), we consider a rotation through an angle of $180\degree$ around the $y$-axis, as implemented by $e^{-i\pi \hat{S}_y}$. This leaves $\hat{S}_y$ unchanged but flips $\hat{S}_z$. Thus the action of this rotation operator on the state $|s,m\rangle_z$---aside from on overall phase factor\footnote{This phase factor appears in \citet[Vol.\ 2, p.\ 1071, Eq.\ C65]{Messiah 1962} and may be proven for instance by appeal to Wigner's explicit formula (see p.\ 1072, Eq.\ C72 in Messiah) for the d-matrix elements. However, this phase factor does not enter into the probabilities and so we do not derive its value here.}
---is to replace $m$ by $-m$:
\begin{equation}
    e^{-i\pi \hat{S}_y} |s,m\rangle_z = (-1)^{s-m}|s,-m\rangle_z.
    \label{WignerD sym 2 proof a}
\end{equation}
Inserting the operators $e^{-i\pi \hat{S}_y}$ and $e^{i\pi \hat{S}_y}$ in the expression for the Wigner d-matrix in Eq.\ (\ref{wigner}),
\begin{equation}
 d^{\,(s)}_{mm'}(\vartheta) =  {_{z\!}}\langle s, m |e^{-i\pi \hat{S}_y}e^{-i \vartheta \hat{S}_{y}}e^{i\pi \hat{S}_y}| s, m' \rangle_{z},
\end{equation}
and using Eq.\ (\ref{WignerD sym 2 proof a}), we arrive at the symmetry in Eq.\ (\ref{WignerD sym 2}):  
\begin{eqnarray}
    d^{\,(s)}_{mm'}(\vartheta) &\!\!=\!\!& {_{z\!}}\langle s, -m |(-1)^{s-m}e^{-i \vartheta \hat{S}_{y}}(-1)^{s-m'}| s, -m' \rangle_{z} \nonumber \\[.3 cm]
    &\!\!=\!\!& (-1)^{m-m'} d^{\,(s)}_{-m-m'}(-\vartheta),
      \label{WignerD sym 2 proof b}
\end{eqnarray}
where in the last step we used that $(-1)^{s-m'}=(-1)^{m'-s}$.

The third symmetry of the Wigner d-matrix, the one in Eq.\ (\ref{WignerD sym 3}), follows as a corollary of the first two, the ones in Eqs.\ (\ref{WignerD sym 1}) and (\ref{WignerD sym 2}), though it can also be established directly through an argument similar to those in Eqs.\ (\ref{WignerD sym 1 proof a})--(\ref{symm of W d-matrices 2}) and Eqs.\ (\ref{WignerD sym 2 proof a})--(\ref{WignerD sym 2 proof a}), involving a rotation around the $x$-axis. 

These three symmetries of the Wigner d-matrix translate into three symmetries of the probabilities in Eq.\ (\ref{prob 3}):
\begin{equation}
\begin{array}{lcll}
\mathrm{Pr}(m_1 \, m_2 | \hat{a}\,\hat{b} ) & \!\!\! =  \!\!\! & \mathrm{Pr}(-m_2  -\!m_1 | \hat{a}\,\hat{b}) & \quad \quad {\mathrm{persymmetry}} \\[.3 cm]
  & \!\!\! =  \!\!\! & \mathrm{Pr}(-m_1  -\!m_2 | \hat{a}\,\hat{b}) & \quad \quad {\mathrm{centrosymmetry}}  \\[.3 cm]
  & \!\!\! =  \!\!\! & \mathrm{Pr}(m_2 \, m_1 | \hat{a}\,\hat{b} ) & \quad \quad {\mathrm{symmetry}}.    
\end{array}
\label{Prob sym}
\end{equation}
This tells us that the probability is unchanged if we either swap $m_1$ and $m_2$ or flip both of their signs. These symmetries, in turn, translate into symmetries of any cell in a correlation array for measurements on the singlet state of two particles with spin $s$. The first line in Eq.\ (\ref{Prob sym}) expresses that any such cell is \emph{persymmetric}, i.e., symmetric across its main anti-diagonal; the second that it is \emph{centrosymmetric}, i.e., symmetric about its center; the third that it is \emph{symmetric}, i.e., symmetric across its main diagonal. Any two of these symmetries imply the third. As we noted in the introduction to Section \ref{2}, in the spin-$\frac12$ case, centrosymmetry implies both symmetry and persymmetry. Since the design of our raffle tickets guarantees centrosymmetry, we did not have to impose any conditions on our raffles to ensure all three symmetries. As we will see in Section \ref{2.2}, however, such conditions are needed for higher-spin cases if we want our raffles to give correlation arrays with the same symmetries as the quantum correlation arrays they are supposed to simulate.

%SUBSECTION 3.2 (labels: 3.2, 3.2.1 through 3.2.4)
\subsection{Designing raffles to simulate the quantum correlations} \label{2.2}
%!TEX root =  ./JanasJanssenCuffaro-August2019.tex

%SUBSECTION 3.2
%\subsection{Designing raffles to simulate the quantum correlations} \label{2.2}

We will now design raffles to simulate the quantum correlations found in measurements on the singlet state of two particles with spin $s \ge \sfrac12$ that we investigated in Section \ref{2.1}. In Section \ref{2.2.1}, to set the stage and fix notation, we give a more formal analysis of the raffles for the spin-$\frac12$ case we introduced in Section \ref{1.4}. In Sections \ref{2.2.2}--{2.2.4} we gradually work our way up to higher-spin cases. 

%SUBSUBSECTION 3.2.1
\subsubsection{Spin-$\frac12$}  \label{2.2.1}

\begin{figure}[h]
 \centering
   \includegraphics[width=6in]{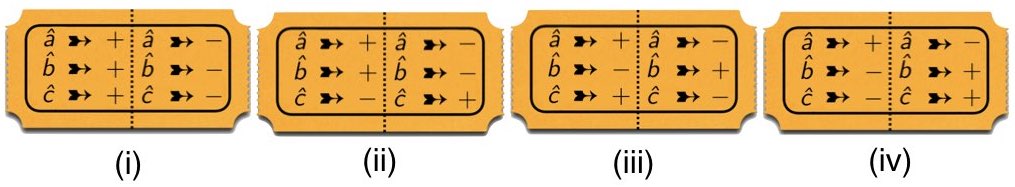} 
   \caption{The four different raffle tickets for three settings and two outcomes (cf.\ Figure \ref{raffle-tickets-3set2out-i-thru-iv}).}
   \label{raffle-tickets-3set2out-i-thru-iv-row}
   \end{figure}

The raffles we considered in Section \ref{1.4} all involved baskets containing a mixture of the four ticket types shown in Figure \ref{raffle-tickets-3set2out-i-thru-iv-row} (see the tree structure in Figure \ref{raffle-tickets-3set2out-i-thru-iv} for their numbering). Let $f^{\mathrm{k}}$ denote the fraction of tickets of type $(\mathrm{k})$ in such a basket (with $\mathrm{k} = \mathrm{i}, \mathrm{ii}, \mathrm{iii}, \mathrm{iv}$). These ticket fractions evidently are non-negative and normalized:
\begin{equation}
f^{\mathrm{k}}\geq 0 \; (\mathrm{k} = \mathrm{i}, \mathrm{ii}, \mathrm{iii}, \mathrm{iv}) \quad\text{and}\quad \sum_{\mathrm{k=i}}^{\mathrm{iv}} f^{\mathrm{k}}=1.
  \label{3-simplex}
\end{equation}

\begin{figure}[h]
 \centering
   \includegraphics[width=4.5in]{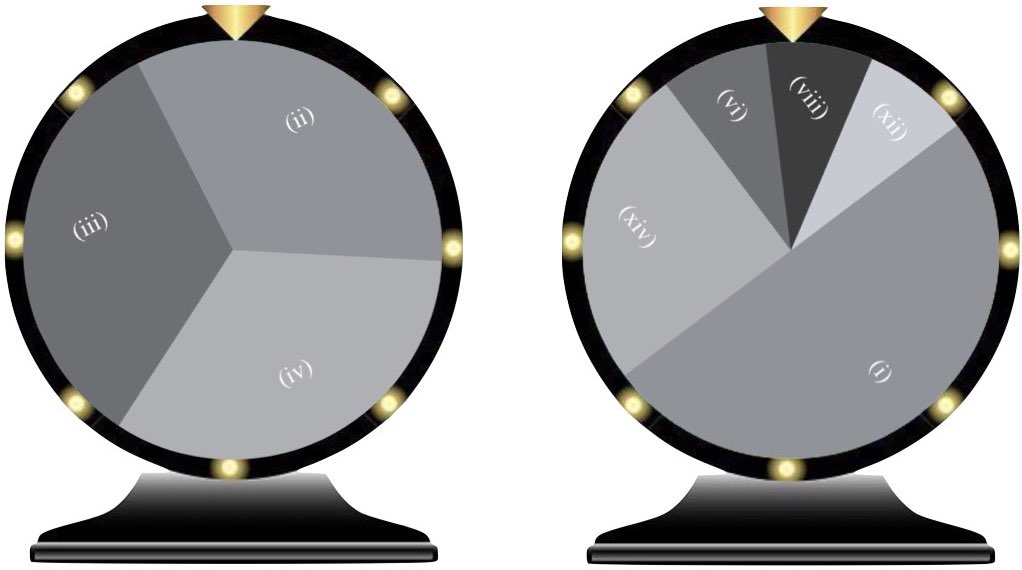} 
   \caption{Wheels of fortune for two raffles. The one on the left is for a raffle with a mix of $\sfrac13$ each of tickets of type (ii), (iii) and (iv) in Figure \ref{raffle-tickets-3set2out-i-thru-iv-row}. This is raffle (b) in Figure \ref{CA-3set2out-raffle-mix} in Section \ref{1.4}, our optimal simulation of the Mermin correlation array in Figure \ref{CA-3set2out-Mermin}. The one on the right is for a 75\%/25\% mix of the admissible raffles ($\alpha$) and ($\beta$) in Figure \ref{admissible-raffles-spin1} in Section \ref{2.2.2}.}
   \label{wheelsoffortune}
   \end{figure}

As we observed in Section \ref{1.4} (see note \ref{dense}), the imagery of a basket with a mix of tickets restricts us to values for $f^{\mathrm{k}}$ that are rational numbers. We will continue to discuss our raffles in terms of baskets of tickets, but we do want to point out that with a simple change of imagery we can accommodate real values as well. Instead of  baskets with tickets, consider wheels of fortune such as those shown in Figure \ref{wheelsoffortune}. These wheels have pie charts printed on them showing the mix of tickets in a particular raffle, each ticket of type $(\mathrm{k})$ occurring with a fraction $f^{\mathrm{k}}$ in the raffle corresponding to a segment $f^{\mathrm{k}} \times 100\%$ of the pie chart. Instead of randomly drawing a ticket from a basket, we would spin the wheel of fortune and pick a ticket of the type the pointer points to when the wheel of fortune comes to rest.   
          
%Pie charts for wheels of fortune for two raffles. The pie chart on the left is for a raffle with a mix of $\sfrac13$ each of tickets of type (ii), (iii) and (iv) in Figure \ref{raffle-tickets-3set2out-i-thru-iv-row}. This is raffle (b) in Figure \ref{CA-3set2out-raffle-mix} in Section \ref{1.4}, our optimal simulation of the Mermin correlation array in Figure \ref{CA-3set2out-Mermin}. The pie chart on the right is for a 75\%/25\% mix of the admissible raffles ($\alpha$) and ($\beta$) in Figure \ref{admissible-raffles-spin1} in Section \ref{2.2.2}. 

When we think of 
\begin{equation}
\vec{f}\equiv (f^{\mathrm{i}},f^{\mathrm{ii}},f^{\mathrm{iii}},f^{\mathrm{iv}})
\label{raffle point f}
\end{equation}
as a point in $\mathbb{R}^{4}$, the constraints in Eq.\ (\ref{3-simplex}) define what is conventionally known as the \emph{3D standard simplex} or \emph{3-simplex} in $\mathbb{R}^{4}$.  Equivalently, this simplex is the convex hull of the four points 
\begin{equation}
\vec{f}_{\mathrm{i}}= 
\begin{pmatrix}
1 \\
0 \\
0 \\
0
\end{pmatrix}, \quad 
\vec{f}_{\mathrm{ii}} = \begin{pmatrix}
0 \\
1 \\
0 \\
0
\end{pmatrix}, \quad
\vec{f}_{\mathrm{iii}} = \begin{pmatrix}
0 \\
0 \\
1 \\
0
\end{pmatrix}, \quad 
\vec{f}_{\mathrm{iv}} = \begin{pmatrix}
0 \\
0 \\
0 \\
1
\end{pmatrix},
\label{single ticket raffles}
\end{equation}
each corresponding to a raffle with just one type of ticket in the basket. These single-ticket raffles are \emph{basic} in the sense that any raffle can be obtained as a mix of them. As the notation already suggests, Eq.\ (\ref{single ticket raffles}) can also be seen as giving the unit vectors of an \emph{orthonormal basis} in which we can expand vectors characterizing arbitrary raffles. The quantity $\vec{f}$ in Eq.\ (\ref{raffle point f}) can then be thought of as the vector
\begin{equation}
\vec{f} =  \sum_{\mathrm{k=i}}^{\mathrm{iv}} f^{\mathrm{k}} \vec{f}_{\mathrm{k}}
\label{expansion of vec f} 
\end{equation}
and the ticket fractions $f^{\mathrm k}$ as its components in this basis.

The 3-simplex is a polytope in 4D Euclidean space and as such cannot be visualized. To circumvent this problem we consider triplets of anti-correlation coefficients rather than quartets of ticket fractions. The anti-correlation coefficient $\chi_{ab}|_{\vec{f}}$ for an arbitrary (mixed or single-ticket) raffle characterized by some vector $\vec{f}$ is defined as (cf.\ Eq.\ (\ref{chi as corr coef}) in Section \ref{1.3}):
\begin{equation}
 \chi_{ab} |_{\vec{f}} \equiv \left. -\frac{\langle X^A_a X^B_b \rangle}{\sigma^A_a \sigma^B_b} \right|_{\vec{f}}.
 \label{chi for f spin 1/2} 
\end{equation}
This anti-correlation coefficient is equal to the weighted average of those for the four single-ticket raffles in Eq.\ (\ref{single ticket raffles}) with the weights given by the ticket fractions $f^{\mathrm{k}}$. In Section \ref{1.4}, we already used this property of our raffles. We will now give a formal proof of it. 
%The result for the spin-$\frac12$ case can easily be generalized to higher-spin cases.
%some wrinkles we will encounter when dealing with raffles for higher-spin cases.  

We start by evaluating the covariance in the numerator on the right-hand side of Eq.\ (\ref{chi for f spin 1/2}) (cf.\ Eq.\ (\ref{prob 2 exp})). By definition,
\begin{equation}
\langle X^A_a X^B_b \rangle \big|_{\vec{f}\,} = \!\! \sum_{m_1, m_2} \! m_1  m_2 \, \mathrm{Pr}(m_1 m_2| \hat{a} \,\hat{b}) \Big|_{\vec{f}} \;, 
 \label{chi for f spin 1/2 a}
\end{equation}
where the outcomes $m_1$ and $m_2$ can only take on the values $\pm \sfrac12$ (if we set $\bbar$ and $\hbar$ equal to 1). It immediately follows from the design of our raffles that the probability of finding some combination of outcomes for some combination of settings in a raffle with a mix of tickets characterized by $\vec{f}$ is the weighted average of those same probabilities in the four basic single-ticket raffles with the weights given by the ticket fractions:
\begin{equation}
\mathrm{Pr}(m_1 m_2| \hat{a} \,\hat{b}) \big|_{\vec{f}\,} = \sum_{\mathrm{k} = \mathrm{i}}^{\mathrm{iv}} f^{\mathrm{k}} \, \mathrm{Pr}(m_1 m_2| \hat{a} \,\hat{b}) 
\Big|_{\vec{f}_{\mathrm{k}}}.
\label{chi for f spin 1/2 b}
\end{equation}
The probabilities on the right-hand side take on a very simple form: they are $\sfrac12$ if the ticket has the combination of outcomes $(X_a = m_1, X_b = m_2)$ and zero if it does not. For example, the tickets in Figure \ref{raffle-tickets-3set2out-i-thru-iv-row} tell us that 
\begin{equation}
\begin{array}{c}
 \mathrm{Pr}\!\left({\textstyle \frac12}, -{\textstyle \frac12} \Big| \hat{a} \,\hat{b}\right) \Big|_{\vec{f}_{\mathrm{i}}} 
 = \mathrm{Pr}\!\left({\textstyle \frac12}, -{\textstyle \frac12} \Big| \hat{a} \,\hat{b}\right) \Big|_{\vec{f}_{\mathrm{ii}}} \! = {\displaystyle{\frac12}}, \\[.6cm]
\mathrm{Pr}\!\left({\textstyle \frac12}, -{\textstyle \frac12} \Big| \hat{a} \,\hat{b}\right) \Big|_{\vec{f}_{\mathrm{iii}}} 
\!\! = \mathrm{Pr}\!\left({\textstyle \frac12}, -{\textstyle \frac12} \Big| \hat{a} \,\hat{b}\right) \Big|_{\vec{f}_{\mathrm{iv}}} \!\! = 0. 
\end{array}
\label{chi for f spin 1/2 ba}
\end{equation}
For this combination of outcomes, Eq.\ (\ref{chi for f spin 1/2 b}) thus gives
\begin{equation}
\mathrm{Pr}\!\left({\textstyle \frac12}, -{\textstyle \frac12} \Big| \hat{a} \,\hat{b}\right) \Big|_{\vec{f}} = \frac12 \big( f^{\mathrm{i}} + f^{\mathrm{ii}} \big).
\end{equation}
As this example illustrates, Eq.\ (\ref{chi for f spin 1/2 b}) is a more formal expression of another feature of our raffles we repeatedly made use of in Section \ref{1.4}: the entries in the correlation array for a mixed raffle are weighted averages of the corresponding entries in the correlation arrays for single-ticket raffles. 

Inserting Eq.\ (\ref{chi for f spin 1/2 b}) into Eq.\ (\ref{chi for f spin 1/2 a}) and changing the order of the summations, we see that the covariance $\langle X^A_a X^B_b \rangle$ in a mixed raffle is likewise the weighted average of that same covariance in the four single-ticket raffles:
\begin{equation}
\langle X^A_a X^B_b \rangle \big|_{\vec{f}} = \sum_{\mathrm{k} = \mathrm{i}}^{\mathrm{iv}} f^{\mathrm{k}} \sum_{m_1, m_2} \! m_1  m_2 \, \mathrm{Pr}(m_1 m_2| \hat{a} \,\hat{b}) 
\Big|_{\vec{f}_{\mathrm{k}}} =  \sum_{\mathrm{k} = \mathrm{i}}^{\mathrm{iv}} f^{\mathrm{k}} \langle X^A_a X^B_b \rangle \big|_{\vec{f}_{\mathrm{k}}}. 
\label{chi for f spin 1/2 c}
\end{equation}

Because the diagonal cells in the correlation arrays for raffles with any mix of tickets of types (i) through (iv) in Figure \ref{raffle-tickets-3set2out-i-thru-iv-row} are the same, the standard deviations in the expression on the right-hand side of Eq.\ (\ref{chi for f spin 1/2 a}) are also the same for all these raffles. For any $\vec{f}$, they are given by Eq.\ (\ref{SD for adm raffle 2}) in Section \ref{1.6} for $s=\sfrac12$:
\begin{equation}
\sigma_a  \big|_{\vec{f}} = \sigma_b  \big|_{\vec{f}} = \sigma_{s = \sfrac12} = \sqrt{\frac{s(s+1)}{3}} = \frac12.
\label{chi for f spin 1/2 d}
\end{equation}

Substituting Eq.\ (\ref{chi for f spin 1/2 c}) into Eq.\ (\ref{chi for f spin 1/2}) and using Eq.\ (\ref{chi for f spin 1/2 d}), we arrive at
\begin{equation}
\chi_{ab} |_{\vec{f}} = -\frac{\langle X^A_a X^B_b \rangle \big|_{\vec{f}}}{\sigma_{s=\sfrac12}^2}  =
\sum_{\mathrm{k} = \mathrm{i}}^{\mathrm{iv}} f^{\mathrm{k}} \left( \! \left. -\frac{\langle X^A_a X^B_b \rangle  }{ \sigma_a \sigma_b }  \right|_{ \vec{f}_{ \mathrm{k} } } \right)
=  \sum_{\mathrm{k} = \mathrm{i}}^{\mathrm{iv}} f^{\mathrm{k}} \, \chi_{ab} |_{\vec{f}_{\mathrm{k}} },
\label{chi for f spin 1/2 e}
\end{equation}
which is what we set out to prove. 

Similar relations hold for $\chi_{ac}$ and $\chi_{bc}$. We combine these three relations and write them in matrix form: 
\begin{equation}
\left. \begin{pmatrix}
\chi_{ab}\\[.2cm]
\chi_{ac}\\[.2cm]
\chi_{bc}
\end{pmatrix} \right|_{\vec{f}} = 
\begin{pmatrix}
   \chi_{ab} |_{\vec{f}_{\mathrm{i}} } \; & \; \chi_{ab} |_{\vec{f}_{\mathrm{ii}} } \; & \; \chi_{ab} |_{\vec{f}_{\mathrm{iii}} } \; & \; \chi_{ab} |_{\vec{f}_{\mathrm{iv}} } \\[.4cm]
     \chi_{ac} |_{\vec{f}_{\mathrm{i}} } & \chi_{ac} |_{\vec{f}_{\mathrm{ii}} } & \chi_{ac} |_{\vec{f}_{\mathrm{iii}} } & \chi_{ac} |_{\vec{f}_{\mathrm{iv}} }  \\[.4cm]
     \chi_{bc} |_{\vec{f}_{\mathrm{i}} } & \chi_{bc} |_{\vec{f}_{\mathrm{ii}} } & \chi_{bc} |_{\vec{f}_{\mathrm{iii}} } & \chi_{bc} |_{\vec{f}_{\mathrm{iv}} }  
\end{pmatrix}
\begin{pmatrix} f^{\mathrm{i}} \\[.2cm]
f^{\mathrm{ii}} \\[.2cm]
f^{\mathrm{iii}} \\[.2cm]
f^{\mathrm{iv}} \end{pmatrix}.
\label{Mapping spin 1/2}
\end{equation}
The $3 \times 4$ matrix $M$ on the right-hand side thus serves as a map from the 3-simplex of possible ticket fractions to the \emph{anti-correlation polyhedron} (see the introduction of Section \ref{2} for a definition). Denoting the vector on the left-hand side as $\vec{\chi}|_{\vec{f}}$, we can write Eq.\ (\ref{Mapping spin 1/2}) more compactly as
\begin{equation}
\vec{\chi}|_{\vec{f}} \equiv M \vec{f}.
\label{Mapping spin 1/2 a}
\end{equation}
The matrix elements of $M$ can be read off of Table \ref{values of chi} in Section \ref{1.4} (or directly from the tickets in Figure \ref{raffle-tickets-3set2out-i-thru-iv-row}: note that $1/\sigma_{s=\sfrac12}^2 = 4$ cancels the 4 in the denominators of the covariances):
\begin{equation}
M = \begin{pmatrix}
    M_{ab}^{\mathrm{i}} & M_{ab}^{\mathrm{ii}} & M_{ab}^{\mathrm{iii}} & M_{ab}^{\mathrm{iv}} \\[.2cm]
    M_{ac}^{\mathrm{i}} & M_{ac}^{\mathrm{ii}} & M_{ac}^{\mathrm{iii}} & M_{ac}^{\mathrm{iv}} \\[.2cm] 
    M_{bc}^{\mathrm{i}} & M_{bc}^{\mathrm{ii}} & M_{bc}^{\mathrm{iii}} & M_{bc}^{\mathrm{iv}}
\end{pmatrix}
= \begin{pmatrix}
    1 & 1 & -1 & -1 \\[.2cm]
    1 & -1 & 1 & -1 \\[.2cm]
    1 & -1 & -1 & 1 
\end{pmatrix}.
\label{Mapping spin 1/2 b}
\end{equation}
$M$ will map the basis vectors in Eq.\ (\ref{single ticket raffles}) for the four single-ticket raffles onto vectors whose components are the columns of $M$:
\begin{equation}
\vec{\chi}|_{\vec{f}_{\mathrm{i}}} = 
\begin{pmatrix}
1 \\
1 \\
1
\end{pmatrix},
\;\;
\vec{\chi}|_{\vec{f}_{\mathrm{ii}}} = 
\begin{pmatrix}
1 \\
-1 \\
-1
\end{pmatrix},
\;\;
\vec{\chi}|_{\vec{f}_{\mathrm{iii}}} =
\begin{pmatrix}
-1 \\
1 \\
-1
\end{pmatrix},
\;\;
\vec{\chi}|_{\vec{f}_{\mathrm{iv}}} = 
\begin{pmatrix}
-1 \\
-1 \\
1
\end{pmatrix}.
\label{Mapping spin 1/2 c}
\end{equation}
These vectors correspond to the vertices (i) through (iv) of the tetrahedron we found in Section \ref{1.4} (see Figure \ref{tetrahedron}).

%SUBSUBSECTION 3.2.2
\subsubsection{Spin-1}  \label{2.2.2}

\begin{figure}[h!]
 \centering
   \includegraphics[width=6in]{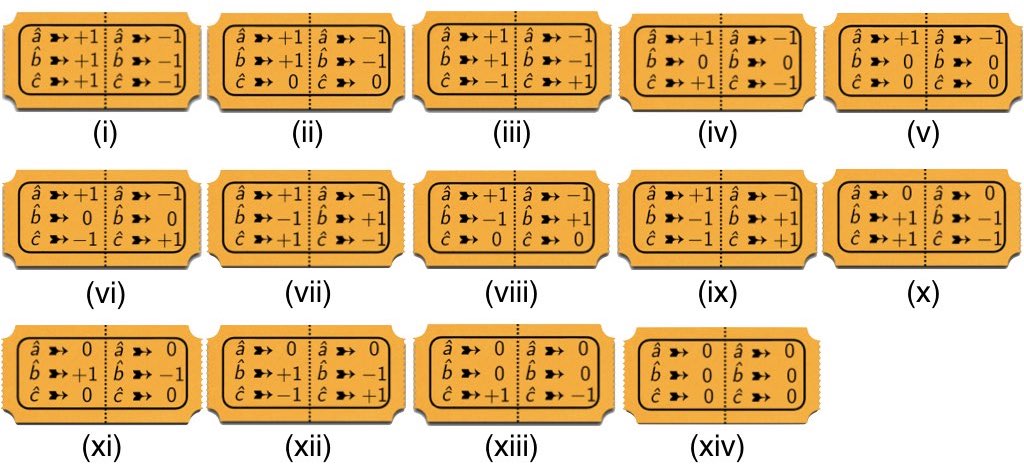} 
   \caption{The fourteen different types of raffle tickets for three settings and three outcomes that are left once the condition is imposed that, when Alice and Bob use the same setting, they should find opposite results if the outcome is $+1$ or $-1$ but the same result if the outcome is $0$.}
   \label{raffle-tickets-3set3out-i-xiv}
\end{figure}

Having reviewed the case of spin-$\frac12$, we now move to new territory and consider the case of spin-$1$. The first change is to the number of tickets. There are now three possible outcomes for the three settings: $1$, $0$ and $-1$ (again setting $\hbar = \bbar =1$). This means that there are now $3^3=27$ ways to specify the outcomes on one side of the ticket, which, as before, determine the outcomes on the other side. Since it is totally random which side of the ticket goes to Alice and which side goes to Bob, it suffices to consider the 14 ticket types labeled $(\mathrm{i})$ through $(\mathrm{xiv})$ shown in Figure \ref{raffle-tickets-3set3out-i-xiv} (to number these we used a tree structure similar to the one in Figure \ref{raffle-tickets-3set2out-i-thru-iv}). 

\begin{figure}[h]
 \centering
   \includegraphics[width=5.5in]{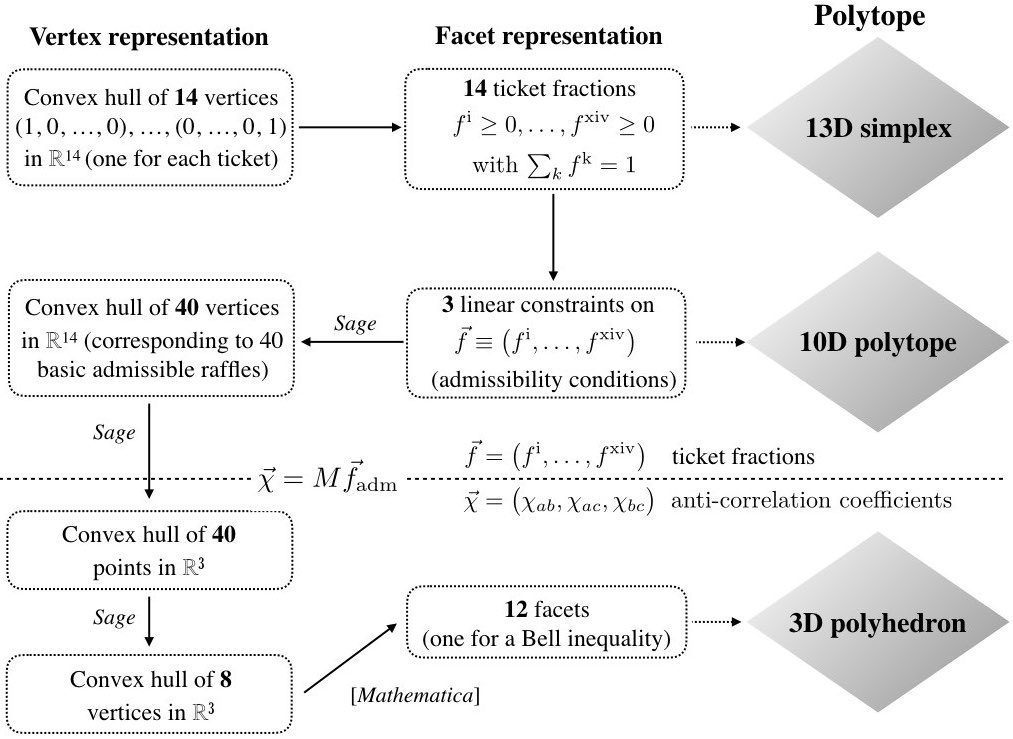} 
   \caption{Flow chart illustrating the construction of the polyhedron characterizing the set of correlations that (a) can be generated with raffles mixing tickets of types (i) through (xiv) in Figure \ref{raffle-tickets-3set3out-i-xiv} and (b) replicate key features of the quantum correlations for measurements on the singlet state of two spin-1 particles.}
   \label{flowchart}
   \end{figure}

A generic raffle is some mix of these 14 ticket types. Geometrically this corresponds to a point in the 13D simplex, defined as the convex hull of the 14 points corresponding to the standard unit basis vectors in $\mathbb{R}^{14}$ (cf.\ Eqs.\ (\ref{3-simplex})--(\ref{expansion of vec f}), with the index k now running from i to xiv). The diagram in Figure \ref{flowchart} provides a flow chart for how to get from this simplex in $\mathbb{R}^{14}$ to a polyhedron in $\mathbb{R}^3$ characterizing the class of quantum correlations found in measurements on the singlet state of two spin-1 particles that can be simulated---albeit, as we will see, imperfectly---with these raffles. 

A polytope can be represented either in terms of its vertices or in terms of its facets. These are called the \emph{V-representation} and the \emph{H-representation}, respectively. The $H$-representation is given in terms of a set of inequalities restricting points to be on one side of some (hyper-)plane (hence the ``$H$'', which stands for ``half-space"). As our flow chart illustrates, we switch back and forth between these two representations as we go from the 13D simplex to the 3D polyhedron. For several computation-intensive
steps we used the open-source mathematical software system SageMath. 

\begin{figure}[h]
 \centering
   \includegraphics[width=4in]{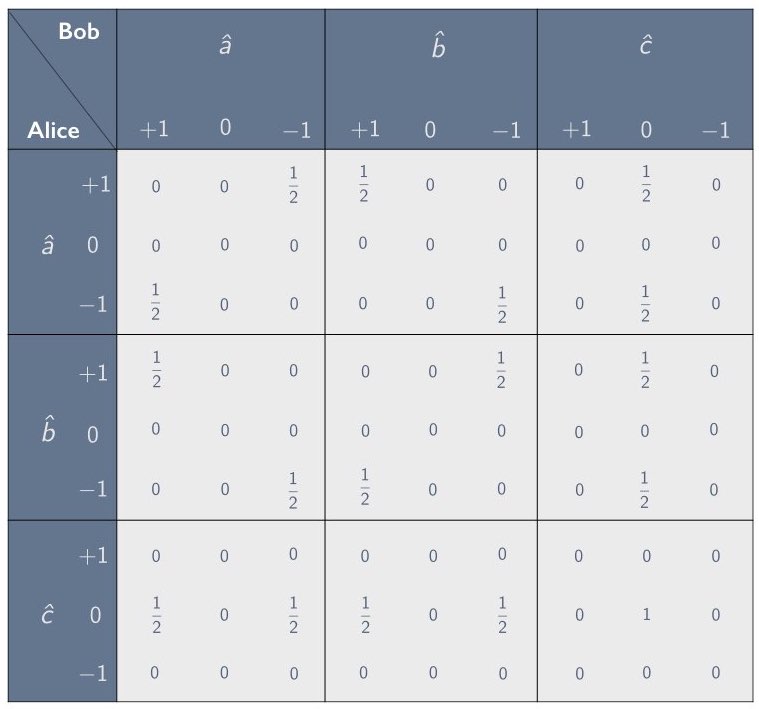} 
   \caption{Correlation array for a single-ticket raffle with tickets of type (vi) (see Figure \ref{raffle-tickets-3set3out-i-xiv}).}
   \label{CA-3set3out-raffle-vi}
\end{figure}

The tickets for the spin-1 case immediately reveal one key difference with the spin-$\frac12$ case that we already drew attention to in Section \ref{1.6}. Our raffle tickets can only have two outcomes for each setting. As soon as there are more than two possible outcomes (and for spin $s$ there are $2s + 1$), it is thus impossible for all outcomes to occur in equal proportion in a single-ticket raffle. Yet this is what Alice and Bob find if they use the same setting in the quantum experiment these raffles are supposed to simulate. In terms of probabilities, the problem is that for spins greater than $\sfrac12$, single-ticket raffles---while non-signaling by construction---do not give uniform marginals, whereas measurements on two particles of arbitrary spin entangled in the singlet state do (as we showed in Section \ref{2.1.4}). The correlation array in Figure \ref{CA-3set3out-raffle-vi} for a single-ticket raffle illustrates the problem. The array is non-signaling but the marginals take on three different values: $0$, $\sfrac12$ and $1$. The solution to this problem is to allow only mixed raffles that give uniform marginals. 

\begin{figure}[h]
 \centering
   \includegraphics[width=5in]{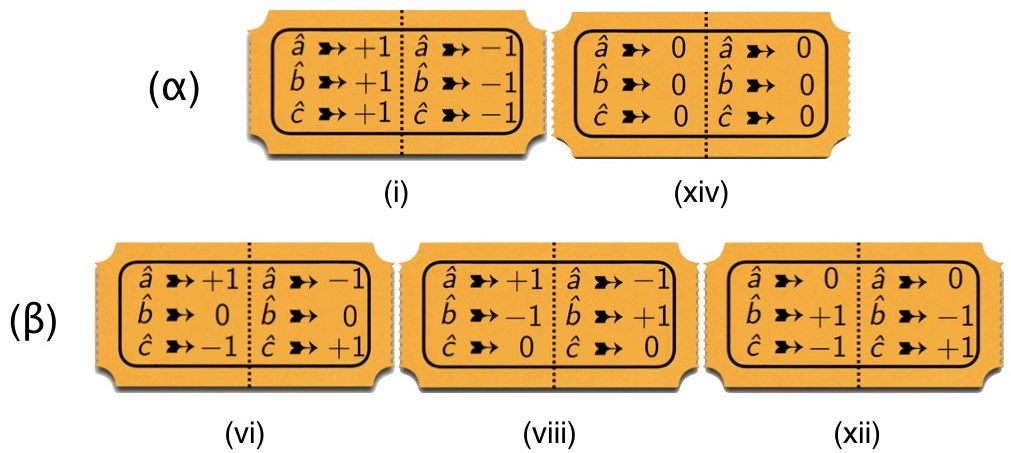} 
   \caption{Two mixed raffles giving uniform marginals: in ($\alpha$) $\sfrac23$ of the tickets are of type (i) and $\sfrac13$ is of type (xiv); ($\beta$) has $\sfrac13$ each of types (vi), (viii) and (xii).}
   \label{admissible-raffles-spin1}
\end{figure}

Before we show how to construct such admissible raffles, we remind the reader of another property of the correlation array in Figure \ref{CA-3set3out-raffle-vi}: the $9 \times 9$ matrix formed by its entries is symmetric. This is a direct consequence of the design of our raffles (for any number of outcomes per setting). Since the correlation array obviously stays the same if Alice and Bob swap both the ticket stubs they receive and the settings they use, its entries for any one of our raffles satisfy  
\begin{equation}
\mathrm{Pr}\!\left(m_1 m_2 \big| \hat{a} \,\hat{b}\right)
=  \mathrm{Pr}\!\left(m_2 m_1 \big| \hat{b} \,\hat{a}\right).
\label{swap symmetry}
\end{equation}

In our raffles for the spin-$\frac12$ case, it followed directly from the symmetry of the $6 \times 6$ matrix for the correlation array as a whole that the $2 \times 2$ matrices formed by the entries of its individual cells are also symmetric and that cells on opposite sides of the diagonal are the same (see, e.g., Figure \ref{CA-3set2out-raffles-i-thru-iv}). For the correlation arrays for the quantum correlations that we are trying to simulate with our raffles, both claims are true for arbitrary spin $s$ (see Figure \ref{CA-cell-spin1-chi} for spin-1; recall that $\chi_{ab} = \cos{\varphi_{ab}} = \cos{\varphi_{ba}} = \chi_{ba}$). Neither one is true, however, for the cells in the correlation array for the single-ticket raffle in Figure \ref{CA-3set3out-raffle-vi}.  In four of the nine cells, the $3 \times 3$ matrices formed by its entries are not symmetric and the matrices for cells on opposite sides of the diagonal are each other's transpose. 

Fortunately, these discrepancies are easily dealt with. First note that the transpose of a symmetric matrix is that matrix itself. Demanding individual cell symmetry thus suffices to ensure that cells on opposite sides of the diagonal are identical. Second, with raffles for the spin-1 case, as we will see, demanding uniform marginals suffices to ensure cell symmetry. In raffles for higher-spin cases, however, cell symmetry calls for additional admissibility conditions (see Section \ref{2.2.3}).

There is one more difference between the spin-$\frac12$ and the spin-1 case that we want to highlight. In the spin-$\frac12$ case, as we saw in Section \ref{1.4},  the probabilities in a cell of a correlation array are fully determined by the anti-correlation coefficient for that cell. In quantum mechanics, this remains true for spin $s > \sfrac12$ even though the dependence of the probabilities on the anti-correlation coefficient is no longer linear (see, e.g., Figure \ref{CA-cell-spin1-chi} in Section \ref{2.1} for $s=1$). As soon as $s > \sfrac12$, however, it is no longer true for the raffles meant to simulate these quantum correlations. In the spin-1 case, as we will see in Section \ref{2.2.3}, it takes two parameters to fix the entries in any off-diagonal cell of a correlation array for an admissible raffle. In these higher spin cases, the same triplet of anti-correlation coefficients will therefore in general correspond to more than one admissible raffle. This should not surprise us. We should not expect the projection from higher-dimensional polytopes of admissible raffles to anti-correlation polyhedra to be injective.    

We are now ready to start constructing raffles for the spin-1 case that give uniform marginals. Figure \ref{admissible-raffles-spin1} shows two examples of such raffles. The first, labeled ($\alpha$) and characterized by the vector 
\begin{equation}
\vec{f}_\alpha = \frac23 \vec{f}_{\mathrm{i}} + \frac13 \vec{f}_{\mathrm{xiv}},
\label{spin 1 raffle alpha}
\end{equation}
is probably the easiest way to ensure uniform marginals. It uses tickets with the same outcomes for all three settings. With two tickets of type (i) and one of type (xiv), all three outcomes occur six times for all three settings. That means that, for all three settings, all three outcomes will be found with equal probability by both Alice and Bob. In the raffle labeled ($\beta$) and characterized by the vector  
\begin{equation}
\vec{f}_\beta =  \frac13 \left( \vec{f}_{\mathrm{vi}} + \vec{f}_{\mathrm{viii}} + \vec{f}_{\mathrm{xii}}\right),
\label{spin 1 raffle beta}
\end{equation}
the outcomes for setting $\hat{a}$ are the same as those in raffle ($\alpha$) while the outcomes for settings $\hat{b}$ and $\hat{c}$ are permutations of those in raffle ($\alpha$). These permutations were chosen so as to make the sum $X_a + X_b + X_c$ vanish on both sides of all three tickets. As we saw in Section \ref{1.6}, this means that the (anti-)correlations produced by this raffle are represented by a point on the elliptope in Figure \ref{elliptope}. 

\begin{figure}[h]
 \centering
   \includegraphics[width=4in]{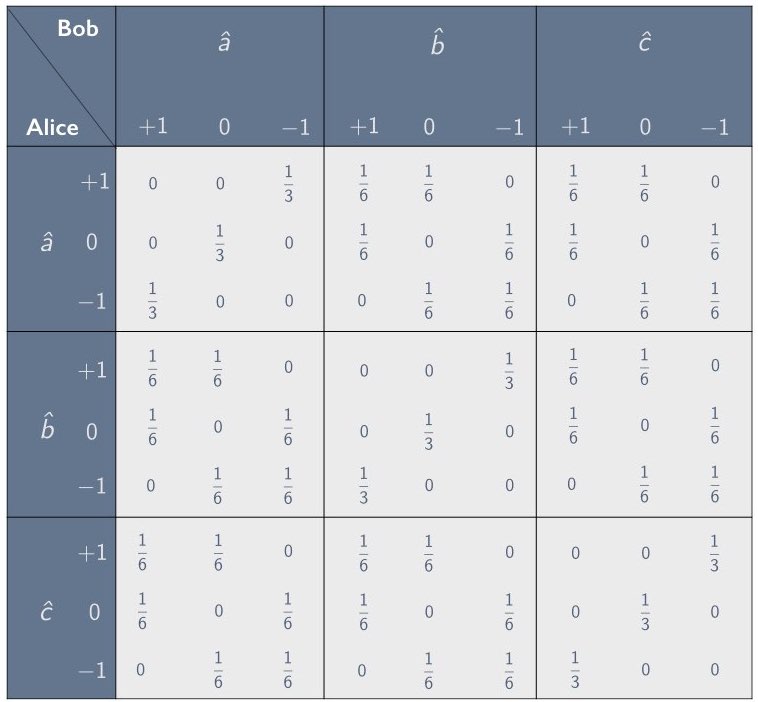} 
   \caption{Correlation array for a raffle with equal numbers of tickets of types (vii), (viii) and (xii) (see Figure \ref{admissible-raffles-spin1}, ($\beta$)).}
   \label{CA-3set3out-raffle-33vi33viii33xii}
\end{figure}

Figure \ref{CA-3set3out-raffle-33vi33viii33xii} shows the correlation array for the mixed raffle ($\beta$). Unlike the correlation array in Figure \ref{CA-3set3out-raffle-vi} (for a single-ticket raffle), it has uniform marginals, as we demanded. The diagonal cells in correlation arrays for raffles that give uniform marginals are all the same. The standard deviations $\sigma_a$, $\sigma_b$ and $\sigma_c$ for such raffles are equal to $\sigma_{s=1} = \sqrt{\sfrac23}$, in accordance with the general formula $\sigma_s^2 = \frac13 s(s+1)$ (see Eq.\ (\ref{SD for adm raffle 2}) in Section \ref{1.6}). This is not true for single-ticket raffles or raffles with an arbitrary mix of ticket types. This provides another reason for restricting ourselves to raffles that give uniform marginals. The projection from $\mathbb{R}^{14}$ to $\mathbb{R}^3$ we are after only works if we can assume that the diagonal cells of the correlation arrays and thus the standard deviations $\sigma_a$, $\sigma_b$ and $\sigma_c$ are the same for all raffles considered.

With the help of the correlation array in Figure \ref{CA-3set3out-raffle-33vi33viii33xii}, we find that
\begin{equation}
\langle X^A_a X^B_b \rangle \big|_{\vec{f}_\beta} = \mathrm{Pr}(1 1| \hat{a} \,\hat{b}) \Big|_{\vec{f}_\beta} + \mathrm{Pr}(-1 -\!1| \hat{a} \,\hat{b}) \Big|_{\vec{f}_\beta} 
= \frac13
\label{covariance ab raffle beta}
\end{equation}
Inserting Eq.\ (\ref{covariance ab raffle beta}) and $\sigma^A_a \sigma^A_b = \sigma_{s=1}^2 = \sfrac23$ into Eq.\ (\ref{chi for f spin 1/2}), we arrive at 
\begin{equation}
 \chi_{ab} |_{\vec{f}_\beta} = \left. -\frac{\langle X^A_a X^B_b \rangle}{\sigma^A_a \sigma^B_b} \right|_{\vec{f}_\beta} = -\sfrac12.
\end{equation}
We similarly find that $ \chi_{ac} |_{\vec{f}_\beta} =  \chi_{bc} |_{\vec{f}_\beta} = -\sfrac12$. The raffle ($\beta$) is thus represented by the point 
\begin{equation}
\left( \chi_{ab}, \chi_{ac}, \chi_{bc} \right)  \big|_{\vec{f}_\beta} = \left(-\sfrac12, -\sfrac12, -\sfrac12 \right)
\label{raffle beta = -1/2^3}
\end{equation}
on the elliptope, where the sum $\chi_{ab} + \chi_{ac} + \chi_{bc}$ has its absolute minimum value of $-\sfrac32$ (cf.\ Section \ref{1.6}). With raffle ($\beta$) we have thus constructed a toy example of a local hidden-variable theory with which we can reach the Tsirelson bound  for this setup. This raffle, however, still fails to reproduce all features of the correlation array for the corresponding quantum case. 

In quantum mechanics, Eq.\ (\ref{raffle beta = -1/2^3}) holds for the correlations found in measurements on the singlet state of two particles of arbitrary (half-)integer spin $s$ if each of  the angles $\varphi_{ab}$, $\varphi_{ac}$ and $\varphi_{bc}$ between the three measuring directions is equal to $120\degree$. In that case the anti-correlation coefficients $\chi_{ab}$, $\chi_{ac}$ and $\chi_{bc}$ are all equal to $\cos 120\degree = -\sfrac12$. By demanding uniform marginals, we made sure that the diagonal cells of the correlation array for raffle ($\beta$) in Figure \ref{CA-3set3out-raffle-33vi33viii33xii} are the same as those of the quantum correlation array for $s=1$ we are trying to simulate. The six off-diagonal cells, however, while identical to each other, differ from the six identical off-diagonal cells in this quantum correlation array. Below we put the entries of two of these off-diagonal cells side-by-side, for raffle ($\beta$) on the left, for the quantum correlations on the right (we obtain the latter by substituting $\chi_{ab} = -\sfrac12$ in the $\hat{a} \hat{b}$ cell in Figure \ref{CA-cell-spin1-chi} in Section \ref{2.1}):  
\begin{equation}
\frac{1}{6}
\begin{pmatrix}
\; 1 & 1 & 0 \; \\[.2cm]
\; 1 & 0 & 1 \; \\[.2cm]
\; 0 & 1 & 1 \;
\end{pmatrix},
\quad\quad
\frac{1}{48}
\begin{pmatrix}
\; 9 & 6 &1 \; \\[.2cm]
\; 6 & 4 & 6 \;\\[.2cm]
\; 1 & 6 & 9 \; 
\end{pmatrix}.
\label{off diag cell quantum v raffle}
\end{equation}

In both these cells, rows and columns sum to $\sfrac13$ and the anti-correlation coefficient is equal to $-\sfrac12$. Both cells are symmetric, persymmetric and centrosymmetric. In the full correlation array, $\chi_{ab} + \chi_{ac} + \chi_{bc} = -\sfrac32$ in both cases. Yet, despite reproducing all these features of the quantum correlations, our raffle still fails to fully simulate the quantum correlations. As Eq.\ (\ref{off diag cell quantum v raffle}) shows, it it is impossible in our raffle for Alice and Bob to find opposite results when using different measurement settings (that would be incompatible with $X^A_a + X^A_b + X^A_c = 0$), whereas in measurements on the singlet state of two spin-1 particles there is a small probability that they do: $\sfrac{1}{48}$ for one of them finding $+1$ and the other one finding $-1$; $\sfrac{1}{12}$ for both of them finding $0$. Finding one such outcome in an actual experiment would thus disprove our local hidden-variable theory even though this theory does not put a tighter bound on the sum of the anti-correlation coefficients than quantum mechanics!

Raffles ($\alpha$) and ($\beta$) are just two examples of raffles that give uniform marginals. We now determine what conditions the vector $\vec{f}$ for some mixed raffle has to satisfy to ensure uniform marginals. As noted above, in the spin-1 case this is the only requirement for a raffle to be \emph{admissible}.
% (in higher-spin cases we need to impose additional symmetry requirements as well). 
Consider the vector giving the ticket fractions for such an admissible raffle:
\begin{equation}
\vec{f}_{\mathrm{adm}} = \sum_{\mathrm{k = i}}^{\mathrm{xiv}} f_{\mathrm{adm}}^{\mathrm{k}} \, \vec{f}_{\mathrm{k}}.
\end{equation}
To ensure that some mixed raffle gives uniform marginals it suffices to require the diagonal cells in its correlation array to have the form
\begin{equation}
\left(
\begin{array}{ccc}
0  & 0 & \sfrac13  \\[.2cm]
0  & \sfrac13  &  0  \\[.2cm]
\sfrac13  &  0  &  0 
\end{array}
\right).
\label{adm spin1 diag}
\end{equation}
Since the correlations produced by our raffles are, by construction, non-signaling, this will automatically take care of the off-diagonal cells. In principle, we thus need to impose the following nine conditions 
\begin{equation}
\mathrm{Pr}(m\,m|\hat{a} \, \hat{a})=\mathrm{Pr}(m\,m|\hat{b} \, \hat{b})=\mathrm{Pr}(m\,m|\hat{c} \, \hat{c})=\sfrac13,
\end{equation} 
for $m=0, \pm 1$. However, we know from normalization and centrosymmetry that
\begin{eqnarray}
1 =  \!  \sum_{m_1, m_2}  \!  \mathrm{Pr}(m_1 m_2|\hat{a}\,\hat{a}) & \!\! = \!\! & \mathrm{Pr}(1\,1|\hat{a}\hat{a})+\mathrm{Pr}(0\,0|\hat{a} \, \hat{a})+\mathrm{Pr}(-\!1 -\!1|\hat{a}\,\hat{a})  \nonumber \\
&  \!\! = \!\!  & 2 \mathrm{Pr}(1\,1|\hat{a}\hat{a})+\mathrm{Pr}(0\,0|\hat{a}\,\hat{a}),
\end{eqnarray}
and similarly for the diagonal cells with settings $\hat{b}\, \hat{b}$ and $\hat{c}\,\hat{c}$. Hence it suffices to impose 
\begin{equation}
\mathrm{Pr}(0\,0|\hat{a}\,\hat{a})=\mathrm{Pr}(0\,0|\hat{b}\,\hat{b})=\mathrm{Pr}(0\,0|\hat{c}\,\hat{c})=\sfrac13.
\label{uniform marginals spin-1}
\end{equation}
These probabilities can be expressed in terms of ticket fractions (exactly how can be seen upon inspection of the tickets in Figure \ref{raffle-tickets-3set3out-i-xiv}):
\begin{equation}
\begin{array}{c}
\mathrm{Pr}(0\,0|\hat{a}\,\hat{a}) =  f^{\mathrm{x}}+f^{\mathrm{xi}}+f^{\mathrm{xii}}+f^{\mathrm{xiii}}+f^{\mathrm{xiv}},  \\[.4cm]
\mathrm{Pr}(0\,0|\hat{b}\,\hat{b})= f^{\mathrm{iv}}+f^{\mathrm{v}}+f^{\mathrm{vi}}+f^{\mathrm{xiii}}+f^{\mathrm{xiv}},  \\[.4cm]
\mathrm{Pr}(0\,0|\hat{c}\,\hat{c})=f^{\mathrm{ii}}+f^{\mathrm{v}}+f^{\mathrm{viii}}+f^{\mathrm{xi}}+f^{\mathrm{xiv}}.
\end{array}
\label{spin 1 constraints}
\end{equation}
The admissibility conditions in this case thus boil down to three linear constraints on the ticket fractions $(f^{\mathrm{i}}, \ldots, f^{\mathrm{xiv}})$. These will restrict the 13D simplex of raffles to a 10D convex polytope of admissible raffles. We used SageMath to compute its vertices. This yields 40 vertices in $\mathbb{R}^{14}$, which we do not reproduce for reasons of space. These corresponds to 40 \emph{basic admissible raffles} (i.e., any admissible raffle can be obtained by mixing these 40).

Like the 13D simplex of arbitrary mixed raffles, the 10D polytope of admissible raffles cannot be visualized as such. As we did in the spin-$\frac12$ case (see Eqs.\ (\ref{chi for f spin 1/2})--(\ref{Mapping spin 1/2 c})), we therefore switch from ticket fractions to anti-correlation coefficients. In other words, we map the 10D polytope to a 3D polyhedron. As before (see Eq.\ (\ref{Mapping spin 1/2 a})), the mapping can compactly be written as
\begin{equation}
\vec{\chi} \big|_{\vec{f}_{\mathrm{adm}}} = M \vec{f}_{\mathrm{adm}}.
\label{mapping spin 1 a}
\end{equation}
What complicates matters in this case is that $M$ is no longer given by $\vec{\chi} \big|_{\vec{f}_{\mathrm{k}}}$, as it was in the spin-$\frac12$ case (see Eq.\ (\ref{chi for f spin 1/2 e})). This is because the standard deviations $\sigma_a$,  $\sigma_b$ and  $\sigma_c$ are different for different single-ticket raffles in the spin-1 case. However, the components of $\vec{\chi} \big|_{\vec{f}_{\mathrm{adm}}}$ can still be written as a sum of  covariances for single-ticket raffles. The $ab$ component, for instance, can be written as (cf.\ Eq.\ (\ref{chi for f spin 1/2 e})):
\begin{equation}
\chi_{ab} \big|_{\vec{f}_{\mathrm{adm}}} = - \frac{1}{\sigma_{s=1}^2} \sum_{\mathrm{k = i}}^{\mathrm{xiv}} f^{\mathrm{k}}_{\mathrm{adm}} \langle X^A_a X^B_b \rangle \Big|_{\vec{f}_{\mathrm{k}}}.
\label{mapping spin 1 b}
\end{equation}
Similar results hold for the other components of $\vec{\chi} \big|_{\vec{f}_{\mathrm{adm}}}$. Comparing Eq.\ (\ref{mapping spin 1 a}) and Eq.\ (\ref{mapping spin 1 b}) and using that $\sigma_{s=1}^2 = \sfrac23$, we see that the components of $M$ in this case are given by 
\begin{equation}
M^{\mathrm{k}}_{ab} = - \frac32 \, \langle X^A_a X^B_b \rangle \Big|_{\vec{f}_{\mathrm{k}}}, \quad \mathrm{with \; k = i \, \ldots \, xiv},
\label{M_ab^i spin-1}
\end{equation}
and similar expressions with $ab$ replaced by $ac$ or $bc$. The covariances for single-ticket raffles on the right-hand side of these expressions are collected in Table \ref{covariances for spin 1}. 
%It may look as if tickets for which all three covariances are zero (i.e., tickets of type (v), (xi), (xiii) and (xiv)) are irrelevant but they serve the purpose of ensuring uniform marginals in some admissible raffles. 

\begin{table}[h]
\centering
\begin{tabular}{|c||c|c|c|}
\hline
\Big. $\!\!\!$ticket$\!\!\!$ & $\!\! \langle X^A_a X^B_b \rangle \!\! $ & $\!\! \langle X^A_a X^B_c\rangle \!\! $  &  $\!\! \langle X^A_b X^B_c \rangle \!\! $ \\[.2cm] 
\hline
$\!\!\!$ (i)$_{[+++]}$ $\!\!\!$ & $-1$ & $-1$ & $-1$ \\[.2cm]
$\!\!\!$ (ii)$_{[++0]}$ $\!\!\!$ & $-1$ & $0$ & $0$ \\[.2cm]
$\!\!\!$ (iii)$_{[++-]}$ $\!\!\!$ & $-1$ & $1$ & $1$ \\[.2cm]
$\!\!\!$ (iv)$_{[+0+]}$ $\!\!\!$ & $0$ & $-1$ & $0$ \\[.2cm]
$\!\!\!$ (v)$_{[+00]}$ $\!\!\!$ & $0$ & $0$ & $0$ \\[.2cm]
$\!\!\!$ (vi)$_{[+0-]}$ $\!\!\!$ & $0$ & $1$ & $0$ \\[.2cm]
$\!\!\!$ (vii)$_{[+-+]}$ $\!\!\!$ & $1$ & $-1$ & $1$ \\[.2cm]
 \hline
\end{tabular}
\;
\begin{tabular}{|c||c|c|c|}
\hline
\Big. $\!\!\!$ ticket $\!\!\!$ & $\!\! \langle X^A_a X^B_b \rangle \!\! $ & $\!\! \langle X^A_a X^B_c\rangle \!\! $  &  $\!\! \langle X^A_b X^B_c \rangle \!\! $ \\[.2cm] 
\hline
$\!\!\!$ (viii)$_{[+-0]}$ $\!\!\!$ & $1$ & $0$ & $0$  \\[.2cm]
$\!\!\!$ (ix)$_{[+--]}$ $\!\!\!$ & $1$ & $1$ & $-1$  \\[.2cm]
$\!\!\!$ (x)$_{[0++]}$ $\!\!\!$ & $0$ & $0$ & $-1$ \\[.2cm]
$\!\!\!$ (xi)$_{[0+0]}$ $\!\!\!$ & $0$ & $0$ & $0$ \\[.2cm]
$\!\!\!$ (xii)$_{[0+-]}$ $\!\!\!$ & $0$ & $0$ & $1$ \\[.2cm]
$\!\!\!$ (xiii)$_{[00+]}$ $\!\!\!$ & $0$ & $0$ & $0$ \\[.2cm]
$\!\!\!$ (xiv)$_{[000]}$ $\!\!\!$ & $0$ & $0$ & $0$ \\[.2cm]
 \hline
\end{tabular}
\caption{Covariances for single-ticket raffles with tickets (i)--(xiv) in Figure \ref{raffle-tickets-3set3out-i-xiv}. The subscript on each ticket number gives the values for the settings $\hat{a}$, $\hat{b}$ and $\hat{c}$ on the left side of a ticket of that type.}
\label{covariances for spin 1}
\end{table} 

Using this table and Eq.\ (\ref{M_ab^i spin-1}) we can write out the $14 \times 3$ matrix $M$ in Eq.\ (\ref{mapping spin 1 a}). Rows in the table multiplied by $-\sfrac32$ turn into columns of $M$:
\setcounter{MaxMatrixCols}{14}
\begin{equation}
M = -\frac32
\begin{pmatrix}
-1 & -1 & -1 & 0 & 0 & 0 & 1 & 1 & 1 & 0 & 0 & 0 & 0 & 0 \; \\[.2cm]
-1 & 0 & 1 & -1 & 0 & 1 & -1 & 0 & 1 & 0 & 0 & 0 & 0 & 0 \; \\[.2cm]
-1 & 0 & 1 & 0 & 0 & 0 & 1 & 0 & -1 & -1 & 0 & 1 & 0 & 0 \;
\end{pmatrix}.
\label{components of M spin 1}
\end{equation}

\begin{figure}[h]
 \centering
   \includegraphics[width=4in]{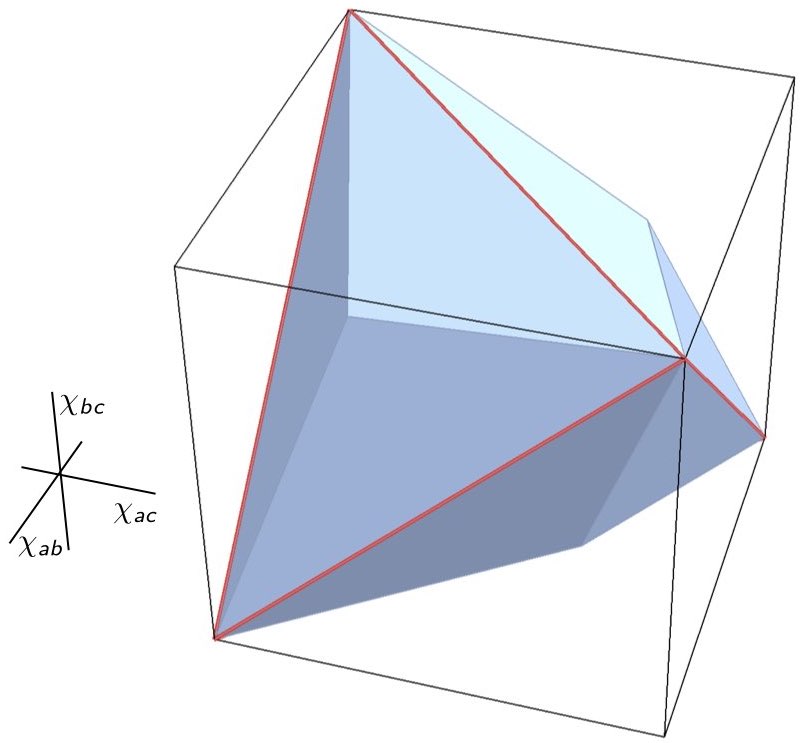} 
   \caption{Anti-correlation polyhedron for raffles simulating quantum correlations for spin-1 (cf.\ the classical tetrahedron in Figure \ref{tetrahedron}).}
   \label{polytope-spin1}
\end{figure}

To illustrate how this mapping works, consider the admissible raffle ($\beta$) introduced above. This raffle is represented by the point
\begin{equation}
(0, 0, 0, 0, 0, f^{\mathrm{vi}}_\beta, 0, f^{\mathrm{viii}}_\beta, 0, 0, 0, f^{\mathrm{xii}}_\beta, 0, 0)
\label{beta point simplex}
\end{equation}
of the 13D simplex, with $f^{\mathrm{vi}}_\beta = f^{\mathrm{viii}}_\beta = f^{\mathrm{viii}}_\beta = \sfrac13$ (see Eq.\ (\ref{spin 1 raffle beta})). We designed this raffle so that it is represented by the point $(-\sfrac12, -\sfrac12, -\sfrac12)$ of the anti-correlation polyhedron (see Eq.\ (\ref{raffle beta = -1/2^3})). We verify that $M$ correctly maps the point in Eq.\ (\ref{beta point simplex}) to the point in Eq.\ (\ref{raffle beta = -1/2^3}). For $\vec{f}_{\mathrm{adm}} = \vec{f}_\beta$, the mapping in Eq.\ (\ref{mapping spin 1 a}) reduces to:
\begin{equation}
\left. \begin{pmatrix}
\chi_{ab}\\[.3cm]
\chi_{ac}\\[.3cm]
\chi_{bc}
\end{pmatrix} \right|_{\vec{f}_\beta} = 
\left. \begin{pmatrix}
M_{ab}^{\mathrm{vi}} & \!\! M_{ab}^{\mathrm{viii}} \!\! & M_{ab}^{\mathrm{xii}} \\[.3cm]
M_{ac}^{\mathrm{vi}} & \!\! M_{ac}^{\mathrm{viii}} \!\! & M_{ac}^{\mathrm{xii}} \\[.3cm]
M_{bc}^{\mathrm{vi}} & \!\! M_{bc}^{\mathrm{viii}} \!\! & M_{bc}^{\mathrm{xii}}
 \end{pmatrix}
\!\! \begin{pmatrix}
f^{\mathrm{vi}} \\[.3cm]
f^{\mathrm{viii}} \\[.3cm]
f^{\mathrm{xii}}
\end{pmatrix} \right|_{\vec{f}_\beta}.
\label{mapping for raffle beta}
\end{equation}
Using  Eq.\ (\ref{components of M spin 1}) for the matrix elements of $M$ and setting these three components of $\vec{f}_\beta$ equal to $\sfrac13$, we confirm that
\begin{equation}
\left. \begin{pmatrix}
\chi_{ab}\\[.3cm]
\chi_{ac}\\[.3cm]
\chi_{bc}
\end{pmatrix} \right|_{\vec{f}_\beta}
=
 \begin{pmatrix}
0 & \!\!\!\! -\sfrac32 \!\!\!\! & 0 \\[.3cm]
-\sfrac32  & 0 & 0 \\[.3cm]
0 & 0 & -\sfrac32 
 \end{pmatrix}
 \begin{pmatrix}
\sfrac13 \\[.3cm]
\sfrac13\\[.3cm]
\sfrac13
\end{pmatrix} 
=
 \begin{pmatrix}
- \sfrac12 \\[.3cm]
- \sfrac12\\[.3cm]
- \sfrac12
\end{pmatrix}.
\label{mapping for raffle beta a}
\end{equation}

More generally, Eq.\ (\ref{mapping spin 1 a}) maps the 10D polytope of admissible raffles to the 3D anti-correlation polyhedron in Figure \ref{polytope-spin1}. This polyhedron is obtained by projecting the 40 vertices MathSage found for us and taking their convex hull. Once again using MathSage, we found that of the 40 points so projected, only 8 are vertices of the polyhedron. Four of them are the vertices of the tetrahedron in Figure \ref{tetrahedron}. The other four are the points $\left(\pm \sfrac12,\pm \sfrac 12,\pm \sfrac12\right)$ in which an odd number of minus signs occur. Raffle ($\beta$) is represented by the one with three minus signs (see Eqs.\ (\ref{raffle beta = -1/2^3}) and (\ref{mapping for raffle beta a})). To construct raffles represented by the three other points, we change the sign of the values $\pm 1$ for one of the three settings for all three ticket types in raffle ($\beta$). If this results in a ticket that is not among the fourteen tickets in Figure \ref{raffle-tickets-3set3out-i-xiv}, we simply switch the left and the right side of the ticket. The proportions of the tickets are kept the same. If we do this for setting $\hat{a}$, we get a new raffle ($\beta^{\prime}$) (with tickets of type (ii), (iv) and (xii)) for which
\begin{eqnarray}
\langle X^A_a X^B_b \rangle  \big|_{\vec{f}_{\beta^\prime}} = - \langle X^A_a X^B_b \rangle  \big|_{\vec{f}_\beta} = \sfrac12 \\[.2cm]
\langle X^A_a X^B_c \rangle  \big|_{\vec{f}_{\beta^\prime}} = - \langle X^A_a X^B_c \rangle  \big|_{\vec{f}_\beta} = \sfrac12 \\[.2cm]
\langle X^A_b X^B_c \rangle \big|_{\vec{f}_{\beta^\prime}}  = \langle X^A_b X^B_c \rangle  \big|_{\vec{f}_\beta} = - \sfrac12
\label{other vertices for spin 1}
\end{eqnarray}
If we do the same thing for setting $\hat{b}$, we get a second new raffle ($\beta^{\prime\prime}$) (with tickets of type (ii), (vi) and (x)), in which $\langle X^A_a X^B_c \rangle$ will be the same as in raffle ($\beta$) while the other two change sign. Finally, if do this for setting $\hat{c}$, we get a third new raffle ($\beta^{\prime\prime\prime}$) (with tickets of type (iv), (viii) and (x)), in which $\langle X^A_a X^B_b \rangle$ will be the same as in raffle ($\beta$) and the other two change sign. In the polyhedron in Figure \ref{polytope-spin1}, the raffles ($\beta^\prime$), ($\beta^{\prime\prime}$)  and  ($\beta^{\prime\prime\prime}$) are thus represented by the points $\left( \sfrac12, \sfrac 12,-\sfrac12\right)$, $\left(\sfrac12, - \sfrac 12, \sfrac12\right)$ and $\left( -\sfrac12, \sfrac 12,\sfrac12\right)$, respectively. 
%One sees in Figure \ref{polytope-spin1} that the four extra points the polyhedron of our raffles for the spin-1 case picks up compared to the tetrahedron for our raffles for the spin-$\frac12$ case lie at the center of each of the four parts into which the elliptope is divided by the tetrahedron. 

%SUBSUBSECTION 3.2.3
\subsubsection{Spin-$\frac32$} \label{2.2.3}

The flow chart in Figure \ref{flowchart} that we used to deal with raffles for the spin-1 case will also guide us in our analysis of raffles for the spin-$\frac32$ case. The number of tickets now jumps from $(3^3 +1)/2 = 14$ to $4^3/2 = 32$. We will only display a subset of these tickets (see Figures \ref{raffles-spin32-tickets-mu} and \ref{raffles-spin32-tickets-nu}). We numbered them using the same convention as before (cf.\ the tree structure in Figure \ref{raffle-tickets-3set2out-i-thru-iv} in Section \ref{1.4}). With 32 tickets, the number of vertices and facets to keep track of is starting to get unwieldy but the spin-$\frac32$ case is still tractable. And there are at least two good reasons for examining it in some detail: it is the simplest case in which cell symmetry calls for separate admissibility conditions; it nicely illustrates the difference between integer and half-integer spin cases in terms of the bound on the sum of the anti-correlation coefficients $\chi_{ab} + \chi_{bc} + \chi_{ac}$ (see Section \ref{1.6}, Eqs.\ (\ref{De Finetti half integer s})--(\ref{Mermin CHSH integer spin})).

\begin{figure}[h]
 \centering
   \includegraphics[width=5in]{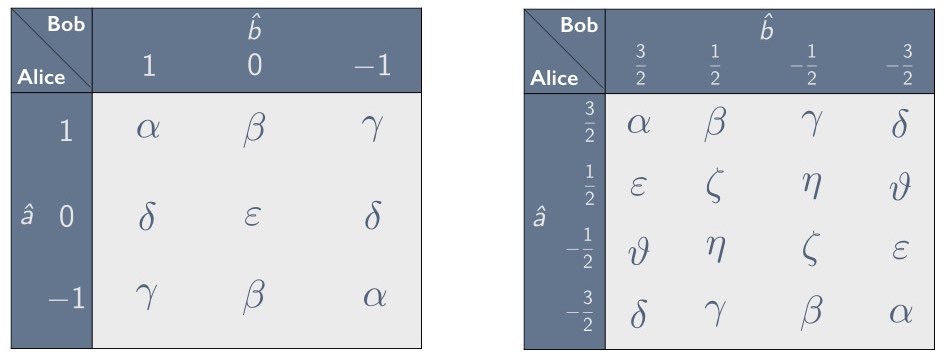} 
   \caption{Cell in the correlation arrays for the spin-1 and spin-$\frac32$ cases showing centrosymmetry.}
   \label{symmetry-spin-1-32}
\end{figure}

Figure \ref{symmetry-spin-1-32} shows a generic off-diagonal cell in the correlation array for two of our raffles, the one on the left for the spin-1 case, the one on the right for the spin-$\frac32$ case. As noted in the introduction to Section \ref{2}, the design of our raffles guarantees that these cells are centrosymmetric (see Eq.\ (\ref{raffle cell centro-symmetry})). Hence, 5 parameters (labeled $\alpha$ through $\varepsilon$) suffice to fix the 9 entries in the cell on the left and 8 parameters (labeled $\alpha$ through $\vartheta$) suffice to fix the 16 entries in the one on the right (these parameters have to satisfy the obvious requirement that the sum of all entries in a cell equals 1). 
%This means that Eq.\ (\ref{raffle cell centro-symmetry}) holds for any single-ticket or mixed raffle whatsoever. Centro-symmetry, in other words, does not require any restrictions on $\vec{f}$.

Requiring uniform marginals further reduces the number of parameters needed to determine the entries in the cells in Figure \ref{symmetry-spin-1-32}. The conditions ensuring uniform marginals in the spin-$\frac32$ case are a straightforward generalization of those in the spin-1 case (cf.\ Eq.\ (\ref{uniform marginals spin-1})):
\begin{equation}
\mathrm{Pr}\!\left({\textstyle{\frac32}} -\!\!{\textstyle{\frac32}} \Big|\hat{a}\hat{a}\!\right) \!\!  \Big|_{\!\vec{f}_{\mathrm{adm}}} \!\!\!
= \mathrm{Pr}\!\left({\textstyle{\frac12}} -\!\!{\textstyle{\frac12}} \Big|\hat{a}\hat{a}\!\right) \!\! \Big|_{\!\vec{f}_{\mathrm{adm}}} \!\!\!
= \mathrm{Pr}\!\left(-\!{\textstyle{\frac12}} \, {\textstyle{\frac12}} \Big|\hat{a}\hat{a}\!\right)  \!\! \Big|_{\!\vec{f}_{\mathrm{adm}}} \!\!\!
= \mathrm{Pr}\!\left(-\!{\textstyle{\frac32}} \, {\textstyle{\frac32}} \Big|\hat{a}\hat{a}\!\right) \!\! \Big|_{\!\vec{f}_{\mathrm{adm}}} \!\!\!
=  \frac14,
\label{uniform marginals spin-32}
\end{equation}
and similarly for the diagonal cells with settings $\hat{b}\,\hat{b}$ and $\hat{c}\,\hat{c}$. As in the spin-1 case, it suffices to impose one of these conditions for each diagonal cell:
\begin{equation}
\mathrm{Pr}\!\left({\textstyle{\frac32}} -\!\!{\textstyle{\frac32}} \Big|\hat{a}\,\hat{a}\right) \!  \Big|_{\vec{f}_{\mathrm{adm}}} \!\!
= \mathrm{Pr}\!\left({\textstyle{\frac32}} -\!\!{\textstyle{\frac32}} \Big|\hat{b}\,\hat{b}\right) \!  \Big|_{\vec{f}_{\mathrm{adm}}} \!\!
= \mathrm{Pr}\!\left({\textstyle{\frac32}} -\!\!{\textstyle{\frac32}} \Big|\hat{c}\,\hat{c}\right) \!  \Big|_{\vec{f}_{\mathrm{adm}}} \!\!
=  \frac14.
\label{uniform marginals spin-32 a}
\end{equation}

Once the conditions for uniform marginals have been imposed, the entries in the cell for the spin-1 case will satisfy
\begin{equation}
\alpha + \beta + \gamma = \alpha + \delta + \gamma = 2\beta + \varepsilon = \sfrac13.
\label{spin 1 relations parameters} 
\end{equation}
It follows that 
\begin{equation}
\gamma = \sfrac13 - \alpha - \beta, \quad \delta = \beta, \quad \varepsilon = \sfrac13 - 2\beta.
\label{spin 1 gamma delta epsilon}
\end{equation}
It thus only takes two independent parameters, $\alpha$ and $\beta$, to fix all entries in the cell. 

Note that this is still one more parameter than is needed to fix all entries in a cell in a correlation array for measurements on the singlet state of two particles of arbitrary (half-)integer spin $s$. The entries in those cells are fixed by (a highly non-linear function of) a single parameter, the angle $\varphi_{ab}$ between the measuring directions. It thus need not surprise us that, for $s \ge 1$, we can no longer perfectly simulate the quantum correlations (see Eq.\ (\ref{off diag cell quantum v raffle})).

Inserting the expressions for $\gamma$, $\delta$ and $\varepsilon$ in Eq.\ (\ref{spin 1 gamma delta epsilon}) in the cell for the spin-1 case in Figure \ref{symmetry-spin-1-32}, we see that this cell is now symmetric as well as centro-symmetric. As we noted in Section \ref{2.2.2}, requiring uniform marginals thus automatically ensures cell symmetry in the spin-1 case. 

In the spin-$\frac32$ case, this is no longer true. Instead of Eq.\ (\ref{spin 1 relations parameters}), the requirement of uniform marginals now gives 
\begin{equation}
\alpha + \beta + \gamma + \delta = \alpha + \varepsilon + \vartheta + \delta = \sfrac14,
\label{symmetry condition cell spin 32}
\end{equation}
from which it follows only that $\beta + \gamma = \varepsilon + \vartheta$. To ensure cell symmetry we need to impose an extra condition. We will set $\beta = \varepsilon$. Since Eq.\ (\ref{symmetry condition cell spin 32}) then entails $\gamma = \vartheta$, this suffices to make the cell on the right in Figure \ref{symmetry-spin-1-32} is symmetric. 

The condition $\beta = \varepsilon$ translates into
\begin{equation}
\left. \mathrm{Pr}\left({\textstyle{\frac32}} {\textstyle{\frac12}} \Big| \hat{a}\hat{b}\right)  \right|_{\vec{f}_{\mathrm{adm}}}
= \left. \mathrm{Pr}\left({\textstyle{\frac12}} {\textstyle{\frac32}} \Big| \hat{a}\hat{b}\right)  \right|_{\vec{f}_{\mathrm{adm}}}.
\label{symmetry condition spin 32} 
\end{equation}
We need three such conditions, one for each of the three off-diagonal cells in the correlation array (recall that the cell symmetry and the symmetry of the correlation array as a whole guarantee that cells on opposite sides of the diagonal are the same). The probabilities in these admissibility conditions are given by half the sum of the ticket fractions $f^{\mathrm{k}}_{\mathrm{adm}}$ for those tickets that have the relevant combination of outcomes (cf.\ Eqs.\ (\ref{chi for f spin 1/2 b})--(\ref{chi for f spin 1/2 ba})). Like the conditions for uniform marginals, these symmetry conditions thus take the form of linear constraints on the ticket fractions $f^\mathrm{k}$ (with k now running from i through xxxii). 

\begin{figure}[h]
 \centering
   \includegraphics[width=3.5in]{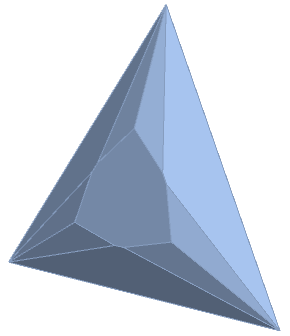} 
   \caption{Facets of the anti-correlation polyhedron for spin-$\frac{3}{2}$. Imagine this figure `glued onto' the four facets of the classical tetrahedron in Figure \ref{tetrahedron}.}
   \label{SpinThreeHalfFace}
\end{figure}

For the spin-$\frac32$ case we have a total of 6 admissibility conditions, 3 to ensure uniform marginals, 3 to ensure cell symmetry. Since there are 32 different tickets, the polytope we start from is the 31D standard simplex in $\mathbb{R}^{32}$ (cf.\ the flow chart in Figure \ref{flowchart}). Upon imposing our 6 admissibility conditions, we arrive at a 25D polytope of admissible raffles in $\mathbb{R}^{32}$. With the help of Sagemath we find that this polytope has a total of 450 (!) vertices. We can map this polytope  in $\mathbb{R}^{32}$ to the 3D anti-correlation polyhedron using a $32 \, \times \, 3$ matrix $M$ with elements of the same form as those of the $14 \, \times \, 3$ matrix for raffles in the spin-1 case (see Eqs.\ (\ref{mapping spin 1 a})--(\ref{components of M spin 1}) and Table \ref{covariances for spin 1}). For our purposes we only need a subset of the 96 elements of this matrix (see Tables \ref{covariances for spin 3/2 raffle mu} and \ref{covariances for spin 3/2 raffle nu}). Applying the mapping given by the matrix $M$ to the 450 vertices of our 25D polytope and using Sagemath to determine which of the resulting 450 points are the vertices of its 3D image, we found a polyhedron in $\mathbb{R}^{3}$ with 40 vertices. We then used the program Mathematica to find the facets of this polyhedron and generate the picture in Figure \ref{SpinThreeHalfFace}. This picture shows the facets we need to ``glue onto" each of the four facets of the tetrahedron in Figure \ref{tetrahedron} to get the full anti-correlation polyhedron in this case.

The polyhedron for raffles (imperfectly) simulating the quantum correlations in the spin-$\frac32$ case picks up 36 extra vertices compared to the tetrahedron for raffles for the spin-$\frac12$ case, 9 for each facet of the tetrahedron. Figure \ref{SpinThreeHalfFace} shows those 9 extra vertices for one of these four facets, 6 of them in pairs that lie so close together that it may look as if there are only 6 rather than 9 points. All 9 vertices lie in the same plane. In the case of the facet of the tetrahedron closest to the point $(-1, -1, -1)$ of the non-signaling cube (see Figure \ref{tetrahedron}) this is the plane where the sum $\chi_{ab} + \chi_{ac} + \chi_{bc}$ has its minimum value. Eq.\ (\ref{Mermin CHSH half-integer spin}) in Section \ref{1.6} tells us that the minimum value of this quantity for $s = \sfrac32$ is
\begin{equation}
%\big( \chi_{ab} + \chi_{ac} + \chi_{bc} \big)_{{\mathrm{minimum \; for}}\; s \,= \, \sfrac32} 
 \frac{1}{8\sigma_{\sfrac32}^2} - \frac32 = \frac{1}{10} - \frac32 = -\frac75.
\end{equation}
where we used Eq.\ (\ref{SD for adm raffle 2}) to set $\sigma_{\sfrac32}^2 = \sfrac54$.

With a little help from the computer, we were able to construct raffles represented by these 9 vertices. Figures \ref{raffles-spin32-tickets-mu} and \ref{raffles-spin32-tickets-nu} show the mix of tickets for two of them, labeled ($\mu$) and ($\nu$). We obtain raffles represented by the other seven vertices by permutations of the outcomes for the three settings $\hat{a}$, $\hat{b}$ and $\hat{c}$ on all tickets in these two raffles (switching left and right sides of tickets if necessary). By changing the sign of the values for one of the settings on all tickets in these nine raffles (again switching sides of tickets if necessary), we can construct raffles for the 27 vertices obtained if the facets in Figure \ref{SpinThreeHalfFace} are ``glued on'' to the other three facets of the tetrahedron (this is the same procedure that we followed for the spin-1 case in Section \ref{2.2.2}).

\begin{figure}[h]
 \centering
   \includegraphics[width=6in]{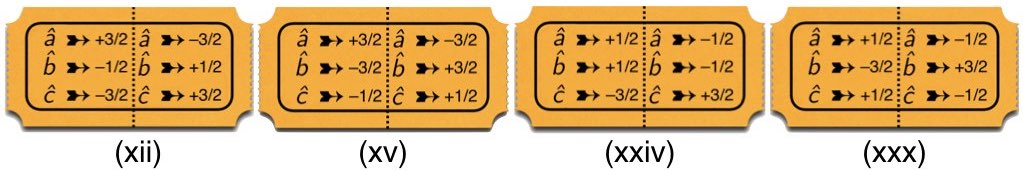} 
   \caption{Admissible raffle ($\mu$) for the spin-$\frac32$ case with equal numbers of tickets of type (xii), (xv), (xxiv) and (xxx).}
    \label{raffles-spin32-tickets-mu}
\end{figure}

Consider Figure \ref{raffles-spin32-tickets-mu} for raffle ($\mu$) characterized by the vector
\begin{equation}
\vec{f}_\mu =  \frac14 \left( \vec{f}_{\mathrm{xii}} + \vec{f}_{\mathrm{xv}} + \vec{f}_{\mathrm{xxiv}} + \vec{f}_{\mathrm{xxx}} \right).
\label{vec spin 32 raffle mu}
\end{equation}

It is easy to see that this raffle yields uniform marginals: For all three settings, all four outcomes occur twice in every set of these four tickets. It is also easy to see that it will be represented by a point on the plane where $\chi_{ab} + \chi_{ac} + \chi_{bc}$ has its minimum value: On both sides of all four tickets, the sum of outcomes for the three settings is $\pm \sfrac12$ resulting in the minimum value of $\sfrac14$ of the expectation value of $(X^A_a + X^A_b + X^A_c)^2$ (cf.\ Section \ref{1.6}). To make sure that the off-diagonal cells in the correlation array for raffle ($\mu$) are symmetric, we check that Eq.\ (\ref{symmetry condition spin 32}) is satisfied, not just for the $\hat{a}\hat{b}$ cell but for the $\hat{a}\hat{c}$ and $\hat{b}\hat{c}$ cells as well. Since
\begin{equation}
\begin{array}{cc}
\mathrm{Pr}\!\left({\textstyle{\frac32}} {\textstyle{\frac12}} \Big| \hat{a}\hat{b}\right) \! \Big|_{\vec{f}_\mu} = {\textstyle{\frac12}} \, f^{\mathrm{xii}},  &
\mathrm{Pr}\!\left({\textstyle{\frac12}} {\textstyle{\frac32}} \Big| \hat{a}\hat{b}\right)  \! \Big|_{\vec{f}_\mu} = {\textstyle{\frac12}} \, f^{\mathrm{xxx}}, \\[.4cm]
 \mathrm{Pr}\!\left({\textstyle{\frac32}} {\textstyle{\frac12}} \Big| \hat{a}\hat{c}\right) \! \Big|_{\vec{f}_\mu}  = {\textstyle{\frac12}} \, f^{\mathrm{xv}}, &
\mathrm{Pr}\!\left({\textstyle{\frac12}} {\textstyle{\frac32}} \Big| \hat{a}\hat{c}\right)  \! \Big|_{\vec{f}_\mu}   = {\textstyle{\frac12}} \, f^{\mathrm{xxiv}}, \\[.4cm]
\mathrm{Pr}\!\left({\textstyle{\frac32}} {\textstyle{\frac12}} \Big| \hat{b}\hat{c}\right) \! \Big|_{\vec{f}_\mu}   = {\textstyle{\frac12}} \, f^{\mathrm{xxx}}, &
\mathrm{Pr}\!\left({\textstyle{\frac12}} {\textstyle{\frac32}} \Big| \hat{b}\hat{c}\right) \! \Big|_{\vec{f}_\mu}  = {\textstyle{\frac12}} \, f^{\mathrm{xxiv}},
\end{array}
\end{equation}
and all four ticket fractions are equal, these symmetry conditions are indeed satisfied.

Using Eq.\ (\ref{mapping spin 1 a}) for the mapping from ticket fractions of admissible raffles to triplets of allowed anti-correlation coefficients, we can find the components of $\vec{\chi}$ for raffle ($\mu$). In this case, Eq.\ (\ref{mapping spin 1 a}) reduces to
\begin{equation}
\left. \begin{pmatrix}
\chi_{ab}\\[.3cm]
\chi_{ac}\\[.3cm]
\chi_{bc}
\end{pmatrix} \right|_{\vec{f}_\mu}
= \left. \begin{pmatrix}
M_{ab}^{\mathrm{xii}} & M_{ab}^{\mathrm{xv}} & M_{ab}^{\mathrm{xxiv}} & M_{ab}^{\mathrm{xxx}} \\[.4cm]
M_{ac}^{\mathrm{xii}} & M_{ac}^{\mathrm{xv}} & M_{ac}^{\mathrm{xxiv}} &  M_{ac}^{\mathrm{xxx}} \\[.4cm]
M_{bc}^{\mathrm{xii}} & M_{bc}^{\mathrm{xv}} & M_{bc}^{\mathrm{xxiv}} &  M_{bc}^{\mathrm{xxx}}
 \end{pmatrix}
\!\! \begin{pmatrix}
f^{\mathrm{xii}} \\[.1cm]
f^{\mathrm{xv}} \\[.1cm]
f^{\mathrm{xxiv}} \\[.1cm]
f^{\mathrm{xxx}}
\end{pmatrix} \right|_{\vec{f}_\mu}. 
\label{mapping for raffle mu}
\end{equation}
The ticket fractions are all equal to $\sfrac14$. The elements of the matrix $M$ are given by
\begin{equation}
M^{\mathrm{k}}_{ab} = -\frac{1}{\sigma_{s = \sfrac32}^2} \langle X^A_a X^B_b \rangle, \quad \mathrm{with \; k  =  i \, \ldots \, xxxii},  
\end{equation}
and similar expressions with $ab$ replaced by $ac$ or $bc$ (cf.\ Eq.\ (\ref{M_ab^i spin-1})). These matrix elements are given by $- 1/\sigma_{s = \sfrac32}^2 = -\sfrac45$ times the relevant entries in Table \ref{covariances for spin 3/2 raffle mu}. 

\begin{table}[h]
\centering
\begin{tabular}{|c||c|c|c|}
\hline
\Big. ticket & $\langle X^A_a X^B_b \rangle$ & $\langle X^A_a X^B_c\rangle$  &  $\langle X^A_b X^B_c \rangle$\\
\hline
\Big. (xii)$_{[\frac32-\frac12-\frac32]}$  & $\sfrac34$ & $\sfrac94$ & $-\sfrac34$ \\[.2cm]
(xv)$_{[\frac32-\frac32-\frac12]}$ & $\sfrac94$ & $\sfrac34$ & $-\sfrac34$ \\[.2cm]
(xxiv)$_{[\frac12\frac12-\frac32]}$ & $-\sfrac14$ & $\sfrac34$ & $\sfrac34$ \\[.2cm]
(xxx)$_{[\frac12-\frac32\frac12]}$ & $\sfrac34$ & $-\sfrac14$ & $\sfrac34$ \\[.2cm]
 \hline
\end{tabular}
\caption{Covariances for single-ticket raffles for the four ticket types in admissible raffle ($\mu$) in Figure \ref{raffles-spin32-tickets-mu}
%The subscript on each ticket number gives the values for the settings $\hat{a}$, $\hat{b}$ and $\hat{c}$ on the left side of a ticket of that type 
(cf.\ Table \ref{covariances for spin 1}).}
\label{covariances for spin 3/2 raffle mu}
\end{table} 

For the components of $\vec{\chi}$ for raffle ($\mu$) we then find:
\begin{equation}
\left. \begin{pmatrix}
\chi_{ab}\\[.3cm]
\chi_{ac}\\[.3cm]
\chi_{bc}
\end{pmatrix} \right|_{\vec{f}_\mu}
= 
\begin{pmatrix}
-\sfrac35 & -\sfrac95 & \sfrac15 & -\sfrac35 \\[.3cm]
-\sfrac95 & -\sfrac35 & -\sfrac35 & \sfrac15 \\[.3cm]
\sfrac35 & \sfrac35 & -\sfrac35 & -\sfrac35 
\end{pmatrix}
\begin{pmatrix}
\sfrac14 \\[.1cm]
\sfrac14 \\[.1cm]
\sfrac14 \\[.1cm]
\sfrac14
\end{pmatrix}
= 
\begin{pmatrix}
-\sfrac{7}{10} \\[.3cm]
-\sfrac{7}{10} \\[.3cm]
0
\end{pmatrix},
\end{equation}
which confirms that $\chi_{ab} + \chi_{ac} + \chi_{bc} =- \sfrac75$, as it should be given the way we constructed this raffle. Raffle ($\mu$) is represented by one of the 9 vertices in the plane of the polyhedron where $\chi_{ab} + \chi_{ac} + \chi_{bc}$ has its minimum value. Through suitable permutation of the outcomes for settings $\hat{a}$,  $\hat{b}$ and  $\hat{c}$, we can create raffles similar to raffle ($\mu$) represented by two other vertices in that same plane: $(0, -\sfrac{7}{10}, -\sfrac{7}{10})$ and $(-\sfrac{7}{10}, 0, -\sfrac{7}{10})$.

%Now consider a raffle ($\mu^\prime$) obtained by replacing all values for $\hat{a}$ on both sides of the four ticket types in raffle ($\mu$) by the values for $\hat{b}$, the values for $\hat{b}$ by those for $\hat{c}$ and the values for $\hat{c}$ by those for $\hat{a}$. The values for $(\chi_{ab}, \chi_{ac}, \chi_{bc})$ for this new raffle ($\mu^\prime$) will be given by the values for $(\chi_{bc}, \chi_{ba} = \chi_{ab}, \chi_{ca} = \chi_{ac})$ for raffle ($\mu$). Raffle ($\mu$) will thus be represented by the point $(0, -\sfrac{7}{10}, -\sfrac{7}{10})$ on the same plane as the point representing raffle ($\mu$). A similar permutation produces a raffle ($\mu^{\prime\prime}$) represented by the point $(-\sfrac{7}{10}, 0, -\sfrac{7}{10})$ on that plane.   

\begin{figure}[h]
 \centering
   \includegraphics[width=5.5in]{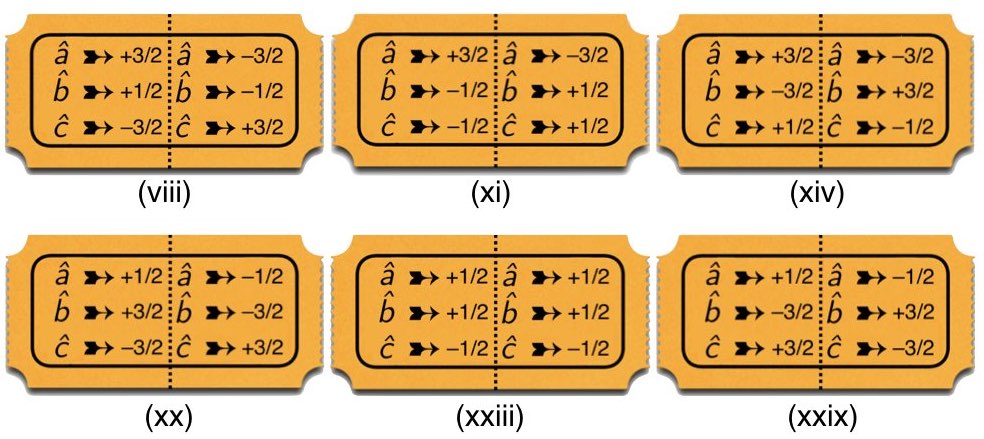} 
   \caption{Admissible raffle ($\nu$) for the spin-$\frac32$ case: a basket with equal numbers of tickets of type (viii), (xi), (xiv), (xx), (xxiii) and (xxix).}
    \label{raffles-spin32-tickets-nu}
\end{figure}

Figure \ref{raffles-spin32-tickets-nu} shows the tickets for a raffle represented by one of the remaining six points in the plane where $\chi_{ab} + \chi_{ac} + \chi_{bc} = -\sfrac75$. This raffle is represented by the vector 
\begin{equation}
\vec{f}_\nu =  \frac16 \left( \vec{f}_{\mathrm{viii}} + \vec{f}_{\mathrm{xi}} + \vec{f}_{\mathrm{xiv}} + \vec{f}_{\mathrm{xx}} + \vec{f}_{\mathrm{xxiii}} + \vec{f}_{\mathrm{xxix}} \right),
\label{vec spin 32 raffle nu}
\end{equation}
As with raffle ($\mu$), the outcomes of both sides of all tickets add up to $\pm \sfrac12$ and both sets of admissibility conditions (uniform marginals and cell symmetry) are met. These last two properties can be verified the same way as in the case of raffle ($\mu$). 

\begin{table}[h]
\centering
\begin{tabular}{|c||c|c|c|}
\hline
\Big. ticket & $\langle X^A_a X^B_b \rangle$ & $\langle X^A_a X^B_c\rangle$  &  $\langle X^A_b X^B_c \rangle$\\
\hline
\Big. (viii)$_{[\frac32\frac12-\frac32]}$ & $-\sfrac34$ & $\sfrac94$ & $\sfrac34$ \\[.2cm]
(xi)$_{[\frac32-\frac12-\frac12]}$ & $\sfrac34$ & $\sfrac34$ & $-\sfrac14$ \\[.2cm]
(xiv)$_{[\frac32-\frac32\frac12]}$ & $\sfrac94$ & $-\sfrac34$ & $\sfrac34$ \\[.2cm]
(xx)$_{[\frac12\frac32-\frac32]}$ & $-\sfrac34$ & $\sfrac34$ & $\sfrac94$ \\[.2cm]
(xxiii)$_{[\frac12\frac12-\frac12]}$ & $-\sfrac14$ & $\sfrac14$ & $\sfrac14$ \\[.2cm]
(xxix)$_{[\frac12-\frac32\frac32]}$ & $\sfrac34$ & $-\sfrac34$ & $\sfrac94$ \\[.2cm]
 \hline
\end{tabular}
\caption{Covariances for single-ticket raffles for the six ticket type in admissible raffle ($\nu$) in Figure \ref{raffles-spin32-tickets-nu} (cf.\ Table \ref{covariances for spin 3/2 raffle mu}).}
\label{covariances for spin 3/2 raffle nu}
\end{table} 

To find the components of $\vec{\chi}$ for this raffle, we once again use Eq.\ (\ref{mapping spin 1 a}), which in this case reduces to:
\begin{equation}
\left. \begin{pmatrix}
\chi_{ab}\\[.3cm]
\chi_{ac}\\[.3cm]
\chi_{bc}
\end{pmatrix} \right|_{\vec{f}_\nu}
= \left. \begin{pmatrix}
M_{ab}^{\mathrm{viii}} & M_{ab}^{\mathrm{xi}} & M_{ab}^{\mathrm{xiv}} & M_{ab}^{\mathrm{xx}} & M_{ab}^{\mathrm{xxiii}} & M_{ab}^{\mathrm{xxix}} \\[.3cm]
M_{ac}^{\mathrm{viii}} & M_{ac}^{\mathrm{xi}} & M_{ac}^{\mathrm{xiv}} & M_{ab}^{\mathrm{xx}} & M_{ab}^{\mathrm{xxiii}} & M_{ac}^{\mathrm{xxix}} \\[.3cm]
M_{bc}^{\mathrm{viii}} & M_{bc}^{\mathrm{xi}} & M_{bc}^{\mathrm{xiv}} & M_{ab}^{\mathrm{xx}} & M_{ab}^{\mathrm{xxiii}} & M_{bc}^{\mathrm{xxix}}
 \end{pmatrix}
\!\! \begin{pmatrix}
f^{\mathrm{viii}} \\
f^{\mathrm{xi}} \\
f^{\mathrm{xiv}} \\
f^{\mathrm{xx}} \\
f^{\mathrm{xxiii}} \\
f^{\mathrm{xxix}}
\end{pmatrix} \right|_{\vec{f}_\nu}.
\label{mapping for raffle nu}
\end{equation}
Evaluating the relevant elements of the matrix $M$ with the help of Table \ref{covariances for spin 3/2 raffle nu} and setting all ticket fractions equal to $\sfrac16$, we arrive at
\begin{equation}
\left. \begin{pmatrix}
\chi_{ab}\\[.3cm]
\chi_{ac}\\[.3cm]
\chi_{bc}
\end{pmatrix} \right|_{\vec{f}_\nu}
= \begin{pmatrix}
\sfrac35 & -\sfrac35 & -\sfrac95 & \sfrac35 & \sfrac15 & -\sfrac35 \\[.3cm]
-\sfrac95 & -\sfrac35 & \sfrac35 & -\sfrac35 & -\sfrac15 & \sfrac35  \\[.3cm]
-\sfrac35 & \sfrac15 & -\sfrac35 & -\sfrac95 & -\sfrac15 & -\sfrac95 
 \end{pmatrix}
\!\! \begin{pmatrix}
\sfrac16 \\
\sfrac16 \\
\sfrac16 \\
\sfrac16 \\
\sfrac16 \\
\sfrac16
\end{pmatrix}
= \begin{pmatrix}
-\sfrac{4}{15} \\[.3cm]
-\sfrac{5}{15}\\[.3cm]
-\sfrac{12}{15}
\end{pmatrix}.
\label{mapping for raffle nu a}
\end{equation}
Note that, once again, $\chi_{ab} + \chi_{ac} + \chi_{bc} =- \sfrac75$, as it should be given the way we constructed this raffle. Through suitable permutation of the outcomes for settings $\hat{a}$,  $\hat{b}$ and  $\hat{c}$, we can create raffles similar to raffle ($\nu$) represented by five other vertices of the facet of the anti-correlation polyhedron where $\chi_{ab} + \chi_{ac} + \chi_{bc} =- \sfrac75$. Since $\sfrac{4}{15}$ and $\sfrac13$ only differ by $\sfrac{1}{15}$, these six vertices can naturally be grouped into three pairs of neighboring points that are hard to tell apart in Figure \ref{SpinThreeHalfFace}:
\begin{equation}
\begin{array}{c}
\big\{ \left( -\sfrac{4}{15}, -\sfrac{1}{3}, -\sfrac{4}{5} \right), \left( -\sfrac{1}{3}, -\sfrac{4}{15}, -\sfrac{4}{5} \right) \big\}, \\[.3 cm]
\big\{ \left( -\sfrac{4}{5}, -\sfrac{4}{15}, -\sfrac{1}{3} \right), \left( -\sfrac{4}{5}, -\sfrac{1}{3}, -\sfrac{4}{15} \right) \big\},  \\[.3 cm]
\big\{ \left( -\sfrac{4}{15}, -\sfrac{4}{5}, -\sfrac{1}{3} \right),  \left( -\sfrac{1}{3}, -\sfrac{4}{5}, -\sfrac{4}{15} \right) \big\}.
\end{array}
\label{six vertices raffle nu}
\end{equation} 

The analysis of raffles ($\mu$) and ($\nu$) allowed us to understand the structure of the facets in Figure \ref{SpinThreeHalfFace} 
%of the polyhedron representing admissible raffles meant to simulate the quantum correlations for measurements on two spin-$\frac32$ particles entangled in the singlet state. 
in full detail.
For higher-spin cases, this becomes impractical. In the next subsection, we will therefore explore alternative methods for dealing with these higher-spin cases.

%SECTION 3.2.4
%!TEX root =  ./JanasJanssenCuffaro-August2019.tex

%SUBSUBSECTION 3.2.4
\subsubsection{Spin-$s$ ($s \ge 2$)}  \label{2.2.4}

\begin{figure}[h]
 \centering
   \includegraphics[width=5.5in]{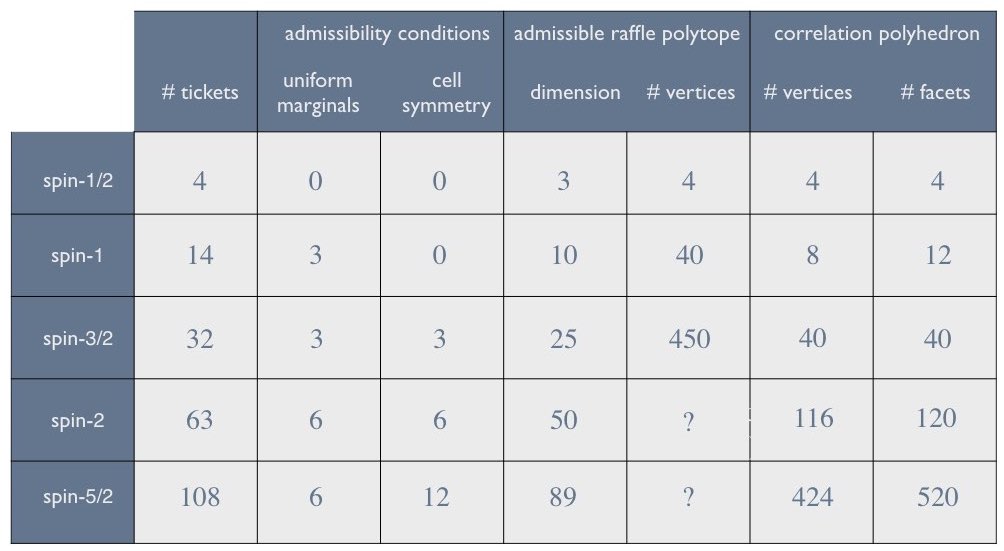} 
   \caption{Number of ticket types, admissibility conditions, and facet/vertex counts for polytopes and polyhedra up to spin-$\frac52$.}
      \label{numberoftickets}
\end{figure}

To simulate the correlations found by measurements on a singlet state of two spin-2 particles, we use  tickets with five outcomes $\pm 2,\pm 1,0$. This results in 63 relevant ticket types, corresponding to a $62$-dimensional simplex of raffles. The conditions for uniform marginals and cell symmetry yield six linear constraints, yielding a 50-dimensional polytope of admissible raffles in 63-dimensional space. Unfortunately, the number of vertices determined by these conditions is daunting: If we consider only the six conditions for uniform marginals, computer calculation produce 553,664 vertices. This is already orders of magnitude larger than the vertex sets for the earlier polytopes. The full case, obtained by including the six conditions for cell symmetry, is presumably even larger but we have not been able to run the vertex enumeration algorithm to completion on a personal computer. Hence we do not know how many vertices the spin-$2$ admissible polytope has, much less the full list of such. This computational obstacle only grows worse as the spin increases, as is evident from the numbers in Figure \ref{numberoftickets}. The enumeration of basic admissible raffles thus becomes intractable for spin $s\geq 2$. As a consequence, the flowchart in Figure \ref{flowchart} comes to a halt and we cannot hope to compute the anti-correlation polyhedron in the way it depicts.

Fortunately, there is an alternative approach which---while it does not yield the full polytope---does allow us to characterize the polyhedron of anti-correlation coefficients. As a first step we observe that, for every spin considered so far, the  polyhedron of anti-correlation coefficients always at least contained the spin-$\frac12$ classical tetrahedron. This is not a  coincidence. For instance, consider the point $\vec{\chi}=(1,1,1)$ where all outcomes are perfectly anti-correlated. There is actually a simple construction of an admissible spin-$s$ raffle with such behavior. To do so, it is convenient to introduce the following notation for the remainder of this section: Let $[m_a,m_b,m_c]$ denote the ticket type for which the outcomes $m_a,m_b,m_c$ for settings $\hat{a},\hat{b},\hat{c}$ appear on one side. As usual, we regard $[-m_a,-m_b,-m_c]$ as equivalent to $[m_a,m_b,m_c]$.

With this notation in mind, consider the spin-$s$ raffle (with $2s+1$ outcomes $m=-s$ to $s$) containing one of each of the tickets $$[s,s,s], [s-1,s-1,s-1], \cdots, [-s+1,-s+1,-s+1], [-s,-s,-s].$$ Since each of these ticket types appears once, the raffle has uniform marginals. If we swap any two settings on every ticket, all tickets are unchanged and therefore the raffle is cell-symmetric. Hence this raffle is admissible. As every outcome on one side of each ticket is strictly anti-correlated with every outcome on the other side, we obtain $\chi_{ab}=\chi_{ac}=\chi_{bc}=1$.

Thus for any spin there is an admissible raffle for which $\vec{\chi}=(1,1,1)$. The other three points then follow by symmetry: We take each ticket type and swap all outcomes for a given setting (e.g., take the outcomes for setting $\hat{a}$ to range from $-s$ to $+s$). This generates three more admissible raffles, characterized by $\vec{\chi}=(-1,-1,1)$, $(-1,1,-1)$, $(1,-1,-1),$ respectively. Hence all four vertices of the classical tetrahedron can be produced by admissible raffles regardless of the number of outcomes. By convexity, it follows that the anti-correlation polyhedron always includes the entire classical tetrahedron.

In general, however, the classical tetrahedron is not the full polyhedron. To establish this, we again recall that the points $\vec{\chi}=(-1,-1,1)$, $(-1,1,-1)$, $(1,-1,-1)$ generate a facet of the classical tetrahedron. This would correspond to the Bell inequality $\chi_{ab}+\chi_{ac}+\chi_{bc}\geq -1$. But, as we have shown for spin $1$ through $\frac32$, there exist admissible raffles that violate this inequality. 
\begin{table}[h]
\centering
\begin{tabular}{|c|c|}
\hline
Ticket  & Ticket \\
fractions & types \\[.1cm]
\hline
$f^1$ & $[0,0,0]$ \\[.2cm]
$f^2$ & $[0,1,-1]$ $[1,0,-1]$ $[1,-1,0]$ \\[.2cm]
$f^3$ & $[0,1,-1]$ $[1,0,-1]$ $[1,-1,0]$ \\[.2cm]
$f^4$ & $[0,1,-1]$ $[1,0,-1]$ $[1,-1,0]$ \\[.2cm]
\hline
\end{tabular}
\caption{Ticket types and ticket fractions for the four ticket groups used to construct admissible raffles which maximally violate the Bell inequality. Each ticket type in a given group occurs with the same frequency in the full raffle.}
\label{Spin2TicketGroups}
\end{table} 

Suppose we now focus on finding admissible raffles which maximally violates this inequality. One approach exploits the observation made in Section \ref{1.6}: To violate the Bell inequality as strongly as possible, we should use ticket types which render $(X_a^A+X_b^A+X_c^A)^2$ as small as possible. For integer spin, this suggests restricting attention to those ticket types whose outcomes sum to zero on the left side (and so also the right side). In the case of spin 2, only 10 out of the 63 tickets to fulfill this criterion. Enforcing cell symmetry on this limited subset of tickets results in the four groups of ticket types shown in Table \ref{Spin2TicketGroups}. For a raffle to be cell-symmetric requires that the fraction of tickets of a given type be the same for all members of a group. We denote these common ticket fractions as $f^1,f^2,f^3,f^4$ for the four respective ticket groups. Enforcing uniform marginals then yields the conditions (cf.\ Eqs.\ (\ref{adm spin1 diag})-(\ref{spin 1 constraints})) 
% \begin{align}
% \text{Pr}(m_1=2|\hat{a})&=&\text{Pr}(m_1=2|\hat{b})=\text{Pr}(m_1=2|\hat{c})= f^3+\frac12 f^4=\frac15, \\
% \text{Pr}(m_1=1|\hat{a})&=&\text{Pr}(m_1=1|\hat{b})=\text{Pr}(m_1=1|\hat{c})= f^2+\frac12 f^1=\frac15,\\
% \text{Pr}(m_1=0|\hat{a})&=&\text{Pr}(m_1=0|\hat{b})=\text{Pr}(m_1=2|\hat{c})= f^1+f^2+f^3=\frac15,
% \end{align}
$$f^1+f^2+f^3=f^2+f^4=f^3+\sfrac12 f^4=\sfrac15.$$ These four equations in three unknowns have a one-dimensional solution set. It can then be shown that the subset of non-negative solutions is generated by convex combinations of the following:

\begin{itemize}
    \item $(f^1,f^2,f^3,f^4)=(\sfrac{1}{10},0,\sfrac{1}{10},\sfrac{2}{10})$, corresponding to a raffle with tickets
    \begin{equation*}
    \begin{array}{*{5}c}
        {[0,\;\, 0,\;\, 0]}  & {[2,0-2]} & {[1,1,-2]} & {[1,-2,1]} & {[2,-1,-1]} \\
        {[0,2,-2]} & {[2,-2,0]} & {[1,1,-2]} & {[1,-2,1]} & {[2,-1,-1]}
    \end{array}
    \end{equation*}
    \item $(f^1,f^2,f^3,f^4)=(0,\sfrac{1}{15},\sfrac{2}{15},\sfrac{1}{15})$, corresponding to a raffle with tickets
    \begin{equation*}
    \begin{array}{*{5}c}
        {[0,1,-1]} & {[0,2,-2]} & {[2,0,-2]} & {[1,1,-2]} & {[1,\;\, -2,\;\, 1]}  \\
        {[1,0,-1]} & {[0,2,-2]} & {[2,-2,0]} & {[1,1,-2]} & {[2,-1,-1]} \\ 
        {[1,-1,0]} & {[2,0-2]} & {[2,-2,0]} & {[1,-2,1]} & {[2,-1,-1]}
    \end{array}
    \end{equation*}
\end{itemize}

% Do not remove the curly brackets. Otherwise, Latex interprets [x] to be a dimension (e.g. [cm]) and will not compile properly.

% \begin{table}[h]
% \centering
% \begin{tabular}{|c|*{5}c|}
%     \hline
%     $(f^1,f^2,f^3,f^4)$ & & & Admissible raffle & & \\
%     \hline \\
%     $(\sfrac{1}{10},0,\sfrac{1}{10},\sfrac{2}{10})$
%         & {[0,0,0]}  & {[0,2,-2]} & {[2,0,-2]} & {[2,-2,0]} & {[1,1,-2]} \\
%         & {[1,1,-2]} & {[1,-2,1]} & {[1,-2,1]} & {[2,-1,-1]} & {[2,-1,-1]} \\ \\
%     \hline \\
%     $(0,\sfrac{1}{15},\sfrac{2}{15},\sfrac{1}{15})$
%         & {[0,1,-1]} & {[1,0,-1]} & {[1,-1,0]} & {[0,2,-2]} & {[0,2,-2]}  \\
%         & {[2,0,-2]} & {[2,0,-2]} & {[2,-2,0]} & {[2,-2,-0]} & {[1,1,-2]} \\ 
%         & {[1,1,-2]} & {[1,-2,1]} & {[1,-2,1]} & {[2,-1,-1]} & {[2,-1,-1]}\\
%     \hline
% \end{tabular}
% \caption{}
% \label{}
% \end{table} 

One may confirm that, in keeping with the ticket outcomes all summing to zero, both raffles map to the point $\vec{\chi}=(-\sfrac12,-\sfrac12,-\sfrac12)$. We have thus gone from a 50-dimensional polytope of admissible raffles to a 1-dimensional subspace of such raffles, all of which maximally violate the Bell inequality.\footnote{These raffles validate that, for spin up to $2$, we can always find an admissible raffle while using only tickets which minimize the magnitude of $X_a^A+X_b^A+X_c^A$. This is actually true in general: For any spin, there is a procedure to construct an admissible raffle using only tickets which minimize $(X_a^A+X_b^A+X_c^A)^2$. This construction, however, is somewhat involved and moreover beyond the scope of the present work, so we do not pursue this line further.}

%For half-integer spin, the smallest magnitude which $X_a^A+X_b^A+X_c^A$ can achieve is now 1/2 rather than 0. This still restricts the set of allowed tickets substantially. For instance, in the case of spin $5/2$ there are six outcomes and therefore $6^3/2=108$ relevant ticket types. However, there are only 27 tickets for which the sum of outcomes has magnitude $1/2$. Imposing the admissibility conditions reduces this 26-dimensional subspace of raffles further. What results is an 11-dimensional polytope of admissible raffles, generated as the convex hull of 288 vertices. Mapping these 288 vertices to their anti-correlation coefficients yields the point set shown in [Figure], all of which lie on the plane $$\chi_{ab}+\chi_{ac}+\chi_{bc} = -\frac{3}{2}+\frac{3}{35}$$ which [as per the deFinetti section] represents the maximum possible violation of the Bell inequality for spin $5/2$. From this point set we obtain 15 vertices, and the convex hull of these 15 vertices is a facet of the anti-correlation polyhedron.

We have thus obtained another vertex of our spin-2 polyhedron, one which lies beyond the facet of the classical tetrahedron opposite the point $\vec{\chi}=(1,1,1)$. By exploiting the tetrahedral symmetry of the anti-correlation coefficients, we obtain three other such vertices that lie beyond the other three facets. Invoking convexity again, we obtain another polyhedron of admissible anti-correlation coefficients which is larger than the classical tetrahedron (though still necessarily bounded by the elliptope) and therefore more closely `approximates' the full anti-correlation polyhedron.

At this point, there is nothing in principle to stop us from pursuing the following strategy, which is essentially the convex hull algorithm of \citet{Lassez and Lassez 1992}. We start with some `approximate' polyhedron, such as the classical tetrahedron, which the true anti-correlation polyhedron contains. We pick one of the facets of the approximate polyhedron, and determine whether any admissible raffles exist which violate the corresponding linear inequality on anti-correlation coefficients. If no such raffle exists, we conclude that the facet under consideration is indeed a facet of the anti-correlation polyhedron and we move on to another facet. If such a raffle does exist, however, we determine one which maximally violates the corresponding inequality. The resulting set of anti-correlation coefficients will be an extreme point of the anti-correlation polyhedron. We then invoke convexity and enlarge our approximate polyhedron to include this new extreme point, thereby obtaining a better approximation of the anti-correlation polyhedron. This algorithm terminates when every facet of the approximate polyhedron is verified to be a facet of the anti-correlation polyhedron, at which point we conclude that the entire polyhedron has been generated.

\begin{figure}[h]
 \centering
   \includegraphics[width=5.5in]{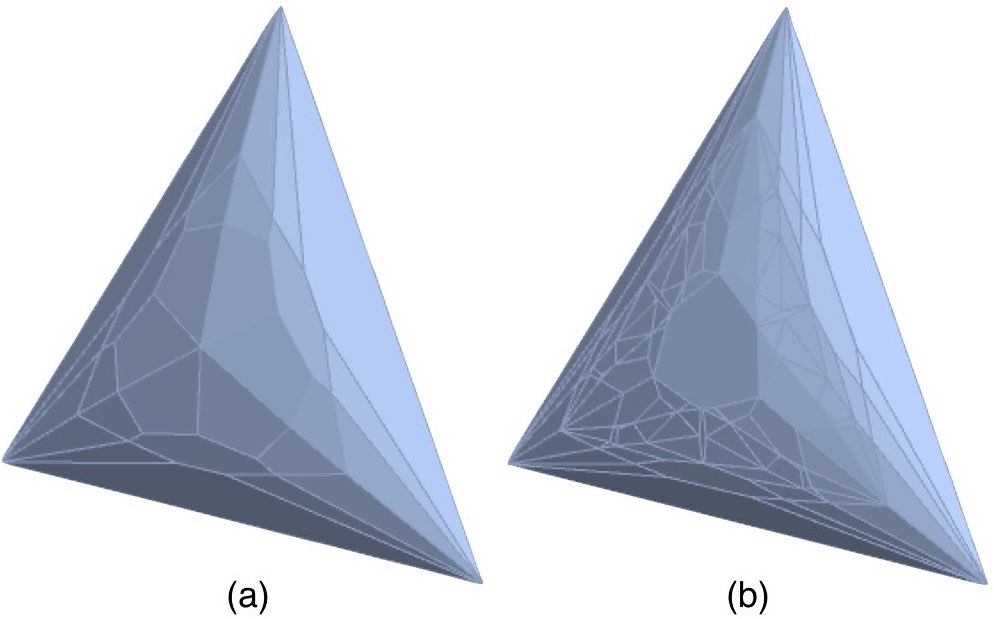} 
   \caption{Facet of polyhedron for (a) spin-$2$ and (b) spin-$\frac52$ (cf.\ Figure \ref{SpinThreeHalfFace}).}
   \label{FacetsSpin2Spin52}
\end{figure}

The hardest step in this method is to determine whether any admissible raffles map to points beyond a given facet. One approach would be to imitate the strategy outlined for the case of $\chi_{ab}+\chi_{ac}+\chi_{bc}$: For each linear inequality, we determine a corresponding linear relation $v_a X_a^A+v_b X_b^A+v_c X_c^A$ and look for admissible raffles using only tickets for which this quantity has small magnitude, thereby ensuring that the facet inequality is violated as strongly as possible. Take, for example, the spin-$2$ case. We have shown that its anti-correlation polyhedron contains the points  $(-1,1,-1)$, $(1,-1,-1)$, and $(-\sfrac12,-\sfrac12,-\sfrac12)$. The plane through these three points is given by $\chi_{ab}+\chi_{ac}+2\chi_{bc}=-2$. We want to determine whether this gives a facet. Observing that 
\begin{equation}
\big\langle \left(X_a^A+2X_b^A+2X_c^A\right)^2\big\rangle  = 18+8 \, (\chi_{ab}+\chi_{ac}+2\chi_{bc}),
\end{equation}
we are led to consider tickets for which $(X_a^A+2X_b^A+2X_c^A)^2$ is small in order to minimize $\chi_{ab}+\chi_{ac}+2\chi_{bc}$. Proceeding in this way, we ultimately obtain all admissible raffles which are characterized by the seven-sided facet at the bottom of Figure \ref{FacetsSpin2Spin52}.

There are, however, several problems with this method. The first is that it is necessarily somewhat tedious: The linear relation to be minimized must be computed for each facet under consideration, and therefore will lead to different sets of tickets in each case. Second, it is not obvious how many tickets one will need to successfully produce admissible raffles. In the case of spin 2, there are 7 tickets for which this expression is zero and 9 for which it has magnitude 1; it turns out that we need to use tickets of both magnitudes to get an admissible raffle. This is still far smaller than the 63 ticket types total, but it is not nearly as attractive as the 10 tickets with outcomes summing to zero in the $\chi_{ab}+\chi_{ac}+\chi_{bc}$ case. Finally, and most importantly, our iterative strategy only needs one admissible raffle that maximally violates the relevant facet inequality. It is therefore altogether excessive to characterize the entire set of such admissible raffles for a given facet.

Rather than pursue this approach further, we will instead exploit the well-known connection between polytopes and \emph{linear programming} \citep{Dantzig and Thapa 1997, Dantzig and Thapa 2003}. We consider a spin-$s$ raffle and let $\Delta^{n-1}$ be the appropriate $(n-1)$-dimensional simplex of raffles, where $n$ is the number of relevant ticket types. Since all the admissibility conditions are all linear in the ticket fractions, we may conveniently express them in the form $B\vec{f}_{\mathrm{adm}}=\vec{b}$; the matrix $B$ and the vector $\vec{b}$ will have as many rows as there are (linearly independent) constraints. Finally, the linear function to be minimized can be expressed as $\vec{c}\cdot \vec{\chi}|_{\vec{f}_{\mathrm{adm}}}$ for some real 3-vector $\vec{c}$. This \emph{objective function} is in terms of the anti-correlation coefficients, but as before the mapping of ticket fractions to anti-correlation coefficients is expressed as $\vec{\chi}|_{\vec{f}_{\mathrm{adm}}}=M\vec{f}_{\mathrm{adm}}$ where $M$ is a $3\times n$ matrix. Hence the objective function may be written in terms of ticket fractions:
\begin{equation}
    \vec{c}\cdot \vec{\chi}|_{\vec{f}_{\mathrm{adm}}}=\vec{c}\cdot (M\vec{f}_{\mathrm{adm}})=(M^\top \vec{c})\cdot \vec{f}_{\mathrm{adm}}.
\end{equation}
To maximally violate a particular facet inequality, then, is equivalent to minimizing this objective function over the set of $\vec{f}_{\mathrm{adm}}\in \Delta^{n-1}$ satisfying $B\vec{f}_{\mathrm{adm}}=\vec{b}$. The problem of finding such an admissible raffle therefore takes the form of a particular \emph{linear program},  and thus our iterative strategy requires the solution of a finite number of linear programs. Such linear programs can be solved via the so-called simplex method at relatively low computational cost. In this way, we implemented our iterative strategy in Mathematica and so obtain an algorithm to compute all vertices of the anti-correlation polyhedron. The resulting local polyhedra, in the case of spin-$2$ as well as spin-$\frac52$, appear in Figure \ref{FacetsSpin2Spin52}. As in Figure \ref{SpinThreeHalfFace}, this picture shows the facets we need to ``glue onto" each of the four facets of the classical tetrahedron to get the full polyhedron.

\begin{figure}[h]
 \centering
   \includegraphics[width=5in]{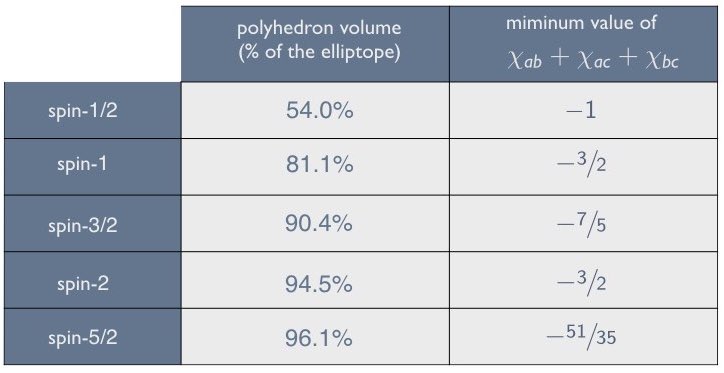} 
   \caption{Polytope volumes and minimal values of the sum of anti-correlation coefficients}
   \label{polytopevolume}
\end{figure}

The anti-correlation polyhedra presented thus far (Figures \ref{tetrahedron}, \ref{polytope-spin1}, and \ref{SpinThreeHalfFace}) suggests that, as the number of outcomes increases, the anti-correlation polyxhedron converges to the full elliptope. This is not really surprising, for if we were to allow for a continuous range of outcomes rather than a discrete set then the full elliptope of correlation matrices is certainly generated. Indeed, for the case of a multivariate Gaussian distribution, the probability distribution is parametrized by the choice of a positive-definite correlation matrix.\footnote{A simpler example is provided by the \emph{3m balance} discussed in Section \ref{1.6.4} (see Figure \ref{3M-balance})).\label{3M convergence}} Thus the failure to obtain the full elliptope rests on the discrete nature of the outcomes.\footnote{One could say that the \emph{discreteness} introduced by quantum mechanics puts some restrictions on the elliptope but that those restrictions are lifted again by the \emph{contextuality} it introduces.\label{discrete and contextual}}  As numerical evidence for convergence we consider the volume of our polyhedra, which can be computed from the list of vertices. These volumes are listed in Figure \ref{polytopevolume} in terms of their fraction of the elliptope volume (which may be shown to be exactly $\pi^2/2$). These volumes are seen to increase monotonically to that of the full elliptope as spin increases, in agreement with the convergence we are seeing in the figures.\footnote{The admissibility conditions on the correlations represented by points in our correlation polyhedra form the main obstacle to a formal proof of this convergence. As we saw in Section \ref{2.2.3}, these conditions are of two kinds: the correlations should have uniform marginals and their correlation arrays should have the same symmetries as the quantum-mechanical ones they are meant to simulate. The convergence issue can meaningfully be studied without these symmetry conditions. The uniform-marginals condition, however, is critical for the construction of our correlation polyhedra. For one thing, as already mentioned in the introduction to Section \ref{2}, we need this condition to ensure that the correlation arrays in some allowed class differ only in their off-diagonal cells.\label{no-convergence-proof}}

%SECTION 4
\section{Correlation arrays, polytopes and the CHSH inequality} \label{3}
%!TEX root =  ./JanasJanssenCuffaro-August2019.tex
%SECTION 4
%\section{Correlation arrays, polytopes and the CHSH inequality} \label{3}
%Four settings and two outcomes per setting
%From three to four different settings:
%A Setup with four settings and two outcomes per setting: the CHSH inequality

In the preceding sections we studied the Mermin setup, which involves two parties and three settings per party. In Section \ref{1}, we analyzed the case with two outcomes per setting; in Section \ref{2}, we extended our analysis to three and more outcomes per setting. In this section, we return to the simple case of two outcomes per setting. However, the two parties now get to choose from two different pairs of settings, rather than from the same triplet of settings. In other words, we replace the Mermin setup by the more common setup for which the CHSH inequality \citep{CHSH} was formulated and tested. As in Section \ref{1}, we focus on correlations found in measurements performed on a pair of spin-$\frac12$ in the singlet state and on raffles designed to simulate these correlations. 

\begin{figure}[h]
 \centering
   \includegraphics[width=6in]{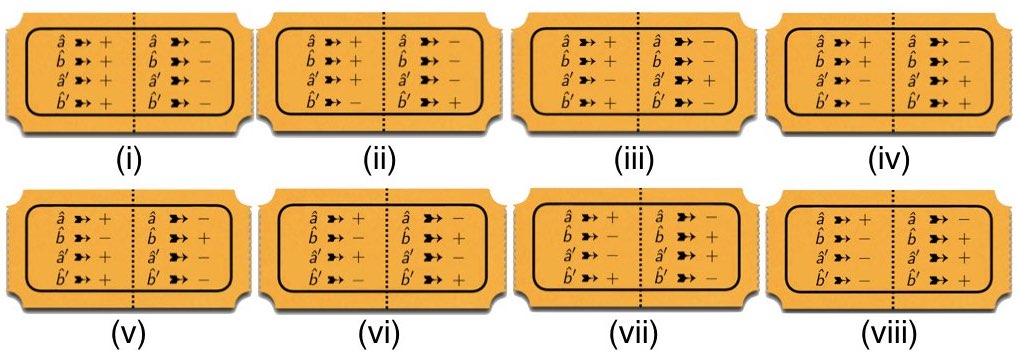} 
   \caption{Raffle tickets for four settings and two outcomes.}
   \label{raffle-tickets-4set2out-i-thru-viii}
\end{figure}

\begin{figure}[h]
 \centering
   \includegraphics[width=3.5in]{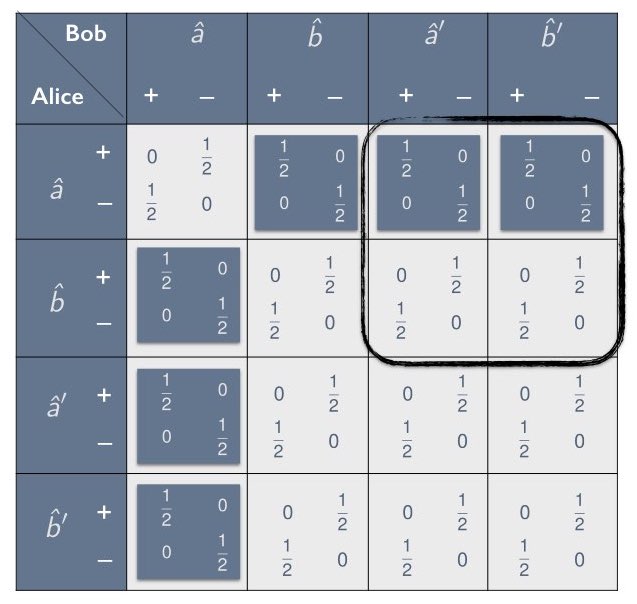} 
   \caption{Correlation array for a raffle with a basket of tickets of type (viii) (see Figure \ref{raffle-tickets-4set2out-i-thru-viii}). Blue-on-white cells show perfect anti-correlation; white-on-blue cells show perfect correlation. The four cells in the upper-right corner are the ones actually probed in tests of the CHSH inequality.}
   \label{CA-4set2out-raffle-viii}
\end{figure}

Our goal in this section is to recover the CHSH inequality and the corresponding Tsirelson bound  in the Bub-Pitowsky-inspired framework developed in Sections \ref{1}--\ref{2}. The key to achieving this objective is to note that the CHSH setup, in which Alice and Bob pick from two \emph{different pairs} of settings, $(\hat{a}, \hat{b})$ and $(\hat{a}', \hat{b}')$, can be treated as a special case of a straightforward generalization of the Mermin setup in which they pick from the \emph{ same quartet} of settings $(\hat{a}, \hat{b}, \hat{a}', \hat{b}')$. The special case of this generalized Mermin setup is that Alice never actually uses the settings $(\hat{a}', \hat{b}')$ and that Bob never actually uses the settings  $(\hat{a}, \hat{b})$. Nothing prevents us, however, from adding cells for the unused combinations of settings to the correlation arrays for the CHSH setup. In this way, the $2 \times 2$ correlation arrays for the CHSH setup turn into $4 \times 4$ correlation arrays (see, e.g., Figure \ref{CA-4set2out-raffle-viii}) that are similar to the $3 \times 3$ ones for the Mermin setup (see, e.g.,  Figures \ref{CA-3set2out-Mermin}, \ref{CA-3set2out-raffles-i-thru-iv} and \ref{CA-3set2out-raffle-mix}). The off-diagonal cells of these $4 \times 4$ correlation arrays can be parametrized by six anti-correlation coefficients, two of which ($\chi_{ab}$ and $\chi_{a'b'}$) do not play a role in the CHSH setup. The CHSH inequality and the Tsirelson bound in this case are conditions on the remaining four, $\chi_{aa'}$, $\chi_{ab'}$, $\chi_{ba'}$ and $\chi_{bb'}$. To derive the CSHS inequality, we use the kind of raffles we introduced in Section \ref{1.4}. As in Section \ref{1.5}, we derive the corresponding Tsirelson bound from the positive semi-definiteness of the anti-correlation matrix $\chi$, which in this case is a symmetric $4 \times 4$ matrix with $1$'s on the diagonal and the six anti-correlation coefficients as its off-diagonal elements. 

Figure \ref{raffle-tickets-4set2out-i-thru-viii} shows the eight different types of raffle tickets for the CHSH setup. Figure \ref{CA-4set2out-raffle-viii} shows the correlation array for a raffle with a basket containing only type-(viii) tickets. Six of the off-diagonal cells in this correlation array show a perfect correlation, the other six show a perfect anti-correlation. This means that the values of the six anti-correlation coefficients parametrizing these twelve cells are:
\begin{equation}
\chi_{ab}^{\mathrm{(viii)}} = \chi_{aa'}^{\mathrm{(viii)}} = \chi_{ab'}^{\mathrm{(viii)}} = -1, \quad \chi_{ba'}^{\mathrm{(viii)}} = \chi_{bb'}^{\mathrm{(viii)}} = \chi_{a'b'}^{\mathrm{(viii)}} = 1.
\label{chi values for ticket (viii)}
\end{equation}
These values can also be read directly off ticket (viii).  

 \begin{table}[h]
\centering
\begin{tabular}{|c||c|c|c|c||c|}
\hline
ticket & \quad $\chi_{aa'}^{\mathrm{(k)}}$ \quad & \quad $\chi_{ab'}^{\mathrm{(k)}}$ \quad & \quad $\chi_{ba'}^{\mathrm{(k)}}$ \quad & \quad $\chi_{bb'}^{\mathrm{(k)}}$ \quad &  $\chi_{aa'}^{\mathrm{(k)}} +\chi_{ab'}^{\mathrm{(k)}}$ \quad \\[.1cm] 
 &  &  &  &  &  \quad $+\chi_{ba'}^{\mathrm{(k)}} - \chi_{bb'}^{\mathrm{(k)}}$ \quad \\[.1cm]
\hline
 (i) & $+1$ & $+1$ & $+1$ & $+1$ & $2$ \\[.1cm]
 (ii) & $+1$ & $-1$ & $+1$ & $-1$ & $2$ \\[.1cm]
 (iii) & $-1$ & $+1$ & $-1$ & $+1$ & $-2$ \\[.1cm]
(iv) & $-1$ & $-1$ & $-1$ & $-1$ & $-2$ \\[.1cm]
 (v) & $+1$ & $+1$ & $-1$ & $-1$ & $2$ \\[.1cm]
 (vi) & $+1$ & $-1$ & $-1$ & $+1$ & $-2$ \\[.1cm]
 (vii) & $-1$ & $+1$ & $+1$ & $-1$ & $2$ \\[.1cm]
(viii) & $-1$ & $-1$ & $+1$ & $+1$ & $-2$ \\[.1cm]
 \hline
\end{tabular}
\caption{Values of four of the anti-correlation coefficients for tickets (i)--(viii) in Figure \ref{raffle-tickets-4set2out-i-thru-viii}. The final column shows the values for a linear combination of these coefficients.}
\label{values of chi-4set2out}
\end{table}
%$\ ($\mathrm{k = i \ldots viii}$), 

In the CHSH setup, data are taken only for the four combinations of settings corresponding to the four cells in the upper-right corner of these $4 \times 4$ correlation arrays. These cells are characterized by four of the six anti-correlation coefficients in Eq.\ (\ref{chi values for ticket (viii)}). Table \ref{values of chi-4set2out} lists their values for all eight tickets in Figure \ref{raffle-tickets-4set2out-i-thru-viii}. The final column gives the value for a linear combination of these four anti-correlation coefficients. Note that for tickets (i), (ii), (v) and (vii), this quantity is equal to 2, while for tickets (iii), (iv), (vi) and (viii), it is equal to $-2$. For a raffle with any mix of tickets of type (i) through (viii), this quantity must therefore lie between $-2$ and $2$:
%(cf.\ Section \ref{1.4}, note \ref{checking averaging of chi}):
\begin{equation}
-2 \le \chi_{aa'} + \chi_{ab'} + \chi_{ba'} - \chi_{bb'} \le 2.
\label{CHSH inequality}
\end{equation}
This is the CHSH inequality \citep[p.\ 68]{Bub 2016}.

We now want to connect the raffles with tickets for the present case of four settings to the raffles in Section \ref{1.4} for the case of three settings. An obvious way to do this is to ignore one of the four settings and have Alice and Bob choose between the remaining three. In that case, there are only three anti-correlation coefficients, which will satisfy the inequality in Eq.\ (\ref{Mermin inequality CHSH-like}) for the Mermin setup. One could object, however, that this way of recovering Eq.\ (\ref{Mermin inequality CHSH-like}) requires that we consider results found for combinations of settings Alice and Bob are not actually using in the CHSH setup. To avoid this objection, suppose Bob's setting $\hat{b}'$ is the same as Alice's setting $\hat{b}$. This means that we restrict ourselves to tickets with the same values for $\hat{b}$ and $\hat{b}'$. These are the tickets (i), (iii), (vi) and (viii) in Figure \ref{raffle-tickets-4set2out-i-thru-viii}. Focusing on the rows for those four tickets in Table \ref{values of chi-4set2out} and using that, in those rows, $\chi_{bb'} = 1$ and $\chi_{ab'} = \chi_{ab}$, we see that the inequality in Eq.\ (\ref{CHSH inequality}) reduces to:
\begin{equation}
-1 \le \chi_{aa'} + \chi_{ab} + \chi_{ba'} \le 3.
\label{CHSH2Bell}
\end{equation}
If $\hat{a}'$ is relabeled $\hat{c}$, this inequality turns into the one in Eq.\ (\ref{Mermin inequality CHSH-like}) for the Mermin setup. Note that this is the form in which \citet{Bell 1964} originally derived the Bell inequality (see Section \ref{1.1}).

As was stressed in Section\ \ref{1.4} for the case of three settings, however, the CHSH inequality is a necessary but not sufficient condition for classical correlations. To obtain a complete characterization of the correlations allowed classically, we once again examine their geometrical representation.

The tickets in Table \ref{values of chi-4set2out} give the coordinates of eight points in the four-dimensional space of anti-correlation coefficients $\chi_{aa'},\chi_{ab'},\chi_{ba'},\chi_{bb'}$. Since the anti-correlation coefficients of a classical mixed state will be a weighted average of those for the classical pure states, we may interpret those eight points as vertices of some convex hull. This is the local polytope, i.e., the set of of anti-correlation coefficients $\chi_{aa'},\chi_{ab'},\chi_{ba'},\chi_{bb'}$ that can be simulated by such raffles.

While we cannot visualize this four--dimensional polytope, some geometric observations can be made. The eight vertices may be viewed as four pairs of antipodal points, with the four line segments between the pairs being mutually orthogonal. As was noted by \citet[p. 5]{Pitowsky 2008}, this polytope is the 4-dimensional octahedron or hyperoctahedron \citep[p.\ 112]{Bub 2016}. It has a total of 16 facets, half of which are given by the inequalities
\begin{eqnarray}
-2 \le \;\, \chi_{aa'} + \chi_{ab'} + \chi_{ba'} - \chi_{bb'} \; \le 2, & & \label{CHSH-ineq 1} \\[.2cm]
-2\le -\chi_{aa'}+\chi_{ab'}-\chi_{ba'}-\chi_{bb'} \; \le  2,  & & \label{CHSH-ineq 2}  \\[.2cm]
-2\le \;\, \chi_{aa'}-\chi_{ab'}-\chi_{ba'}-\chi_{bb'} \; \le  2, & & \label{CHSH-ineq 3}  \\[.2cm]
-2\le  -\chi_{aa'}-\chi_{ab'}+\chi_{ba'}-\chi_{bb'} \; \le  2. & & \label{CHSH-ineq 4}
\end{eqnarray}
Eq.\ (\ref{CHSH-ineq 1}) is just the CHSH inequality stated above when Alice and Bob use settings $(\hat{a},\hat{b})$ and $(\hat{a}',\hat{b}')$, respectively. We may obtain the others similarly by reversing some of the settings. For instance, Eq.\ (\ref{CHSH-ineq 2}) is the CHSH inequality if Bob were to use setting $-\hat{b}$ instead of $\hat{b}$. For our purposes, we will regard each of them as a CHSH inequality.

For the remaining eight facets, note that all coordinates of the eight vertices are $\pm 1$, with $+1$ occurring an even number of times. Hence these are eight of the sixteen vertices for a four-dimensional hypercube, which is the non-signaling polytope in the CHSH setup. It bounds the local polytope, thus providing the remaining eight facets which \citet[p.\ 3, Eq.\ (2)]{Pitowsky 2008} refers to as ``trivial'':
\begin{equation}
-1\leq \chi_{aa'}\leq 1,\quad -1\leq \chi_{ba'}\leq 1,\quad -1\leq \chi_{ab'}\leq 1,\quad -1\leq \chi_{bb'}\leq 1.
\end{equation}
In the case of the Mermin example, the four facets of the classical tetrahedron already sufficed to restrict the correlations to the non-signaling cube. This is not sufficient in the present setup. For instance, the values $\chi_{aa'}=2,\chi_{ab'}=\chi_{ba'}=\chi_{bb'}=0$ are allowed by the CHSH inequalities but lie outside the non-signaling hypercube. It is thus necessary to include the hypercube facets explicitly.

The parts of the local polytope which lie on one of these hypercube facets are particularly notable. If we restrict our attention to the facet $\chi_{bb'}=1$, for instance, then we are looking at what part of the local polytope falls in the region parametrized by $\chi_{aa'}$, $\chi_{ab'}$ and $\chi_{ba'}$. Since each of these coefficients has magnitude 1, we are working with a copy of the non-signaling cube for three outcomes. The inequalities in Eqs. (\ref{CHSH-ineq 1})--(\ref{CHSH-ineq 4})) then simplify to
\begin{eqnarray}
-1 \le \;\, \chi_{aa'} + \chi_{ab'} + \chi_{ba'}\; \le 3, & & \\[.2cm]
-1\le -\chi_{aa'}+\chi_{ab'}-\chi_{ba'} \; \le  3, & & \\[.2cm]
-1\le \;\, \chi_{aa'}-\chi_{ab'}-\chi_{ba'} \; \le  3, & & \\[.2cm]
-1\le  -\chi_{aa'}-\chi_{ab'}+\chi_{ba'} \; \le 3. & &
\end{eqnarray}
Aside from having $\hat{a}',\hat{b}'$ instead of $\hat{c},\hat{b}$, this system of inequalities is identical to those that generated the classical tetrahedron in the Mermin setup (see Eqs.\ (\ref{Mermin inequality CHSH-like (i)})--(\ref{Mermin inequality CHSH-like (iv)})). Hence the local polytope in the CHSH setup contains a copy of the classical tetrahedron, occurring on the $\chi_{bb'}=1$ facet of the non-signaling hypercube; this facet is itself a copy of the non-signaling cube.
%(This is actually true in general: The sixteen cells of the 4-dimensional octahedron are regular tetrahedra, as may be shown by considering the case where the CHSH inequality reaches its maximum value.) 

We now consider the analogous quantum story, i.e., the correlations of the singlet state for the CHSH setup. Cells in the correlation array for the CHSH setup are no different from cells in the correlation array for the Mermin setup given in Section \ref{1.5}. The anti-correlation coefficients for any combination of measurement directions $\vec{e}_a$ and $\vec{e}_b$ are given by  
\begin{equation}
\chi_{ab} =-\langle \hat{S}_{1a}\hat{S}_{2b}\rangle_{00} =  \cos\varphi_{ab}
\end{equation}
(see Eq.\ (\ref{chi2angle}) with the standard deviations $\sigma_{1a}$ and $\sigma_{2b}$ set equal to 1). 

We now introduce the anti-correlation matrix
\begin{equation}
\chi=
\begin{pmatrix}
\chi_{aa} & \chi_{ab} & \chi_{aa'} & \chi_{ab'} \\[.2cm]
\chi_{ba} & \chi_{bb} & \chi_{ba'} & \chi_{bb'} \\[.2cm]
\chi_{a'a} & \chi_{a'b} & \chi_{a'a'} & \chi_{a'b'} \\[.2cm]
\chi_{b'a} & \chi_{b'b} & \chi_{b'a'} & \chi_{b'b'}.
\end{pmatrix}=
\begin{pmatrix}
1 & \cos \varphi_{ab} & \cos \varphi_{aa'} & \cos \varphi_{ab'} \\[.2cm]
\cos \varphi_{ab} & 1 & \cos \varphi_{ba'} & \cos \varphi_{bb'} \\[.2cm]
\cos \varphi_{aa'} &\cos \varphi_{ba'} & 1 & \cos \varphi_{a'b'} \\[.2cm]
\cos \varphi_{ab'} & \cos \varphi_{bb'} & \cos \varphi_{a'b'} & 1.
\end{pmatrix}.
\label{CHSH-matrix}
\end{equation}
Using that $\cos{\varphi_{ab}} = \vec{e}_a \cdot \vec{e}_b$ etc.\ and that $\vec{e}_a = (a_x, a_y, a_z)$, etc., we can factorize this matrix as (cf.\ Eqs.\ (\ref{QM10})--(\ref{QM12})):
%\begin{eqnarray}
%\chi 
%& \!\! = \!\! &
%\begin{pmatrix}
%\vec{e}_a \! \cdot  \vec{e}_a &  \vec{e}_a \! \cdot  \vec{e}_b  &   \vec{e}_a \! \cdot  \vec{e}_{a'} & \vec{e}_a \! \cdot  \vec{e}_{b'} \\
%\vec{e}_b \! \cdot  \vec{e}_a & \vec{e}_b \! \cdot  \vec{e}_b & \vec{e}_b \! \cdot  \vec{e}_{a'} & \vec{e}_b \! \cdot  \vec{e}_{b'} \\
%\vec{e}_{a'} \! \cdot  \vec{e}_a & \vec{e}_{a'} \! \cdot  \vec{e}_b & \vec{e}_{a'} \! \cdot  \vec{e}_{a'} & \vec{e}_{a'} \! \cdot  \vec{e}_{b'} \\
%\vec{e}_{b'} \! \cdot  \vec{e}_a & \vec{e}_{b'} \! \cdot  \vec{e}_b & \vec{e}_{b'} \! \cdot  \vec{e}_{a'} & \vec{e}_{b'} \! \cdot  \vec{e}_{b'}
%\end{pmatrix} \nonumber \\
%& \!\! = \!\!  &
%\begin{pmatrix}a_x & a_y & a_z \\ b_x & b_y & b_z \\ a'_x & a'_y & a'_z \\ b'_x & b'_y & b'_z \end{pmatrix}
%\begin{pmatrix} a_x & b_x & a'_x & b'_x \\ a_y & b_y & a'_y & b'_y  \\ a_z & b_z & a'_z & b'_z.  \end{pmatrix}
%\equiv T^\top T.
%\end{eqnarray}
\begin{equation}
\chi = 
\begin{pmatrix}
a_x & a_y & a_z \\[.2cm]
b_x & b_y & b_z \\[.2cm] 
a'_x & a'_y & a'_z \\[.2cm] 
b'_x & b'_y & b'_z 
\end{pmatrix}
\begin{pmatrix} 
a_x & b_x & a'_x & b'_x \\[.2cm]
a_y & b_y & a'_y & b'_y  \\[.2cm]
a_z & b_z & a'_z & b'_z.  
\end{pmatrix}
\equiv L^\top L.
\end{equation}
This factorization implies that, given any vector $\vec{v}=(v_a,v_b,v_c)^\top$, one has 
\begin{equation}
\vec{v}^\top \! \chi\vec{v} = \vec{v}^\top L^\top L \vec{v} = (L\vec{v})^\top L \vec{v} \geq 0,
\end{equation}
where in the last step we used that this quantity is the length squared of the vector $L\vec{v}$. Hence $\chi$ is positive semi-definite (cf.\ Eq.\ (\ref{inf the 4})).

The set of such $4 \times 4$ matrices---that is, those which are symmetric, have $1$'s on the diagonal and are positive semi-definite---is conventionally known as the 4-elliptope. The terminology is an obvious generalization of the elliptope for the $3 \times 3$ case and for the purposes of this section we refer to the latter as the 3-elliptope. The 4-elliptope, being parametrized by six anti-correlation coefficients, is a six-dimensional set; like the 3-elliptope, it is moreover convex (since any weighted average of two positive semi-definite matrices is itself positive semi-definite). As noted before, however, the values of $\chi_{ab}$ and $\chi_{a'b'}$ are not provided in the context of the CHSH inequality. As such we must project this convex set to the four-dimensional space of the coefficients $\chi_{a'b'}$, $\chi_{aa'}$, $\chi_{ab'}$ and $\chi_{ba'}$. The resulting four-dimensional convex set, a shadow of the 4-elliptope, represents the class of correlations that Alice and Bob can obtain by measuring the singlet state using the settings $(\hat{a}, \hat{b})$ and $(\hat{a}', \hat{b}')$, respectively.\footnote{In the present case, we can further note that the four columns of $L$ are unit vectors in three-dimensional space. As such they must be linearly dependent, which implies that zero is an eigenvalue of $L$ (and thus of $\chi$ as well). Hence $\chi$ is not positive definite and must lie in the five-dimensional boundary of the 4-elliptope. However, this restriction plays no role upon projecting to the shadow of the 4-elliptope and so will not be elaborated upon further.}

Like the local polytope, neither the 4-elliptope nor its projected shadow can be visualized in its totality, yet some geometrical conclusions can still be drawn. For instance, since these  anti-correlation coefficients are cosines, their magnitude does not exceed 1. Hence, the set of quantum correlations is contained within the non-signaling hypercube. Moreover, this set contains the local polytope. To see this, consider a configuration in which the unit vectors for all four measurement settings lie along the same line, i.e., are either parallel or anti-parallel. This configuration can be realized in eight ways. Then all the cosines are $\pm 1$, with $+1$ occurring an even number of times. This corresponds directly to the eight vertices of the local polytope (i.e., the rows of Table\ \ref{values of chi-4set2out}). Since the 4-elliptope (and therefore its shadow) is convex, we conclude that the set of quantum correlations does indeed include the local polytope.

It remains to show that the set of quantum correlations is larger than the local polytope. Recall that we can use $-\hat{S}_{2b}$ as a stand-in for $\hat{S}_{1b}$ when evaluating the expectation value $ \langle \hat{S}_{1a}\hat{S}_{1b}\rangle_{00}$ (see the end of  Section \ref{1.5}):
\begin{equation}
\chi_{ab} =-\langle \hat{S}_{1a}\hat{S}_{2b}\rangle_{00} = \langle \hat{S}_{1a}\hat{S}_{1b}\rangle_{00}.
\end{equation}
Now note that
\begin{equation}
\Big\langle \Big(\hat{S}_{1a'}-\frac{1}{\sqrt{2}}\hat{S}_{1a}-\frac{1}{\sqrt{2}}\hat{S}_{1b}\Big)^{\!2}\Big\rangle_{00}
= 2-\sqrt{2}\chi_{aa'}-\sqrt{2}\chi_{a'b}+\chi_{ab}\geq 0.
\label{CHSH-Tsirelson 1}
\end{equation}
Similarly,
\begin{equation}
\Big\langle \Big( \hat{S}_{1b'}-\frac{1}{\sqrt{2}}\hat{S}_{1a}+\frac{1}{\sqrt{2}}\hat{S}_{1b} \Big)^{\!2} \Big\rangle_{00}
= 2-\sqrt{2}\chi_{ab'}+\sqrt{2}\chi_{bb'}-\chi_{ab}\geq 0.
\label{CHSH-Tsirelson 2}
\end{equation}
Adding these two inequalities, we arrive at
\begin{equation}
2-\sqrt{2}\chi_{aa'}-\sqrt{2}\chi_{a'b} + 2-\sqrt{2}\chi_{ab'}+\sqrt{2}\chi_{bb'} \geq 0,
\label{CHSH-Tsirelson 3}
\end{equation}
which can be rewritten as
\begin{equation}
\chi_{aa'}+\chi_{a'b}+\chi_{ab'}-\chi_{bb'} \leq 2\sqrt{2}.
\label{CHSH-Tsirelson 4}
\end{equation}
This is the Tsirelson bound for the CHSH setup \citep[p.\ 68]{Bub 2016}. 

\begin{figure}[h]
\centering
    \includegraphics[width=1.5in]{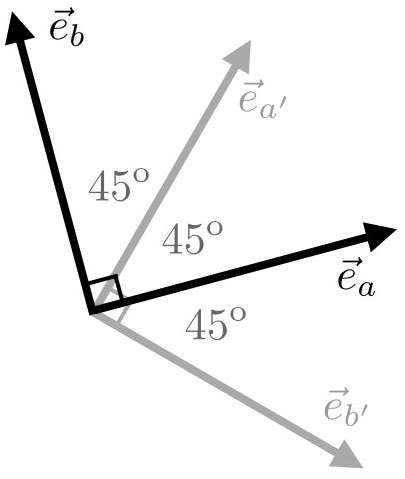}
 \caption{Measurement directions for maximum violation of the CHSH inequality.}
   \label{vectorsCHSH}
\end{figure}

To reach the Tsirelson bound we need both of the expectation values in Eqs.\ (\ref{CHSH-Tsirelson 1})--(\ref{CHSH-Tsirelson 2}) to vanish. This occurs when the unit vectors for the measurement settings of Alice and Bob are related as 
\begin{equation}
\vec{e}_{a'}=\frac{1}{\sqrt{2}}\vec{e}_a+\frac{1}{\sqrt{2}}\vec{e}_b,\quad \vec{e}_{b'}= \frac{1}{\sqrt{2}}\vec{e}_a- \frac{1}{\sqrt{2}}\vec{e}_b. 
\end{equation}
Squaring the first of these relations, we find that
\begin{equation}
1 = \left( \vec{e}_{a'} \right)^2 = \left( \frac{1}{\sqrt{2}}\vec{e}_a+\frac{1}{\sqrt{2}}\vec{e}_b \right)^2 = \frac12 + \frac12 + \vec{e}_{a} \! \cdot \vec{e}_{b},
\end{equation}
from which it follows that $\vec{e}_{a} \perp \vec{e}_{b}$. The situation is thus as shown in Figure \ref{vectorsCHSH}. It follows that $\varphi_{ab} =  \varphi_{a'b'} = 90^{\mathrm{o}}$, $\varphi_{aa'} =  \varphi_{ab'} = \varphi_{ba'} = 45^{\mathrm{o}}$ and $\varphi_{bb'} =  135^{\mathrm{o}}$. Hence
\begin{equation*}
\chi_{ab}=\chi_{a'b'}= \cos{90^{\mathrm{o}}} = 0,
\end{equation*}
\begin{equation}
\chi_{aa'}=\chi_{ba'}=\chi_{ab'}= \cos{45^{\mathrm{o}}} = \frac{1}{\sqrt{2}},
\end{equation}
\begin{equation*}
\chi_{bb'}= \cos{135^{\mathrm{o}}} = - \frac{1}{\sqrt{2}}.
\end{equation*}
In this case we therefore have
\begin{equation}
\chi_{aa'}+\chi_{ab'}+\chi_{ab'}-\chi_{bb'} =2\sqrt{2}.
\label{CHSH-Tsirelson 5}
\end{equation}
This represents the maximum violation of the CHSH inequality and we see once again that the set of quantum correlations is larger than the local polytope.

As was noted in Section \ref{1.4} for the three-settting case, however, Tsirelson's bound is a necessary but not sufficient condition for quantum correlations. To see this presently, note that \emph{every} possible expectation value of the form Eqs. (\ref{CHSH-Tsirelson 1})--(\ref{CHSH-Tsirelson 2}) ought to be non-negative. But each such expectation gives rise to another linear inequality on the six anti-correlation coefficients. For instance, if Bob were to instead use $(-\hat{a}',-\hat{b}')$ then the inequality obtained is 
\begin{equation}
\chi_{aa'}+\chi_{ab'}+\chi_{ab'}-\chi_{bb'}\geq -2\sqrt{2}.
\label{CHSH-Tsirelson 6}
\end{equation}
The anti-correlation coefficients therefore satisfy not one but infinitely many linear bounds, all of which must be satisfied if the coefficients arise quantum-mechanically. Hence Tsirelson's bound is evidently not sufficient to characterize the set of quantum correlations in the CHSH setup and moreover no finite list of such inequalities will suffice either. The set of quantum correlations in the CHSH setup is therefore not a polytope and is instead some smooth convex body.

As was the case for the 3-elliptope in relation to the three-setting case, however, this still leaves the possibility to characterize the set of quantum correlations by \emph{nonlinear} inequalities. To find these it is useful to relate the set of quantum correlations for the four-setting case to those for the three-setting case. For instance, suppose Alice and Bob are allowed to use three of the four settings, e.g., Alice and Bob measure using settings $(\hat{a},\hat{b},\hat{a}')$. The results of Section \ref{1.5} then apply and therefore the anti-correlation coefficients $(\chi_{ab},\chi_{aa'},\chi_{ba'})$ must lie within the corresponding 3-elliptope:
\begin{equation}
1-\chi_{ab}^2-\chi_{aa'}^2-\chi_{ba'}^2+2\chi_{ab}\chi_{aa'}\chi_{ba'}\geq 0.
\label{3-elliptope}
\end{equation}
In mathematical terms, this is an application of Sylvester's criterion: A matrix is positive semi-definite if and only if none of its principal minors are negative. It is then useful to rewrite the 3-elliptope in the equivalent form\footnote{Aside from an overall square root, this form of the inequality (for three rather than four random variables) can be found in \citet[p.\ 486]{Yule 1896}. Cf.\ Eq.\ (\ref{lastminute}) in Section \ref{1.6}.}
\begin{equation}
 |\chi_{aa'}\chi_{ba'}-\chi_{ab}| \leq \sqrt{1-\chi_{aa'}^2}\sqrt{1-\chi_{ba'}^2}.
\end{equation}
The coefficient $\chi_{ab}$ cannot be deduced from Alice and Bob's measurements in the CHSH setup. Nevertheless this quantity does exist and the last inequality signifies that it cannot differ too much from $\chi_{aa'}\chi_{ba'}$. 
A similar calculation shows that, if Alice and Bob instead only made measurements on the settings $(\hat{a},\hat{b},\hat{b}'),$ then
\begin{equation}
 |\chi_{ab'}\chi_{bb'}-\chi_{ab}| \leq \sqrt{1-\chi_{ab'}^2}\sqrt{1-\chi_{bb'}^2}.
\end{equation}
We thus have bounds on how much $\chi_{ab}$ can differ from the values of $\chi_{aa'}\chi_{ba'}$ and $\chi_{ab'}\chi_{bb'}$. The triangle inequality then bounds how much these two quantities can differ from each other:
\begin{eqnarray}
|\chi_{aa'}\chi_{ba'}-\chi_{ab'}\chi_{bb'}|
&=& |(\chi_{aa'}\chi_{ba'}-\chi_{ab})+(\chi_{ab}-\chi_{ab'}\chi_{bb'})| \nonumber \\[.2cm]
&\leq & |\chi_{aa'}\chi_{ba'}-\chi_{ab}|+|\chi_{aa'}\chi_{ba'}-\chi_{ab}| \nonumber \\[.2cm]
&\leq & \sqrt{1-\chi_{aa'}^2}\sqrt{1-\chi_{ba'}^2}+\sqrt{1-\chi_{ab'}^2}\sqrt{1-\chi_{bb'}^2}.
\end{eqnarray}
Any set of anti-correlation coefficients obtained from measurements on the singlet state in the CHSH setup must satisfy this inequality, first obtained by \citet{Landau 1988}. The 3-elliptope therefore proves useful even when characterizing quantum correlations in the non-visualizable case of four settings.\footnote{It should be noted that we have only established that this this is a necessary condition for quantum correlations, i.e., if $\chi_{aa'}$, $\chi_{ab'}$, $\chi_{ba'}$ and $\chi_{bb'}$ do not satisfy Landau's inequality then one cannot find $\chi_{ab},\chi_{a'b'}$ such that $\chi$ is positive semi-definite (and therefore these are not quantum correlations). The converse claim, i.e., that Landau's inequality  is also a sufficient condition for quantum correlations, is beyond the scope of this paper and will not be addressed further.}

As an application of these results we consider again the special case where Alice and Bob share one setting, e.g. $\hat{b}'=\hat{b}$. Then $\chi_{bb'}=1$ and $\chi_{ab'}=\chi_{ab}$, so the second 3-elliptope inequality is then fulfilled trivially (both sides vanish identically) and Landau's result collapses to the 3-elliptope inequality. If we further relabel Bob's setting $\hat{a}'\to\hat{c}$, then this 3-elliptope takes the form
\begin{equation}
 |\chi_{ac}\chi_{bc}-\chi_{ab}| \leq \sqrt{1-\chi_{ac}^2}\sqrt{1-\chi_{bc}^2}.
\end{equation}
But this is the same 3-elliptope as considered originally in the Mermin example. Geometrically this means that the shadow of the 4-elliptope contains a copy of the 3-elliptope, occurring on the $\chi_{bb'}=1$ facet of the non-signaling hypercube.  This is as it should be:  The scenario where Alice and Bob, respectively, use settings $(\hat{a},\hat{b})$ and $(\hat{b},\hat{c})$ is exactly the setup originally employed by \citet{Bell 1964}. Combining this with the corresponding results for the local polytope and the non-signaling hypercube, we conclude that we recover the Mermin setup of Section \ref{1.4}--\ref{1.5} (including the entirety of Figure \ref{elliptope}) is indeed recovered when $\hat{b}'=\hat{b}$.

To concretely illustrate how the non-signaling hypercube, the set of quantum correlations and the local polytope are related, we consider the family of anti-correlation coefficients given by 
\begin{equation}
\chi_{aa'}=\chi_{ab'}=\chi_{ba'}=-\chi_{bb'}=-t
\end{equation}
for $0\leq t\leq 1$. The case $t=0$, where all correlation coefficients vanish, can be simulated classically (and therefore quantum mechanically), for instance by a raffle with basket containing all ticket types in equal proportion. By contrast, the case $t=1$ corresponds to the PR box shown in Figure \ref{CA-PRbox} in Section \ref{1.2} and should not be realizable in quantum mechanics despite being non-signaling. (More precisely, it is one of the vertices of the non-signaling hypercube which is not also a vertex of the local polytope).

The questions are then for what range of $t$ can such correlation coefficients be simulated classically or quantum-mechanically. For the classical case we observe that 
\begin{equation}
|\chi_{aa'}+\chi_{ab'}+\chi_{ba'}-\chi_{bb'}| =2t.
\end{equation}
Since the CHSH inequality bounds this magnitude by $2$, we conclude that $t\leq 1/2$ is the classical bound. For the quantum case we appeal to Landau's inequality, which takes the form
\begin{eqnarray}
|\chi_{aa'}\chi_{ba'}-\chi_{ab'}\chi_{bb'}| & \!\! = \!\! & 2t^2 \nonumber \\[.1cm]
& \!\!  \leq \!\! & \sqrt{1-\chi_{aa'}^2}\sqrt{1-\chi_{ba'}^2}+\sqrt{1-\chi_{ab'}^2}\sqrt{1-\chi_{bb'}^2}  \\[.2cm]
&  \!\! = \!\!  & 2(1-t^2). \nonumber
\end{eqnarray}
We therefore conclude that $t\leq 1/\sqrt{2}$, corresponding to Tsirelson's bound, i.e., the maximal violation of the CHSH inequality, marks the boundary between quantum and non-quantum correlations along this family of anti-correlation coefficients.% The relation between these cases is illustrated in [Figure].

%SECTION 5
\section{Interpreting quantum mechanics} \label{4}
%!TEX root =  ./JanasJanssenCuffaro-August2019.tex

\subsection{The story so far}
\label{4.1}

In Section \ref{1} we introduced the concept of a correlation array---a concise representation of the statistical correlations between separated parties in the context of a given experimental setup. We focused primarily on setups involving two parties, Alice and Bob, who are each given one of two correlated systems and are asked to measure them using one of the three settings $\hat{a}$, $\hat{b}$ and $\hat{c}$. Such a setup can be characterized using a 3$\times$3 correlation array in which each cell corresponds to one of the nine possible combinations for Alice's and Bob's setting choices. In Section \ref{1.3} we showed how to parameterize the cells in such a correlation array by means of an anti-correlation coefficient, defined as the negative of the expectation value of the product of Alice's and Bob's random variables, divided by the product of their standard deviations (see Eq.\ \eqref{chi as corr coef}). For example, when there are two possible outcomes per measurement, a symmetric 3$\times$3 correlation array with zeroes along the diagonal can be parameterized using three anti-correlation coefficients $\chi_{ab}$, $\chi_{ac}$ and $\chi_{bc}$, as depicted in Figure \ref{CA-3set2out-non-signaling-chis}. One of the correlation arrays describable in this way is the correlation array for the Mermin setup given in Figure \ref{CA-3set2out-Mermin}.

We considered local-hidden variable models for 3$\times$3 correlation arrays of this kind in Section \ref{1.4}. We imagined, in particular, modeling such arrays with mixtures of raffle tickets like the ones in Figure \ref{raffle-tickets-3set2out-i-thru-iv}, and for such models we derived the following constraints on the anti-correlation coefficients $\chi_{ab}$, $\chi_{ac}$ and $\chi_{bc}$:\footnote{These are identical to the inequalities given in Eqs.\ (\ref{Mermin inequality CHSH-like (i)}--\ref{Mermin inequality CHSH-like (iv)}) of Section \ref{1.4}.}
\begin{align}
\label{repeatInequalities1}
-1 \leq \chi_{ab} + \chi_{ac} + \chi_{bc} \leq 3, \\[.3cm]
\label{repeatInequalities2}
-1 \leq \chi_{ab} - \chi_{ac} - \chi_{bc} \leq 3, \\[.3cm]
\label{repeatInequalities3}
-1 \leq \chi_{ab} + \chi_{ac} - \chi_{bc} \leq 3, \\[.3cm]
\label{repeatInequalities4}
-1 \leq \chi_{ab} - \chi_{ac} + \chi_{bc} \leq 3.
\end{align}
Together these four linear inequalities are necessary and sufficient to characterize the space of possible statistical correlations realizable in any such model. This space can be visualized as the tetrahedron in Figure \ref{tetrahedron}; i.e., for any given point $(\chi_{ab}, \chi_{ac}, \chi_{bc})$, it is contained in the convex set represented by the tetrahedron if and only if it satisfies all four of Eqs.\ (\ref{repeatInequalities1}--\ref{repeatInequalities4}). In Section \ref{1.5} we showed that the convex set characterizing the allowable \emph{quantum} correlations for 3$\times$3 setups of this kind is a superset of those allowed in a local-hidden variables model. It can be characterized by the \emph{non}-linear inequality\footnote{This equation is identical to Eq.\ \eqref{QM14} from Section \ref{1.5}.}
\begin{align}
\label{repeatElliptopeEqn}
1 - \chi^2_{ab} - \chi^2_{ac} - \chi^2_{bc} + 2\chi_{ab}\chi_{ac}\chi_{bc} \geq 0,
\end{align}
whose associated inflated tetrahedron or elliptope is shown in Figure \ref{elliptope}.

Our work is both continuous with and extends that of Pitowsky. Pitowsky, in turn building on the work of George Boole \citep{Pitowsky 1994}, also considers the distinction between quantum and classical theory in light of the inequalities that characterize the possibility space of relative frequencies for a given classical event space. Pitowsky describes a general algorithm for determining these inequalities: Given the logically connected events $E_1, \dots E_n$, write down the propositional truth table corresponding to them and then take each row to represent a vector in an $n$-dimensional space. Their convex hull yields a polytope, and the sought-for inequalities characterize the facets of this polytope. Alternately, if we already know the inequalities we can then determine the polytope associated with them.

In our own case the event space associated with a 3$\times$3 correlation array for a setup involving two possible outcomes per measurement yields an easily visualisable three-dimensional representation of possible correlations between events for both a quantum and a local-hidden variables model. Moreover in the quantum case we showed that the resulting representation remains three-dimensional even when we transition to setups involving more than two outcomes---indeed we showed in Section \ref{2.1.5} that it is in every case the very same elliptope as the one we derived in Section \ref{1.5} for two outcomes (i.e., for spin-$\frac12$) and which we depicted in Figure \ref{elliptope}. In the local-hidden variables case (where we model correlations with raffles) the local polytopes characterizing the space of possible correlations for setups with more than two possible outcomes per measurement are of much higher dimension than three. In part through considering only those raffles that have a hope of recovering the quantum set, we showed in Section \ref{2.2} how to project these higher-dimensional polytopes down to three-dimensional anti-correlation polyhedra (see Figure \ref{flowchart}).\footnote{We call them polyhedra rather than polytopes since they are always three-dimensional.} We showed that with increasing spin these polyhedra become further and further faceted and correspondingly more and more closely approximate the full quantum elliptope (see Figure \ref{polytopevolume})---though actually computing these polyhedra becomes more and more intractable as the number of possible outcomes per setting increases. Finally, in addition to providing an easily visualisable representation in three dimensions of the quantum and local-hidden variable correlations associated with a 3$\times$3 Mermin-style setup, we showed how the correlation array formalism for this case can be straightforwardly extended so as to provide useful insight into the more familiar correlational space associated with CHSH-style setups, if the latter are characterized using 4$\times$4 correlation arrays and parameterized using six anti-correlation coefficients (see Section \ref{3}).

As Pitowsky observes \citeyearpar[p. IV]{Pitowsky 1989b},\footnote{We previously noted Pitowsky's observation in Section \ref{0}, where we quoted him.} linear inequalities such as those characterizing the facets of our polytopes have been an object of study for probability theorists since at least the 1930s. And although they were (re)discovered in a context far removed from these abstract mathematical investigations, the various versions of Bell's inequality are all inequalities of just this kind. \emph{Non}-linear inequalities like the one in Eq.\ \eqref{repeatElliptopeEqn}, on the other hand, are not. Nevertheless, equations like this one have also been an object of study for probability theorists. Drawing directly on their work, we showed in Section \ref{1.6.1} how one can derive an equation analogous to Eq.\ \eqref{repeatElliptopeEqn} characterizing the quantum elliptope from general statistical considerations concerning three balanced random variables $X_a$, $X_b$ and $X_c$ (for the meaning of \emph{balanced}, see the definition numbered (\ref{def balanced}) in Section \ref{1.3}). Specifically, we derived a constraint on the correlation coefficients $\overline{\chi}_{ab}$, $\overline{\chi}_{ac}$ and $\overline{\chi}_{bc}$ that is of exactly the same form as Eq.\ \eqref{repeatElliptopeEqn} (which, recall, constrains the \emph{anti}-correlation coefficients $\chi_{ab}$, $\chi_{ac}$ and $\chi_{bc}$):\footnote{A correlation coefficient $\overline{\chi}_{\alpha\beta}$ is just the negative of its corresponding anti-correlation coefficient.}$^,$\footnote{The following equation is identical to Eq.\ \eqref{inf the 5} of Section \ref{1.6.1}.}
\begin{equation}
1 - \overline{\chi}_{ab}^2 - \overline{\chi}_{ac}^2 - \overline{\chi}_{bc}^2 + 2 \, \overline{\chi}_{ab} \, \overline{\chi}_{ac} \, \overline{\chi}_{bc} \ge 0.
\label{repeat inf the 5}
\end{equation}

In Sections \ref{1.6.2} and \ref{1.6.3} we took up the questions, respectively, of how to model this general statistical constraint quantum-mechanically and in a local-hidden variables model, noting that the general derivation of Eq.\ \eqref{repeat inf the 5} relies essentially on the fact that we can consider a linear combination of the three random variables $X_a$, $X_b$ and $X_c$ in order to determine the expectation value of its square:\footnote{This equation is identical to Eq.\ \eqref{inf the 1} of Section \ref{1.6.1}.}
\begin{equation}
\Big\langle \Big( v_a \frac{X_a}{\sigma_a} + v_b \frac{X_b}{\sigma_b} + v_c \frac{X_c}{\sigma_c} \Big)^{\!2} \Big\rangle \ge 0.
\label{repeat inf the 1}
\end{equation}
To model such a relation with local-hidden variables, however, we require a joint probability distribution over $X_a$, $X_b$ and $X_c$. This in turn actually entails a tighter bound on the correlation coefficients than the one given by Eq.\ \eqref{repeat inf the 5}. Namely, it entails the analogue of the CHSH inequality for our setup, which should be unsurprising given the classical assumptions we began with. Thus, while the elliptope equation given by Eq.\ \eqref{repeat inf the 5} indeed constrains correlations between local-hidden variables in the setups we are considering, those correlations do not saturate that elliptope. In the case where there are only two possible values corresponding to each of the three random variables, the subset of the elliptope achievable is just the tetrahedron given in Figure \ref{tetrahedron}. For more than two values per variable the situation is more complicated: When the number of possible values, $n$, per variable is odd, one can actually reach the Tsirelson bound for this setup---the minimum value of 0 in Eq.\ \eqref{repeat inf the 5}---while when the number of possible values, $n$, per variable is even, one reaches the bound only in the limit as $n \to \infty$ (see Eqs.\ (\ref{Mermin CHSH half-integer spin}--\ref{Mermin CHSH integer spin})). But in either case---whether one reaches the Tsirelson bound or not---it appears that one requires a number of possible values $n \to \infty$ per random variable in order to saturate the volume of the elliptope in its entirety.\footnote{For further discussion, see Section \ref{2.2.4} and in particular the caveat contained in note \ref{no-convergence-proof}.}

From a slightly different point of view we can understand this as follows. Think of an arbitrary linear combination of the variables $X_a$, $X_b$ and $X_c$ as a vector $\mathbf{X}$ in a vector space. (Note that it follows from this that each of the variables $X_a$, $X_b$, $X_c$ is itself trivially also a vector.) And let $\varphi_{ab}$, $\varphi_{ac}$ and $\varphi_{bc}$ represent the ``angles'' between such vectors \citep[cf.][Section 4]{De Finetti 1937}. The correlation coefficient $\overline{\chi}_{\alpha\beta}$ may then be defined as the inner product of the vectors $X_\alpha$ and $X_\beta$, yielding (for instance) the natural property that two vectors are uncorrelated whenever they are orthogonal. As we explained in Section \ref{1.6.4}, from this point of view we can interpret Eq.\ \eqref{repeat inf the 5} geometrically as a constraint on the angles $\varphi_{\alpha\beta}$ between such vectors.

To express this mathematically is one thing. It is another to give a model for it. Note that such a model \emph{need not be classical.} De Finetti's own interpretation of the probability calculus, for instance, was not.\footnote{De Finetti distinguished between coherent degrees of belief in---and therefore probabilities associated with---verifiable as opposed to unverifiable events. This has consequences for his theory of probability. For instance if $A$ and $B$ are verifiable but not jointly verifiable they are not subject to the inequality $P(A) + P(B) - P(A\& B) \leq 1$. See \citet{Berkovitz 2012, Berkovitz 2019} for further discussion.} Any underlying model for these correlations that is classical, however, presupposes the existence of a joint distribution over the individual random variables $X_a$, $X_b$ and $X_c$. From this it follows that the correlations realizable in such a model cannot saturate the full volume of the elliptope expressed by Eq.\ \eqref{repeat inf the 5} except in the limit as the number of possible values corresponding to each of the random variables goes to infinity.

As we explained in Section \ref{1.6.2}, there are a number of challenges which need to be overcome in order to provide a quantum-mechanical model for the general statistical constraint expressed in Eq.\ \eqref{repeat inf the 5}. The most important of these is that in quantum mechanics one cannot consistently assume a joint probability distribution over incompatible observables, such as one would have to do in order to non-ambiguously define a vector $\mathbf{X}$ by taking a linear combination over the quantum analogues of $X_a$, $X_b$ and $X_c$. Since the sum of any two Hermitian operators is also Hermitian, however, then given three observables represented by, say, the operators $\hat{S}_a$, $\hat{S}_b$ and $\hat{S}_c$, one can always also consider the observable represented by the operator $\hat{S} \equiv \hat{S}_a + \hat{S}_b + \hat{S}_c$. As von Neumann observed already in 1927,\footnote{See note \ref{Myrvold 2} in Section \ref{1.6.2}.} quantum mechanics allows us to assign in this way a value to the sum of three variables without assigning values to all of them individually. From this it follows, not only that the elliptope equation constrains the possible correlations in the setups we are considering, but also that it tightly constrains them. The quantum-mechanical correlations in these setups, that is, saturate the full volume of the elliptope. Moreover we saw in Section \ref{1.6.2} how, in virtue of certain other assumptions we needed to model the constraint quantum-mechanically, Eq.\ \eqref{repeat inf the 5}---the equation we derived \emph{from without}---reduces to Eq.\ \eqref{repeatElliptopeEqn}---the equation we derived \emph{from within} quantum mechanics.

The remainder of this section is devoted to the philosophical conclusions that can be drawn from the foregoing. Below, in Section \ref{4.2} we will comment on the nature of our derivation of the space of quantum correlations for the setups we have considered. We will note that our derivation evinces aspects of both the principle-theoretic and the constructive approaches to physics, and that in our own derivation and generally in the practice of theoretical physics, both work together to yield understanding of the physical world. In Section \ref{4.3} we will argue that the insight yielded by our own investigation is that the fundamental novelty of the quantum mode of description can be located in the kinematics rather than in the dynamics of the theory. This distinction---between the kinematical and dynamical parts of a theory---is one we take to be of far more significance than the distinction between principle-theoretic and constructive approaches that has been the object of so much recent attention. In Section \ref{4.3a} we consider examples, from the history of quantum theory, of puzzles solved as a direct result of the changes to the kinematical framework introduced by quantum mechanics. We close, in Section \ref{4.4}, with the topic of measurement. We conclude that there are yet philosophical puzzles to be resolved concerning the quantum-mechanical account of measurement, though we locate these puzzles elsewhere than is standardly done.

Before moving on to Section \ref{4.2} we want to comment on the interpretation of the distinction between principle-theoretic and constructive approaches that figures prominently within it. The idea of such a distinction dates back to a popular article Einstein published in the London Times \citep{Einstein 1919} shortly after the Eddington-Dyson eclipse expeditions had (practically overnight) turned him into an international celebrity. The distinction Einstein drew there has since taken on a life of its own, both in the historical and in the foundational physics literature. The account of this distinction, which we give in the next section, is meant to more closely reflect the latter literature (especially the literature on quantum foundations). It is not meant to reflect what Einstein intended by the distinction either in 1919 or in his later career.\footnote{For the views of one of us on what Einstein meant by this distinction and how it captures Einstein's own scientific methodology, see \citet[sec.\ 3.5, pp.\ 38--41]{Janssen 2009}, \citet[p.\ 16, pp.\ 26--28]{Janssen and Lehner 2014} and \citet[Ch.\ 3, especially p.\ 102 and pp.\ 119--120]{Duncan and Janssen 2019}.}

The account that we give of the distinction is also different from certain others whose interpretations of quantum mechanics are close to ours on the phylogenetic tree we mentioned in Section \ref{0}. For instance, on our reading of him (based on his unpublished monograph), Bill Demopoulos uses the label ``constructive'' to refer to particular dynamical hypotheses concerning the micro-constituents of matter, and uses the label ``principle-theoretic'' to refer to the specific structural constraints that a theory imposes on the representations it allows. In contrast, our own way of using the label ``constructive'' is broader than this; a constructive characterization may involve the kinematical features of a theory \citep[cf.][]{Janssen 2009}, and a principle-theoretic characterization may include dynamical posits \citep[see][especially Section 12.4]{Koberinski and Mueller 2018}. In the next section we will be speaking about constructive and principle-theoretic \emph{derivations} in particular. What is essential about the former kind of derivation is that it begins from an internal perspective---it is a derivation \emph{from within} quantum theory of some aspect of the world that it describes, while what is essential about the latter kind of derivation is that it begins from an external perspective---it is a derivation \emph{from without} (i.e., from a more general mathematical framework) of some aspect of the quantum world.

Jeff Bub and Itamar Pitowsky also distinguish principle-theoretic from constructive approaches in their \citeyear{Bub and Pitowsky 2010} paper. In their case it is actually not clear to us which of the two senses of the distinction given above is the one they really intend, and at various times they seem to be appealing to both (see especially Section 2 of their paper), although in fairness they appear to do so consistently. This slippage is in any case understandable: The idea that the kinematical core of a theory constrains all of its representations is easily mistaken for the idea that this core constitutes a characterizing principle for the theory. In our own discussion we will endeavor to be careful in distinguishing the former from the latter. But regardless of what one makes of the distinction between constructive and principle-theoretic approaches, we take this distinction to be of relatively minor importance. As we will see further below, the more important distinction to bear in mind when interpreting a physical theory, as one of us pointed out in the context of special relativity, is the distinction between the kinematics and the dynamics of that theory \citep[p.\ 38]{Janssen 2009}.

\subsection{From within and from without}
\label{4.2}

Our derivation of the space of possible quantum correlations in the 2-party, 3-parameter, Mermin-style setup illustrates the interplay between principle-theoretic and constructive approaches that is typical of the actual practice and methodology of theoretical physics \citep[compare, e.g.,][]{Sainz et al 2018}. Our goal was to carve out the space of quantum correlations so as to gain insight into what distinguishes quantum from classical theory. Accordingly, guided by the work of probability theorists and statisticians like De Finetti, Fisher, Pearson and Yule, we associated vectors with random variables and derived a constraint on the angles between such vectors, Eq.\ \eqref{repeat inf the 5}, which has the same form as the constraint on anti-correlation coefficients that characterizes the quantum correlational space of our Mermin-style setup. But it would be wrong to stop here. In and of itself Eq.\ \eqref{repeat inf the 5} is just an abstract equation; it neither explains the space of quantum correlations, nor what distinguishes that space from the corresponding classical space. To gain insight into these matters we needed to model the angle inequality both in quantum theory and in a local-hidden variables model.

In the case of a local-hidden variables model, the classical assumptions that underlie the vectors constrained by Eq.\ \eqref{repeat inf the 5} entail a tighter bound on the correlations between them than what is given by the inequality itself. Specifically, assigning a value to the sum of three variables classically requires that we assign a value to all three of them individually. And because of this, the correlations in a local-hidden variables model cannot saturate the full space described by Eq.\ \eqref{repeat inf the 5}, unless the number $n$ of possible values for a random variable goes to infinity---unless, that is, the range of possible values for a random variable is actually continuous (see Section \ref{2.2.4} for further discussion, as well as note \ref{discrete and contextual}).

In quantum theory, in contrast, this classical presupposition regarding a sum of random variables does not apply. We can indeed still take a sum of three random variables in quantum theory, but we do not need to assign a value to each of them individually in order to do so. As a result, the constraint expressed by the quantum version of the inequality turns out to be tight---quantum correlations, that is, saturate the full volume of the elliptope---regardless of the number of possible values we can assign to the random variables in a particular setup. In this sense Eq.\ \eqref{repeatElliptopeEqn}---a constraint on expectation values---expresses an essential structural aspect of the quantum probability space. Moreover a visual comparison of the quantum elliptope with the various polyhedra we derived for local-hidden variable models vividly demonstrates the way that their respective probabilistic structures differ. This, finally, motivates us to think of quantum mechanics as a theory that is, at its core, \emph{about} probabilities. But this should not be misunderstood. What is being expressed here is the thought that the conceptual \emph{novelty} of quantum theory consists precisely in the way that it departs from the assumptions that underlie classical probability spaces.

One of the strengths of principle-theoretic approaches to physics is that they give us insight into the multi-faceted nature of the objects of a theory.\footnote{See the end of Section \ref{4.1} for a discussion of the way that our characterization of the principle-theoretic and constructive approaches differs from other ways in which they have been characterized in the literature.} A formal framework is set up, for example the $C^*$-algebraic framework of \citet[]{CBH}, one of the minimalist operationalist frameworks of states, transformations and effects discussed in \citet[]{Myrvold 2010}, ``general probabilistic'' frameworks \citep[]{Koberinski and Mueller 2018}, ``informational'' and/or ``computational'' frameworks \citep[]{Chiribella and Ebler 2019}, ``operator tensor'' formulations \citep[]{Hardy 2012} and so on.\footnote{For more on all of these and other related topics, see the collection of essays edited by \citet[]{Chiribella and Spekkens 2016}.} Each such framework focuses on a particular aspect of quantum phenomena, for example on distant quantum correlations, quantum measurement statistics, quantum dynamics and so on. In the language of a given framework one then posits a principle (or a small set of them), e.g., ``no signaling'' \citep[]{Popescu and Rohrlich 1994}, ``no restriction'' \citep[]{Chiribella et al 2010}, ``information causality'' \citep[]{Pawlowski et al 2009} or what have you. These principles carve up the conceptual space of a given framework into those theories that satisfy them with respect to the phenomena considered, and those that do not. The correlations predicted by quantum theory, for instance, satisfy the information causality principle, but any theory that allows correlations above the Tsirelson bound corresponding to the CHSH inequality does not \citep[for discussion, see][]{Cuffaro 2018}.

It may sometimes even be possible to uniquely characterize a theory in a given context---to fix the point in a framework's conceptual space that is occupied by the theory---and if the principles from which such a unique characterization follows are sufficiently compelling in that context, then situating the theory within it adds to our understanding both of the theory and of the phenomena described by it \citep[cf., e.g.,][]{Masanes et al 2019}.\footnote{One of us has expressed previously in print the contention that only constructive approaches to physics can yield explanatory content \citep[]{Balashov and Janssen 2003, Janssen 2009}. All three of us are now of the opinion that both principle-theoretic and constructive approaches can be explanatory.} We are not, of course, claiming that this or that abstract characterizing principle exhausts all that there is to say about quantum theory. But by situating quantum theory within the abstract space provided by a formal framework we subject it to a kind of ``theoretical experiment''. Just as with an actual experiment, which we set up to determine this or that property of a physical system, in the course of which we control (i.e., in our lab) parameters that we deem irrelevant to or that interfere with our determination of the particular property of interest, in our theoretical experiments we likewise abstract away from features of quantum theory that are irrelevant to or obfuscate our characterization of it \emph{as a theory of information processing of a particular sort}, or \emph{as a particular kind of $C^*$ algebra}, or \emph{as a theory of probabilities} and so on. Quantum theory can be thought of as each of these things. Insofar as it occupies a particular position (or region) within the conceptual space of these respective frameworks, it can be characterized from each of these points of view. And within each perspective within which it can be so characterized, there are constraints on what a quantum system can be from that perspective. It is these constraints which our theoretical experiments set out to discover. And it is these constraints which convey to us information about what quantum theory \emph{is} and how the systems it describes \emph{actually behave} under that mode of description.

The value of the principle-theoretic approach is, moreover, not limited to this descriptive role. Principle-theoretic approaches are also instrumental for the purposes of theory development. For instance in the course of setting up a conceptual framework in which to situate quantum theory, we might consider it more natural to relax rather than maintain one or more of the principles that characterize quantum theory in that framework (cf. \citealt[]{Hardy 2007}). In this way we feel our way forward to new physics. Even, that is, if we do not expect that they themselves will constitute new physical theories, the formal frameworks we set up enable forward theoretical progress by helping us to grasp the descriptive limits of our existing theories and to get a sense of what we may find beyond them.

And yet, earlier we stated our conviction that, ``at its core'', quantum mechanics is fundamentally \emph{about} probabilities. How can this univocal statement be consistent with the claim we have just been making regarding the essentially perspectival nature of the insights obtainable through a principle-theoretic approach? In fact it would be wrong to describe the interpretation of quantum theory that we have been advancing in this paper as a principle-theoretic one.\footnote{Ours is not a principle-theoretic interpretation on the way that we have expounded that term here. As discussed at the end of Section \ref{4.1}, our own usage of the term is intended to reflect its usage in the contemporary literature on quantum foundations. Our interpretation could, though, be seen as a principle-theoretic one in the sense in which (for instance) Bill Demopoulos uses that term.} As we have described them above, principle-theoretic approaches offer perspectives on quantum theory (or on some aspect of it) that are essentially external: One first sets up a formal framework which in itself has little to do with quantum theory; next one seeks to motivate and define a principle or set of them with which to pinpoint quantum theory within that framework. But what does it mean to pinpoint quantum theory within a framework? Generally this means matching the set of phenomena circumscribed by the principle(s) with the set of phenomena predicted by quantum mechanics, i.e., with those obtained via a derivation \emph{from within}. In this way one tests that the set of phenomena captured by a set of principles really is the one predicted by quantum mechanics, that these characterizing principles really do constitute a perspective on the theory. A principle-theoretic approach to understanding quantum theory, therefore, is not wholly external. But on the approach just outlined the internal perspective only becomes relevant at the end of the procedure, as a way to gauge the success of one's theoretical experiment.

For us this latter step was very far from trivial. Indeed it was only through it that we were able to gain full insight into the aspect of quantum phenomena that we were seeking to understand. To recapitulate: We first set up a generalized framework for characterizing correlations and within this framework we considered the angle inequality relating correlation coefficients for linear combinations of random variables expressed by Eq.\ \eqref{repeat inf the 5}. We thus began our derivation \emph{from without}. We then asked whether one could view this as an expression of the fundamental nature of the correlations between random variables in either a local-hidden variables model such as our raffles, or in a quantum model. That is, we asked whether the correlations in either case saturate the elliptope described by Eq.\ \eqref{repeat inf the 5}. To answer this question we then took a constructive step in both cases: We gave both a local-hidden variables and a quantum model for the general constraint expressed by Eq.\ \eqref{repeat inf the 5}. And by proceeding in this way \emph{from within} both frameworks we were able to show that, as a consequence of the assumptions underlying the framework of classical probability theory, the angle inequality and its corresponding elliptope cannot be seen as a fundamental expression of the nature of correlations in a local-hidden variables model, for there are further constraints that need to be satisfied in such a model in order to saturate the elliptope. As for the mathematical framework of quantum theory, we saw how it is able to succeed, where a local-hidden variables model cannot, in entirely filling up the elliptope. Finally by considering \emph{how} it is capable of doing this we are able to understand what the essential distinction between quantum and classical theory is.

\subsection{The new kinematics of quantum theory}
\label{4.3}

What, then, is the essential distinction between quantum and classical theory? In the end we saw that the key assumption we needed to derive the quantum version of the angle inequality is one which follows straightforwardly from the Hilbert space formalism of quantum mechanics. The Hilbert space formalism, however, applies universally to all quantum systems. Our case studies were limited to a relatively small number of particular experimental setups---the Mermin-inspired setups we considered in Sections \ref{1} and \ref{2} and the CHSH-like setup we considered in Section \ref{3}. They were also limited in terms of the quantum states measured in those setups. But we see now that the wider significance of our analyses of these case studies is not likewise limited. For the key feature of the quantum formalism that these special but informative case studies point us to is in fact a fully general one; it expresses quantum theory's kinematical core \citep[cf.][Section 1.2]{Janssen 2009}.

As mentioned in Section \ref{0}, our interpretation of quantum theory owes much to the work of Jeff Bub, Bill Demopoulos, Itamar Pitowsky and others who have proceeded from similar motivations. In their \citeyear{Bub and Pitowsky 2010} paper, Bub and Pitowsky characterize their interpretation of quantum theory as both principle- and information-theoretic (pp. 445--446), arguing both that the Hilbert space structure of quantum theory is derivable on the basis of information-theoretic constraints, and that quantum theory should in this sense be thought of as being all about information \citep[cf.][]{Bub 2004}. Interpreted as some sort of ontological claim, the latter is surely false. If, instead, one interprets this as a claim about where the conceptual novelty of quantum theory is located \citep[cf.][]{Demopoulos 2018}, namely in the structural features of its kinematical core, then we take this claim to be correct, even if we prefer to speak of \emph{probability} rather than of information (see Section \ref{0}).

There is a common viewpoint on interpretation that holds that what it means to interpret a theory is to ask the question: ``What would the world be like (in a representational sense) if the theory were literally true of it?'' \citep[cf., for instance,][p. 153]{Caulton 2015}. We reject this as exhaustive of what it means to interpret a theory \citep[cf. also][]{Curiel 2019}, and rather affirm that often the more interesting interpretational question is the one which asks what the world must be like (not necessarily in a representational sense) in order for a given theory to be of use to us; i.e., to be effective in describing and structuring our experience and in enabling us to speak objectively about it to one another. Note, on the one hand, the realist commitment implicit in this question. But note, on the other hand, that the question does not presuppose the literal or even the approximately literal truth of the theory being considered. For even classical mechanics, superseded as it has been by quantum mechanics, is of use to us in this sense. And it is a meaningful question to ask how this constrains our possible conceptions of the world.

Such a question can be answered in a number of ways. One might begin, for instance, by positing a priori constraints on what an underlying ontological picture of the world must be like in a general sense, e.g., that it must be some kind of particle ontology \citep[cf.][]{Albert 2018}. The descriptive success of quantum mechanics (and, correspondingly, the descriptive failure of classical mechanics) would then entail a number of constraints on this general ontological picture, in particular that it must be fundamentally non-local. Alternately (i.e., rather than positing a general ontological picture of the world a priori) one might choose instead to focus more directly on the relation between the formalism of the theory and the phenomena it describes. What aspect of the formalism, one might ask from this point of view, is key to enabling quantum theory to be successful in describing phenomena and coordinating our experience, and what does that tell us about the world? A natural way of illuminating this question is to compare quantum with classical modes of description---to consider what is \emph{novel} in the quantum as compared with the classical mode of description---and to consider how this allows quantum theory to succeed where classical theory cannot. We take the investigations in the prior sections of this paper to have shown that this novel content can be located in the kinematical core of quantum theory, in the structural constraints that quantum theory places on our representations of the physical systems it describes.

In classical mechanics, an observable $A$ is represented by a function on the phase space of a physical system: $A = f(p, q)$ where $p$ and $q$ are the system's momentum and position coordinates within its phase space. Points in this space can be thought of as ``truthmakers'' \citep[]{Bub and Pitowsky 2010} for the occurrence or non-occurrence of events related to the system in the sense that specifying a particular $p$ and $q$ fixes the values assigned to every observable defined over the system in question. With each observable $A$ one can associate a Boolean algebra representing the possible yes or no questions that can be asked concerning that observable in relation to the system. And because one can simultaneously assign values to every observable given the state specification $(\mathbf{q}, \mathbf{p}) \equiv ((x_i, y_i, z_i), (p_{x_i}, p_{y_i}, p_{z_i}))$, one can embed the Boolean algebras corresponding to each of them within a global Boolean algebra that is the union of them all. In general there is no reason to think of observables as representing the properties of a physical system within this framework. But because we can fix the value of every observable associated with the system \emph{in advance} given a specification of the system's state---because the union of the Boolean algebras corresponding to these observables is itself representable as a Boolean algebra---it is in this case conceptually unproblematic to treat these observables as though they do represent the properties of the system, properties that are possessed by that system irrespective of how we interact (or not) with it.

In quantum mechanics an observable, $A$, is represented by a Hermitian operator, $\hat{A}$ (whose spectrum can be discrete, continuous or a combination of both) acting on the Hilbert space associated with a physical system, with the possible values for $A$ given by the eigenvalues of $\hat{A}$: $\{a: \hat{A}| \psi \rangle = a| \psi \rangle\}$. Unlike the case in classical mechanics, the quantum state specification for a physical system, $| \psi \rangle$, cannot be thought of as the truthmaker for the occurrence or non-occurrence of events related to it, for specifying the state of a system at given moment in time does not fix in advance the values taken on at that time by every observable associated with the system. First, the state specification of a system yields, in general, only the probability that a given observable associated with it will take on a particular value when selected. Second, and more importantly, the Boolean algebras corresponding to the observables associated with the system cannot be embedded into a larger Boolean algebra comprising them all. Thus one can only say that \emph{conditional} upon the selection of the observable $A$, there will be a particular probability for that observable to take on a particular value. At the same time no one of the individual Boolean sub-algebras of this larger non-Boolean structure yields what would be regarded, from a classical point of view, as a complete characterization of the properties of the system in question. As we will see in Section \ref{4.4}, this does not preclude a different kind of completeness from being ascribed to the quantum description of a system. But because our characterization is not \emph{classically} complete, it is no longer unproblematic to take the observable $A$ as a stand-in for one of the underlying properties of the system, even in the case where quantum mechanics predicts a particular value with certainty conditional upon a particular measurement.

To put it a different way: Because classical-mechanical observables can be set down in advance, irrespective of the nature of the interaction with the system from which they result, they can straightforwardly be taken to represent ``beables'' \citep[see][sec.\ 2]{Bell 1984} with respect to a given state specification. Quantum-mechanical observables cannot be, or at any rate there can be no direct, unproblematic, inference from observable to beable within quantum theory---something more, some further argument must be given. As for us, we have yet to see a convincing argument to this effect. We rather take quantum theory to be telling us that there can be no ground in the classical sense of a fully determinate globally Boolean noncontextual assignment of values to all of the observables relevant to a given system \citep[cf.][Section 9]{Pitowsky 1994}.

In the context of space-time theories, Minkowski space-time encodes generic constraints on the space-time configurations allowed by any specific relativistic theory compatible with its kinematics. These constraints are satisfied as long as all of the observables are represented by mathematical objects that transform as tensors (or spinors) under Lorentz transformations. Analogously, in quantum mechanics, Hilbert space encodes generic constraints on the possible values of observables as well as on the correlations between such values that are allowed within any specific quantum theory compatible with its kinematics. These constraints are satisfied as long as all of the observables are represented by Hermitian operators acting on Hilbert space. In the case of Minkowski space-time, the determination of the particular tensor (or spinor) representative of a given transformation is the province of the dynamics, not the kinematics, of the specific relativistic theory in question. Likewise, determining the particular self-adjoint operator representative of a given action on a system is a province of the dynamics, not the kinematics, of the specific quantum theory in question.

Just as in special relativity, the kinematical part of quantum theory is a comparatively small one. The lion's share (and more) of the practice of quantum theory is concerned with determining the dynamical aspects of particular systems of interest. And yet, conceptually, the kinematics of quantum theory may justifiably be regarded as its most important part; it constitutes the ``operating system'' upon which the dynamics of particular physical systems can be seen as ``applications'' being run \citep[p. 2]{Nielsen and Chuang 2016}.

\subsection{Examples of problems solved by the new kinematics}
\label{4.3a}

As in the transition from 19th-century ether theory to special relativity \citep[see][]{Janssen 2009}, one can find examples in the transition from the old to the new quantum theory of puzzles
solved as a direct result of changes in the basic kinematical framework. Unsurprisingly, given our characterization of the ``big discoveries'' of Heisenberg and Schr\"odinger in Section \ref{0}, these examples are easier to come by in the early history of matrix mechanics than in the early history of wave mechanics, but they can be found in both.

The basic idea of the paper with which \citet{Heisenberg 1925} laid the foundation of matrix mechanics was not to repeal the laws of classical mechanics but to reinterpret them \citep[p.\ 139]{Janssen 2019}. This is clearly expressed in the title of the paper: ``Quantum-theoretical reinterpretation (\emph{Umdeutung}) of kinematical and mechanical relations.'' Heisenberg replaced the real numbers $p$ and $q$ by non-commuting arrays of numbers soon to be recognized as matrices and then as operators. These operators, $\hat{p}$ and $\hat{q}$, satisfy the same relations as $p$ and $q$ (e.g., the functional dependence of the Hamiltonian on these variables will remain the same) but they are subject to the commutation relation, $[\hat{q} \, , \, \hat{p}] = i\hbar$, the quantum analogue, as \citet{Dirac 1926} realized early on, of Poisson brackets in classical mechanics.

In the final section of the \emph{Dreim\"annerarbeit}, the joint effort of Max Born, Werner Heisenberg and Pascual Jordan that consolidated matrix mechanics, the authors (or rather Jordan who was responsible for this part of the paper) showed that the new formalism automatically yields both terms of a famous formula for energy fluctuations in black-body radiation \citep[pp.\ 375--385]{dreimaenner}.\footnote{For a detailed reconstruction of Jordan's argument, see \citet{Duncan and Janssen 2008}. The ensuing debate over this reconstruction \citep[see, especially,][]{BCM 2017} does not, as far as we can tell, affect our use of this example in the present context.} \citet{Einstein 1909a, Einstein 1909b} had derived this formula from little more than the connection between entropy and probability expressed in the formula $S = k \ln{W}$ carved into Boltzmann's tombstone and Planck's law for black-body radiation. One of its two terms suggested waves, the other particles. Einstein had argued in 1909 that the latter called for a modification of Maxwell's equations \citep[pp.\ 120--126]{Duncan and Janssen 2019}. He had contemplated such drastic measures before when faced with the tension between Maxwell's equations and the relativity principle. The new kinematics of special relativity had resolved that tension. Jordan now showed that the tension between Maxwell's equations and Einstein's fluctuation formula could also be resolved by a change in the kinematics.

Instead of a cavity with electromagnetic waves obeying Maxwell's equations, Jordan considered a simple model, due to Paul \citet{Ehrenfest 1925}, of waves in a string fixed at both ends. This string can be replaced by an infinite number of uncoupled harmonic oscillators. Quantizing those oscillators, using the basic commutation relation $[\hat{q} \, , \, \hat{p}] = i\hbar$, and calculating the fluctuation of the energy in a small segment of the string in a narrow frequency interval, Jordan recovered both the wave and the particle term of Einstein's formula. Using classical kinematics, one only finds the wave term. As Jordan concluded:  
\begin{quote}
The reasons for the occurrence of a term not delivered by the classical theory are obviously closely related to the reasons for the occurrence of the zero-point energy [of the harmonic oscillator, which itself follows directly from the commutation relation for position and momentum]. In both cases, the basic difference between the theory attempted here and the one attempted so far  [i.e., classical theory with the restrictions imposed on it in the old quantum theory] \emph{lies not in a disparity of the mechanical laws but in the kinematics characteristic for this theory}. One could even see in [this fluctuation formula], into which no mechanical principles whatsoever even enter, one of the most striking examples of the difference between quantum-theoretical kinematics and the one used hitherto \citep[p.\ 385; our emphasis and our translation, quoted in part in Janssen, 2009, p.\ 50]{dreimaenner}.
\end{quote}

Our second example turns on the quantum-mechanical treatment of orbital angular momentum, which proceeds along the exact same lines as the treatment of intrinsic or spin angular momentum underlying the quantum-mechanical analysis of the experiments we have been studying in Sections \ref{1}--\ref{3}. We already alluded to this example at the end of Section \ref{1.6.2}. It is the problem of the electric susceptibility of diatomic gases such as hydrogen chloride.\footnote{For a detailed analysis of this episode, see \citet{Midwinter and Janssen 2013}.} One of the two terms in the so-called Langevin-Debye formula for this quantity comes from the alignment of the molecule's permanent dipole with the external field. This term decreases with increasing temperature as the thermal motion of these dipoles frustrates their alignment. This makes it at least intuitively plausible that only the lowest energy states of the molecule contribute to the susceptibility. This is indeed what the classical theory predicts. In the old quantum theory, however, this feature was lost. This is a direct consequence of the way in which angular momentum was quantized. The length $L$ of the angular momentum vector could only take on values $l \hbar$  in the old quantum theory, where $l$ is an integer greater than 1. The value $l=0$ was ruled out for the same reason that it was ruled out for the hydrogen atom: an orbit with zero angular momentum would have to be a straight line going back and forth through the nucleus! Hence $l \ge 1$ for all states contributing to the susceptibility. This led to the strange situation, as \citet[p.\ 325]{Pauli 1921} noted in one of his early papers, that there are ``{\it only such orbits present that according to the classical theory do not give a sizable contribution to the electrical polarization}'' (emphasis in the original). Fortunately, the allowed orbits (or energy states) with $l \ge 1$ do give a sizable contribution. Unfortunately, this contribution is almost five times too large.

The quantization of angular momentum in the new quantum mechanics was worked out in the \emph{Dreim\"annerarbeit} mentioned above \citep[pp.\ 364--374]{dreimaenner}. The upshot was that the correct quantization of angular momentum leads to the eigenvalues $l(l+1)$ for $\hat{L}^2$, where the allowed integer values of $l$ start at 0 rather than 1 (cf.\ Eq.\ \eqref{state dfn} in Section \ref{2.1}). This new quantization rule for angular momentum follows directly from the basic commutation relation for position and momentum.

Pauli and his former student Lucy Mensing showed how this new quantization rule solved the puzzle of the electric susceptibility of diatomic gases. As in classical theory, only the lowest ($l=0$) state contributes to the susceptibility, the contributions of all other terms sum to zero (and this depends delicately on the exact quantization rule). As \citet[p.\ 512]{Mensing and Pauli 1926} noted with palpable relief: ``{\it Only the molecules in the lowest state will therefore give a contribution to the temperature-dependent part of the dielectric constant}" (emphasis, once again, in the original). The new quantum theory thus reverted to the classical theory in this respect. In a note to \emph{Nature} on the topic, Van Vleck made the same point: ``The remarkable result is obtained that only molecules in the state of lowest rotational energy make a contribution to the polarisation. This corresponds very beautifully to the fact that in the classical theory only molecules with [the lowest energy] contribute to the polarisation" \citep[p.\ 227]{Van Vleck 1926}.

Van Vleck expanded on this comment when interviewed in 1963 by his former PhD student Thomas S.\ Kuhn for the Archive for History of Quantum Physics (AHQP):
\begin{quote}
I showed that [the Langevin-Debye formula for susceptibilities] got restored in quantum mechanics, whereas in the old quantum theory, it had all kinds of horrible oscillations \ldots\  you got some wonderful nonsense, whereas it made sense with the new quantum mechanics. I think that was one of the strong arguments for quantum mechanics. One always thinks of its effect and successes in connection with spectroscopy, but I remember Niels Bohr saying that one of the great arguments for quantum mechanics was its success in these non-spectroscopic things such as magnetic and electric susceptibilities.\footnote{Cf.\ the opening sentence of the preface of his classic text on magnetic and electric susceptibilities \citep{Van Vleck 1932} quoted in note \ref{Van Vleck} in Section \ref{1.6.2}, the book that earned him the informal title of ``father of modern magnetism'' \citep[p.\ 139]{Midwinter and Janssen 2013}.}
\end{quote}
Van Vleck was so taken with this result that it features prominently in his Nobel lecture in 1977 \citep[p.\ 138]{Midwinter and Janssen 2013}. The important point for our purpose is that this is another example of a problem that was solved by a change in the kinematics rather than the dynamics.

The two examples given so far both turned on the commutation relation  $[\hat{q} \, , \, \hat{p}] = i\hbar$ at the heart of matrix mechanics. Our third and last example turns on a key feature of wave mechanics. As we noted in Section \ref{0}, Schr\"odinger, unlike Heisenberg, may not have emphasized that his new theory provided a new framework for doing physics but  this is, of course,  as true for wave mechanics as it is for matrix mechanics. An obvious example of a change in the basic framework for doing physics that emerged from the development of wave mechanics rather than matrix mechanics is the introduction of quantum statistics, especially Bose-Einstein statistics, which preceded the formulation of wave mechanics. We close this subsection with a less obvious but informative example.\footnote{For a detailed analysis of this episode, see \citet[Sec.\ 6.3 and Appendix A]{Duncan and Janssen 2014, Duncan and Janssen 2015, Duncan and Janssen 2019}.}

In the same year that saw the appearance of Bohr's atomic model, Johannes \citet{Stark 1913} discovered the effect named after him, the splitting of spectral lines due to an external electric field, the analogue to the effect discovered by Pieter Zeeman in 1896, the splitting of spectral lines due to a magnetic field. It was not until two key contributions---one by a physicist, Arnold Sommerfeld, in late 1915; one by an astronomer, Karl Schwarzschild, in early 1916---that there was any hope of accounting for the Stark effect on the basis of the old quantum theory, the extension, mainly due to Sommerfeld, of Bohr's original ideas. Sommerfeld's key contribution to the explanation of the Stark effect was to introduce (even though he did not call it that) \emph{degeneracy}, the notion that the same energy level can be obtained with different combinations of quantum numbers. External fields will lift this degeneracy and result in a splitting of the spectral lines associated with transitions between these energy levels. Schwarzschild's key contribution was to bring the advanced techniques developed in celestial mechanics to bear on the analysis of the miniature planetary systems representing atoms in the old quantum theory.  Once those two ingredients were available, \citet{Schwarzschild 1916} and Paul \citet{Epstein 1916}, an associate of Sommerfeld, quickly and virtually simultaneously derived formulas for the line splittings in the Stark effect in hydrogen that were in excellent agreement with the experimental data.  

Even though some energy states and some transitions between them had to be ruled out rather arbitrarily and even though there was no convincing explanation for the polarizations and relative intensities of the components into which the Stark effect split the spectral lines, this was seen as a tremendous success for the old quantum theory. As Sommerfeld exulted in the conclusion of the first edition of \emph{Atombau und Spektrallinien} (Atomic structure and spectral lines), which became known as the ``the bible of atomic theory'' \citep[pp. 255--256]{Eckert 2013}: ``the theory of the Zeeman effect and especially the theory of the Stark effect belong to the most impressive achievements of our field and form a beautiful capstone on the edifice of atomic physics'' \citep[pp.\ 457--458]{Sommerfeld 1919}.

Even in the case of the Stark effect (to say nothing of the Zeeman effect), Sommerfeld's jubilation would prove to be premature. In addition to the limitations mentioned above, there was a more subtle but insidious difficulty with Schwarzschild and Epstein's result. To find the line splittings of the Stark effect, they had to solve the so-called Hamilton-Jacobi equation, familiar from celestial mechanics, for the motion of an electron around the nucleus of a hydrogen atom immersed in an external electric field. This could only be done in coordinates in which the Hamilton-Jacobi equation for this problem is \emph{separable}, i.e., in coordinates in which the equation splits into three separate equations, one for each of the three degrees of freedom of the electron. Similar problems in celestial mechanics made it clear that they needed so-called parabolic coordinates for this purpose. These then also were the coordinates in which Schwarzschild and Epstein imposed the quantum conditions to select a subset of the orbits allowed classically. As long as there is no external electric field, it was much simpler to do the whole calculation in polar coordinates. Letting the strength of the external field go to zero, one would expect that  the quantized orbits found in parabolic coordinates reduce to those found in polar coordinates. This turns out not to be the case. The energy levels are the same in both cases but the orbits are not. Both Sommerfeld and Epstein recognized that this is a problem (Schwarzschild died the day his paper appeared in the proceedings of the Berlin academy). As \citet[p.\ 507]{Epstein 1916} put it:
\begin{quote}
Even though this does not lead to any shifts in the line series, the notion that a preferred direction introduced by an external field, no matter how small, should drastically alter the form and orientation of stationary orbits seems to me to be unacceptable \citep[quoted in][p.\ 251]{Duncan and Janssen 2015}.
\end{quote}    
The old quantum theory simply did not have the resources to tackle this problem and nothing was done about it.

The Stark effect in hydrogen was one of the first applications of Schr\"odinger's new wave mechanics. The calculation is actually very similar to the one in the old quantum theory. This is no coincidence. An important inspiration for Schr\"odinger's wave mechanics was Hamilton's optical-mechanical analogy \citep{Joas and Lehner 2009}. So it is not terribly surprising that Hamilton-Jacobi theory informed the formalism Schr\"odinger came up with. The time-independent Schr\"odinger equation was actually modeled on the Hamilton-Jacobi equation. The  time-independent Schr\"odinger equation for an electron in a hydrogen atom in an external electric field is, once again, most easily solved in parabolic coordinates. Independently of one another, \citet{Schroedinger 1926} and \citet{Epstein 1926} did this calculation shortly after wave mechanics arrived on the scene. To first order in the strength of the electric field, this calculation yields the same splittings as the old quantum theory. However, as both Schr\"odinger and Epstein emphasized, no additional restrictions on states or transitions between states are necessary and the theory also correctly predicts the polarizations and intensities of the various Stark components. What Schr\"odinger and Epstein did not mention, however, was that wave mechanics also solves the problem of the non-uniqueness of orbits of the old quantum theory. Although physicists at the time lacked the mathematical tools to express this---and the problem, it seems, quickly got lost in the waves of excitement about the new theory\footnote{In the case of special relativity, it also took some time for physicists to recognize that some puzzles had been resolved by the new kinematics. In the case of the Trouton-Noble experiment, \citet{Butler 1968} was the first to show that the torque on a moving capacitor that the experimenters had been looking for in 1903 was nothing but an artifact of how one slices Minkowski space-time when defining the momentum and angular momentum of spatially extended systems \citep[p.\ 45; see also Teukolsky, 1996]{Janssen 2009}.}---the problematic non-uniqueness of orbits in the old quantum theory turns into the completely innocuous non-uniqueness of bases of wave functions in an instantiation of Hilbert space \citep[sec.\ 5, pp.\ 76--77]{Duncan and Janssen 2014}.

Both Heisenberg and Schr\"odinger recognized the problematic nature of the old quantum theory's electron orbits, which had been imported from celestial mechanics along with the mathematical machinery to analyze atomic structure and atomic spectra \citep[p.\ 171]{Janssen 2019}. An area in which the trouble with orbits had become glaringly obvious by the early 1920s was optical dispersion, the study of the dependence of the index of refraction on the frequency of the refracted light. Heisenberg's \emph{Umdeutung} paper builds on a paper he co-authored with Hans Kramers, Bohr's right-hand man in Copenhagen, on Kramers's new quantum theory of dispersion \citep{Kramers and Heisenberg 1925}. Taking his cue from this theory, Heisenberg steered clear of orbits altogether in his \emph{Umdeutung} paper and focused instead on observable quantities such as frequencies and intensities of spectral lines \citep[pp.\ 134--142]{Duncan and Janssen 2007, Janssen 2009}. The quantities with which he replaced position and momentum were not, in his original scheme, themselves observable. Instead they functioned as auxiliary quantities that allowed him to calculate the values of (indirectly) observable quantities such as energy levels and transition probabilities. Schr\"odinger did not get rid of orbits as radically as Heisenberg. His wave functions can be seen as a new way to characterize atomic orbits once we have come to recognize that they are the manifestation of an underlying wave phenomenon. Comparing these different responses to the trouble with orbits in the old quantum theory, we see the beginnings of the two main lineages of the genealogy we proposed in Section \ref{0} to classify different interpretations of quantum mechanics.

\subsection{Measurement}
\label{4.4}

Consider a measurement device that has been set up to assess the spin state of an ensemble of electrons that has been prepared in a particular way. For instance, imagine we have prepared a uniform ensemble of electrons in the superposition state (cf. Section \ref{1.5}):
\begin{align}
  \label{eqn:pm-superpos}
  | \psi \rangle = \alpha | + \rangle_z \, + \, \beta | - \rangle_z.
\end{align}
We direct the electrons one at a time toward the device, which we have prepared so that it will measure their spin in the $z$-direction. We observe the results of our experiment and see that in each case, the spin state of the electron is recorded as having a definite value of either up ($+$) or down ($-$) along the $z$-axis, and further that the distribution of results is such that an electron's spin is recorded as up with a relative frequency that tends toward $|\alpha|^2$ and as down with a relative frequency that tends toward $|\beta|^2$. What is the explanation?

Here is an attempt. The quantum-mechanical state description assigns a probability to the outcome of a measurement that is given (in the case of a projective measurement in the $z$-basis) by:
\begin{align}
\label{eqn:outcome-prob}
\mathrm{Pr}(m|\hat{z}) = \, _{z\!}\langle \psi | \hat{P}_m | \psi \rangle_{\!z},
\end{align}
where
\begin{align}
\label{eqn:projection}
\hat{P}_m \equiv |m\rangle_{\!z} \, _{z\!}\langle m|
\end{align}
is the projection operator corresponding to the outcome $m$. For the current example involving a uniform ensemble of electrons in the state given by Eq.\ \eqref{eqn:pm-superpos} this entails that:
\begin{align}
  \mathrm{Pr}(+| \hat{z}) & = \Big(\alpha^*\,_{z\!}\langle + | \,+\, \beta^*\,_{z\!}\langle - |\Big)\Big(| + \rangle_{\!z}\,_{z\!}\langle + | \Big) \Big(\alpha| + \rangle_{\!z} \,+\, \beta| - \rangle_{\!z}\Big) \nonumber \\[.3cm]
  & = \alpha^*\,_{z\!}\langle + |\Big(\alpha| + \rangle_{\!z} \,+\, \beta| - \rangle_{\!z}\Big) = \alpha^*\alpha = |\alpha|^2,
\end{align}
and similarly for $\mathrm{Pr}(-| \hat{z})$. This agrees with the statistics actually observed. In the more general case of a non-uniform ensemble described by the \emph{density operator}\footnote{As we noted in Section \ref{0}, density operators were first introduced by \citet{von Neumann 1927b}.}
\begin{align}
  \label{eqn:densityop}
  \hat{\rho} = \sum_i| \psi \rangle_{\!i} \, _{i\!}\langle \psi |,
\end{align}
the probability of the outcome $m$, in the case of a projective measurement in the $z$-basis, is given by
\begin{align}
  \label{eqn:probdensityop}
  \mathrm{Pr}(m|\hat{z}) = \mbox{Tr}(\hat{\rho} \hat{P}_m).
\end{align}
Gleason's theorem \citeyearpar{Gleason 1957} tells us that quantum mechanics' assignment of probabilities is \emph{complete} in the sense that every probability measure on the Boolean sub-algebras associated with the observables of a system is representable by means of a density operator in the manner just described.\footnote{Gleason's proof assumes that measurements are represented as projections and is valid for Hilbert spaces of dimension $\geq 3$. \citet[]{Busch 2003} proves an analogous result for the more general class of positive operator valued measures (POVMs, or ``effects'') which is valid for Hilbert spaces of dimension $\geq 2$. An extended discussion of the issue of completeness in relation to Gleason's theorem may be found in \citet[]{Demopoulos 2018}.}

The account of a quantum-mechanical measurement given above will be criticized. What has been given, it will be maintained, is merely a recipe for recovering the statistics associated with such a measurement. All we learn from this recipe is that, and how, the quantum formalism may be used to calculate the probabilities that will be observed upon interacting the system of interest with a device we set up to measure one of its dynamical parameters. No account has been given here of how the measurement interaction itself allows for this, however. And this is what is demanded by our objector.

Now consider again a measurement in the $z$-basis on an electron that is part of a uniform ensemble of systems prepared in the state described by Eq.\ \eqref{eqn:pm-superpos}. This state description is non-classical. However given such a measurement one knows---even before it has interacted with the measurement device---that conditional upon that measurement, we can consider each electron as a member, not of the uniform ensemble that has actually been prepared, but rather of a non-uniform ensemble whose relative proportion of systems in the states $| + \rangle_{z}$ and $| - \rangle_{z}$ is $|\alpha|^2$ and $|\beta|^2$, respectively. That is, conditional upon a $z$-basis measurement, the observed statistics will not be distinguishable from those that would be observed from a $z$-basis measurement on an ensemble characterized by the density operator for the mixed state
\begin{align}
\label{eqn:mixedstate}
\hat{\rho} = |\alpha|^2| + \rangle_{\!z} \, _{z\!}\langle + | ~+~ |\beta|^2| - \rangle_{\!z} \, _{z\!}\langle - |.
\end{align}
Because of this we can simulate the observed statistics, conditional upon such a measurement, with a local-hidden variables model similar to the raffles we used in the previous sections of this paper. Unlike those raffles, the phenomena we are simulating here are not correlations, thus our tickets will not need to have two halves like the ones depicted in Figure \ref{raffle-tickets-3set2out-i-thru-iv}. In the current scenario we can make do with a basket of raffle tickets inscribed with a single symbol, either ``$|+\rangle$'' or ``$|-\rangle$'', whose relative proportions in the basket are $|\alpha|^2$ and $|\beta|^2$, respectively. Thus, we have here an account of how, through measuring a system in a given basis, our characterization of the system transitions from a quantum to an effectively classical description. Moreover if one repeats this procedure sufficiently many times, for measurements in the $z$ and possibly also in other measurement bases, one can convince oneself that the statistics yielded by these measurements accord with one's expectations given the initial quantum-mechanical description of the ensemble, i.e., the description of it as a uniform ensemble of electrons in the state given by Eq.\ \eqref{eqn:pm-superpos}.

Again this will be criticized. This explanation of the measurement process, it will be objected, is no explanation at all. Our measurement seems almost magical on the account just given, a black box whose inner workings we do not grasp. But it is a goal of physical inquiry to open all such boxes, and it will be demanded of us that we open this one as well.

In the present instance this demand is completely legitimate, for so far we have told you nothing of the details of the measurement interaction outlined above. Obviously, though, there are many good reasons to want to be informed of such details. If the measurement statistics do not accord with our expectations, for instance, we will want to examine the inner workings of the measurement device in more detail to see whether it is functioning properly. Even when we have full confidence in a particular device, we might still want information about its inner workings so that we can reproduce the experiment in another physical location with different equipment. Or maybe we simply want to understand its inner workings for understanding's sake. These are all legitimate reasons to demand a deeper explanation of the measurement interaction described above. And within quantum theory it is always possible to give you such a description, i.e., to describe how a particular measurement device dynamically interacts with a given system of interest, gives rise to an entangled state of the system of interest and apparatus, and yields probabilities for the state of the measuring device that will be found upon its being assessed.

To come back to our running example: Rather than considering our system of interest to be a member of a uniform ensemble of electrons in the state given by Eq.\ \eqref{eqn:pm-superpos}, one can instead describe our system of interest as a member of a uniform ensemble of composite systems, where the state of each member is describable by the entangled superposition:
\begin{align}
  \label{eqn:compound}
  \alpha |+ \rangle_{Mz} |+ \rangle_{Sz} \, + \; |-  \rangle_{Mz} | - \rangle_{Sz}
\end{align}
with $| + \rangle_{Sz}$, $|- \rangle_{Sz}$ (where $S$ stands for the system of interest) representing the two possible spin-$z$ states of the electron, and $| + \rangle_{Mz}$, $| - \rangle_{Mz}$ representing the corresponding two possible magnetic field orientations of the DuBois magnets used in the apparatus. In this way we move back ``the cut'' \citep[see][sec. 10.4]{Bub 2016}: the dividing line between, on the one hand, our quantum description of the system we are measuring, and on the other hand, our description of the instrument we are using to assess that system's state. That part of the measurement phenomenon which, on our earlier analysis, was the instrument of measurement is now, on this more detailed analysis, part of the (quantum) system measured. But as before, if one considers measuring this system in (for instance) the basis
\begin{align}
  \label{eqn:zzbasis}
  \mathcal{B}_{zz} \equiv \{|+\rangle_{Mz}|+\rangle_{Sz},~ |+\rangle_{Mz}|-\rangle_{Sz},~ |-\rangle_{Mz}|+\rangle_{Sz},~ |-\rangle_{Mz}|-\rangle_{Sz}\},
\end{align}
one can treat the expected statistics, conditional upon that choice of basis, as arising from measurements on a \emph{non}-uniform ensemble of composite systems for which the proportion of such systems in the state $| + \rangle_{Mz}| + \rangle_{Sz}$ is $|\alpha|^2$ and the proportion of such systems in the state $| - \rangle_{Mz}| - \rangle_{Sz}$ is $|\beta|^2$. Similarly to before, one can simulate these statistics using a raffle with tickets marked as either ``$|+\rangle|+\rangle$'' or ``$|-\rangle|-\rangle$'', with proportions $|\alpha|^2$ and $|\beta|^2$, respectively. In other words we see, in more detail now, how the measurement interaction gives rise to an effectively classical description of the statistics observed. Note, though, that in this particular case the more detailed analysis of the interaction yields the very same expected statistics as the less detailed analysis. In the scenario we are imagining, this is as it should be, for we were not looking for a different result but merely for a deeper understanding of the interaction between the electron and the measuring device.

It is not the case, however, that any cut we impose on a given phenomenon will be compatible with any other. Consider, for instance, two identically prepared ensembles of electrons, both in the state given by Eq.\ \eqref{eqn:pm-superpos}. Imagine that we subject the electrons in the first ensemble to a $z$-basis measurement while we subject the electrons in the second ensemble to an $x$-basis measurement. If we now examine the $z$-component of spin for the electrons in both ensembles (through a \emph{further} measurement in the $z$-basis in both cases), we will see that the statistics yielded by the first ensemble are incompatible with the statistics yielded by the second. And similarly, if we take two identical ensembles of compound systems, each in the state given by Eq.\ \eqref{eqn:compound}, and subject the first to a measurement in the basis
\begin{align}
  \label{eqn:xzbasis}
  \mathcal{B}_{xz} \equiv \{|+\rangle_{Mx}|+\rangle_{Sz},~ |+\rangle_{Mx}|-\rangle_{Sz},~ |-\rangle_{Mx}|+\rangle_{Sz},~ |-\rangle_{Mx}|-\rangle_{Sz}\},
\end{align}
while we subject the second to a measurement in the basis $\mathcal{B}_{zz}$, then our statistics for $| + \rangle_{Mz}| + \rangle_{Sz}$ and $| - \rangle_{Mz}| - \rangle_{Sz}$ (which, again, we will have to determine through a \emph{further} measurement in the basis $\mathcal{B}_{zz}$) will not be compatible with one another.

Corresponding to any particular cut that we impose on a particular phenomenon is a particular experimental arrangement, and with it a different physical interaction through which we assess the state of the system being probed \citep[cf.][pp. 392--393]{Bohr 1958}.\footnote{Bohr writes: ``In the treatment of atomic problems, actual calculations are most conveniently carried out with the help of a Schr\"odinger state function, from which the statistical laws governing observations obtainable under specified conditions can be deduced by definite mathematical operations. It must be recognized, however, that we are here dealing with a purely symbolic procedure, the unambiguous physical interpretation of which in the last resort requires a reference to a complete experimental arrangement. Disregard of this point has sometimes led to confusion, and in particular the use of phrases like `disturbance of phenomena by observation' or `creation of physical attributes of objects by measurements' is hardly compatible with common language and practical definition.''} Corresponding to this new cut---to this new subdivision of the measurement phenomenon---is a different description which will in general be incompatible with the first (cf. \citet{Frauchiger and Renner 2018} and especially \citeauthor[]{Bub 2016}'s response in Section 10.4 of the revised edition of \emph{Bananaworld}). As we saw in Section \ref{4.3}, quantum theory presents us with a fundamentally non-Boolean kinematical structure of possibilities. Upon this structure, we impose a particular Boolean frame. We do this through the need to express our experience of the result of a particular measurement---an experience of events that either do or do not occur, and which together fit into a consistent picture of the phenomenon in the particular measurement context being considered \citep[cf.][p. 701]{Bohr 1935}.\footnote{Bohr writes: ``While, however, in classical physics the distinction between object and measuring agencies does not entail any difference in the character of the description of the phenomena concerned, its fundamental importance in quantum theory, as we have seen, has its root in the indispensable use of classical concepts in the interpretation of all proper measurements, even though the classical theories do not suffice in accounting for the new types of regularities with which we are concerned in atomic physics.'' Compare also \citet[p. 293]{Bohr 1937}: ``the requirement of communicability of the circumstances and results of experiments implies that we can speak of well defined experiences only within the framework of ordinary concepts''.} In this way we partition the quantum-theoretical description into a quantum (non-Boolean) part, and a classical (Boolean) part. The latter is what we leave out of the quantum description. But it is left out by stipulation. The cut is movable. It is something that we impose upon our description of nature. Importantly, however, along with every cut comes a particular measurement context, and a particular measurement interaction corresponding to that context. And yet there is \emph{something} that we may call perspective-independent within quantum theory: This is its kinematical core, the fundamental structural constraints that quantum theory places on the possible representations of the physical systems it describes.

Our example of the measurement interaction given in Eq.\ \eqref{eqn:compound} could of course be given in even more detail. More of the components of the Stern-Gerlach device being used and of the dynamical interactions occurring between them and between them and the electron can be included in our description of the experiment---in fact one can include as many of these components as one likes. Moreover if there is an external system being used to assess the state of the Stern-Gerlach device after its interaction with the electron (your eyes, your ears, or even your nose, for instance), these can in principle be included in a dynamical description of the measurement as well. Indeed, quantum mechanics can be used to describe the interaction between \emph{any} two systems, one of which is to be called the ``system of interest'', the other the ``measuring device'', \emph{irrespective} of the level of internal complexity of either of them. This description will be of essentially the same form that it took in the simple examples given above. And in all cases the quantum description of an interaction will give us the answer to how, conditional upon it, the observed statistics can effectively be treated as classical.

What of the universe as a whole? There are areas of physics (notably cosmology) in which we aim to describe the universe in its totality as well as the dynamical evolution of that totality. Even putting cosmology to one side, is it not the goal of fundamental physics, generally speaking, to yield up a \emph{total} description of whatever aspect of the world is being considered? In order to do that, however, one would seem to require, not an account of this or that particular measurement (however detailed it may be), but rather an account of the measurement process in \emph{in general}. Second, if we are to provide a total description of reality, the scope of quantum description in the case of a measurement cannot be limited to the system of interest alone. Rather, the measurement apparatus itself should be included in one's quantum description of the interaction. It is true that it has been shown above how to do this to some extent, yet on the account of a measurement interaction given it is still the case that the emergence of a particular probability distribution is always conditional upon the particular (classical) assessment that we make. No matter how far we push back the cut, \emph{some} cut must always remain on this account. But this, it will be objected, is unacceptable; it cannot constitute a total description of reality.

The first demand---that an account of measurement must take the form of a general dynamical account \citep[cf.][]{Ghirardi 2018}---is a demand we reject. There is no dynamical process of measurement in general. There are only particular measurements. And in every particular case quantum mechanics provides, as we have illustrated, the \emph{general scheme} through which a dynamical account of that measurement process can be given. Quantum mechanics provides, that is, the tools we need in order to give an account of how the particular measurement apparatus in question dynamically interacts with a particular system of interest so as to give rise to a combined system in an entangled superposition yielding probabilities for the state of the measuring device that will be found upon assessing it.

As for the second objection: This, we maintain, misunderstands the nature of the cut upon which the quantum-mechanical assignment of conditional probabilities is based. For asserting the necessity of such a cut does not amount to the claim that measurement necessarily involves an interaction with some ``classical physical system'', where by this we imagine something large or heavy or both. Indeed, an \emph{atomic} system can in many cases serve very well as a measuring apparatus \citep[see, for instance,][]{Bacciagaluppi 2017}. The claim being made here, rather, is a logical one. Specifically, the claim is that in order to represent the assessment of a system's state, one needs to distinguish between that assessment and the system being assessed. This is true regardless of the measurement interaction in question, and indeed it is true even if the measurement scenario imagined is one in which it is the state of the entire universe being assessed, say, by a supreme being. It is still the case that this supreme being must distinguish, in its description of its measurement, its assessment of that measurement from the system it is measuring. And there is no reason to stop there; for there is nothing to stop one from considering the supreme being and the universe as together comprising a single physical system (supposing that the supreme being exists somehow in space and time); and in that case one still needs to distinguish one's assessment of that larger system from the system being assessed. There is no ``view from nowhere'' within quantum mechanics with respect to its account of observation. Nor should there be.

Consider, by way of analogy, the claim one might make in the context of classical physics that one can measure the length of a given body with a rod, or the lifetime of a given particle with a clock. Now to conduct an accurate measurement, the rod must be rigid, the clock ideal. And a legitimate demand one might make in this instance is that the existence of such rigid rods and ideal clocks be substantiated. Einstein accepted this, and in his debate with Weyl over the issue, appealed to the identical spectral lines manifested by atoms of the same kind as compelling evidence for the existence of such ideal instruments \citep[see][]{Giovanelli 2014}. Now the further objection was not made, though it conceivably could have been, that in connecting the theory up with our experience in this way, it is still presupposed, in every particular case, that somehow a rod or a clock has been determined to be a suitable one, and to complete the theory we require an account of how such a determination can be possible. In the context of special relativity this objection is easily dismissed as an extra-physical, purely philosophical concern. And yet, the analogous question in the case of quantum mechanics is not so readily dismissed.\footnote{See \citet{Pokorny et al 2019} for an investigation into the existence of ideal quantum measurements, and see \citet[]{Cabello 2019} for discussion of the quantum correlations that can be realized with ideal and non-ideal measurements.}

The issue, it seems, is the intrinsic randomness of the theory. The dynamical account of a given measurement that is provided by quantum mechanics ultimately ends in probabilities; it does not end in definite outcomes. And yet when one assesses the state of a given system the result is in every case a definite outcome. What, one will ask, is to be made of the definite character of these particular outcomes as contrasted with the apparent indefiniteness we attach to the description of a quantum state, and how is it that the former can be seen as arising from the latter? For someone motivated by this worry, appealing to the quantum-mechanical account of the dynamics of a particular measurement, as we did above, is a \emph{non sequitur}; the quantum-mechanical account of a measurement, no matter how deep or encompassing one makes it, in the end can only yield indefiniteness; it can in general only assign a probability to a particular measurement outcome. But it is an account of the mechanism through which a particular definite outcome emerges from this indefiniteness which is now being demanded.

\citet[]{Bub and Pitowsky 2010} refer, with irony, to this as the ``big'' measurement problem. We will dispense with the irony. On our view it would be better to call it the \emph{superficial} measurement problem. If one compares a uniform ensemble of quantum systems in the state given by Eq.\ \eqref{eqn:compound} with a basket of raffle tickets in which the proportion of tickets marked ``$|+\rangle|+\rangle$'' and ``$|-\rangle|-\rangle$'' is respectively $|\alpha|^2$ and $|\beta|^2$, the important conceptual difference between the two cases is \emph{not} that the outcome obtained for a particular experimental run in one but not the other scenario is determined stochastically. For this is in fact true of both the quantum ensemble and the raffle. To be sure, in the case of the raffle, we can always interpret away this indeterminism. As mentioned earlier, the complete classical state specification (which by assumption our raffle contestant has no access to) for a given ticket is the truthmaker for the occurrence or non-occurrence of an event in the sense that it fixes the value of every yes-or-no question one may ask of the system. In the case of the quantum system this is not true. A quantum state assignment fixes in advance only the probability that a selected observable will take on a particular value when we query the system concerning it (i.e., when the operator representing the observable is applied to the state vector describing the system). But rather than add further structure to the quantum formalism so as to make possible the same sort of interpretation that seems so straightforward in the classical case, we rather elect to take as true what the kinematical core of quantum theory is telling us: that the world is fundamentally nondeterministic, that there is no further story to tell about how a particular definite outcome emerges as the result of a given measurement; that measurement outcomes are intrinsically random---in general only determinable probabilistically.

The \emph{profound} problem of measurement is not this. Nor is it quite what \citet[]{Bub and Pitowsky 2010} refer to as the ``small'' measurement problem: the problem of how to dynamically account for the effective emergence of a globally Boolean macrostructure of events out of a globally non-Boolean microstructure underlying them. Unlike the ``small'' problem, the profound problem of measurement cannot be resolved by considering the dynamics of decoherence alone, nor is it truly dynamical in nature at all. The profound problem of measurement stems, rather, from the fact that of the many classical probability distributions that are implicit in the quantum state description, the one that emerges in a given scenario is always conditional upon the choice that we make from among the many possible measurements performable on the system. In other words it is the---in part physical and in large part \emph{philosophical}---problem to account for the fact that, owing to the nature of the non-Boolean kinematical structure of quantum mechanics, \emph{only some} of the classical possibility distributions implicit in the quantum state are actualized in the context of a given measurement, and moreover which of them are actualized is always conditional upon that measurement context.

An ensemble of quantum systems prepared in the state given by Eq.\ \eqref{eqn:pm-superpos}, for example, yields a particular classical probability distribution over the outcomes $| + \rangle_z$ and $| - \rangle_z$ when the systems from the ensemble interact with a Stern-Gerlach apparatus whose DuBois magnets are oriented along the $z$-direction. If we interact the ensemble with an apparatus whose magnets are oriented along the $x$-direction, however, we require a different probability distribution to describe the measurement statistics that ensue, which is in general incompatible with the first. Note that this problem is not resolved by including aspects of the measuring apparatus (or indeed all of it) in our quantum-mechanical description of the experimental setup as we did above. For given the entangled superposition in Eq.\ \eqref{eqn:compound}, we are still left with the choice of whether to measure the combined system in the basis $\mathcal{B}_{zz}$ or in some other basis. Quantum mechanics does not make this choice for us. It is up to us. This is the profound problem of measurement.

And yet to think of it as a problem pertaining to the quantum-mechanical \emph{account} of a measurement is misleading. Given a particular measurement context, quantum mechanics provides us with all of the resources we need in order to account for the dynamics of the measurement interaction between the system of interest and measurement device, and through this account we explain why a particular classical probability distribution is applicable given that measurement context, despite the non-classical nature of the quantum state description. Quantum mechanics does not tell you, however, which of the many possible measurements on a system you should apply in a given case. From the point of view of the theory the choices you make or do not make are up to you.

%SECTION 6
\section{Conclusion} \label{5}
%!TEX root =  ./JanasJanssenCuffaro-August2019.tex

%SECTION 6 -- Conclusion

We noted in Section \ref{0} that for Heisenberg, quantum mechanics' significance lay in its provision of a new framework for doing physics, one that was sorely needed in light of the persistent failures of classical mechanics and the old quantum theory of Bohr and Sommerfeld to deal with the puzzling (mostly spectroscopic) experimental data it was confronted with in the first two decades of the last century \citep{Duncan and Janssen 2019}. Heisenberg's core insight into quantum mechanics' significance is one that we and the others close to us on the phylogenetic tree of interpretations share. In the body of this paper we saw a number of concrete examples vividly illustrating the essential differences between the quantum and the classical kinematical framework, how those differences are manifested in the correlations between and in the dynamics of quantum systems, and finally how the quantum-kinematical framework enables us to learn about the specifics of particular systems through measurement. In this final section we present our view in a nutshell.

Quantum mechanics is about probabilities. The kinematical framework of the theory is probabilistic in the sense that the state specification of a given system yields, in general, only the probability that a selected observable will take on a particular value when we query the system concerning it. Quantum mechanics' kinematical framework is also non-Boolean: The Boolean algebras corresponding to the individual observables associated with a given system cannot be embedded into a global Boolean algebra comprising them all, and thus the values of these observables cannot (at least not straightforwardly) be taken to represent the properties possessed by that system in advance of their determination through measurement. It is in this latter---non-Boolean---aspect of the probabilistic quantum-kinematical framework that its departure from classicality can most essentially be located.

Despite this character, we have seen above how the quantum-mechanical framework provides a recipe\footnote{Cf. \citet[]{Wallace 2019}, who argues that the Everett interpretation provides a general ``recipe'' for interpreting quantum theory (see also note \ref{wallace-recipe} in Section \ref{0}).} through which one can acquire information concerning particular systems by classical means. Given an ensemble of quantum systems either prepared uniformly in a particular state $| \psi \rangle$ or as a mixture of states $| \psi \rangle_i$ (described by the density operators $\hat{\rho} = | \psi \rangle\langle \psi |$ and $\hat{\rho} = \sum_i| \psi \rangle_{\!i} \, _{i\!}\langle \psi |$, respectively), and conditional upon a particular classically describable assessment of one of the parameters of the systems in that ensemble---conditional, that is, upon a particular Boolean frame that we impose on those systems---the information we obtain from our assessment can always be (re)described as having arisen from an ensemble of classical systems (like the raffles in our examples) with a certain distribution of values for the parameter in question. Further, the particular distribution observed can be predicted from the quantum state.

This recipe does not solve the \emph{profound} problem of measurement; i.e., the problem to account for how it is that only some of the classical probability distributions implicit in the quantum state description are actualized in the context of a given measurement. But even without providing an answer to this question, we see how the kinematical core of quantum mechanics provides us with all of the tools we need to give an account of the dynamics of a particular measurement interaction, and through this explain why a particular classical probability distribution can be used to characterize the statistics observed within that measurement context, despite the non-classical nature of the quantum state description.

It may be objected that the world we experience does not consist in probability distributions. Its objects include this table, that banana and the other dynamical objects we observe and interact with, both in the kitchens of the world and outside of them, every day. These objects will not be found within the quantum-kinematical framework, nor will the recipe just mentioned yield them up in and of itself. Conditional upon a given measurement, however, that recipe will allow one to transition from the quantum description of a system to the classical description of the observations which ensue. And from there we already know how to use classical theory to construct, from these observations, the familiar objects of our world.

As our examples have demonstrated, quantum theory is successful where classical theory falls short in its description of physical phenomena, and its advent has uncovered aspects of our world that were before then veiled in darkness. But besides these particular lessons there is a wider moral that we can glean from the new kinematical framework of quantum theory, and in particular by considering how it differs from classical theory. The logical framework of classical physics is a globally Boolean structure. Through it a global noncontextual assignment of values to the observables associated with physical systems becomes possible. Because of this, these value assignments may unproblematically be thought of as the underlying properties of the physical systems they have been assigned to. This allows us to speak of a world that exists in a particular way irrespective of our particular interactions with it. Quantum mechanics, however, shows us that this classical description is valid only up to a certain point, and that the logical structure of the world \emph{as it presents itself to us} is globally non-Boolean. Whatever else we may discover in the course of the future development of physical theory, this is a non-trivial fact that we have discovered about the world. Moreover it is a fact that will remain with us \citep[cf.][p. 98]{Pitowsky 1994}. It is, further, a non-trivial fact that we can learn about our world, despite this non-Boolean character, through classical means \citep[cf.][p. 293]{Bohr 1937}.\footnote{Bohr writes: ``the proper r\^ole of the indeterminacy relations consists in assuring quantitatively the logical compatibility of apparently contradictory laws which appear when we use two different experimental arrangements, of which only one permits an unambiguous use of the concept of position, while only the other permits the application of the concept of momentum.''}

It will be objected that what we have just called ``facts about the world'' are really only relational facts about our connection to the world \citep[cf.][]{Healey 2016}. This is entirely correct. But that, we maintain, is how it should be. For we are entangled with the world, and our concepts both of the world and of ourselves are only marginals of that true entangled description. That description, along with its many seemingly incompatible aspects, arises out of and is made possible through the non-Boolean probabilistic structure of the quantum-mechanical kinematical core.

Quantum theory provides us with an \emph{objective} description of a given system. This description is valid \emph{irrespective} of one's particular choices and irrespective of one's particular interests in making those choices. At the same time the description that quantum theory provides to us of a given system's dynamical state is unlike the corresponding description given to us by classical theory. In quantum theory, what is exhibited to us through the quantum state description is not the set of dynamical properties, in the classical sense, of the system of interest. What is exhibited, rather, is the structure of, interrelations between, and interdependencies among the possible perspectives one can take on that system. In this way quantum theory informs us regarding the structure of the world---a world that \emph{includes} ourselves---and of our place within that structure.

\section*{Acknowledgments}

Thanks, above all, to Jeff Bub for his enthusiastic support of our efforts and for his patience in explaining his information-theoretic interpretation of quantum mechanics to us. The three of us came to this project from different directions. Janssen (a historian of science) is a recovering Everettian who has been defending Bub's information-theoretic interpretation with the zeal of the converted. Cuffaro (a philosopher of science) has been working on information-theoretic interpretations for some time and has tried to keep Janssen's proselytizing at bay. Janas (a theoretical physicist) was and remains a Bohm sympathizer but got intrigued by Pitowsky's polytopes. One thing the three of us share is a broadly Kantian outlook. This paper has been in the making for some time and we have presented preliminary results in a variety of settings. We thank Laurent Taudin for drawing chimps and bananas both for more recent presentations and for this paper. For their feedback on our presentations, we are grateful to members of the Physics Interest Group (PIG) of the Minnesota Center for Philosophy of Science; to students at the University of Minnesota and Washburn High in Minneapolis; to audiences at a physics colloquium at Minnesota State University Mankato, a lunchtime talk at the Center for Philosophy of Science at the University of Pittsburgh, and seminars in the School of Mathematics at the University of Minnesota and at the \emph{Max-Planck-Institut f\"ur Wissenschaftsgeschichte} in Berlin; to participants in a workshop on interpreting quantum mechanics at \emph{Politecnico di Milano}; and, finally, to participants in the 2019 edition of the conference series, \emph{New Directions in the Foundations of Physics}, in Viterbo, Italy. Especially Wayne Myrvold's intervention in the Q\&A following our presentation in Viterbo, our only joint performance, greatly improved this paper (see notes \ref{Myrvold 1} and \ref{Myrvold 2} in Section \ref{1.6}). For further discussion we thank Tony Duncan, Sam Fletcher, Louisa Gilder, Peter Grul, Luc Janssen, Christian Joas, David Kaiser, Jim Kakalios, Alex Kamenev, Marius Krumm, Christoph Lehner, Charles Marcus, Markus M\"uller, Max Niedermayer, Sergio Pernice, J\"urgen Renn, Serge Rudaz, David Russell, Rob ``Ryno'' Rynasiewicz, Rob Spekkens and Brian Woodcock.  Instead of dedicating our paper to them we would have loved to discuss it with Bill Demopoulos and Itamar Pitowsky. Cuffaro gratefully acknowledges support from the Rotman Institute of Philosophy and from the Foundational Questions Institute (FQXi). Janssen gratefully acknowledges support from the \emph{Alexander von Humboldt Stiftung} and the \emph{Max-Planck-Institut f\"ur Wissenschaftsgeschichte}. Janas thanks the staff of Al's Breakfast in Dinkytown.

\renewcommand{\refname}{Bibliography}


\begin{thebibliography}{}

\bibitem[Accardi and Fedullo(1982)]{Accardi and Fedullo 1982}  Accardi, L.\ and Fedullo, A.\ (1982). On the statistical meaning of complex numbers in quantum mechanics. \emph{Lettere al Nuovo Cimento} 34: 161--172.

\bibitem[Acu\~na(2014)]{Acuna 2014} Acu\~na, P.\ (2014). On the empirical equivalence between special relativity and Lorentz's ether theory. {\em Studies in History and Philosophy of Modern Physics} 46: 283--302.

\bibitem[Albert(1992)]{Albert 1992} Albert, D.\ Z.\ (1992). \emph{Quantum Mechanics and Experience.} Cambridge, MA: Harvard University Press.

\bibitem[Albert(2018)]{Albert 2018} Albert, D.\ Z.\ (2018). Quantum's leaping lizards (Review of \citet{Becker 2018}) \emph{The New York Review of Books} 65, no.\ 7 (April 19, 2018).

\bibitem[Aldrich(1998)]{Aldrich 1998} Aldrich, J.\ (1998). Doing least squares: perspectives from Gauss and Yule. \emph{International Statistical Review} 66: 61--81.

\bibitem[Bacciagaluppi(2017)]{Bacciagaluppi 2017} Bacciagaluppi, G.\ (2017). Bohr's slit and Hermann's microscope. In G.\ Bacciagaluppi and E.\ Crull, Eds., \emph{Grete Hermann: Between Physics and Philosophy} (pp. 135--148). Dordrecht: Springer.

\bibitem[Bacciagaluppi, Crull and Maroney(2017)]{BCM 2017} Bacciagaluppi, G., Crull, E.\ and Maroney, O.\ J.\ E.\ (2017). Jordan's derivation of blackbody fluctuations. \emph{Studies in History and Philosophy of Modern Physics} 60: 23--34.

\bibitem[Balashov and Janssen(2003)]{Balashov  and Janssen 2003} Balashov, Y.\ and Janssen, M.\ (2003). Presentism and relativity.  {\it British Journal for the Philosophy of Science}
 54: 327--346.

\bibitem[Ball(2018)]{Ball 2018} Ball, P.\ (2018). \emph{Beyond Weird: Why Everything You Thought You Knew About Quantum Physics is Different}. London: The Bodley Head.

%\bibitem[Barrett \emph{et al.}(2005)]{Barrett 2005} Barrett, J., Linden, N., Massar, S., Pironio, S., Popescu, S., and Roberts, D.\ (2005). Non-local correlations as an information theoretic resource. \emph{Physical Review} A 71: 022101.
%xxx--yyy see bananaworld, p. 120, note 2.

\bibitem[Baym(1969)]{Baym 1969} Baym, G.\ (1969). \emph{Lectures on Quantum Mechanics.} Reading, MA: Addison-Wesley.

\bibitem[Becker(2018)]{Becker 2018} Becker, A.\ (2018). \emph{What is Real? The Unfinished Quest for the Meaning of Quantum Physics}. New York: Basic Books.

\bibitem[Bell(1964)]{Bell 1964} Bell, J.\ S.\ (1964). On the Einstein-Podolsky-Rosen paradox. \emph{Physics} 1: 195--200. Reprinted in \citet[pp.\ 14--21]{Bell 1987}.

\bibitem[Bell(1984)]{Bell 1984} Bell, J.\ S.\ (1984). Beables for quantum field theory. CERN-TH.4035/84. Reprinted in \citet[pp.\ 159--166]{Bell 1987}

\bibitem[Bell(1987)]{Bell 1987} Bell, J.\ S.\ (1987). \emph{Speakable and Unspeakable in Quantum Mechanics}. Cambridge: Cambridge University Press. Page references to second edtion (2004).

\bibitem[Bell(1990)]{Bell 1990} Bell, J.\ S.\ (1990). Against measurement. \emph{Physics World} 8: 33--40.

\bibitem[Ben-Menahem(2017)]{Ben-Menahem 2017} Ben-Menahem, Y.\ (2017). The PBR theorem: Whose side is it on? \emph{Studies in History and Philosophy of Modern Physics} 57: 80--88.

\bibitem[Berkovitz(2012)]{Berkovitz 2012} Berkovitz, J.\ (2012). The world according to de {F}inetti. In Y.\ Ben-Menahem and M.\ Hemmo, Eds., \emph{Probability in Physics} (pp. 249--280). Heidelberg: Springer.

\bibitem[Berkovitz(2019)]{Berkovitz 2019} Berkovitz, J.\ (2019). On de {F}inetti's instrumentalist philosophy of probability. \emph{European Journal for the Philosophy of Science} 9: 25.

\bibitem[Bingham(2000)]{Bingham 2000} Bingham, N.\ H.\ (2000). Studies in the history of probability and statistics XLVI. Measure into probability: From Lebesgue to Kolmogorov. \emph{Biometrika} 87: 145--156.

\bibitem[Birkhoff and Kreyszig(1984)]{Birkhoff and Kreyszig 1984} Birkhoff, G.\ and Kreyszig, E.\ (1984). The establishment of functional analysis. \emph{Historia Mathematica} 11: 258--321.

\bibitem[Bohr(1935)]{Bohr 1935} Bohr, N.\ (1935). Can quantum-mechanical description of physical reality be considered complete? \emph{Physical Review} 48: 696--702.

\bibitem[Bohr(1937)]{Bohr 1937} Bohr, N.\ (1937). Causality and complementarity. \emph{Philosophy of Science} 4: 289--298.

\bibitem[Bohr(1958)]{Bohr 1958} Bohr, N.\ (1958). Quantum physics and philosophy. In R.\ Klibansky, Ed., \emph{Philosophy in the Mid-Century: A Survey} (pp.\ 308--314). Firenze: La Nuova Italia Editrice. Page reference to reprint in: J.\ Kalckar, Ed., \emph{Niels Bohr: Collected Works}, Vol.\ 7 (pp.\ 385--394). Amsterdam: Elsevier.

%\bibitem[Bohm(1951)]{Bohm 1951} Bohm, D.\ (1951). \emph{Quantum Theory.} Englewood Cliffs, NJ: Prentice-Hall.
%pp. 614--623

%\bibitem[Born(1971)]{Born 1971} Born, M., Ed.\  (1971). \emph{The Born-Einstein Letters.} New York: Walker.

\bibitem[Born, Heisenberg and Jordan(1926)]{dreimaenner} Born, M., Heisenberg, W.\ and Jordan, P.\ (1926). Zur Quantenmechanik II. \emph{Zeitschrift f\"{u}r Physik}  35: 557--615. English translation in \citet[pp.\ 321--385]{Van der Waerden}.

\bibitem[Bowler(2003)]{Bowler 2003} Bowler, P.\ J.\ (2003). \emph{Evolution. The History of an Idea}. 3rd revised and expanded edition. Berkeley, Los Angeles, London: University of California Press.

\bibitem[Bravais(1846)]{Bravais 1846} Bravais, A.\ (1846). Analyse math\'ematique sur les probabilit\'es des erreurs de situation d'un point. \emph{M\'emoires de l'Acad\'emie Royale des Sciences de l'Institut de France} 9: 255--332.

\bibitem[Brown(2005)]{Brown 2005} Brown, H.\ R.\ (2005). \emph{Physical Relativity. Space-time Structure from a Dynamical Perspective}, Oxford: Oxford University Press.

\bibitem[Brown and Timpson(2006)]{Brown and Timpson 2006}  Brown, H.\ R.\ and Timpson, C.\ G.\ (2006). Why special relativity should not be a template for a fundamental reformulation of quantum mechanics. In W.\ Demopoulos and I.\ Pitowsky, Eds., \emph{Physical Theory and its Interpretation} (pp.\ 29--42). Berlin: Springer.  

\bibitem[Bub(2004)]{Bub 2004} Bub, J.\ (2004). Why the quantum? \emph{Studies in History and Philosophy of Modern Physics} 35: 241--266.

\bibitem[Bub(2010)]{Bub 2010} Bub, J.\ (2010). Von Neumann's `no hidden variables' proof: A re-appraisal.  \emph{Foundations of Physics} 40: 1333--1340.

\bibitem[Bub(2016)]{Bub 2016} Bub, J.\ (2016). \emph{Bananaworld. Quantum Mechanics for Primates}. Oxford: Oxford University Press. Slightly revised paperback edition: 2018.

\bibitem[Bub(2019)]{Bub 2019} Bub, J.\ (2019). Interpreting the quantum world: old questions and new answers (essay review of \citet{Freire 2015}, \citet{Becker 2018} and \citet{Ball 2018}). \emph{Historical Studies in the Natural Sciences} 49: 226--239.

\bibitem[Bub and Demopoulos(2010)]{Bub and Demopoulos 2010} Bub, J.\ and Demopoulos, W.\ (2010). Itamar Pitowsky 1950--2010. \emph{Studies in History and Philosophy of Modern Physics} 41: 85.

\bibitem[Bub and Pitowsky(2010)]{Bub and Pitowsky 2010} Bub, J.\ and Pitowsky, I.\ (2010). Two dogmas about quantum mechanics.  In \citet[pp.\ 433--459]{Many Worlds 2010}.

\bibitem[Bub and Bub(2018)]{Bub and Bub 2018} Bub, T.\ and Bub, J.\ (2018). \emph{Totally Random. Why Nobody Understands Quantum Mechanics. A Serious Comic on Entanglement.} Princeton: Princeton University Press.

\bibitem[Busch(2003)]{Busch 2003} Busch, P.\ (2003). Quantum states and generalized observables: A simple proof of Gleason's theorem. \emph{Physical Review Letters} 91: 120403.

\bibitem[Butler(1968)]{Butler 1968} Butler, J.\ W.\ (1968). On the Trouton-Noble experiment. \emph{American Journal of Physics} 36: 936--941.

\bibitem[Cabello(2019)]{Cabello 2019} Cabello, A.\ (2019). Quantum correlations from simple assumptions. \emph{Physical Review A} 100: 032120.

\bibitem[Caulton(2015)]{Caulton 2015} Caulton, A. (2015). The role of symmetry in the interpretation of physical theories. \emph{Studies in History and Philosophy of Modern Physics} 52: 153--162.

\bibitem[Chiribella, D'Ariano and Perinotti(2010)]{Chiribella et al 2010} Chiribella, J., D'Ariano, G.\ M.\ and Perinotti, P.\  (2010). Probabilistic theories with purification. \emph{Physical Review A} 81:
062348.

\bibitem[Chiribella and Ebler(2019)]{Chiribella and Ebler 2019} Chiribella, J.\ and Ebler, D.\, Quantum speedup in the identification of cause-effect relations. \emph{Nature Communications} 10: 1472.

\bibitem[Chiribella and Spekkens(2016)]{Chiribella and Spekkens 2016} Chiribella, G.\ and Spekkens, R.\ (2016). \emph{Quantum Theory: Informational Foundations and Foils}. Dordrecht: Springer.

\bibitem[Cirel'son(1980)]{Cirel'son 1980} Cirel'son, B.\ S.\ (1980). Quantum generalizations of Bell's inequality. \emph{Letters in Mathematical Physics} 4: 93--100.

\bibitem[Clauser and Freedman(1972)]{Clauser and Freedman 1972} Clauser, J.\ F.\ and Freedman, S.\ J.\ (1972). Experimental test of local hidden-variable theories. \emph{Physical Review Letters} 28: 938--941.

\bibitem[Clauser, Horne, Shimony and Holt(1969)]{CHSH} Clauser, J.\ F., Horne, M.\ A., Shimony, A.\ and Holt, R.\ A.\ (1969). Proposed experiment to test local hidden-variable theories. \emph{Physical Review} 23: 880--884.

\bibitem[Clifton, Bub and Halvorson(2003)]{CBH} Clifton, R., Bub, J.\ and Halvorson, H.\ (2003). Characterizing quantum theory in terms of information-theoretic constraints. \emph{Foundations of Physics} 33: 1561--1591.

\bibitem[Cuffaro(2018)]{Cuffaro 2018} Cuffaro, M.\, E.\ (2018). Information causality, the Tsirelson bound, and the `being-thus' of things. \emph{Studies in History and Philosophy of Modern Physics}. In press.

\bibitem[Cuffaro and Fletcher(2018)]{Cuffaro and Fletcher 2018} Cuffaro, M.\ E.\  and Fletcher, S.\ C., Eds.\ (2018). \emph{Physical Perspectives on Computation, Computational Perspectives on Physics.} Cambridge: Cambridge University Press.

\bibitem[Curiel(2019)]{Curiel 2019} Curiel, E.\ (2019). Schematizing the observer and the epistemic content of theories. Preprint, {arXiv:1903.02182}.

\bibitem[Dantzig and Thapa(1997)]{Dantzig and Thapa 1997}  Dantzig, G.\ B.\ and Thapa, M.\ N.\ (1997). \emph{Linear programming 1: Introduction}. New York: Springer.

\bibitem[Dantzig and Thapa(2003)]{Dantzig and Thapa 2003} Dantzig, G.\ B.\ and Thapa, M.\ N.\ (2003). \emph{Linear programming 2: Theory and Extensions.} New York: Springer.

\bibitem[De Finetti(1930)]{De Finetti 1930} De Finetti, B.\ (1930). Spazi astratti metrici ($\mathcal{D}_M$). \emph{Atti della Pontificia Accademia delle Scienze Nuovi Lincei} 83: 248--256.

\bibitem[De Finetti(1937)]{De Finetti 1937} De Finetti, B.\ (1937). A proposito di ``correlazione." \emph{Supplemento Statistico ai Nuovi problemi di Politica, Storia ed Economia}, 3:  41--57. Page reference to English translation in \emph{Journal \'Electronique d'Histoire des Probabilit\'es et de la Statistique} (www.jehps.net) 4 (4) (2008): 1--15.

\bibitem[Demopoulos(2010)]{Demopoulos 2010}  Demopoulos, W.\ (2010). Effects and propositions. \emph{Foundations of Physics} 40: 368--389.

\bibitem[Demopoulos(2018)]{Demopoulos 2018} Demopoulos, W.\ (2018). \emph{On Theories}. Unpublished book manuscript.

\bibitem[Deza and Laurent(1997)]{Deza and Laurent 1997} Deza, M.\ M.\ and Laurent, M.\  (1997). \emph{Geometry of Cuts and Metrics}. Berlin: Springer.

\bibitem[Dieks(2017)]{Dieks 2017} Dieks, D.\ (2017). Von Neumann's impossibility proof: Mathematics in the service of rhetorics. \emph{Studies in History and Philosophy of Modern Physics} 60: 136--148.

\bibitem[Dirac(1926)]{Dirac 1926} Dirac, P.\ A.\ M.\ (1926). The fundamental equations of quantum mechanics. \emph{Proceedings of the Royal Society of London} A109: 642--653. Reprinted in \citet[pp.\ 307--320]{Van der Waerden}

\bibitem[Dirac(1927)]{Dirac 1927} Dirac, P.\ A.\ M.\  (1927).  The physical interpretation of the quantum dynamics. \emph{Proceedings of the Royal Society of London. Series A} 113: 621--641. 

\bibitem[Dirac(1930)]{Dirac 1930} Dirac, P.\ A.\ M.\  (1930). \emph{Principles of Quantum Mechanics.} Oxford: Clarendon.

\bibitem[Doob(1934)]{Doob 1934} Doob, J.\ L.\ (1934). Probability and statistics. \emph{Transactions of the American Mathematical Society} 36: 759--775.

\bibitem[Duncan and Janssen(2007)]{Duncan and Janssen 2007} Duncan, A.\ and Janssen, M.\  (2007). On the verge of \emph{Umdeutung} in Minnesota: Van Vleck and the correspondence principle. 2 Pts. \emph{Archive for History of Exact Sciences} 61: 553--624, 625--671.

\bibitem[Duncan and Janssen(2008)]{Duncan and Janssen 2008} Duncan, A.\ and Janssen, M.\   (2008). Pascual Jordan's resolution of the conundrum of the wave-particle duality of light. \emph{Studies in History and Philosophy of Modern Physics} 39: 634--666.

\bibitem[Duncan and Janssen(2013)]{Duncan and Janssen 2013} Duncan, A.\ and Janssen, M.\  (2013). (Never) mind your $p$'s and $q$'s: Von Neumann versus Jordan on the foundations of quantum theory. \emph{The European Physical Journal H} 38: 175--259.

\bibitem[Duncan and Janssen(2014)]{Duncan and Janssen 2014} Duncan, A.\ and Janssen, M.\  (2014). The trouble with orbits: the Stark effect in the old and the new quantum theory. \emph{Studies in History and Philosophy of Modern Physics} 48: 68--83.

\bibitem[Duncan and Janssen(2015)]{Duncan and Janssen 2015} Duncan, A.\ and Janssen, M.\  (2015).  The Stark effect in the Bohr-Sommerfeld theory and in Schr\"odinger's wave mechanics. In F.\ Aaserud and H.\ Kragh, Eds., \emph{One Hundred Years of the Bohr Atom} (pp. 217--271). Copenhagen: Det Kongelige Danske Videnskabernes Selskab.

\bibitem[Duncan and Janssen(2019)]{Duncan and Janssen 2019} Duncan, A.\ and Janssen, M.\  (2019). \emph{Constructing Quantum Mechanics}, Vol.\ 1, \emph{The Scaffold, 1900--1923}. Oxford: Oxford University Press.

\bibitem[Eckert(2013)]{Eckert 2013} Eckert, M.\ (2013). \emph{Arnold Sommerfeld. Science, Life and Turbulent Times.} New York: Springer,.

\bibitem[Ehrenfest(1925)]{Ehrenfest 1925} Ehrenfest, P.\ (1925). Energieschwankungen im Strahlungsfeld oder Kristallgitter bei Superposition quantisierter Eigenschwingungen. \emph{Zeitschrift f\"ur Physik}, 34, 362--373.

\bibitem[Einstein(1909a)]{Einstein 1909a} Einstein, A.\ (1909a). Zum gegenw\"artigen Stand des Strahlungsproblems. \emph{Physikalische Zeitschrift} 10: 185--193. Reprinted in Einstein (1987--2018, Vol.\ 2, Doc.\ 56).

\bibitem[Einstein(1909b)]{Einstein 1909b} Einstein, A.\ (1909b). \"Uber die Entwicklung unserer Anschauungen \"uber das Wesen und die Konstitution der Strahlung. \emph{Physikalische Zeitschrift} 10: 817--825. Reprinted in Einstein (1987--2018, Vol.\ 2, Doc.\ 60).

\bibitem[Einstein(1919)]{Einstein 1919} Einstein, A.\ (1919). Time, space, and gravitation. \emph{The London Times}, November 28, 1919. Reprinted in Einstein (1987--2018, Vol.\ 7, Doc.\ 26).
%M.\ Janssen \emph{et al}., Eds. \emph{The Collected Papers of Albert Einstein}. Vol.\ 7. \emph{The Berlin Years: Writings, 1918--1921} (Doc.\ 26). Princeton: Princeton University Press, 2002. 
Reprinted under a different title (What is the theory of relativity?) in A.\ Einstein, \emph{Ideas and Opinions} (pp.\ 227--232). New York: Crown Publishers, 1954.

\bibitem[Einstein(1987--2018)]{Einstein 1987--2018} Einstein, A.\ (1987--2018). \emph{The Collected Papers of Albert Einstein} (15 Vols).  J.\ Stachel, \emph{et al.} (Eds.). Princeton: Princeton University Press.

\bibitem[Einstein, Podolsky and Rosen(1935)]{EPR} Einstein, A., Podolsky, B.\ and Rosen, N.\ (1935). Can quantum-mechanical description of physical reality be considered complete? \emph{Physical Review} 47: 777--780.

\bibitem[Eisberg and Resnick(1985)]{Eisberg and Resnick 1985} Eisberg, R.\ and Resnick, R.\ (1985). \emph{Quantum Physics of Atoms, Molecules, Solids, Nuclei, and Particles}. 2nd ed.\ (1st ed.: 1974). New York: Wiley.   

\bibitem[Epstein(1916)]{Epstein 1916} Epstein, P.\ (1916). Zur Theorie des Starkeffektes. \emph{Annalen der Physik} 50: 489--521.

\bibitem[Epstein(1926)]{Epstein 1926} Epstein, P.\ (1926). The Stark effect from the point of view of Schr\"odinger's quantum theory. \emph{Physical Review} 28: 695--710.

%\bibitem[Fine(1982a)]{Fine 1982a}  Fine, A.\ (1982a). Hidden variables, joint probability, and the Bell inequalities. \emph{Physical Review Letters} 48:
% (5): 291--294.

%\bibitem[Fine(1982b)]{Fine 1982b}  Fine, A.\ (1982b). Joint distributions, quantum correlations, and commuting observables. \emph{Journal of Mathematical Physics}  23: 
%(7): 1306--1310.

\bibitem[Fisher(1915)]{Fisher 1915}  Fisher, R.\ A.\ (1915). Frequency distribution of the values of the correlation coefficient in samples from an indefinitely large population. \emph{Biometrika} 10: 507--521.

\bibitem[Fisher(1924)]{Fisher 1924}  Fisher, R.\ A.\ (1924). The distribution of the partial correlation coefficient. \emph{Metron} 3: 329--332. 

\bibitem[Forster(1942)]{Forster 1942} Forster, E.\ M.\ (1942). \emph{Virginia Woolf. The Rede Lecture 1941}. Cambridge: Cambridge University Press.

\bibitem[Franklin \emph{et al.}(2008)]{Franklin et al 2008}  Franklin, A., Edwards, A.\ W.\ F., Fairbanks, D.\ J., Hartl, D.\ L.\ and Seidenfeld, T.\ (2008).  \emph{Ending the Mendel-Fisher Controversy}. Pittsburgh: University of Pittsburgh Press.

\bibitem[Frauchiger and Renner(2018)]{Frauchiger and Renner 2018} Frauchiger, D.\ and Renner, R.\ (2018). Quantum theory cannot
consistently describe the use of itself. \emph{Nature Communications} 9: 3711.

\bibitem[Freire(2015)]{Freire 2015} Freire, O.\ Jr.\ (2015). \emph{The Quantum Dissidents: Rebuilding the Foundations of Quantum Mechanics (1950--1990)}. Heidelberg: Springer.

\bibitem[Fristedt and Gray(1997)]{Fristedt and Gray 1997}  Fristedt, B.\ E.\ and Gray, L.\ F.\ (1997). \emph{A Modern Approach to Probability Theory}. Boston, Basel, Berlin: Birkh\"auser. 
%Fristedt, Bert E., Gray, Lawrence F. 

\bibitem[Ghirardi(2018)]{Ghirardi 2018} Ghirardi, G.\ (2018). Collapse Theories. In E.\ N.\ Zalta (ed.), \emph{The Stanford Encyclopedia of Philosophy} (Fall 2018 Edition). URL =
\textless\url{plato.stanford.edu/archives/fall2018/entries/qm-collapse/}\textgreater

\bibitem[Gilder(2008)]{Gilder 2008} Gilder, L.\ (2008). \emph{The Age of Entanglement. When Quantum Physics Was Reborn.} New York: Knopf. Reprinted: New York: Vintage, 2009.

\bibitem[Giovanelli(2014)]{Giovanelli 2014} Giovanelli, M.\ (2014). `But one must not legalize the mentioned sin': Phenomenological vs.\ dynamical treatments of rods and clocks in Einstein's thought. \emph{Studies in History and Philosophy of Modern Physics} 48: 20--44.

\bibitem[Gleason(1957)]{Gleason 1957} Gleason, A.\ M.\ (1957). Measures on the closed subspaces of a Hilbert space. \emph{Journal of Mathematics and Mechanics} 6: 885--893.

\bibitem[Greenberger, Horne and Zeilinger(2014)]{GHZ} Greenberger, D.\ M., Horne, M.\ A.\ and Zeilinger, A. (1998). Going beyond Bell's theorem. In \citet[69--72]{Kafatos 1989}

%\bibitem[Griffiths(2005)]{Griffiths 2005} Griffiths, D.~J.\ (2005). \emph{Introduction to Quantum Mechanics}. 2nd ed. Upper Saddle River, NJ: Pearson Prentice Hall.

\bibitem[Hald(1998)]{Hald 1998} Hald, A.\ (1998).\emph{A History of Mathematical Statistics from 1750 to 1930}. New York: Wiley.

\bibitem[Handsteiner \emph{et al.}(2017)]{Handsteiner 2017} Handsteiner, J.\ \emph{et al.} (2017). Cosmic Bell test: measurement settings from milky way stars. \emph{Physical Review Letters} 118: 060401-1--7.

\bibitem[Hardy(2007)]{Hardy 2007} Hardy, L.\ (2007). Towards quantum gravity: A framework for probabilistic theories with non-fixed causal structure. \emph{Journal of Physics A} 30: 3081.

\bibitem[Hardy(2012)]{Hardy 2012} Hardy, L.\ (2012). The operator tensor formulation of quantum theory. \emph{Philosophical Transactions of the Royal Society A} 370: 3385--3417.

\bibitem[Harrigan and Spekkens(2010)]{Harrigan and Spekkens 2010} Harrigan, N.\ and Spekkens, R.\ W.\ (2010). Einstein, incompleteness, and the epistemic view of quantum states. \emph{Foundations of Physics} 40: 125--157.

\bibitem[Healey(2016)]{Healey 2016} Healey, R.\ (2016). A pragmatist view of the metaphysics of entanglement. \emph{Synthese} (in press). Advance access version available via \emph{Springer Netherlands online}.

\bibitem[Healey(2019)]{Healey 2019} Healey, R.\ (2019). The mysteries of quantum mechanics in graphic-novel form (review of Bub and Bub, 2018). \emph{Physics Today}  72 (9): 56--57. 

\bibitem[Heisenberg(1925)]{Heisenberg 1925} Heisenberg, W.\ (1925). \"Uber die quantentheoretische Umdeutung kinematischer und mechanischer Beziehungen. \emph{Zeitschrift f\"ur Physik} 33: 879--893. English translation in \citet[pp.\ 261--276]{Van der Waerden}.

%B.\ L.\ van der Waerden, Ed.\ (1968). \emph{Sources of Quantum Mechanics} (pp.\ 261--276) New York: Dover.

\bibitem[Janssen(2009)]{Janssen 2009} Janssen, M.\ (2009). Drawing the line between kinematics and dynamics in special relativity. \emph{Studies in History and Philosophy of Modern Physics} 40: 26--52.

\bibitem[Janssen(2019)]{Janssen 2019} Janssen, M.\ (2019). Arches and scaffolds: bridging continuity and discontinuity in theory change. In A.\ C.\ Love and W.\ C.\ Wimsatt, Eds., \emph{Beyond the Meme. Development and Structure in Cultural Evolution.} (pp.\ 95--199). Minneapolis: University of Minnesota Press.

\bibitem[Janssen and Lehner(2014)]{Janssen and Lehner 2014} Janssen, M.\ and Lehner, C., Eds.\ (2014). \emph{The Cambridge Companion to Einstein}. Cambridge: Cambridge University Press.

\bibitem[Joas and Lehner(2009)]{Joas and Lehner 2009} Joas, C.\ and Lehner, C.\ (2009). The classical roots of wave mechanics: Schr\"odinger's transformations of the optical-mechanical analogy. \emph{Studies in History and Philosophy of Modern Physics} 40: 338--351.

\bibitem[Jordan(1927a)]{Jordan 1927a} Jordan, P.\ (1927a).  \"Uber eine neue Begr\"undung der Quantenmechanik. \emph{Zeitschrift f\"ur Physik} 40: 809--838. 

\bibitem[Jordan(1927b)]{Jordan 1927b} Jordan, P.\  (1927b).  \"Uber eine neue Begr\"undung der Quantenmechanik II. \emph{Zeitschrift f\"ur Physik} 44: 1--25.

\bibitem[Kafatos(1989)]{Kafatos 1989} Kafatos, M., Ed., (1989).  \emph{Bell's Theorem, Quantum Theory and Conceptions of the Universe}. Dordrecht: Kluwer. 

\bibitem[Kaiser(2011)]{Kaiser 2011} Kaiser, D.\ (2011). \emph{How the Hippies Saved Physics. Science, Counterculture, and the Quantum Revival.} New York: Norton.

\bibitem[Kakalios(2009)]{Kakalios 2009} Kakalios, J.\ (2009). \emph{The Physics of Superheroes}. 2nd ed. New York: Gotham Books

\bibitem[Kakalios(2018)]{Kakalios 2018} Kakalios, J.\ (2018). A graphic tale of entanglement (review of Bub and Bub, 2018). \emph{Physics World} 31 (12): 37--38. 

\bibitem[Kendall(1961)]{Kendall 1961} Kendall, M.~G.\ (1961). \emph{A Course in the Geometry of $n$ Dimensions.} New York: Hafner. 

\bibitem[Koberinski and M\"uller(2018)]{Koberinski and Mueller 2018} Koberinski, A.\ and M\"uller, M.\ (2018). Quantum theory as a principle theory: Insights from an information theoretic reconstruction. In \citet[pp. 257--281]{Cuffaro and Fletcher 2018}. 

\bibitem[Kramers and Heisenberg(1925)]{Kramers and Heisenberg 1925} Kramers, H.\ A.\ and Heisenberg, W.\ (1925). \"{U}ber die Streuung von Strahlung durch Atome. \emph{Zeitschrift f\"{u}r Physik} 31: 681--707. English translation in \citet[pp. 223--252]{Van der Waerden}.

\bibitem[Landau(1988)]{Landau 1988} Landau, L.\ J.\ (1988). Empirical two-point correlation functions. \emph{Foundations of Physics} 18: 440--460.

\bibitem[Lassez and Lassez(1992)]{Lassez and Lassez 1992} Lassez, C.\ and Lassez, J.-L.\ (1992). Quantifier elimination for conjunctions of linear constraints via a convex hull algorithm. In B.\ R.\ Donald, D.\ Kapur, J.\ L.\ Mundy, Eds., \emph{Symbolic and Numerical Computation for Artificial Intelligence} (pp.\ 103--119). San Diego: Academic Press, 1992.

\bibitem[Laurence, Hwang and Barone(2008)]{Laurence 2008} Laurence, P., Hwang, T-H.\ and Barone, L.\ (2008). Geometric properties of multivariate correlation in de Finetti's approach to insurance theory. \emph{Journal \'Electronique d'Histoire des Probabilit\'es et de la Statistique} (www.jehps.net) 4 (2): 1--13.

%\bibitem[Lehner(2014)]{Lehner 2014} Lehner, C.\ (2014). Einstein's realism and his critique of quantum mechanics. In M.\ Janssen and C.\ Lehner, \emph{The Cambridge Companion to Einstein} (pp.\ 306--353). Cambridge: Cambridge University Press.

\bibitem[Masanes, Galley and M\"uller(2019)]{Masanes et al 2019} Masanes, L., Galley, T.~D. and M\"uller, M.~P.\ (2019). The measurement postulates of quantum mechanics are operationally redundant. \emph{Nature Communications} 10: 1361.

\bibitem[McGrayne(2011)]{McGrayne 2011}  McGrayne, S.\ B.\ (2011). \emph{The Theory That Would Not Die: How Bayes' Rule Cracked the Enigma Code, Hunted Down Russian Submarines, \& Emerged Triumphant from Two centuries of Controversy}. New Haven: Yale University Press. 

\bibitem[Mensing and Pauli(1926)]{Mensing and Pauli 1926} Mensing, L.\ and Pauli, W.\ (1926). \"Uber die Dielektrizit\"atskonstante von Dipolgasen nach der Quantenmechanik. \emph{Physikalische Zeitschrift} 27: 509--512.

\bibitem[Mermin(1981)]{Mermin 1981} Mermin, N.\ D.\ (1981). Quantum mysteries for everyone. \emph{Journal of Philosophy} 78: 397--408. Page references to reprint in \citet[pp.\ 81--94]{Mermin 1990}.

\bibitem[Mermin(1988)]{Mermin 1988} Mermin, N.\ D.\ (1988). Spooky actions at a distance: mysteries of the quantum theory.  Page references to reprint in \citet[pp.\ 110--176]{Mermin 1990}.

\bibitem[Mermin(1990)]{Mermin 1990} Mermin, N.\ D.\ (1990). \emph{Boojums All the Way Through. Communicating Science in a Prosaic Age.} Cambridge: Cambridge University Press.

\bibitem[Messiah(1962)]{Messiah 1962} Messiah, A.\ (1962). \emph{Quantum Mechanics.} 2 Vols. New York: Wiley. 

\bibitem[Midwinter and Janssen(2013)]{Midwinter and Janssen 2013} Midwinter, C.\ and Janssen, M.\ (2013). Kuhn losses regained: Van Vleck from spectra to susceptibilities. In M.\ Badino and J.\ Navarro, Eds., \emph{Research and Pedagogy: A History of Early Quantum Physics Through Its Textbooks} (pp.\ 137--205). Berlin: Edition Open Access.

\bibitem[Myrvold(2010)]{Myrvold 2010} Myrvold, W.\ (2010). From physics to information theory and back. In A.\ Bokulich and G.\ Jaeger, Eds., \emph{Philosophy of Quantum Information and Entanglement} (pp.\ 181--207). Cambridge: Cambridge University Press.

\bibitem[Nielsen and Chuang(2016)]{Nielsen and Chuang 2016} Nielsen, M.\ A.\ and Chuang, I.\ L.\ (2016). \emph{Quantum Computation and Quantum Information.} 10th ann.\ ed.\ Cambridge: Cambridge University Press.

\bibitem[Pauli(1921)]{Pauli 1921} Pauli, W.\ (1921). Zur Theorie der Dieelektrizit\"atskonstante zweiatomiger Dipolgase. \emph{Zeitschrift f\"ur Physik} 6: 319--327.

\bibitem[Pauli(1994)]{Pauli 1994} Pauli, W.\ (1994). \emph{Writings on Physics and Philosophy.} C.\ P.\ Enz and K.\ von Meyenn, Eds. Berlin: Springer.

\bibitem[Pawlowski \emph{et al}(2009)]{Pawlowski et al 2009} Paw{\l}owski, M., Paterek, T., Kaszlikowski, D., Scarani, V., Winter, A.\ and {\.Z}ukowski, M.\  (2009). Information causality as a physical principle. \emph{Nature} 461: 1101--1104.

\bibitem[Pearson(1896)]{Pearson 1896} Pearson, K.\ (1896). Mathematical contributions to the theory of evolution. III. Regression, heredity and panmixia. \emph{Philosophical Transactions of the Royal Society, London, Series A} 186: 343--414.

\bibitem[Pearson(1916)]{Pearson 1916} Pearson, K.\ (1916). On some novel properties of partial and multiple Correlation in a universe of manifold characteristics. \emph{Biometrika} 11: 231--238.

\bibitem[Pearson(1921)]{Pearson 1921} Pearson, K.\ (1921). Notes on the history of correlation. \emph{Biometrika} 13: 25--45. Page reference to reprint in: E.\ S.\ Pearson and M.\ G.\ Kendall, Eds., \emph{Studies in the History of Statistics and Probability} (pp.\ 185--205). London: Griffin.  

\bibitem[Pitowsky(1986)]{Pitowsky 1986} Pitowsky, I.\ (1986). The range of quantum probabilities. \emph{Journal of Mathematical Physics} 27: 1556--1566.

\bibitem[Pitowsky(1989a)]{Pitowsky 1989a} Pitowsky, I.\ (1989a). From George Boole to John Bell: The Origins of Bell's inequalities. In \citet[pp.\ 37--49]{Kafatos 1989}.

\bibitem[Pitowsky(1989b)]{Pitowsky 1989b} Pitowsky, I.\ (1989b).   (Lecture Notes in Physics, No.\ 321). Heidelberg: Springer.

\bibitem[Pitowsky(1991)]{Pitowsky 1991} Pitowsky, I.\ (1991). Correlation polytopes, their geometry and complexity. \emph{Mathematical Programming} A 50: 395--414.

\bibitem[Pitowsky(1994)]{Pitowsky 1994} Pitowsky, I.\ (1994). George Boole's `conditions of possible experience' and the quantum puzzle. \emph{The British Journal for the Philosophy of Science} 45 95--125.

\bibitem[Pitowsky(2006)]{Pitowsky 2006} Pitowsky, I.\ (2006). Quantum mechanics as a theory of probability.  In W.\ Demopoulos and I.\ Pitowsky, Eds., \emph{Physical Theory and its Interpretation. Essays in Honor of Jeffrey Bub} (pp. 213--240). New York: Springer.

\bibitem[Pitowsky(2008)]{Pitowsky 2008} Pitowsky, I.\ (2008). Geometry of quantum correlations. \emph{Physical Review} A 77: 062109. 
%xxx--yyy see bananaworld, p. 120, note 1.

\bibitem[Pokorny \emph{et al.}(2019)]{Pokorny et al 2019} Pokorny, F., Zhang, C., Higgins, G., Cabello, C., Kleinmann, M.\ and Hennrich, M.\  (2019). Tracking the dynamics of an ideal quantum measurement. \emph{arXiv}: 1903.10398v1.

\bibitem[Popescu(2016)]{Popescu 2016} Popescu, S.\ (2016). Foreword. In \citet[pp.\ v--vi]{Bub 2016}

\bibitem[Popescu and Rohrlich(1994)]{Popescu and Rohrlich 1994} Popescu, S.\ and Rohrlich, D.\ (1994). Quantum nonlocality as an axiom. \emph{Foundations of Physics} 24: 370--385.

\bibitem[Pusey, Barrett and Rudolph(2012)]{PBR} Pusey, M.\ F., Barrett, J.\ and Rudolph, T.\ (2012). On the reality of the quantum state. \emph{Nature Physics} 8: 475--478

\bibitem[Quine(1951)]{Quine 1951} Quine, W.\ V.\ O.\ (1951). Two dogmas of empiricism. \emph{The Philosophical Review} 60: 20--43.

\bibitem[Rastall(1995)]{Rastall 1995} Rastall, P.\ (1995). Locality, Bell's theorem, and quantum mechanics. \emph{ Foundations of Physics} 15: 963--972.

%\bibitem[Renn(2007)]{Renn 2007} Renn, J., Ed.\ (2007). \emph{The Genesis of General Relativity.} 4 Vols. New York, Berlin: Springer.

\bibitem[Sainz \emph{et al.}(2018)]{Sainz et al 2018} Sainz, A.\ B., Guryanova, Y\, Ac\'in, A., and Navascu\'es, M.\  (2018). Almost-quantum correlations violate the no-restriction hypothesis. \emph{Physical Review Letters} 120: 200402.

\bibitem[Saunders \emph{et al.}(2010)]{Many Worlds 2010} Saunders, S., Barrett, J., Kent, A.\ and D.\ Wallace, D.,Eds.\ (2010). \emph{Many Worlds? Everett, Quantum Theory, and Reality}. Oxford: Oxford University Press.

%\bibitem[Schlosshauer (2007)]{Schlosshauer 2007} Schlosshauer, M.\ (2007). \emph{Decoherence and the Quantum-to-Classical Transition.} Berlin: Springer.

\bibitem[Schr\"odinger(1926)]{Schroedinger 1926} Schr\"odinger, E.\ (1926).  Quantisierung als Eigenwertproblem.  Dritte Mitteilung: St\"orungstheorie, mit Anwendung auf den Starkeffekt der Balmerlinien. \emph{Annalen der Physik}  80: 437--490. 

%\bibitem[Schr\"odinger(1935)]{Schroedinger 1935} Schr\"odinger, E.\ (1935). Discussion of probability relations between separated systems. \emph{Mathematical Proceedings of the Cambridge Philosophical Society} 31: 555--563.

\bibitem[Schwarzschild(1916)]{Schwarzschild 1916} Schwarzschild, K.\ (1916). Zur Quantenhypothese. \emph{K\"oniglich Preussischen Akademie der Wissenschaften} (Berlin)  \emph{Sitzungsberichte}: 548--568.

\bibitem[Sommerfeld(1919)]{Sommerfeld 1919} Sommerfeld, A.\ (1919). \emph{Atombau und Spektrallinien.} Braunschweig: Vieweg.

\bibitem[Spekkens(2007)]{Spekkens 2007} Spekkens, R.\ W.\, Evidence for the epistemic view of quantum states: A toy theory. \emph{Physical Review A} 75: 032110.

%\bibitem[Stachel(1994)]{Stachel 1994} Stachel, J.\ (1994). Changes in the concepts of space and time brought about by relativity. In C.\ C.\ Gould and R.\ S.\ Cohen, Eds., \emph{Artifacts, Representation and Social Practice} (pp.\ 141--162). Dordrecht: Kluwer.

\bibitem[Stark(1913)]{Stark 1913} Stark, J.\ (1913). Beobachtungen \"uber den Effekt des elektrischen Feldes auf Spektrallinien. {\it K\"oniglich Preussischen Akademie der Wissenschaften} (Berlin). {\it Sitzungsberichte}: 932-946.

\bibitem[Stone(1932)]{Stone 1932} Stone, M.\ H.\  ( 1932). \emph{Linear Transformations in Hilbert Space}. New York: American Mathematical Society.

\bibitem[Tanabashi \emph{et al.}(2018)]{Tanabashi et al 2018} Tanabashi, M.\ \emph{et al}.\ (2018). Review of Particle Physics (Particle Data Group). \emph{Physical Review} D 98, 030001.

\bibitem[Teukolsky(1996)]{Teukolsky 1996} Teukolsky, S.\ A.\ (1996). The explanation of the Trouton-Noble experiment revisited. \emph{American Journal of Physics}   64: 1104--1106. 

\bibitem[Timpson(2010)]{Timpson 2010} Timpson, C.\ (2010). Rabid dogma? Comments on Bub and Pitowsky [2010]. In \citet[pp.\ 460--466]{Many Worlds 2010}.

\bibitem[van der Waerden(1968)]{Van der Waerden} Van der Waerden, B.\ L., Ed.\ (1968). \emph{Sources of quantum mechanics.} New York: Dover.

\bibitem[van Fraassen(1991)]{Van Fraassen 1991} Van Fraassen, B.\ C.\ (1991). \emph{Quantum Mechanics. An Empiricist View.} Oxford: Clarendon.

\bibitem[Van Vleck(1926)]{Van Vleck 1926} Van Vleck, J.\ H.\ (1926). Magnetic susceptibilities and dielectric constants in the new quantum mechanics. \emph{Nature} 118: 226--227.

\bibitem[Van Vleck(1928)]{Van Vleck 1928} Van Vleck, J.\ H.\ (1928). The new quantum mechanics. \emph{Chemical Reviews} 5: 467--507.

\bibitem[Van Vleck(1932)]{Van Vleck 1932} Van Vleck, J.\ H.\ (1932). \emph{The Theory of Electric and Magnetic Susceptibilities}. Oxford University Press, Oxford.

\bibitem[von Mises(1928)]{von Mises 1928} Von Mises, R.\ (1928). \emph{Wahrscheinlichkeit, Statistik und Wahrheit}. Vienna: Springer.  

\bibitem[von Neumann(1927a)]{von Neumann 1927a} Von Neumann, J.\ (1927a). Mathematische Begr\"undung der Quantenmechanik.  \emph{K\"onigliche Gesellschaft der Wissenschaften zu G\"ottingen. Mathematisch-physikalische Klasse. Nachrichten} 1--57.

\bibitem[von Neumann(1927b)]{von Neumann 1927b} Von Neumann, J.\ (1927b). Wahrscheinlichkeitstheoretischer Aufbau der Quantenmechanik.  \emph{K\"onigliche Gesellschaft der Wissenschaften zu G\"ottingen. Mathematisch-physikalische Klasse. Nachrichten} 245--272.

\bibitem[von Neumann(1927c)]{von Neumann 1927c} Von Neumann, J.\ (1927c). Thermodynamik quantenmechanischer Gesamtheiten. \emph{K\"onigliche Gesellschaft der Wissenschaften zu G\"ottingen. Mathematisch-physikalische Klasse. Nachrichten} 273--291.

\bibitem[von Neumann(1932)]{von Neumann 1932} Von Neumann, J.\ (1932). \emph{Mathematische Grundlagen der Quantenmechanik.} Berlin: Springer.

%\bibitem[Wallace(2016)]{Wallace 2016} Wallace, D.\ (2016). What is orthodox quantum mechanics? \emph{arXiv}: 1604.05973v1.

\bibitem[Wallace(2019)]{Wallace 2019} Wallace, D.\ (2019). On the plurality of quantum theories: Quantum theory as a framework, and its implications for the quantum measurement problem. Preprint, \emph{PhilSci Archive}. To be published in: S. French and J. Saatsi, Eds., \emph{Realism and the Quantum}. Oxford, Oxford University Press, forthcoming.

\bibitem[Wigner(1931)]{Wigner 1931} Wigner, E.\ (1931). \emph{Gruppentheorie und ihre Anwendung auf die Quantenmechanik der Atomspektren}. Braunschweig: Vieweg. 
%Reprinted in facsimile by Authority of the Alien Property Custodian  in 1944.
%under License No.\ A-281.

\bibitem[Yule(1897)]{Yule 1896} Yule, G.\ U.\ (1897). On the significance of Bravais' formulae for regression, \&c., in the case of skew correlation. \emph{Proceedings of the Royal Society of London} 60: 477--489. 
%(issues: 359Ð367). Udny Yule.
%In 1846 he published a paper on the statistical concept of correlation, and arrived at a definition of the correlation coefficient before Karl Pearson.
%Stephen Mack Stigler. 

  

\end{thebibliography}
\end{document}